\documentclass[aip,preprint]{revtex4-1}

\usepackage{graphicx}
\usepackage{amssymb}
\usepackage{amsthm}
\usepackage{braket}
\usepackage{amsmath}
\usepackage{subfig}

\newcommand{\tr}{\operatorname{tr}}

\begin{document}

\title{On the Von Neumann Entropy of Graphs}

\author{Giorgia Minello}
\affiliation{Dipartimento di Scienze Ambientali, Informatica e Statistica, Universit\`{a} Ca' Foscari Venezia}

\author{Luca Rossi\footnote{Corresponding author}}
\affiliation{School of Engineering and Applied Science, Aston University, UK}

\author{Andrea Torsello}
\affiliation{Dipartimento di Scienze Ambientali, Informatica e Statistica, Universit\`{a} Ca' Foscari Venezia}

\begin{abstract}
{The von Neumann entropy of a graph is a spectral complexity measure that has recently found applications in complex networks analysis and  pattern recognition. Two variants of the von Neumann entropy exist based on the graph Laplacian and normalized graph Laplacian, respectively. Due to its computational complexity, previous works have proposed to approximate the von Neumann entropy, effectively reducing it to the computation of simple node degree statistics. Unfortunately, a number of issues surrounding the von Neumann entropy remain unsolved to date, including the interpretation of this spectral measure in terms of structural patterns, understanding the relation between its two variants, and evaluating the quality of the corresponding approximations.

In this paper we aim to answer these questions by first analysing and comparing the quadratic approximations of the two variants and then performing an extensive set of experiments on both synthetic and real-world graphs. We find that 1) the two entropies lead to the emergence of similar structures, but with some significant differences; 2) the correlation between them ranges from weakly positive to strongly negative, depending on the topology of the underlying graph; 3) the quadratic approximations fail to capture the presence of non-trivial structural patterns that seem to influence the value of the exact entropies; 4) the quality of the approximations, as well as which variant of the von Neumann entropy is better approximated, depends on the topology of the underlying graph.}
{Graph, Entropy.}
\\
\end{abstract}

\maketitle

\section{Introduction}
Complex networks provide a natural way to model the underlying structure of a large number of biological, social, and technological systems. Examples of such systems include metabolic networks~\cite{jeong2000large}, brain networks~\cite{bullmore2009complex}, social networks~\cite{chorley2016pub}, collaboration networks~\cite{lima2014coding}, and transport networks~\cite{guimera2005worldwide}. The ability to measure the complexity of these networks plays a central role in the analysis of the corresponding systems. Intuitively, the complexity of a network should capture the level of organization of its structural features, e.g., the scaling behaviour of its degree distribution. To this end, a number of entropic complexity measures have been proposed in the past years~\cite{bonchev2005quantitative,dehmer2008information,passerini2009quantifying,anand2009entropy,anand2011shannon,escolano2012heat}.

The von Neumann entropy of a network was introduced by Braunstein et al.~\cite{braunstein2006some} and then analyzed further in a number of later works~\cite{passerini2009quantifying,anand2009entropy,du2010note,anand2011shannon,de2016interpreting,dairyko2017note,simmons2018quantum}. The intuition behind this measure is that of associating graphs to density matrices and measuring the complexity of the graphs in terms of the von Neumman entropy of the corresponding density matrices. This in turn is based on the mapping between quantum states and the combinatorial graph Laplacian proposed by Braunstein et al.~\cite{braunstein2006some}. In~\cite{passerini2009quantifying}, Passerini and Severini briefly investigated the use of the normalized Laplacian, although their analysis mainly focused on the unnormalized version. In both cases, a necessary step is the computation of the eigenvalues of the (normalized) graph Laplacian. This has computational complexity quadratic in the number of nodes of the network, thus making the application to large networks unfeasible.

Han et al.~\cite{han2012graph} sought to overcome this by looking at the second order polynomial approximation of the Shannon entropy. They considered the von Neumann entropy obtained from the normalized graph Laplacian and they showed that its quadratic approximation can be computed in terms of degree statistics. A similar result was obtained by Lockhart et al.~\cite{lockhart2016edge} for the graph Laplacian. With this approximation to hand, the von Neummann network entropy has found applications in the analysis of several real-world networks~\cite{han2012graph,ye2014approximate,rossi2017measuring} as well as in pattern recognition~\cite{bai2013graph,han2015generative}. More recently, Simmons et al. showed that the von Neumann entropy can be used as a measure of graph centralization~\cite{simmons2018quantum}, i.e., the extent to which a graph is organized around a number of central nodes. Unfortunately, due to the spectral nature of this measure, it remains unclear how different structural patterns influence its value. Despite several attempts, a general structural interpretation of the von Neumann entropy remains an open problem.

In this paper, our aim is three-fold. We intend to: 1) shed light on the relation between the structure of a network and its von Neumann entropy, both for the version based on the graph Laplacian and the normalized Laplacian, thus also 2) deepening our understanding of the difference between these two entropies; 3) evaluate the quality of the quadratic approximation. Han et al.~\cite{han2012graph} also briefly analysed the accuracy of the quadratic approximation, but only for the version of the von Neumann entropy based on the normalized Laplacian. As explained in Section~\ref{sec:experiments}, their analysis is also strongly influenced by the use of datasets with graphs of varying size, whereas our experimental evaluation is on datasets of fixed graph size. Finally, in the present work we are particularly interested in looking at how different edges contribute to the overall graph entropy, revealing additional inaccuracies introduced by the quadratic approximation.

Our analysis shows that:
\begin{itemize}
\item the two versions of the von Neumann entropy based on the Laplacian and normalized Laplacian (respectively) are connected to the presence of similar structural patterns, although with some significant differences;
\item the correlation between these two entropic measures ranges from weakly positive to strongly negative, depending on the underlying graph structure;
\item the quadratic approximations fail to explain the presence of non-trivial structures observed when the growth is driven by the exact entropies;
\item the quality of the approximations, as well as which variant of the von Neummann entropy is better approximated, depends on the topology of the underlying graph;
\end{itemize}

The remainder of this paper is organized as follows: Section~\ref{sec:background} introduces the necessary mathematical and physical background. Section~\ref{sec:entropy} introduces the quadratic approximation of the two variants of the von Neumann entropy considered in this paper. In Section~\ref{sec:experiments} we empirically compare the exact and approximated entropies. Finally, Section~\ref{sec:conclusion} discusses the results of our investigation and concludes this paper.

\section{Background}\label{sec:background}
\subsection{Quantum states and density matrices}
In quantum mechanics, a system can be either in a pure state or a mixed state. Using the Dirac notation, a pure state is represented as a complex-valued column vector $\Ket{\psi_i}$. A mixed state, on the other hand, is a statistical ensemble of pure states $\Ket{\psi_i}$, each with probability $p_i$. Density matrices are trace-one positive semidefinite matrices introduced to describe mixed state systems~\cite{neumann2013mathematische}. For such a system, $\rho = \sum_i p_i \Ket{\psi_i}\Bra{\psi_i}$, where $\Ket{\psi_i}$ is a pure state and $p_i$ is the probability associated to it. Density matrices play a pivotal role in quantum mechanics and are linked with the observables of quantum systems, e.g., the expectation value of the measurement of an observable $O$ is $\langle O \rangle = \tr(\rho O)$.

\subsection{The von Neumann entropy}
Given a quantum mechanical system described by a density matrix $\rho$, its von Neumann entropy~\cite{neumann2013mathematische} is defined as
\begin{equation}\label{eq:entropy_tr}
S(\rho) = -\tr(\rho \ln \rho),
\end{equation}
where $\tr$ denotes the trace operator and $\ln$ denotes the matrix logarithm. The von Neumann entropy of $\rho$ can also be computed as the Shannon entropy of the spectrum of $\rho$, i.e.,
\begin{equation}\label{eq:entropy_eig}
S(\rho) = -\sum_{i=1}^n \lambda_i \ln \lambda_i
\end{equation}
where $\lambda_i$ denotes the $i$-th eigenvalue of $\rho$, with the convention $0 \ln 0 = 0$.

The von Neumann entropy measures the maximum amount of classical information that we can extract from a mixture of pure states~\cite{vedral2006introduction}. It has also been extensively used in the literature to study correlated systems and to define entanglement and distinguishability measures~\cite{nielsen2002quantum,ohya2004quantum,majtey2005jensen}. Finally note that the von Neumann entropy of a pure state $\rho = \Ket{\psi_i}\Bra{\psi_i}$ is always zero. On other hand, a mixed state always has non-zero entropy. Therefore, the von Neumann entropy $S(\rho)$ can also be seen as a measure of how close $\rho$ is to being a pure state.

\subsection{Graph density matrices}
Let $G$ be an undirected graph with vertex set $V$ and edge set $E \subseteq V \times V$. Recall that the adjacency matrix of the graph $G$ is the symmetric matrix with elements
\begin{equation}
A_{uv}=\left\{\begin{array}{l}
1~~$if$~(u,v)\in E\\
0~~$otherwise$
\end{array}
\right.
\end{equation}
Let $D$ be the diagonal matrix with elements $d_u = \sum_{v=1}^n A(u,v)$, where $d_u$ is the degree of the node $u$. Then $L=D-A$ is the graph Laplacian, the combinatorial analogue of the Laplace-Beltrami operator~\cite{jost2011riemannian}.

Braunstein et al.~\cite{braunstein2006some} proposed to use the graph Laplacian to map graphs to quantum states. More specifically, let $G$ be a graph with Laplacian $L$, then its density matrix is defined as $\rho(L) = \frac{L}{\tr(L)} = \frac{L}{2m}$, where $m$ denotes the number of edges of $G$. Passerini and Severini~\cite{passerini2009quantifying} proposed an alternative version of the von Neumann entropy for graphs based on the normalized Laplacian $\mathcal{L} = D^{-1/2}LD^{-1/2}$. Given a graph $G$ with $n$ nodes and normalized Laplacian $\mathcal{L}$, they define the density matrix of $G$ as $\rho(\mathcal{L}) = \frac{\mathcal{L}}{\tr(\mathcal{L})} = \frac{\mathcal{L}}{n}$.

\subsection{The von Neumann entropy of a Graph}
With the density matrix of a graph to hand, one can compute its von Neumann entropy using either Eq.~\ref{eq:entropy_tr} or Eq.~\ref{eq:entropy_eig}. In the remainder of this paper, we refer to the von Neumann entropies computed on $\rho(L)$ and $\rho(\mathcal{L})$ as the \emph{Laplacian entropy} and \emph{normalized Laplacian entropy}, respectively.

A number of previous works have made steps toward a general interpretation of the Laplacian entropy, although this remains an open problem~\cite{passerini2009quantifying,du2010note,anand2009entropy,anand2011shannon,de2016interpreting,dairyko2017note}. Passerini and Severini~\cite{passerini2009quantifying} have observed that the Laplacian entropy of a graph tends to grow with the
number of connected components, long paths and nontrivial symmetries. They have also shown that the Laplacian entropy of a graph $G$ is upper bounded by $\ln(n-1)$, where $n$ denotes the number of nodes of $G$, and that this upper bound is saturated by both complete graphs and regular graphs (for large $n$), suggesting that the Laplacian entropy can be interpreted as a measure of regularity. Du et al.~\cite{du2010note} proved that the same bound holds also for Erd\"os-R\'enyi random graphs, highlighting a connection between randomness and regularity. In~\cite{anand2009entropy}, the authors showed that for scale free networks the Laplacian entropy
of a graph is linearly related to the Shannon entropy of the graph ensemble~\cite{anand2009entropy}. More in general, Anand et al. observed in~\cite{anand2011shannon} that for graphs with heterogeneous degree distributions there exists a correlation between these entropies.

De Beaudrap et al. have shown that the Laplacian entropy of a graph can be interpreted as a measure of the amount of entanglement between a system corresponding to the vertices and a system corresponding to the edges of the graph~\cite{de2016interpreting}. This in turn allows them to identify cospectral graphs (i.e., graphs having the same graph spectrum) as graphs with local unitarily equivalent pure states~\cite{de2016interpreting}. Finally, Dairyko et al.~\cite{dairyko2017note} show that adding an edge to a graph can result in a decrease of its Laplacian entropy, i.e., the Laplacian entropy does not satisfy the subadditivity property~\cite{de2016spectral}. More recently, Simmons et al.~\cite{simmons2018quantum} have proved that the Laplacian entropy of a graph is related to both the graph Theil index and the graph Jain fairness index, highlighting an interesting connection between the Laplacian entropy and the level of centralization across a graph.

\section{Quadratic approximation of the von Neumann entropy}\label{sec:entropy}
While the von Neumann entropy entropy of a graph has found many applications in the analysis of real-world networks~\cite{han2012graph,ye2014approximate,bai2013graph,han2015generative}, a major drawback of this entropic measure is the fact that it requires the computation of the eigenvalues of the (normalized) graph Laplacian. This has computational complexity which is quadratic in the number of nodes of the network, thus making the application to large networks unfeasible.

For this reason, a number of researchers resorted to a quadratic approximation of the entropy~\cite{han2012graph,ye2014approximate,lockhart2016edge}. Although this only captures simple degree statistics of the graph, Han et al.~\cite{han2012graph} show that for Erd\"os-R\'enyi, scale-free, and Delaunay graphs this is a sufficiently good approximation. However their analysis is limited to the normalized Laplacian entropy, and does not consider the unnormalized version. In fact, to the best of our knowledge, no previous study has investigated the difference between the Laplacian and the normalized Laplacian entropies. Interestingly, despite a lack of evidence suggesting that one formulation should be preferred to the other, most works in the literature make use of the normalized version~\cite{han2012graph,ye2014approximate,bai2013graph,han2015generative}.

One of the main aims of this paper is indeed that of shedding light on the differences between these two formulations. To this end, we rewrite the Shannon entropy $-\sum_i\lambda_i \ln(\lambda_i)$ using the second order polynomial approximation $k\sum_i\lambda_i(1-\lambda_i)$, where the value of $k$ depends on the dimension of the simplex. Given a graph $G$, let $\rho(G)$ denote its associated density matrix, i.e., $\rho(G)$ is either $\rho(L)$ or $\rho(\mathcal{L})$. We obtain
\begin{equation}\label{eq:quadratic_approximation}
S(\rho) = -\tr\left(\rho \ln \rho\right) \approx \tr\left(\rho(I_n - \rho)\right),
\end{equation}
where $n$ is the number of nodes of $G$, $I_n$ is the $n \times n$ identity matrix and we the ignored the node set size-dependent factor $\frac{|V|\ln(|V|)}{|V|-1}$.

In the next subsections we look at the specific form of these approximations in the case of the Laplacian and normalize Laplacian entropies. We also derive the expressions for the change in approximated entropy when a single edge is added to the graph, which in turn allows us to shed light on the type of structures that lead to maximal entropy changes.

\subsection{Laplacian}
We start by considering the Laplacian entropy. Recall that in this case $\rho(L) = \frac{L}{2m}$, where $m$ denotes the number of edges of $G$. Using simple algebra, we can rewrite Eq.~\ref{eq:quadratic_approximation} as
\begin{equation}\label{eq:laplacian_approximation}
S(\rho(L)) \approx 1 - \frac{1}{2m} - \frac{1}{4m^2} \sum_{v\in V} d_v^2
\end{equation}
In other words, the quadratic approximation of the Laplacian entropy can be expressed in terms of simple degree statistics. More interestingly, this allows us to probe into the behaviour of the (approximated) Laplacian entropy as the edge set of the graph grows. This was already investigated numerically in Passerini and Severini~\cite{passerini2009quantifying}, but the quadratic approximation allows us to get a deeper analytical insight, although dependent on the approximation.

Let $\Delta(\rho(L)) = S(\rho(L_{\cup (x,y)})) - S(\rho(L))$ be the increment in entropy when a new edge is added to a graph $G$. From Eq.~\ref{eq:laplacian_approximation}, we see that
\begin{align}\label{eq:laplacian_increment}
\Delta(\rho(L)) \approx& \frac{1}{2m} +\frac{1}{4m^2} \sum_{v \in V } d_v^2 - \frac{1}{2(m+1)} -  \frac{1}{4(m+1)^2} \left( \sum_{v \ne x,y } d_v^2  + (d_x + 1)^2 + (d_y + 1)^2 \right) \nonumber \\
=& \frac{1}{2m(m+1)} - \frac{1}{4m^2(m+1)^2} \Bigg( (m + 1)^2 \sum_{v \in V } d_v^2 -  m^2 \left( \sum_{v \ne x,y } d_v^2  + (d_x + 1)^2 + (d_y + 1)^2 \right) \Bigg) \nonumber \\
=& \frac{1}{2m(m+1)} - \frac{1}{4m^2(m+1)^2} \Bigg( (2m + 1) \sum_{v \in V } d_v^2 -  2m^2 (d_x + d_y + 1) \Bigg) \nonumber \\
=& \frac{1}{2m(m+1)} - \frac{2m+1}{(m+1)^2} \left( 1 - \frac{1}{2m} - \tilde{S}(\rho(L))\right) - \frac{d_x + d_y + 1}{2(m+1)^2} \nonumber \\
=&  -\frac{d_x + d_y}{2(m+1)^2} - \frac{1 + (2m+1) (1-\tilde{S}(\rho(L)) }{(m+1)^2},
\end{align}
where $\tilde{S}(\rho(L))$ denotes the approximated Laplacian entropy. Eq.~\ref{eq:laplacian_increment} indicates that edges connecting low degree nodes produce the maximum increment in the graph entropy, while connecting high degree nodes has the opposite effect. This in turn suggests that highly regular graphs with low average degree will be assigned higher values of the approximated Laplacian entropy. This is also the case for the exact version of the Laplacian entropy, as shown in Section~\ref{sec:experiments}.

Note, however, that this does not explain the emergence of structures such as long paths, connected components, and non-trivial symmetries observed by Passerini and Severini in the Laplacian entropy~\cite{passerini2009quantifying} and confirmed in our experimental evaluation. Indeed, the quadratic approximation provides an interesting but incomplete picture of the structural patterns captured by the Laplacian entropy.

\subsection{Normalized Laplacian}
We now consider the normalized Laplacian entropy. In this case $\rho(\mathcal{L}) = \frac{\mathcal{L}}{n}$, where $n$ denotes the number of nodes of $G$. We can rewrite Eq.~\ref{eq:quadratic_approximation} as
\begin{equation}\label{eq:nlaplacian_approximation}
S(\rho(\mathcal{L})) \approx 1-\frac{1}{n}-\frac{1}{n^2}\sum_{(u,v) \in E} \frac{1}{d_u d_v},
\end{equation}
as previously observed by Han et al.~\cite{han2012graph}. As for the Laplacian entropy, the quadratic approximation is based on simple degree statistics. Unlike the approximated Laplacian entropy, however, Eq.~\ref{eq:nlaplacian_approximation} shows that the approximated normalized Laplacian entropy is defined in terms of degree statistics for pairs of nodes that are connected by edges.

As in the previous subsection, we now turn our attention to the increment in entropy when the edge set of $G$ grows. Let $\Delta(\rho(\mathcal{L})) = S(\rho(\mathcal{L}_{\cup (x,y)})) - S(\rho(\mathcal{L}))$ denote this increment. Then let $N_x$ and $N_y$ denote set of vertices connected to $x$ and $y$ in $G$ (before introducing the edge $(x,y)$), respectively. We have that
\begin{align}\label{eq:nlaplacian_increment}
\Delta(\rho(\mathcal{L})) \approx& -\frac{1}{n^2} \Bigg( \sum_{v \in N_x} \frac{1}{d_v(d_x+1)} + \sum_{v \in N_y} \frac{1}{d_v(d_y+1)} + \frac{1}{(d_x+1)(d_y+1)} \nonumber \\
& -\sum_{v \in N_x} \frac{1}{d_vd_x} - \sum_{v \in N_y} \frac{1}{d_vd_y} \Bigg) \nonumber \\
=& -\frac{1}{n^2} \Bigg( \frac{d_x-(d_x+1)}{d_x(d_x+1)}\sum_{v \in N_x}\frac{1}{d_v}+\frac{d_y-(d_y+1)}{d_y(d_y+1)}\sum_{v \in N_y}\frac{1}{d_v} + \frac{1}{(d_x+1)(d_y+1)} \Bigg) \nonumber \\
=& \frac{1}{n^2} \Bigg( \frac{1}{d_x(d_x+1)}\sum_{v \in N_x}\frac{1}{d_v}+\frac{1}{d_y(d_y+1)}\sum_{v \in N_y}\frac{1}{d_v} - \frac{1}{(d_x+1)(d_y+1)} \Bigg) \nonumber \\
=& \frac{1}{n^2} \Bigg( \frac{1}{(d_x+1)H(d_{N_x})}+\frac{1}{(d_y+1)H(d_{N_y})}-\frac{1}{(d_x+1)(d_y+1)} \Bigg),
\end{align}
where $H(d_{N_x})$ and $H(d_{N_y})$ denote the harmonic means of the degrees of the vertices in $N_x$ and $N_y$, respectively. Compared to Eq.~\ref{eq:laplacian_increment}, Eq.~\ref{eq:nlaplacian_increment} shows a more complex relation between the node degrees and the graph entropy. The third term of the last line of Eq.~\ref{eq:nlaplacian_increment} drives the entropy change in the opposite direction of Eq.~\ref{eq:laplacian_increment}, as maximizing (minimizing) the entropy requires establishing connections between high (low) degree nodes. The first two terms, on the other hand, highlight the importance of the neighbourhood of the nodes being connected, with the connection of pairs of low degree nodes with low average degree neighbourhoods yielding the maximum increment in the entropy of the graph.

\subsection{Discussion}
The analysis of the quadratic approximations of the two entropies suggests that these may be only weakly correlated, if not perhaps negatively correlated, depending on the topology of the underlying graphs. Note that simply looking at Eq.~\ref{eq:laplacian_approximation} and Eq.~\ref{eq:nlaplacian_approximation} one may conclude that the correlation between the quadratic approximations of the Laplacian and normalized Laplacian entropies should be negative. However the actual relation is more subtle, and it is better understood through Eqs.~\ref{eq:laplacian_increment} and~\ref{eq:nlaplacian_increment}. In fact, note that while Eq.~\ref{eq:laplacian_approximation} involves a summation over the nodes of the graph, Eq.~\ref{eq:nlaplacian_approximation} involves a summation over its edges, thus making the relation between the two quantities more complex. Indeed, the negative correlation suggested by Eqs.~\ref{eq:laplacian_approximation} and~\ref{eq:nlaplacian_approximation} is also observed when examining Eqs.~\ref{eq:laplacian_increment} and~\ref{eq:nlaplacian_increment}, though the second pair of equations reveals a more subtle relation between the two entropies, with the degree distribution of the nodes neighbourhoods playing an important role.

As for the exact version of the entropies, it is harder to draw any conclusion on their relation as we do not know what type of structural information (beyond simple degree statistics) is being lost in the approximation. In the next section we aim to answer the following questions: 1) are the Laplacian and normalized Laplacian capturing similar structural patterns? and 2) can we rely on the quality of their quadratic approximations when the high computational complexity of the exact version becomes an issue? To answer this questions, in the next section we run an extensive set of numerical experiments on both synthetic and real-world graphs.

\begin{figure}[t!]
\centering
\includegraphics[width=0.15\textwidth]{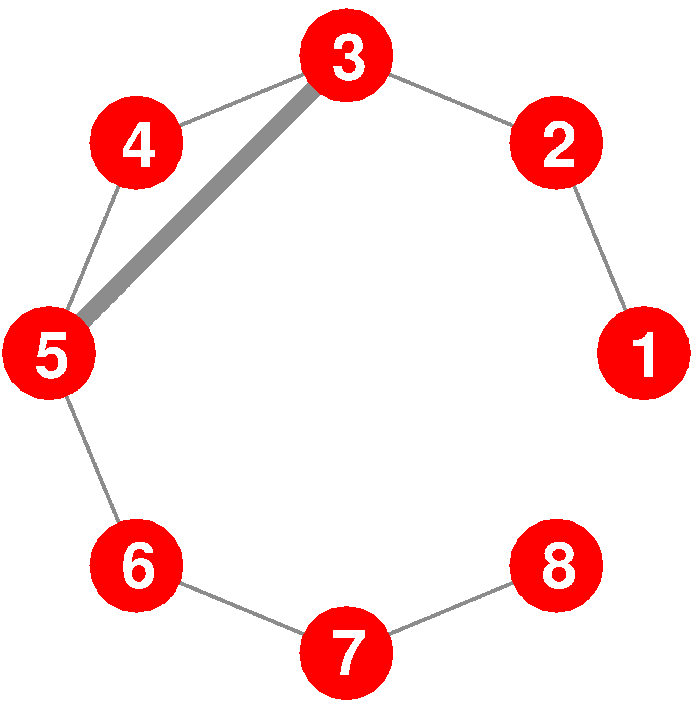}
\hspace{0.15in}
\includegraphics[width=0.15\textwidth]{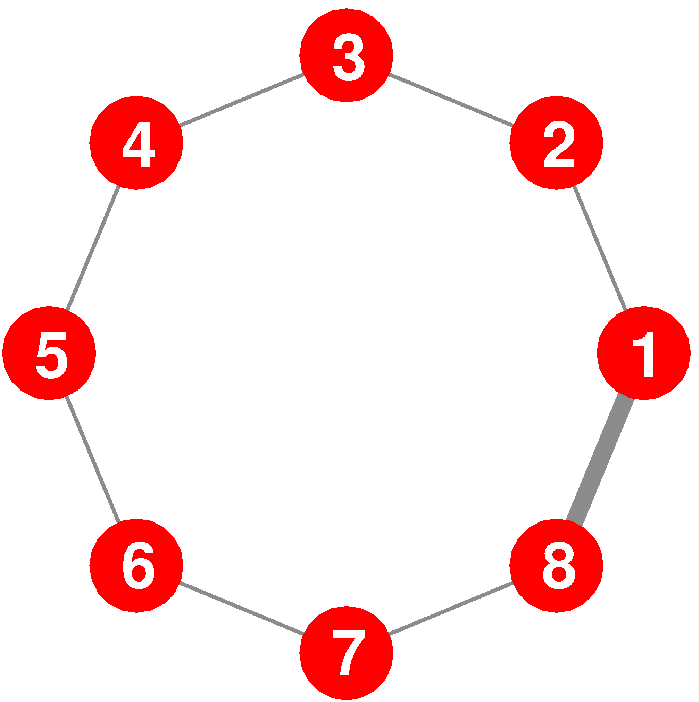}
\hspace{0.15in}
\includegraphics[width=0.15\textwidth]{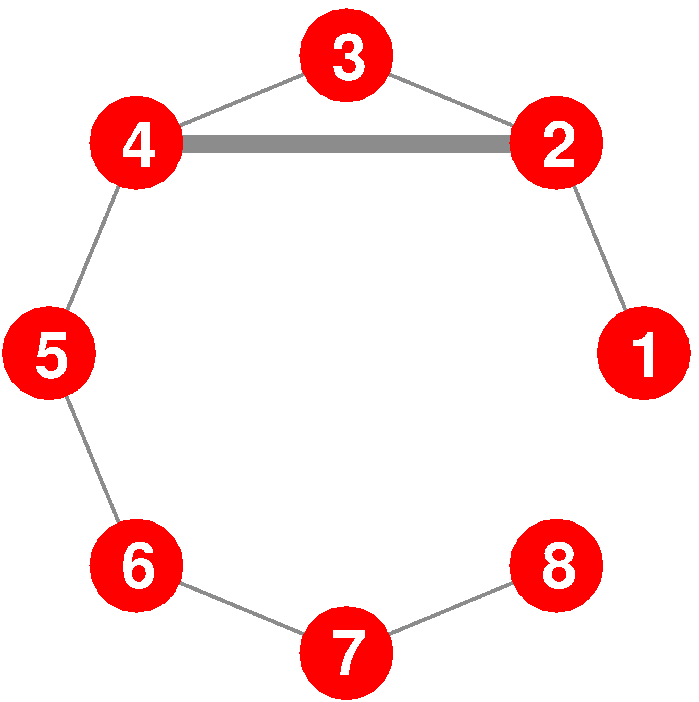}
\hspace{0.15in}
\includegraphics[width=0.15\textwidth]{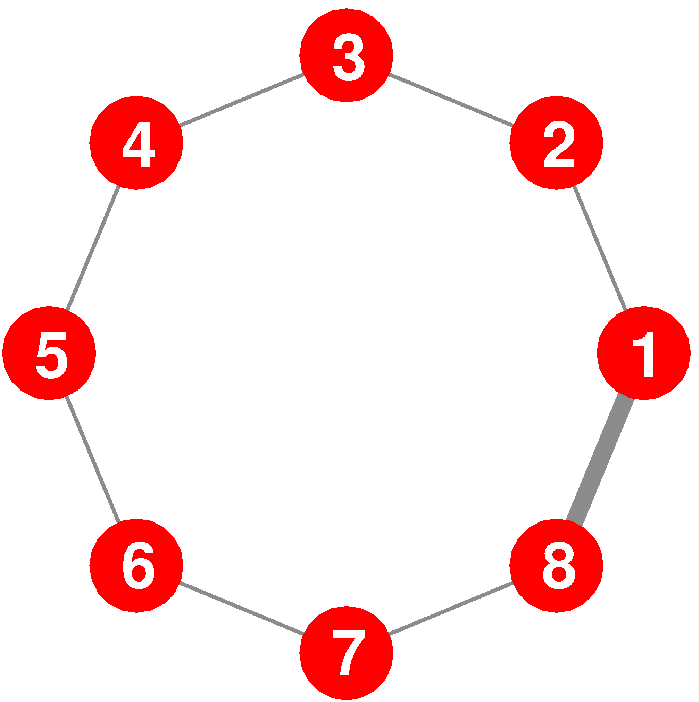}\\
\vspace{0.15in}

\includegraphics[width=0.15\textwidth]{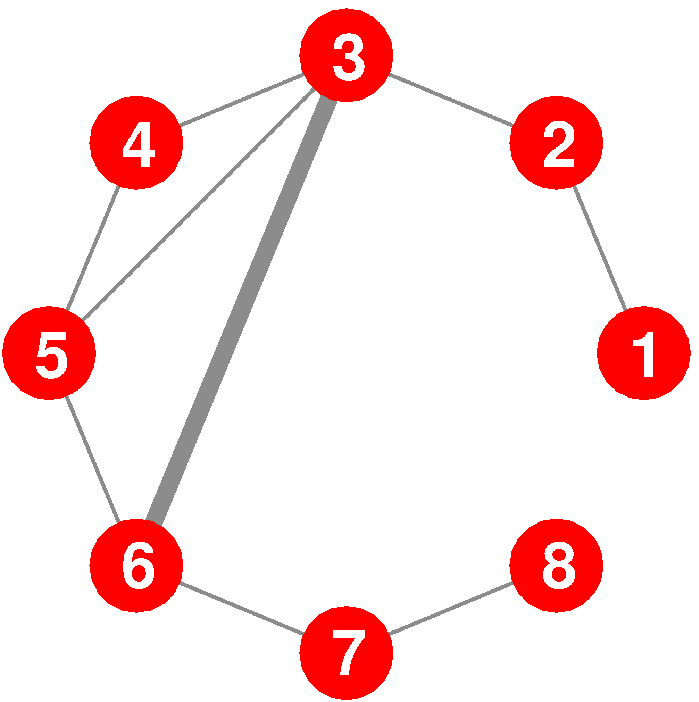}
\hspace{0.15in}
\includegraphics[width=0.15\textwidth]{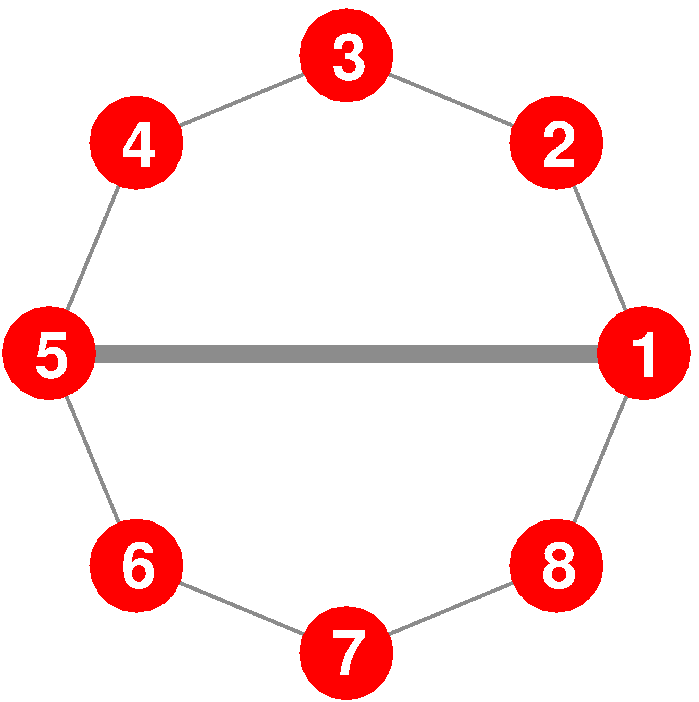}
\hspace{0.15in}
\includegraphics[width=0.15\textwidth]{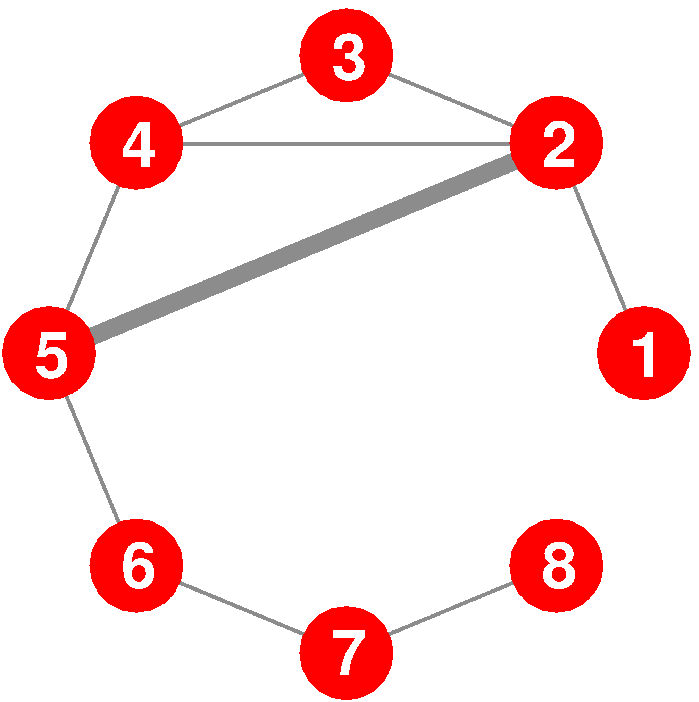}
\hspace{0.15in}
\includegraphics[width=0.15\textwidth]{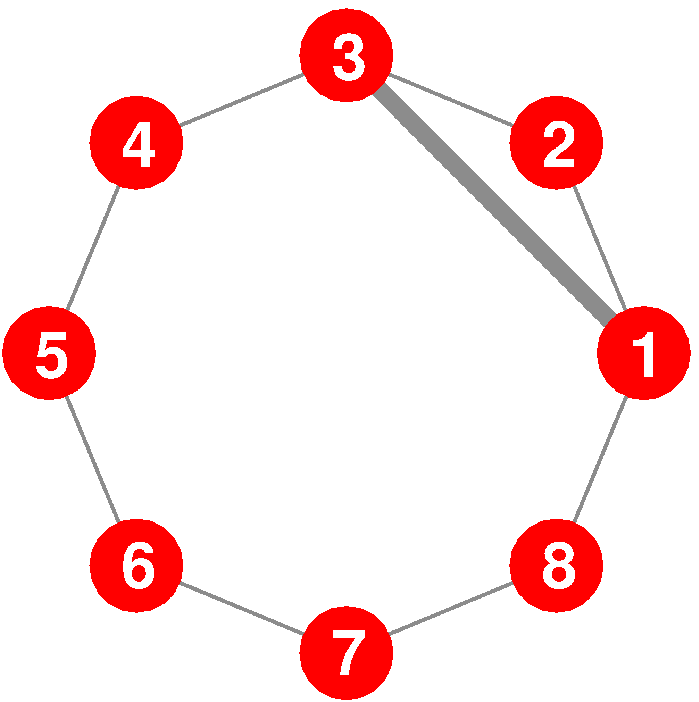}\\

\vspace{0.2in}
\includegraphics[width=0.15\textwidth]{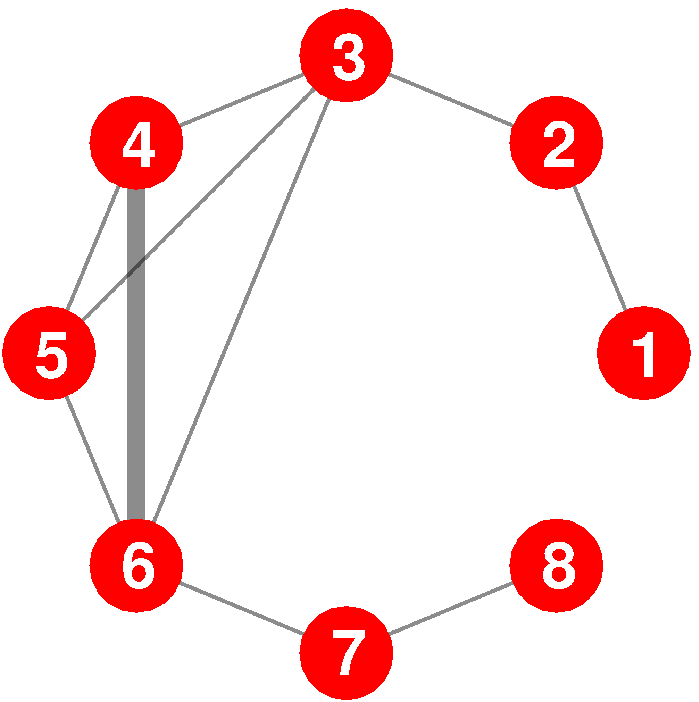}
\hspace{0.15in}
\includegraphics[width=0.15\textwidth]{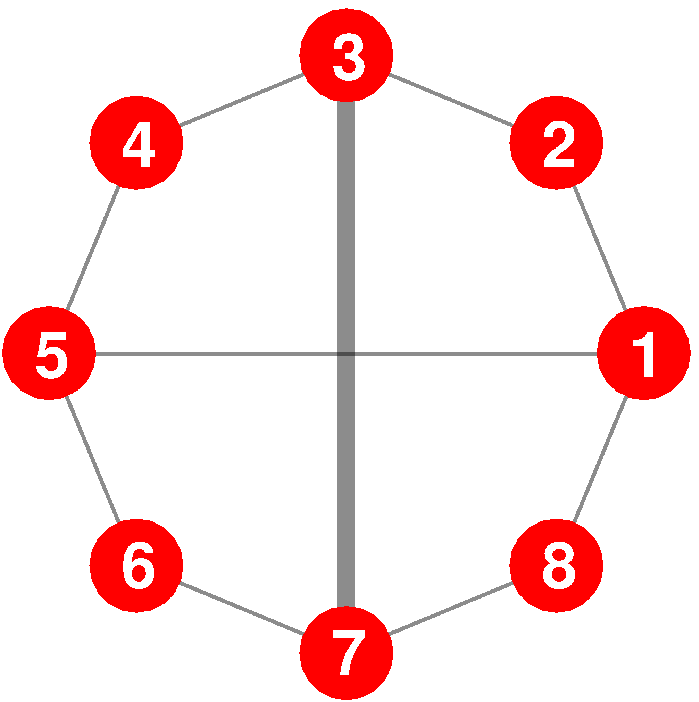}
\hspace{0.15in}
\includegraphics[width=0.15\textwidth]{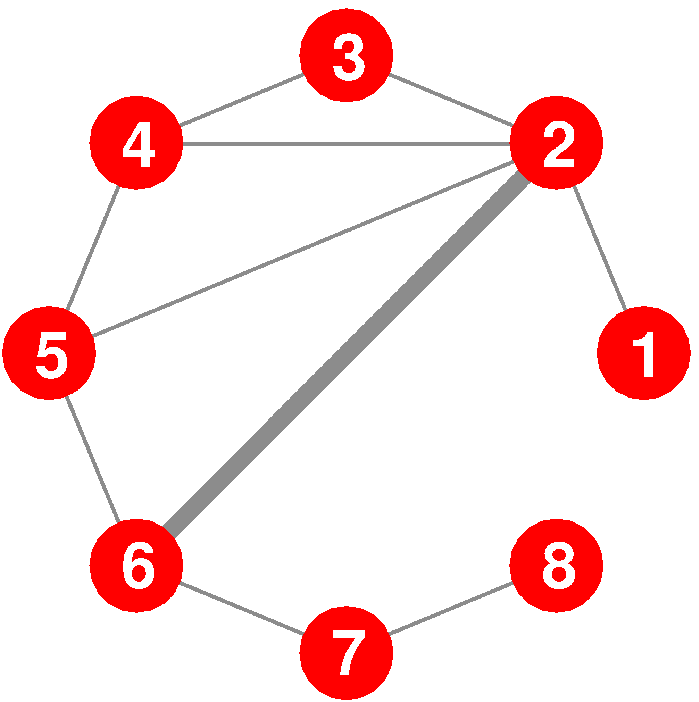}
\hspace{0.15in}
\includegraphics[width=0.15\textwidth]{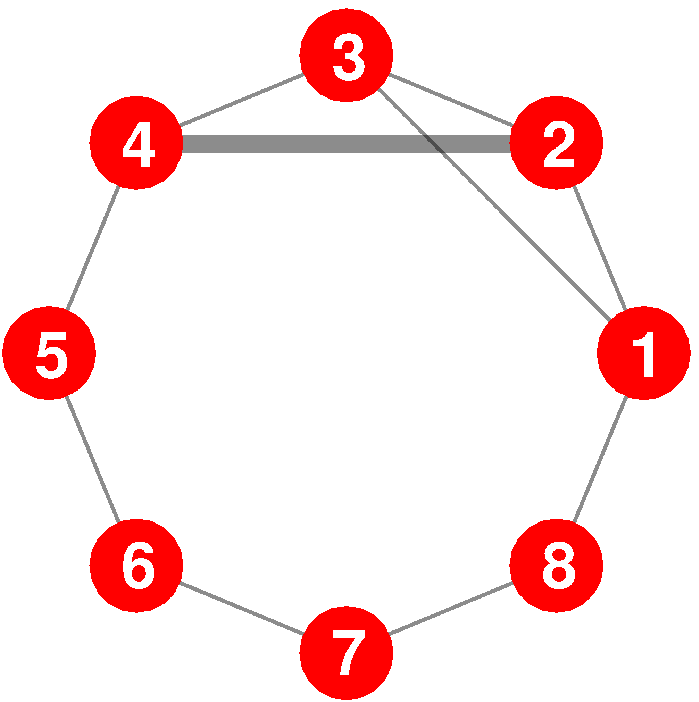}\\

\vspace{0.2in}
\subfloat[Min LE]{\includegraphics[width=0.15\textwidth]{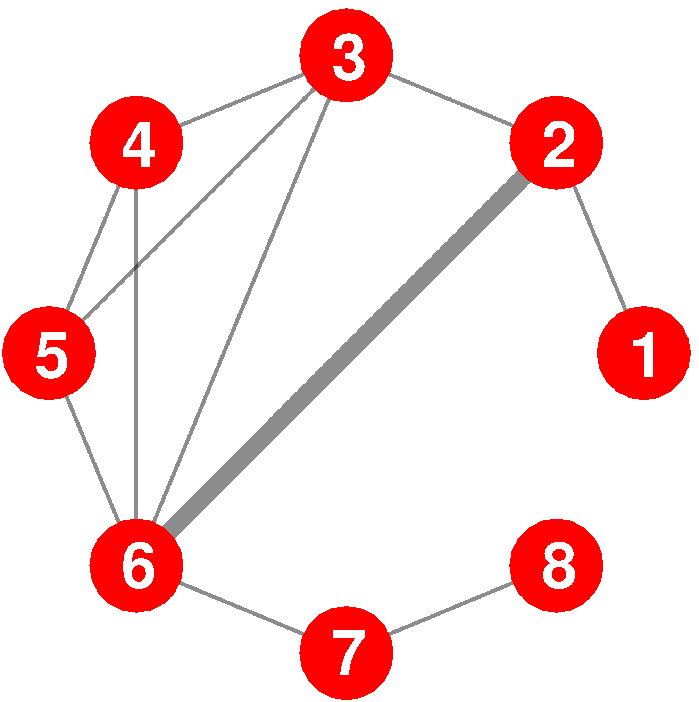}}
\hspace{0.15in}
\subfloat[Max LE]{\includegraphics[width=0.15\textwidth]{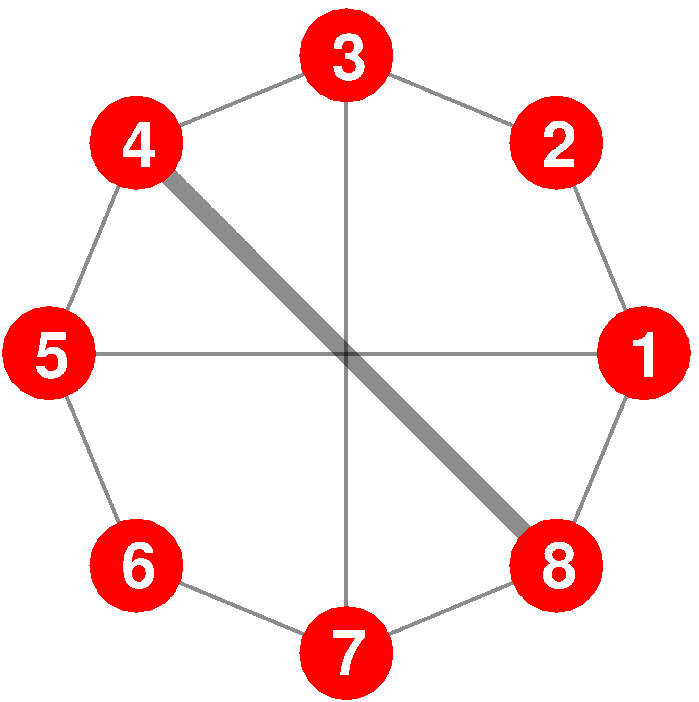}}
\hspace{0.15in}
\subfloat[Min ALE]{\includegraphics[width=0.15\textwidth]{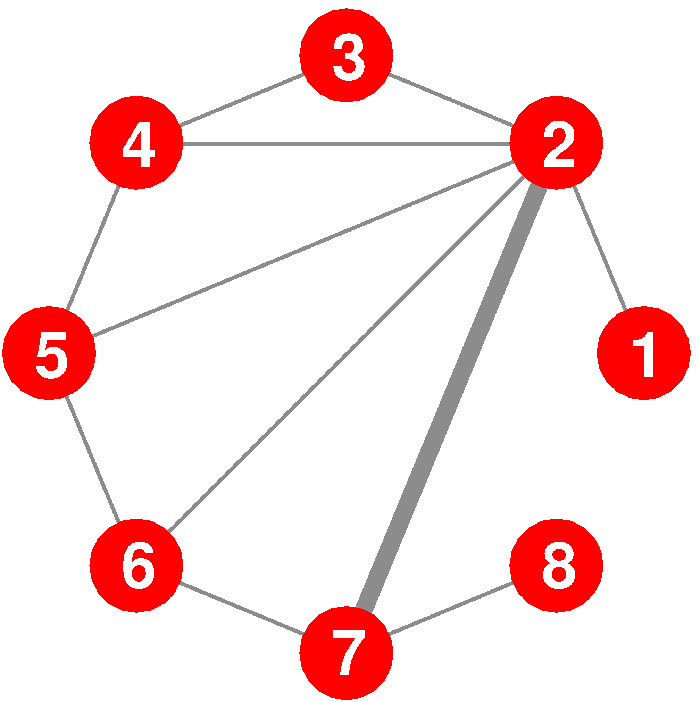}}
\hspace{0.15in}
\subfloat[Max ALE]{\includegraphics[width=0.15\textwidth]{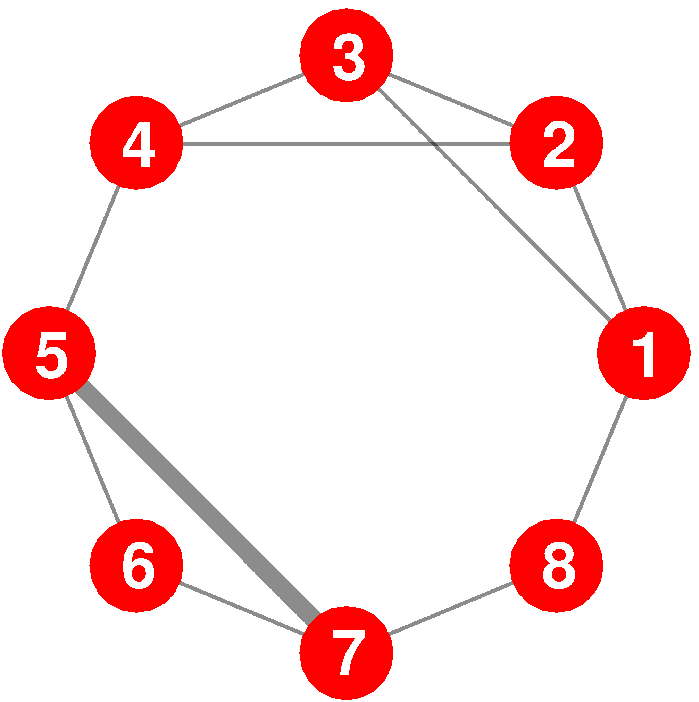}}
\caption{Evolution of the edge structure of a path graph over 8 nodes when we iteratively add edges that (a) minimize and (b) maximize the Laplacian entropy (LE). (c) and (d) show similar results for the approximate Laplacian entropy (ALE).}
\label{fig:lucatoy}
\end{figure}

\section{Experiments}\label{sec:experiments}
In the previous sections we have introduced the concepts of (normalized) Laplacian entropy of a graph and its quadratic approximation. This in turn provided us with a partial intuition of the relation between graph structure and entropy. In this section we aim to validate these initial intuitions with an extensive set of experiments and to investigate further the relation between the normalized and unnormalized Laplacian entropies, as well as the quality of their quadratic approximation.

\begin{figure}[t!]
\centering
\includegraphics[width=0.15\textwidth]{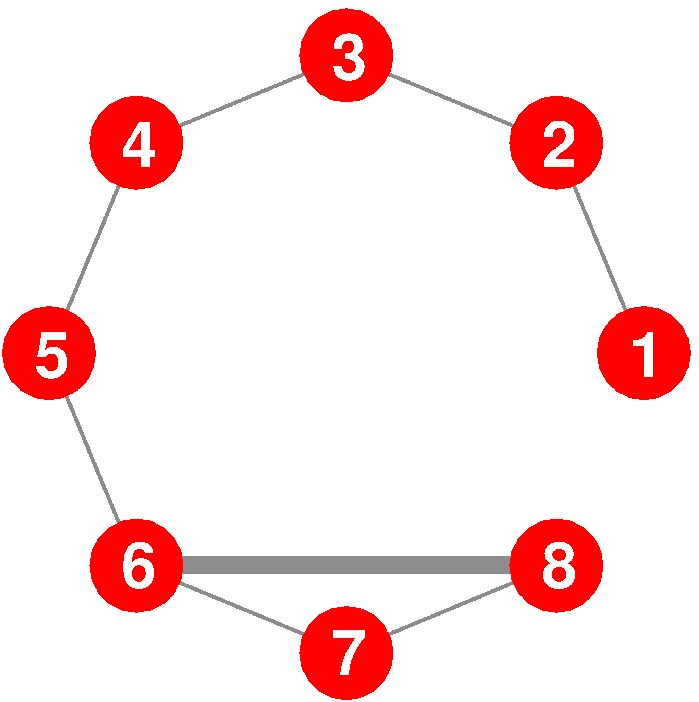}
\hspace{0.15in}
\includegraphics[width=0.15\textwidth]{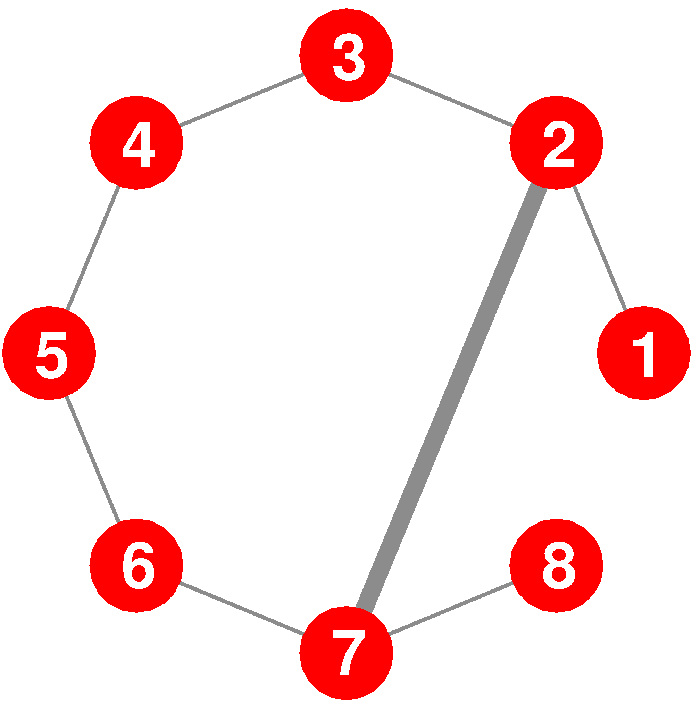} 
\hspace{0.2in}
\includegraphics[width=0.15\textwidth]{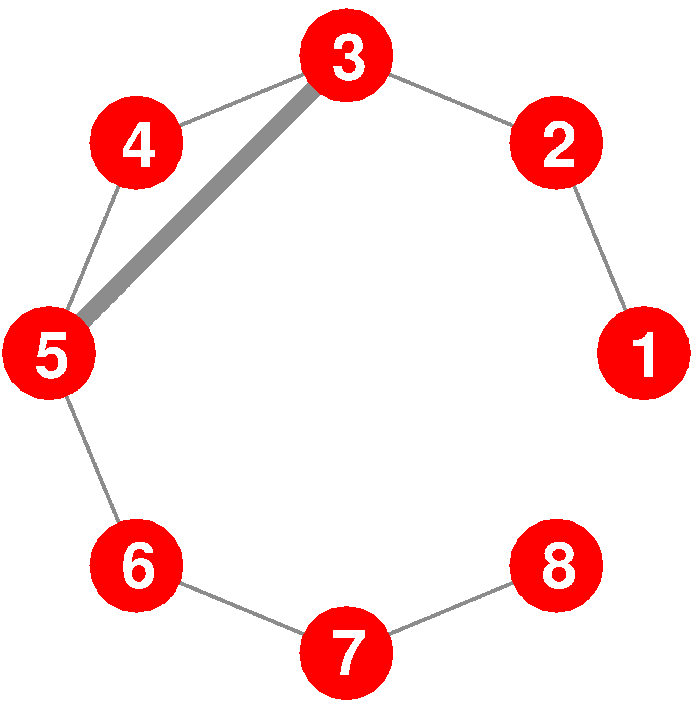}
\hspace{0.15in}
\includegraphics[width=0.15\textwidth]{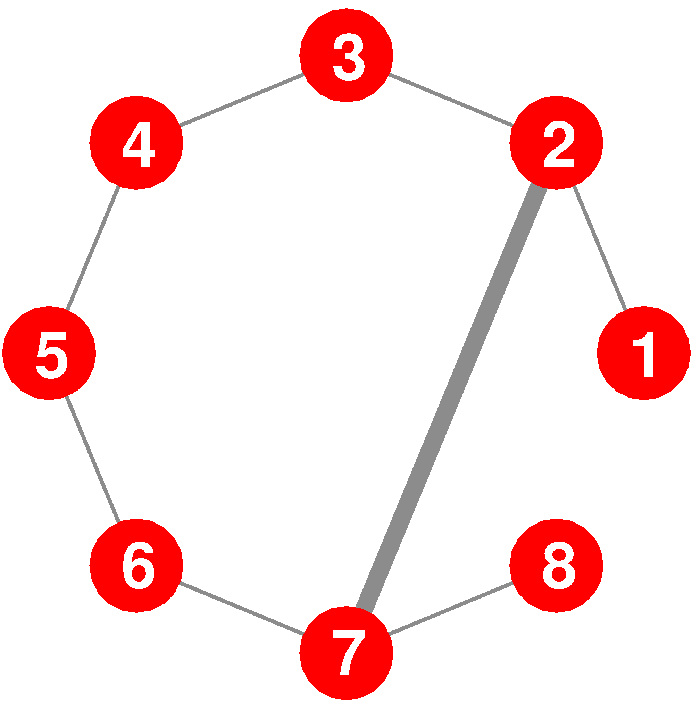}\\

\vspace{0.2in}
\includegraphics[width=0.15\textwidth]{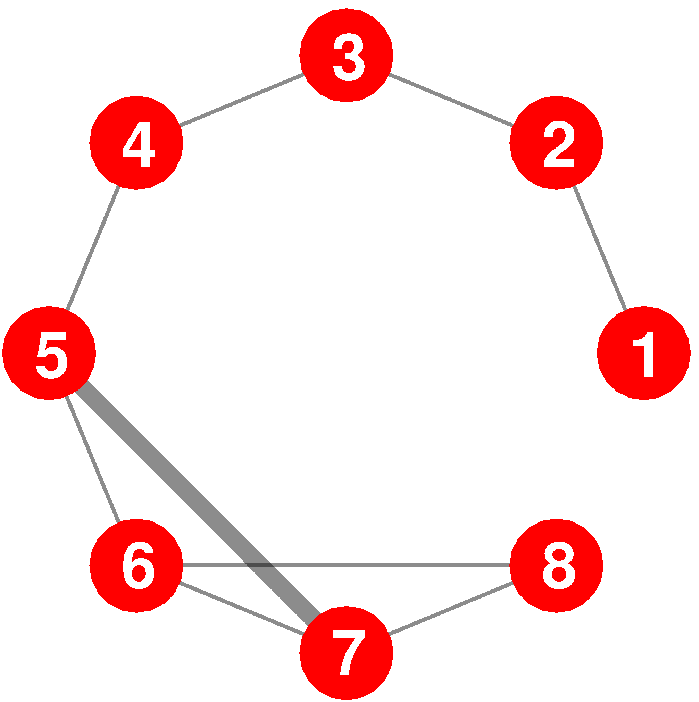}
\hspace{0.15in}
\includegraphics[width=0.15\textwidth]{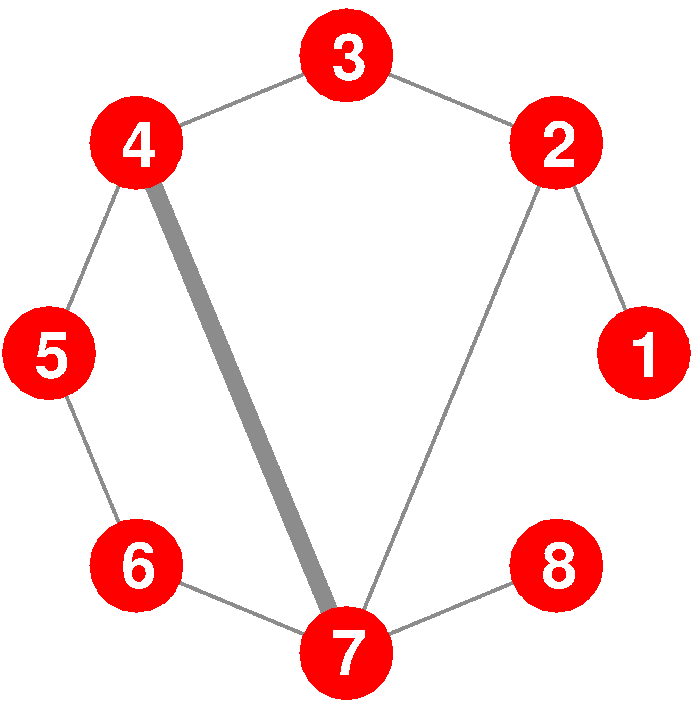}
\hspace{0.15in}
\includegraphics[width=0.15\textwidth]{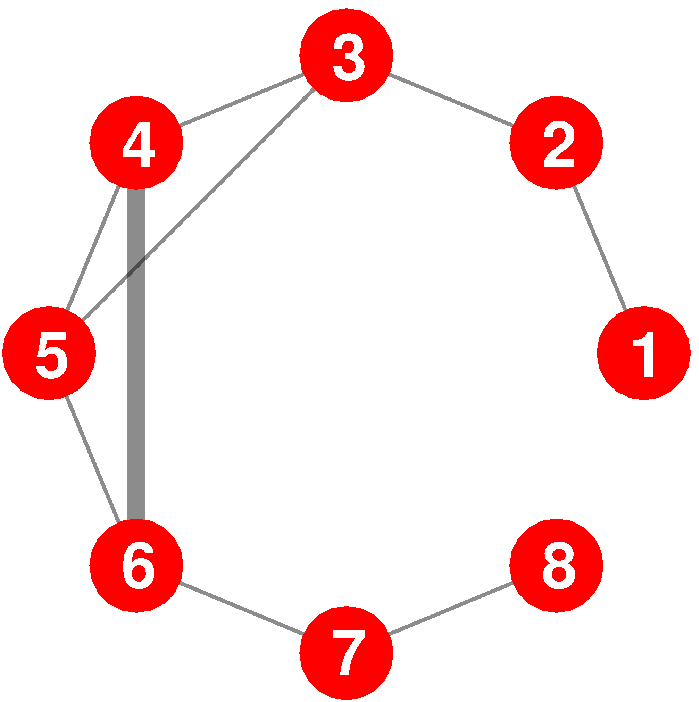}
\hspace{0.15in}
\includegraphics[width=0.15\textwidth]{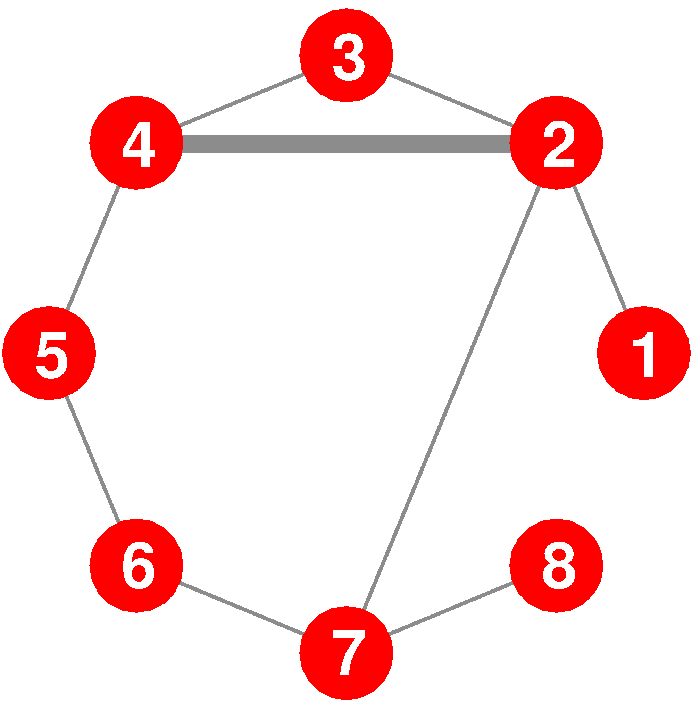}\\

\vspace{0.2in}
\includegraphics[width=0.15\textwidth]{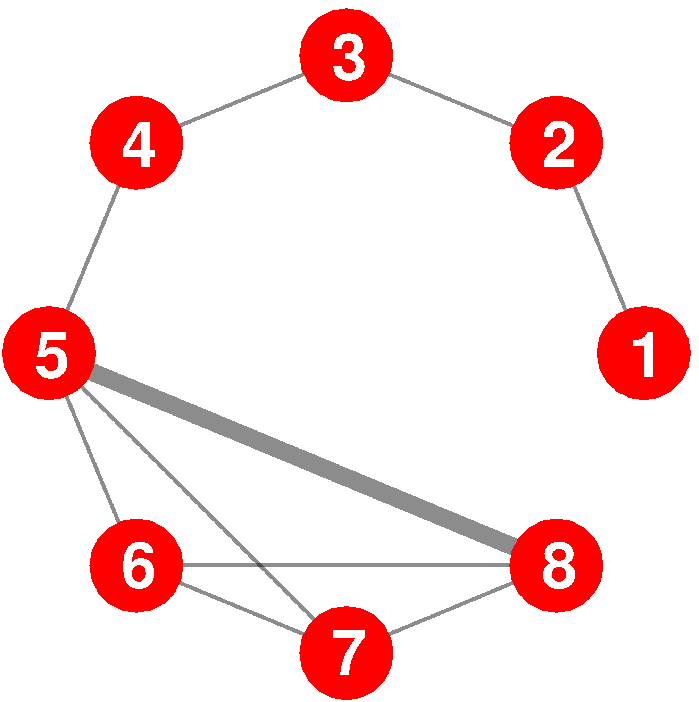}
\hspace{0.15in}
\includegraphics[width=0.15\textwidth]{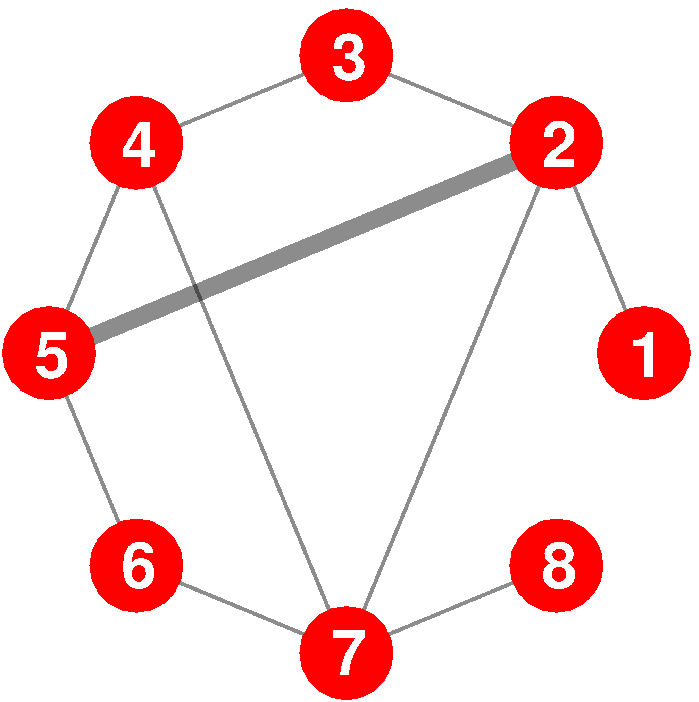}
\hspace{0.15in}
\includegraphics[width=0.15\textwidth]{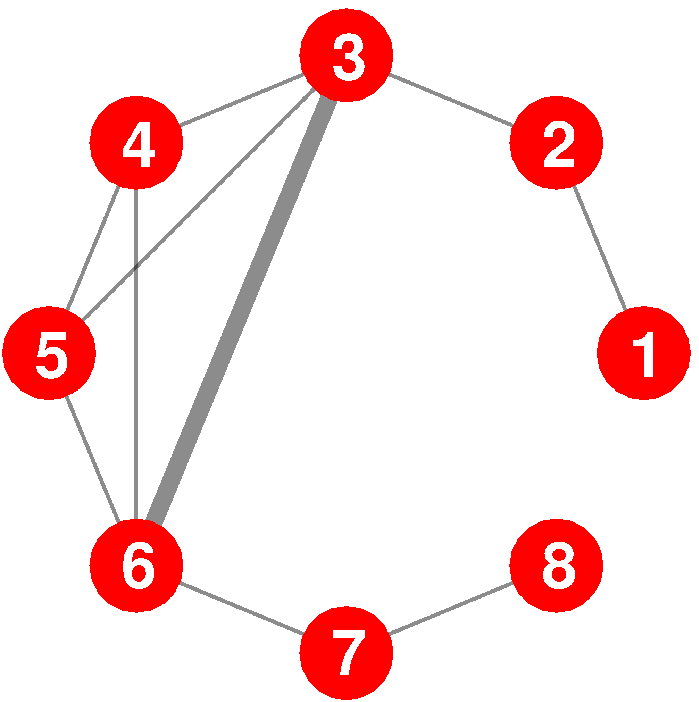}
\hspace{0.15in}
\includegraphics[width=0.15\textwidth]{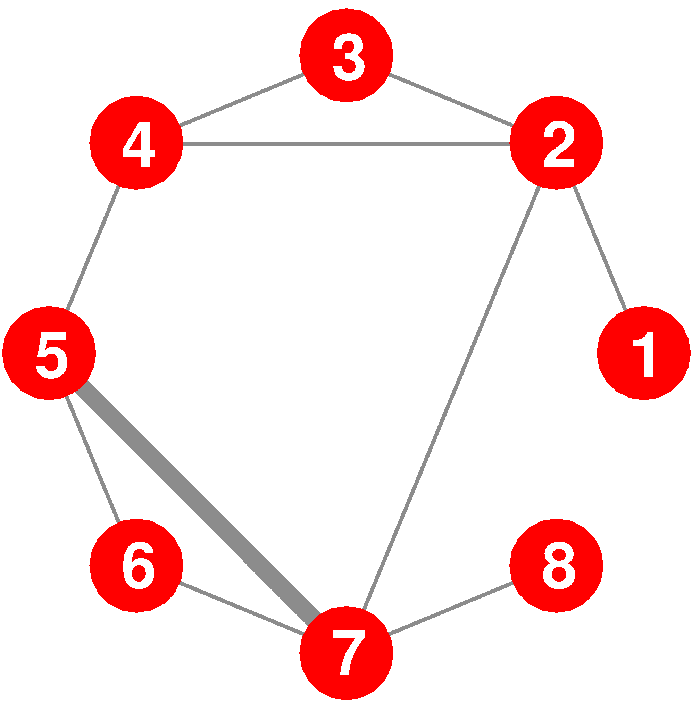}\\

\vspace{0.2in}
\subfloat[Min NLE]{\includegraphics[width=0.15\textwidth]{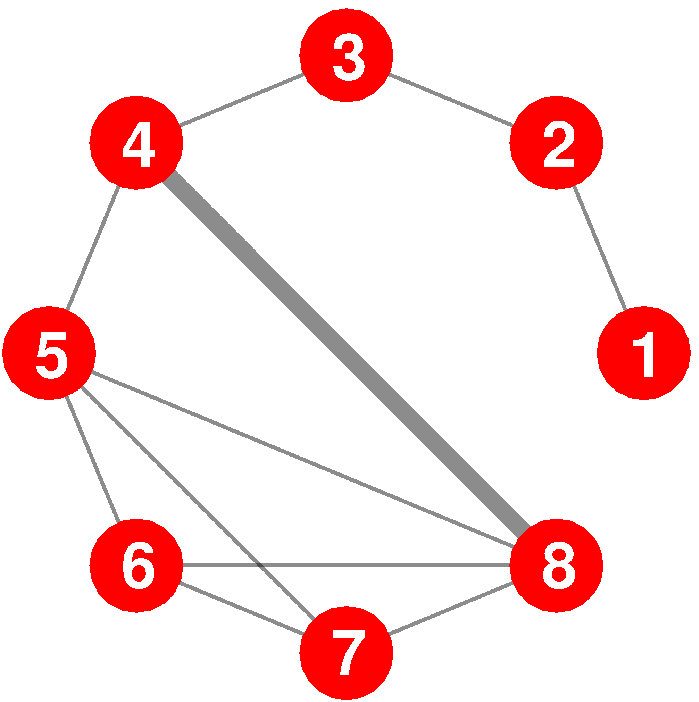}}
\hspace{0.15in}
\subfloat[Max NLE]{\includegraphics[width=0.15\textwidth]{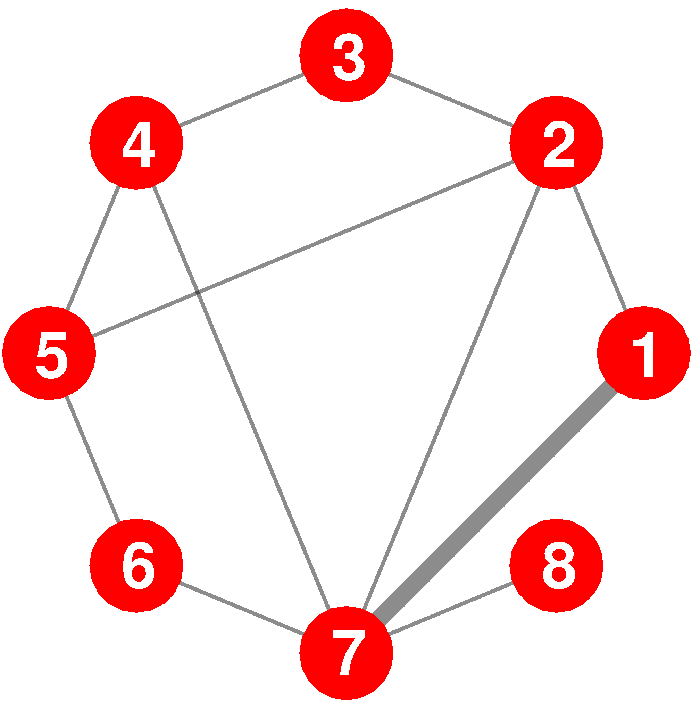}}
\hspace{0.15in}
\subfloat[Min ANLE]{\includegraphics[width=0.15\textwidth]{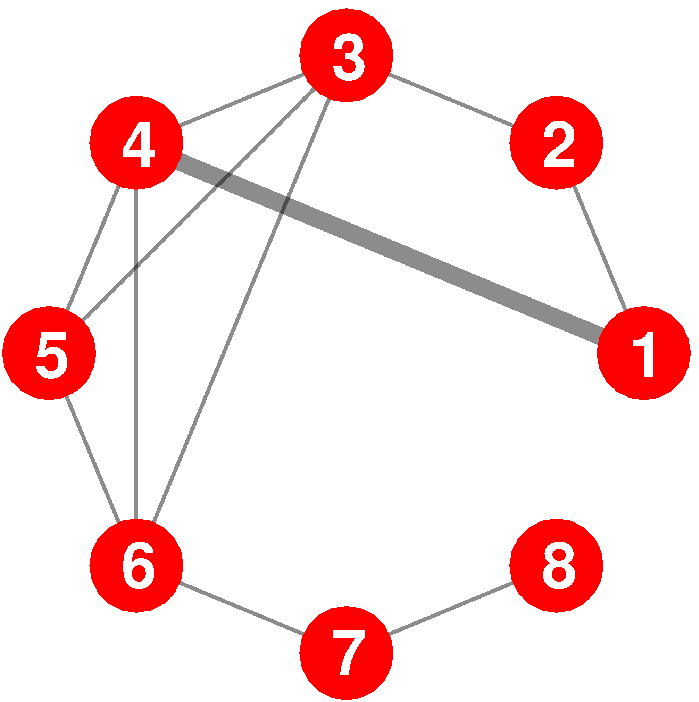}}
\hspace{0.15in}
\subfloat[Max ANLE]{\includegraphics[width=0.15\textwidth]{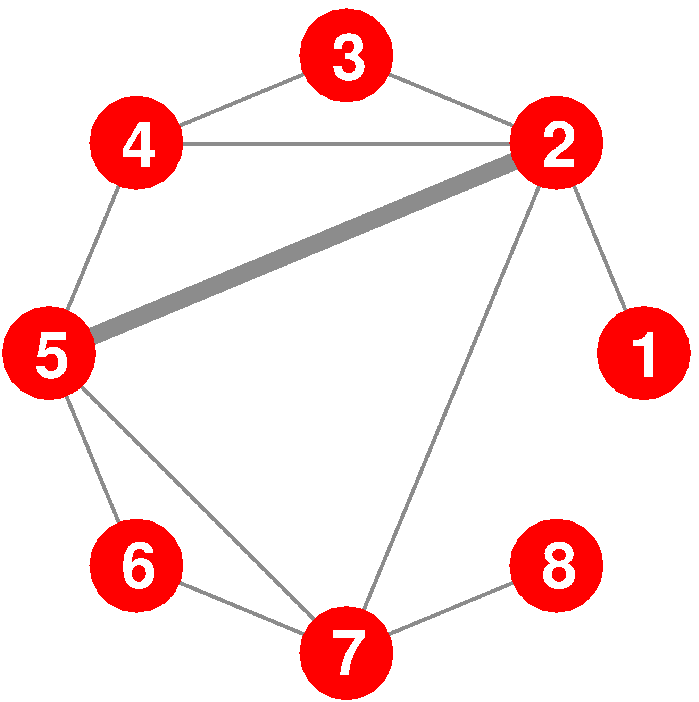}}
\caption{Evolution of the edge structure of a path graph over 8 nodes when we iteratively add edges that (a) minimize and (b) maximize the normalized Laplacian entropy (NLE). (c) and (d) show similar results for the approximate normalized Laplacian entropy (ANLE).}
\label{fig:giorgiatoy}
\end{figure}

\subsection{Entropy-driven Graph Evolution}
We commence by investigating how the structure of a graph changes as we add new nodes and edges to it. To this end, we introduce a simple growth model where new connections are established if they maximise (minimise) the graph entropy. We perform the same experiment for both the Laplacian and the normalized Laplacian entropy, as well as their quadratic approximations.

\subsubsection{Edge Growth Model}
We first consider the case where the number of nodes is fixed and new edges are iteratively added to the graph. Fig.~\ref{fig:lucatoy}  shows the first four stages of the evolution of a graph with eight nodes where the growth process is driven by the Laplacian entropy. Each column of Fig.~\ref{fig:lucatoy} corresponds to a different choice of the process (maximization or minimization) and entropy (exact or approximated). Similarly, Fig.~\ref{fig:giorgiatoy} shows the results for the normalized Laplacian entropy.

Figs.~\ref{fig:lucatoy}(c) and (d) confirm what already observed when looking at the quadratic approximation of the Laplacian entropy. Indeed, edges that maximize the approximate Laplacian entropy are edges that connect low degree nodes, as shown in Fig.~\ref{fig:lucatoy}(d). In contrast to Fig.~\ref{fig:lucatoy}(b), where we maximize the exact entropy, in Fig.~\ref{fig:lucatoy}(d) all pairs of nodes with minimum degree sum have the same probability of being connected. This is not the case in Fig.~\ref{fig:lucatoy}(b), where, given two pairs of nodes with equal sum of their degrees, the pair of nodes with the highest geodesic distance\footnote{Recall that the geodesic distance between two nodes $u$ and $v$ is the number of edges in the shortest path connecting $u$ and $v$.} leads to a higher increment in the entropy. As a result of this, each new edge added in the Fig.~\ref{fig:lucatoy}(b) seems to act as an axis of symmetry. While it would be tempting to argue that the latter is evidence of structural symmetries being picked up by the exact Laplacian entropy as opposed to its approximated version, a quick numerical investigation proves that this hypothesis is incorrect. 

\begin{figure}[t!]
\centering
\subfloat[\hspace{0.1in} Erd\"os-R\'enyi]{\includegraphics[width=.32\textwidth]{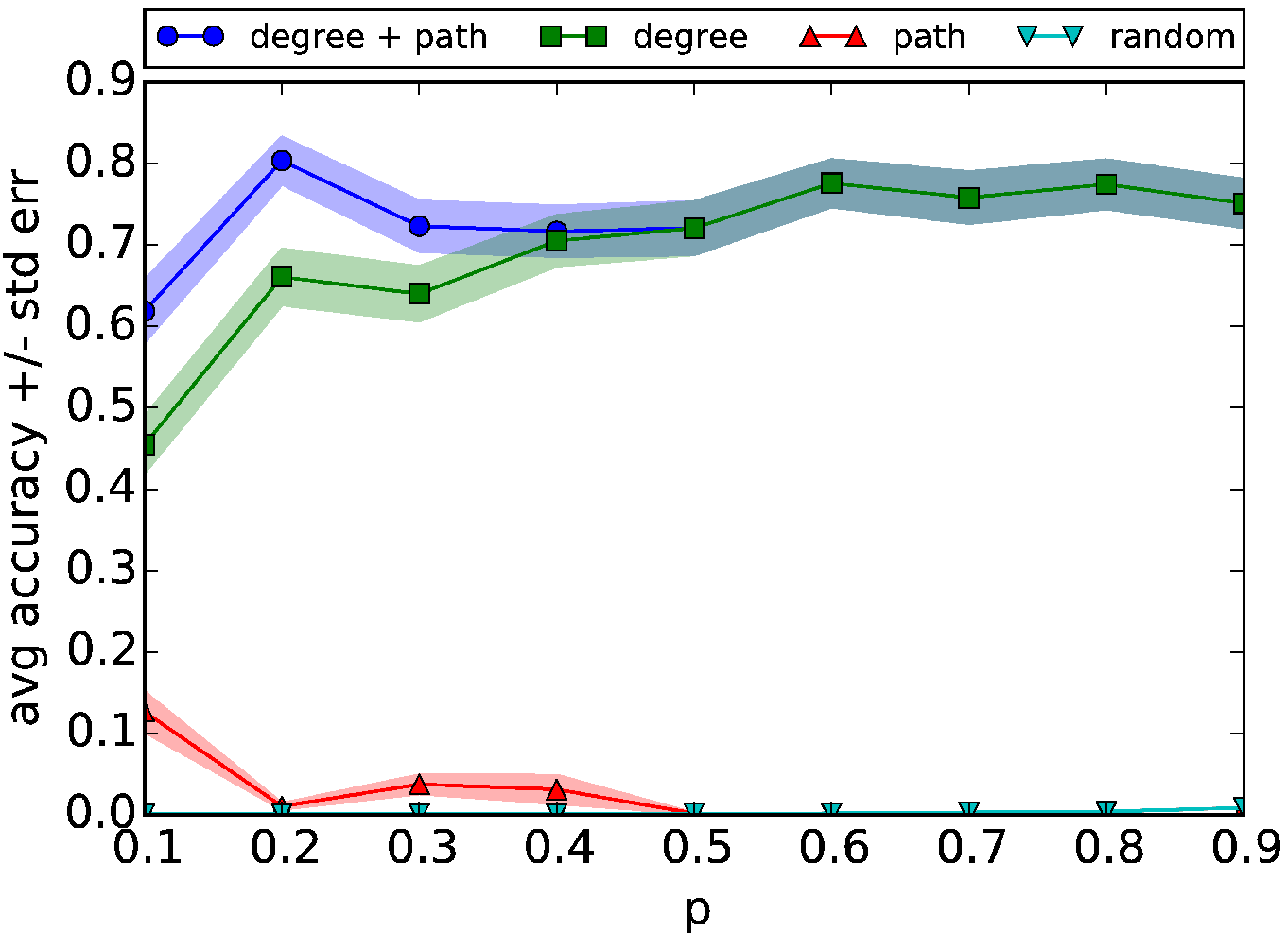}}
\hspace{0.01in}
\subfloat[\hspace{0.1in} Scale-free]{\includegraphics[width=.32\textwidth]{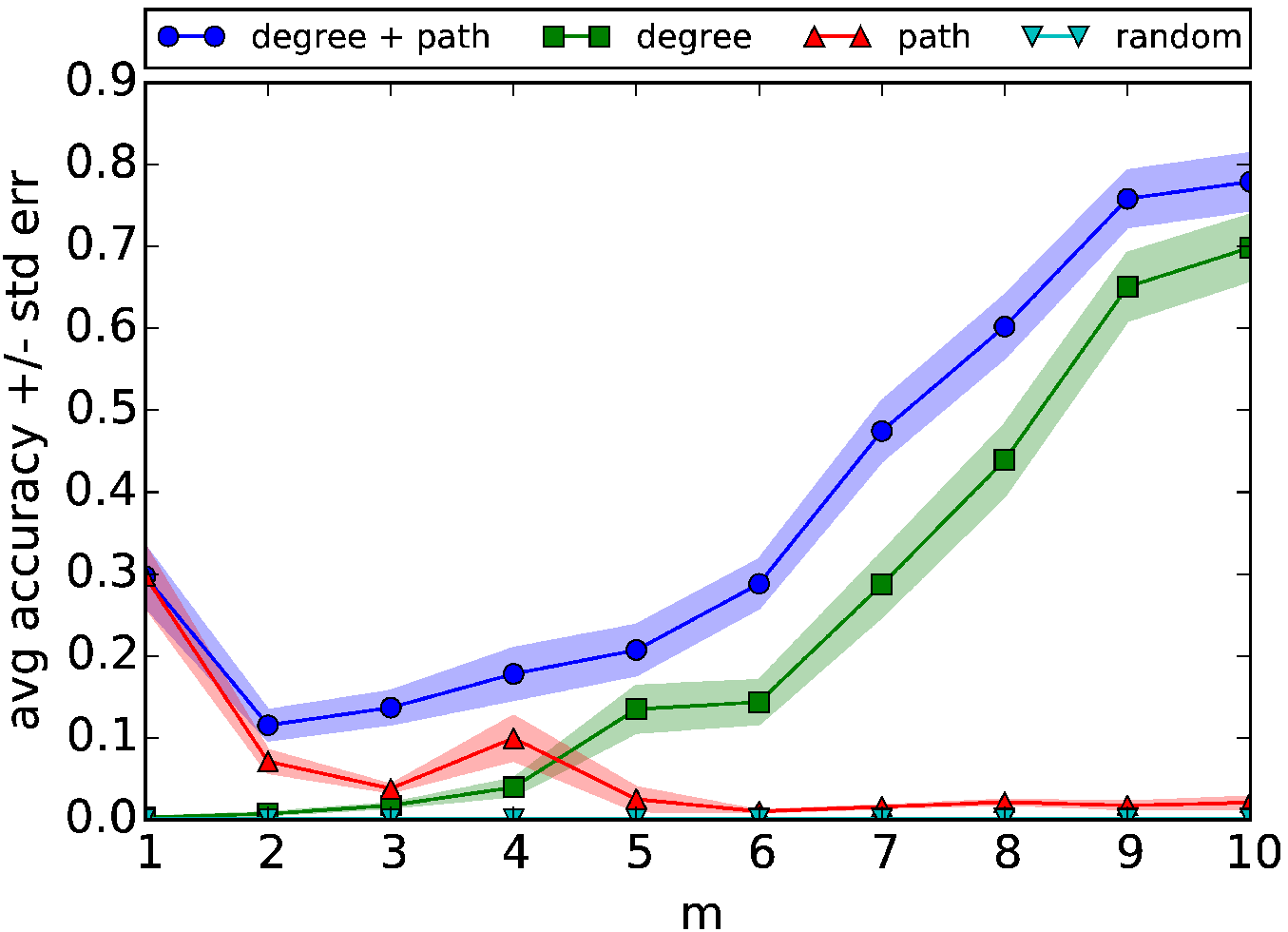}}
\hspace{0.01in}
\subfloat[\hspace{0.1in} Small world]{\includegraphics[width=.32\textwidth]{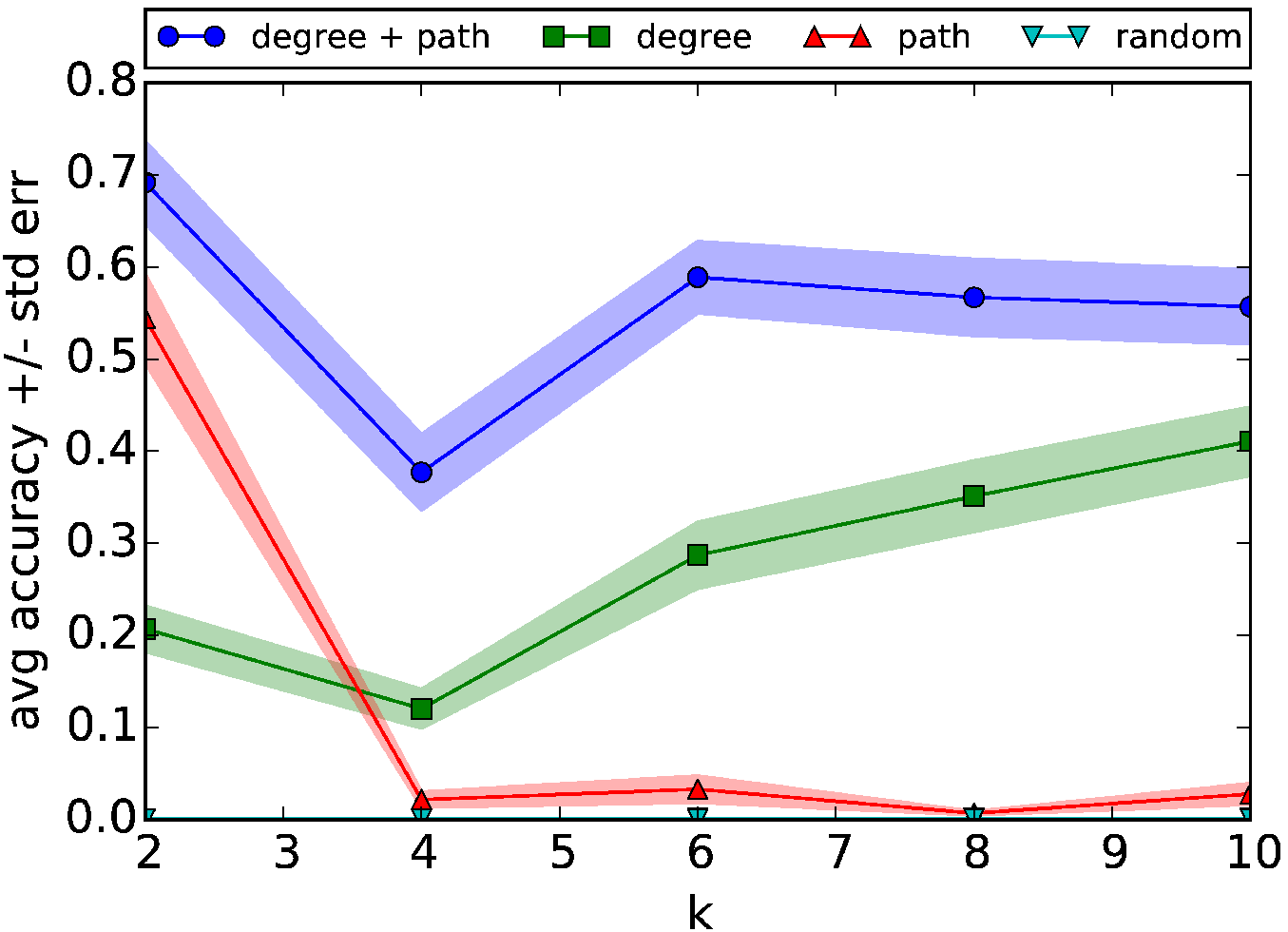}}
\hspace{0.01in}
\caption{Average accuracy when predicting the edge whose addition maximizes the Laplacian entropy when using four different heuristics, i.e., choose (1) the pair of nodes with minimum degree sum; (2) the pair of nodes with maximum geodesic distance; (3) the pair of nodes with minimum degree sum and maximum geodesic distance; (4) a pair of nodes $u$ and $v$ picked at random, with $(u,v) \notin E$.}
\label{fig:heuristic}
\end{figure}

It is also not true, for a general graph, that the pair of nodes being connected is always one with minimum degree sum and maximum distance. However the distance between the nodes being connected clearly plays a role, together with a number of yet unspecified structural properties. To show this, we perform the following experiment. Starting from a random graph $G$, we use four different heuristics to predict what edge will lead to the maximum increment of the Laplacian entropy. Each of the four heuristics selects the pairs of nodes that optimize the following measures, respectively: (1) the pair of nodes with minimum degree sum (which corresponds to the structural information contained in the approximated Laplacian entropy); (2) the pair of nodes with maximum geodesic distance; (3) the pair of nodes with minimum degree sum and maximum geodesic distance; (4) a pair of nodes $u$ and $v$ picked at random, with $(u,v) \notin E$. The prediction accuracy of a heuristic is computed as the fraction of edges it identified correctly. Fig.~\ref{fig:heuristic} shows the results for the different random graph models, the Erd\"os-R\'enyi model, the Watts-Strogatz model, and the Preferential Attachment model~\cite{barabasi1999emergence} (see Section~\ref{sec:graph_model} for a detailed description of the models and their parameters). In all cases, the addition of the path length information leads to a significant increment in the accuracy of the heuristic which solely looks for the pair of nodes with minimum degree sum. In other words, both degree statistics (captured by the quadratic approximations) and path length information are important structural patterns captured by the exact version of the Laplacian entropy. Note also that as the graphs becomes denser (which, given a fixed number of nodes, for the three random models considered correspond to increasing values of $p$, $m$, and $k$, respectively) the path length information loses importance. This is due to the fact that for sufficient higher densities all pairs of nodes lie at the same distance from each other.

\begin{figure}[t!]
{\includegraphics[width=0.22\textwidth]{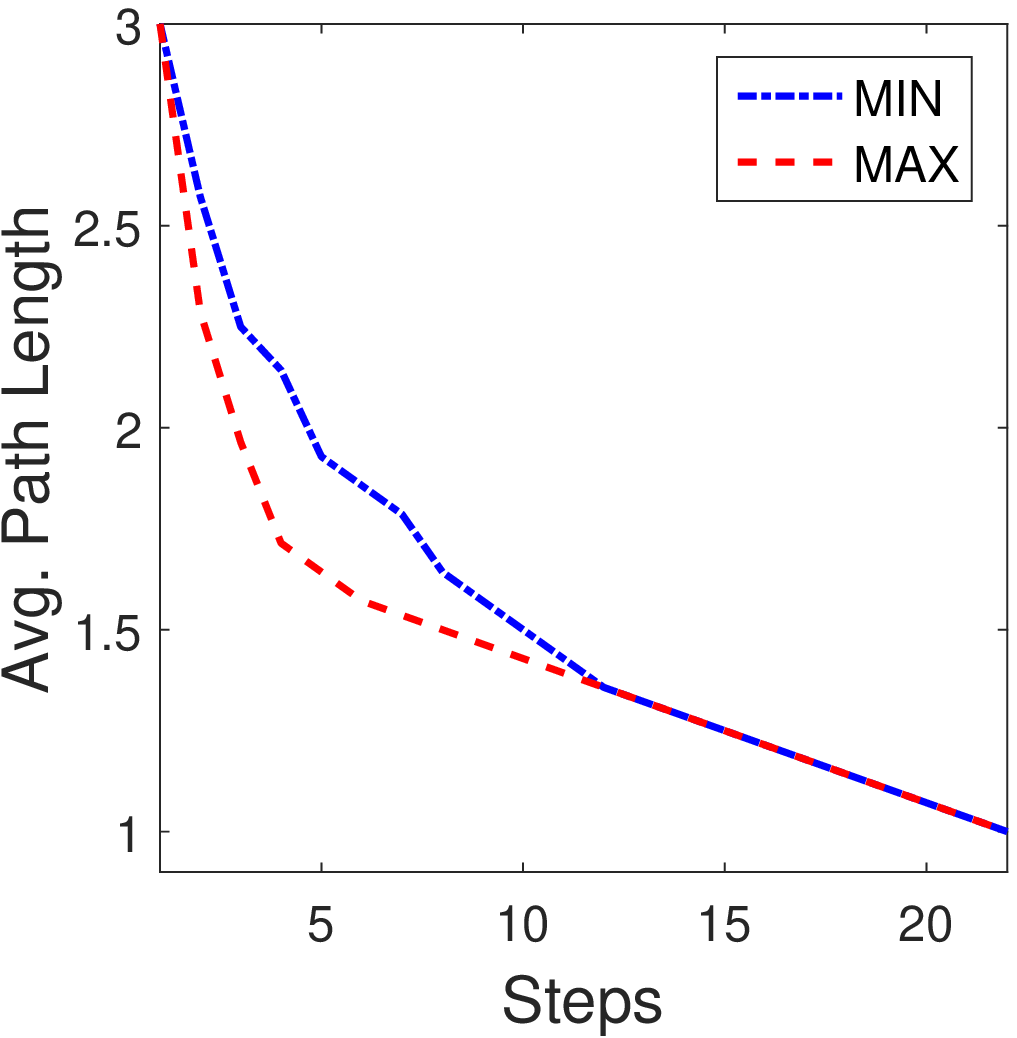}}
{\includegraphics[width=0.22\textwidth]{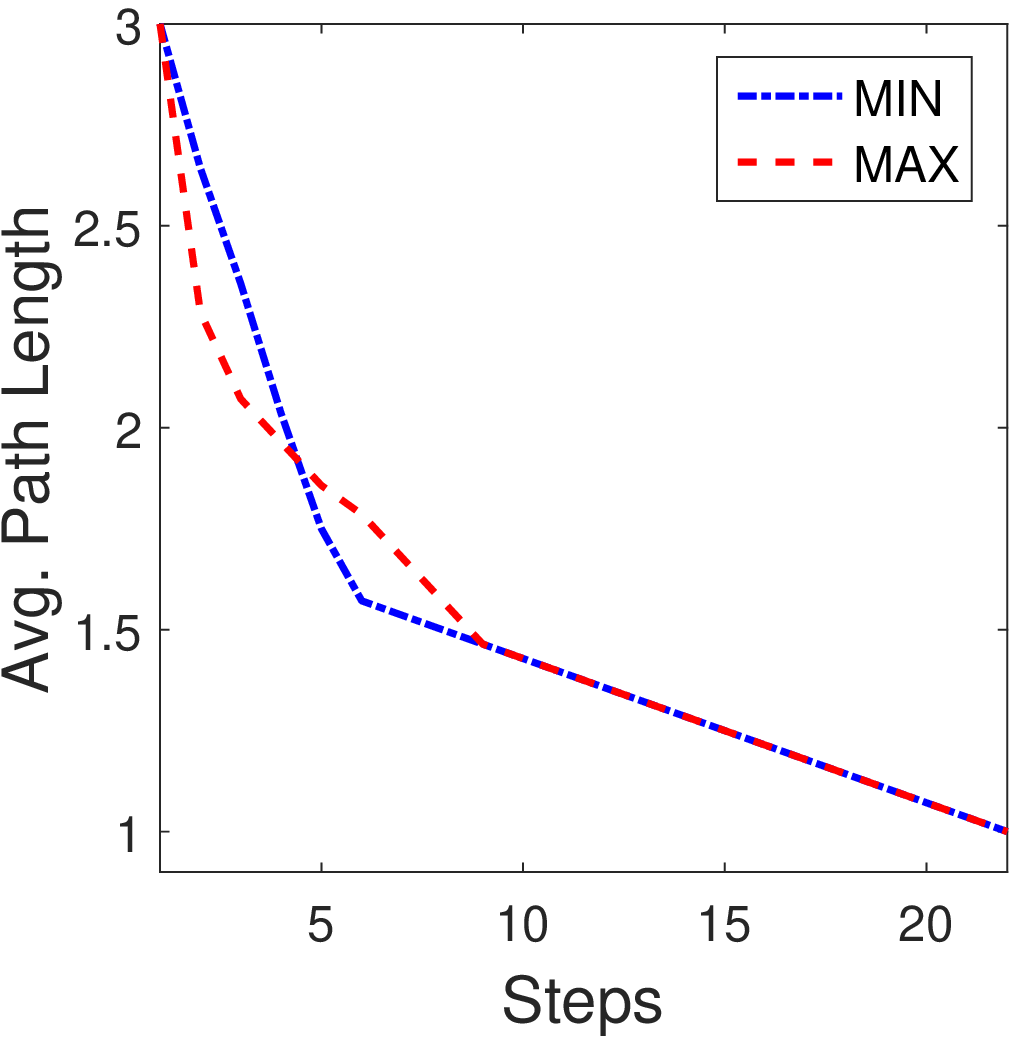}}
{\includegraphics[width=0.22\textwidth]{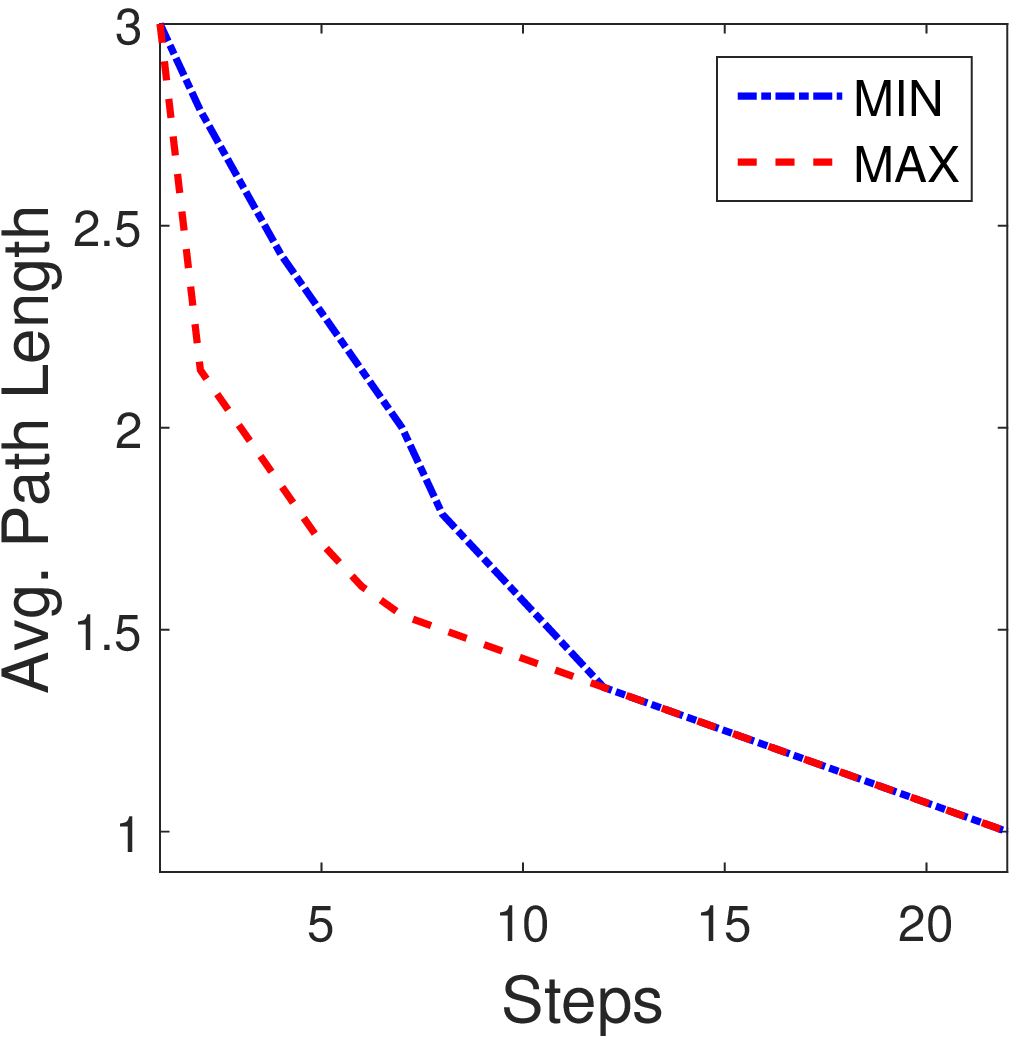}}
{\includegraphics[width=0.22\textwidth]{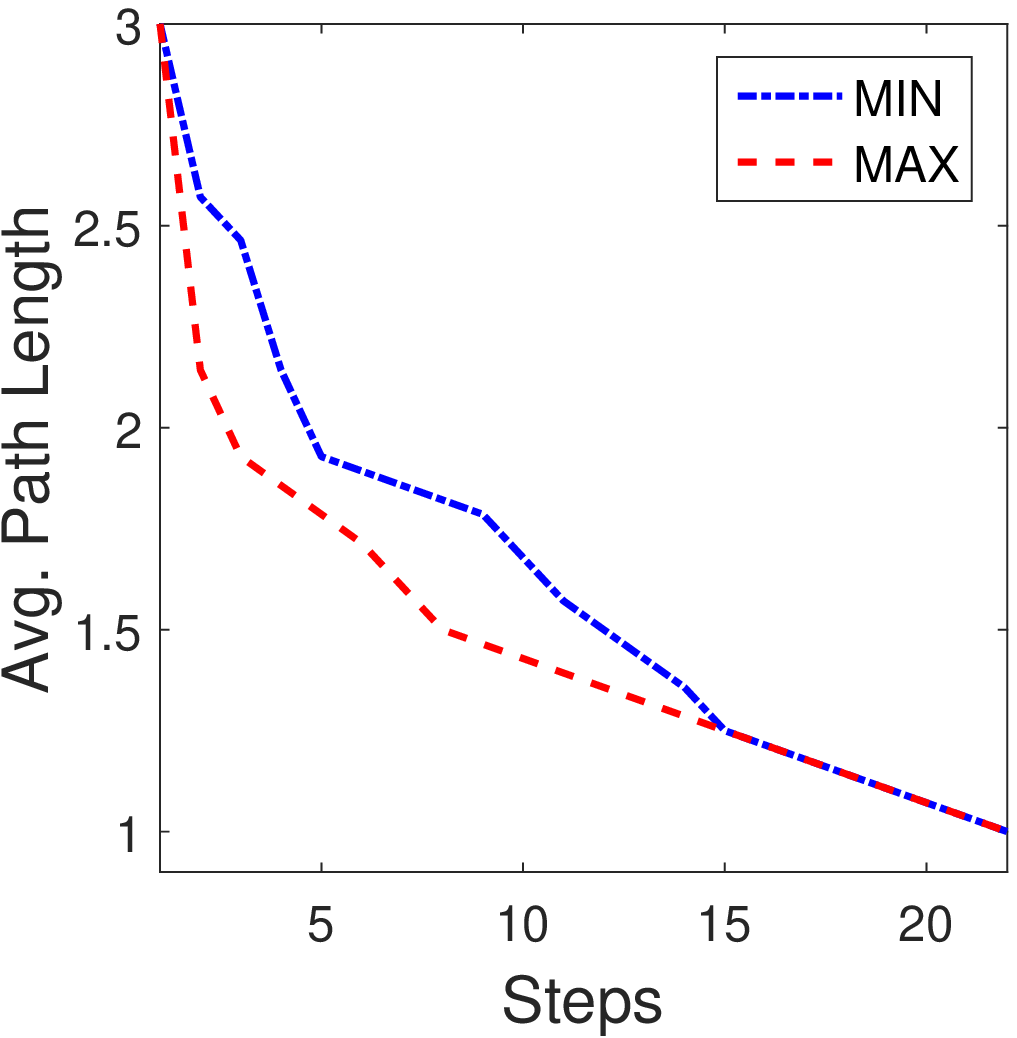}}


{\includegraphics[width=0.22\textwidth]{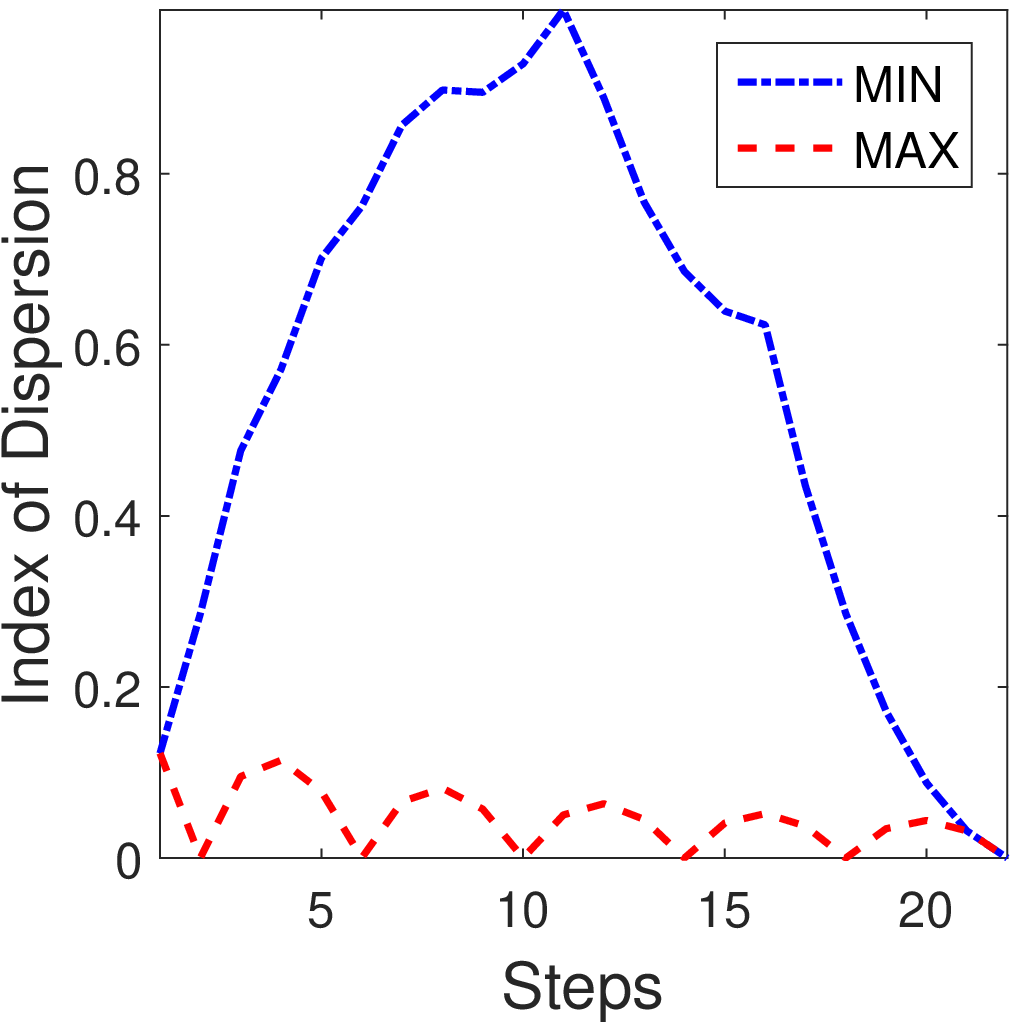}}
{\includegraphics[width=0.22\textwidth]{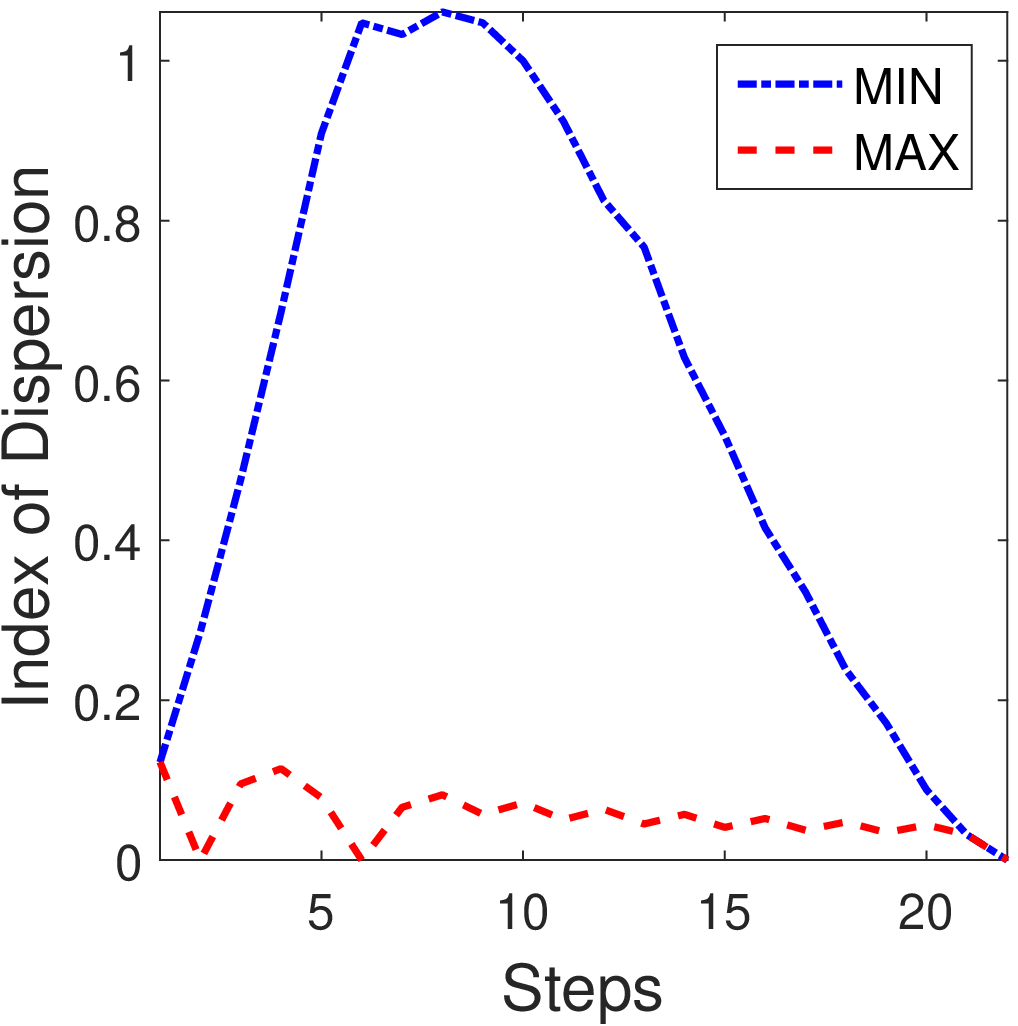}}
{\includegraphics[width=0.22\textwidth]{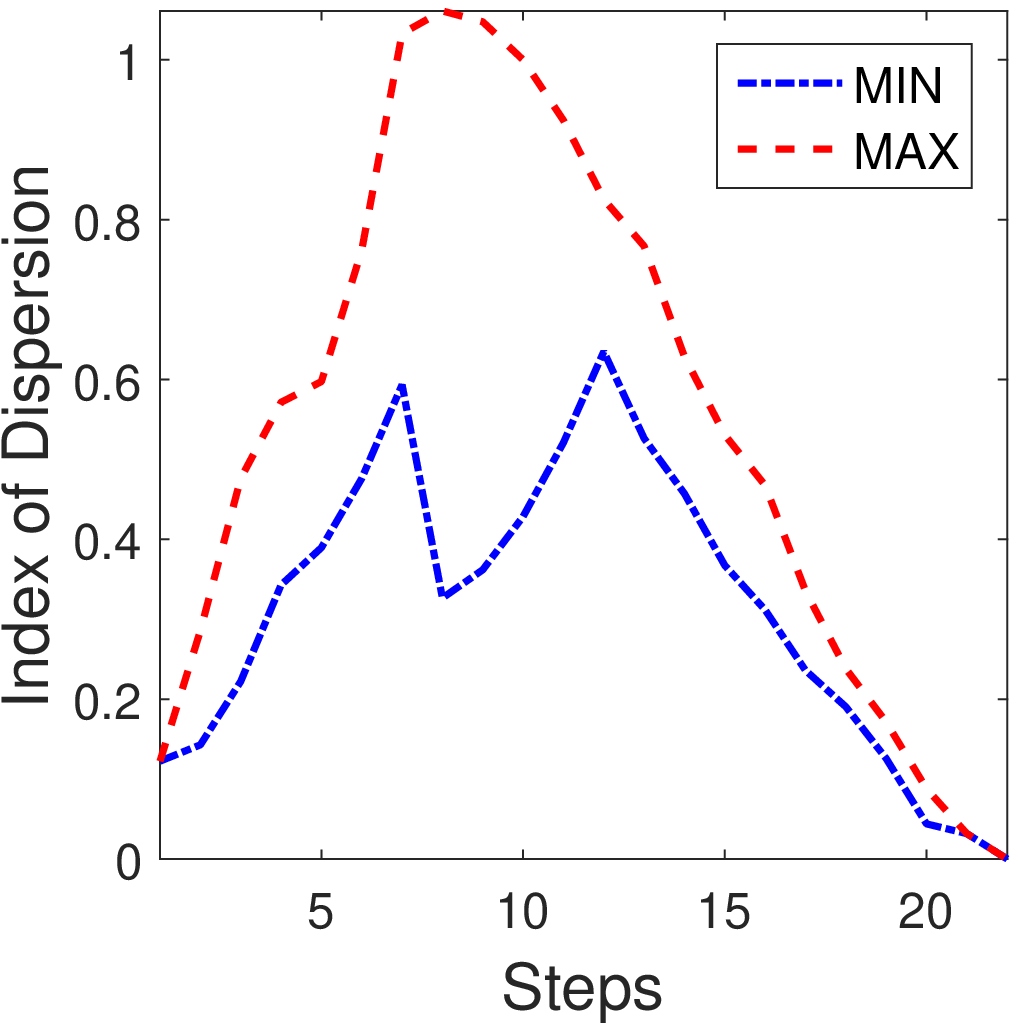}}
{\includegraphics[width=0.22\textwidth]{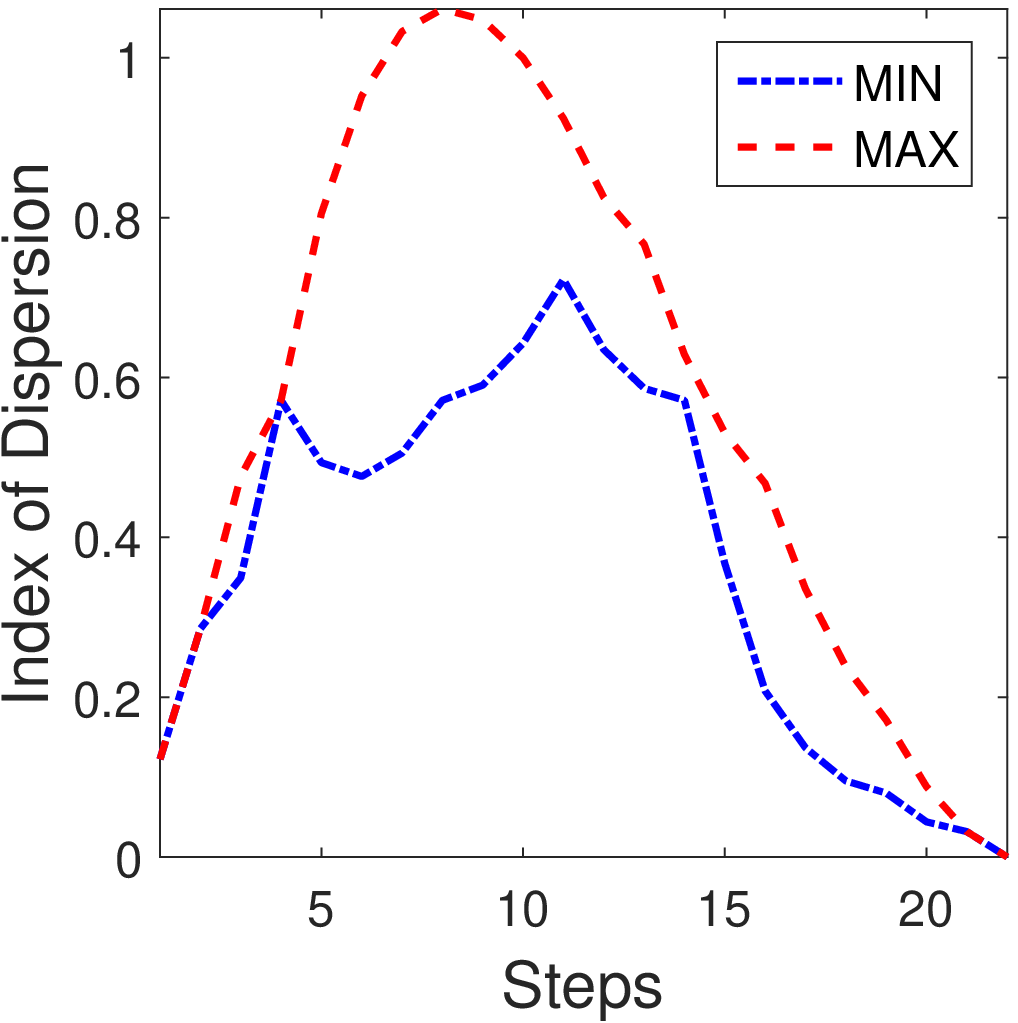}}

\subfloat[LE]{\includegraphics[width=0.22\textwidth]{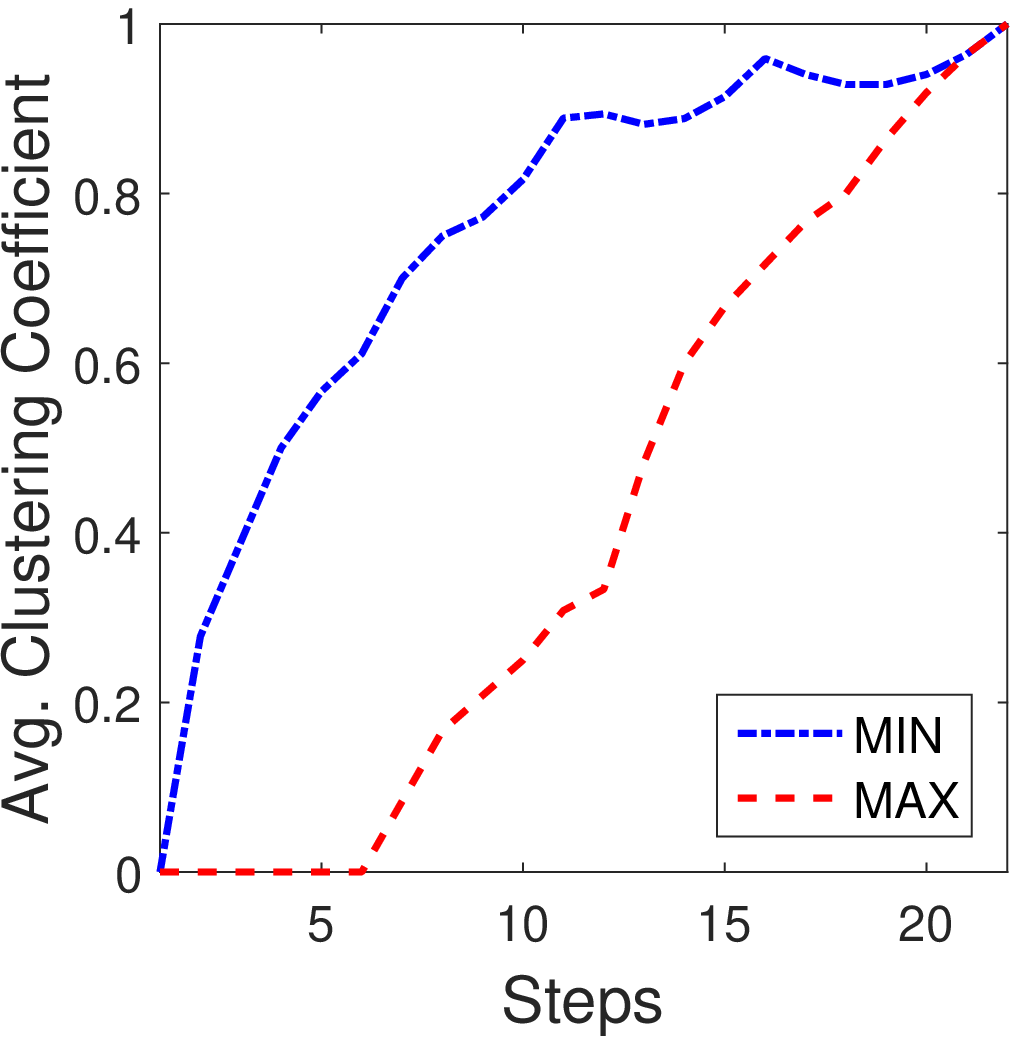}}
\subfloat[ALE]{\includegraphics[width=0.22\textwidth]{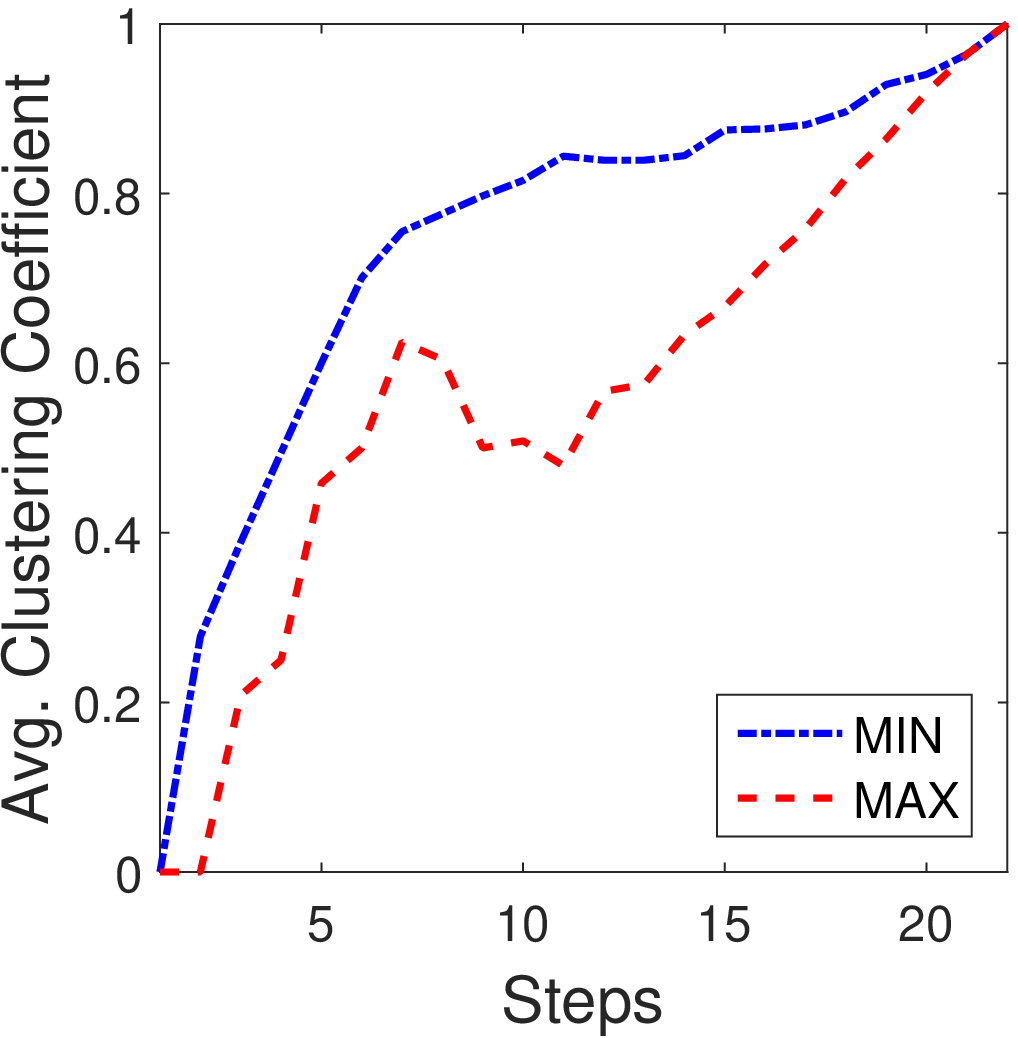}}
\subfloat[NLE]{\includegraphics[width=0.22\textwidth]{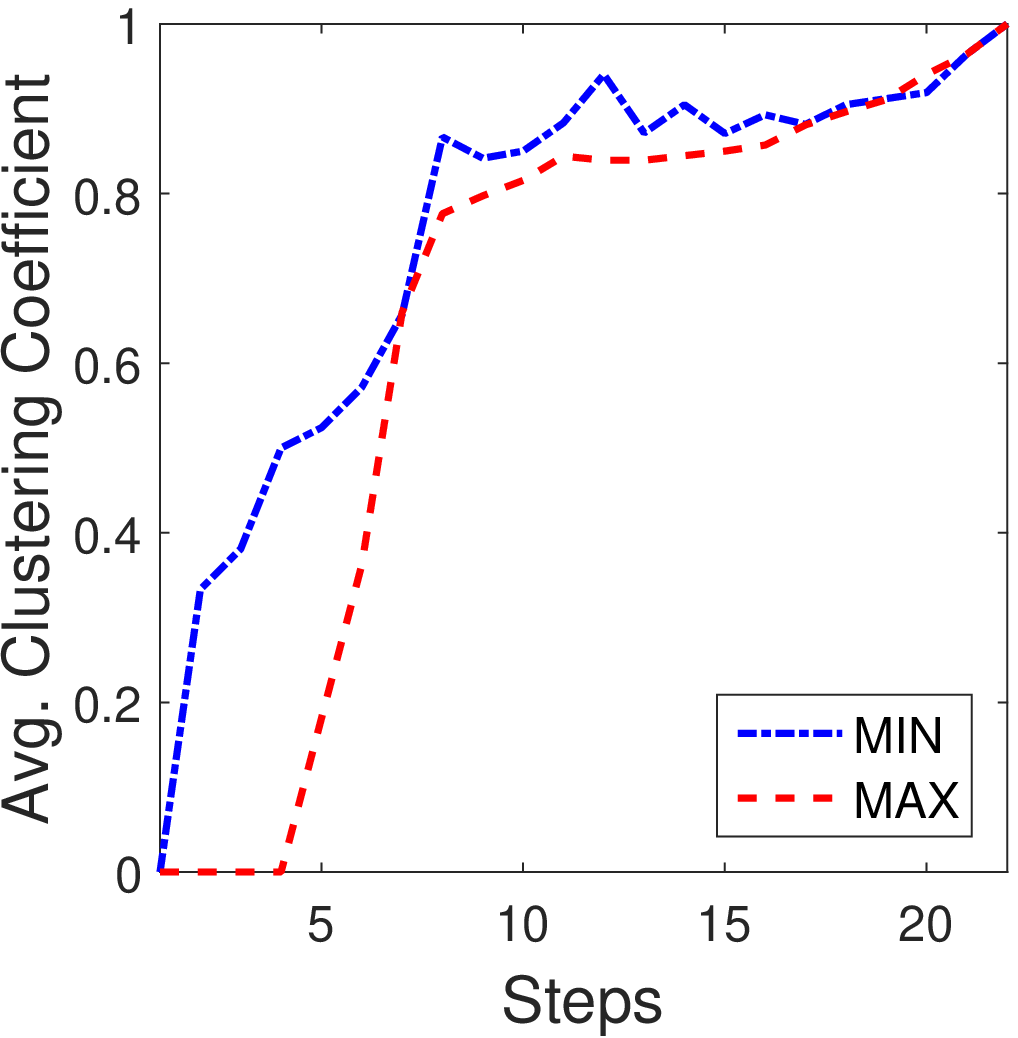}}
\subfloat[ANLE]{\includegraphics[width=0.22\textwidth]{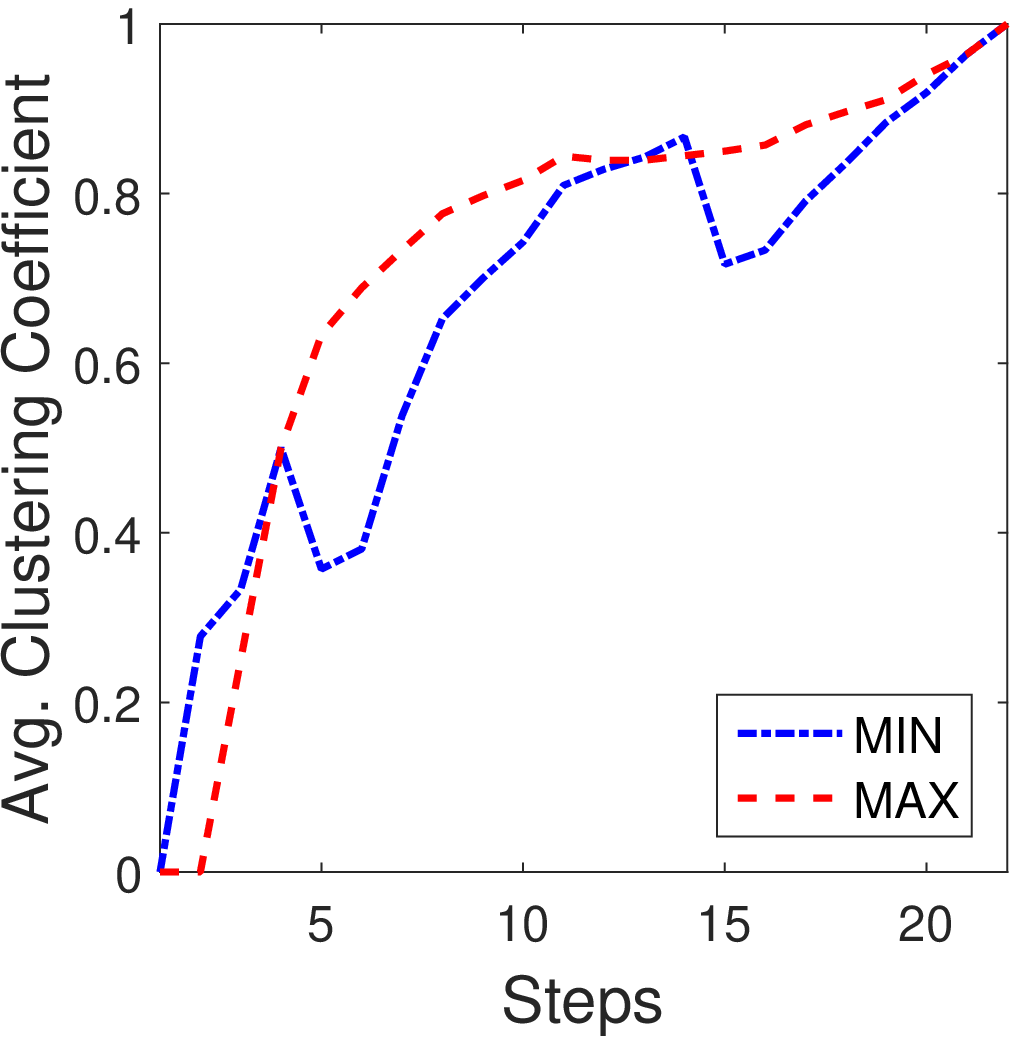}}

\caption{Top-to-bottom: Average path length, index of dispersion, and average clustering coefficient at different steps of the temporal evolution of the graphs in Fig.~\ref{fig:lucatoy}.}
\label{fig:stats_toy}
\end{figure}

We also compute a number of statistics that capture different structural properties of the graph during its evolution, namely the average shortest path length, the index of dispersion of the degree distribution, and the average clustering coefficient, as shown in Fig.~\ref{fig:stats_toy}. Recall that the index of dispersion of a distribution measures the ratio of its variance to its mean, and the clustering coefficient quantifies the degree to which nodes in a graph tend to cluster together. Fig.~\ref{fig:stats_toy} highlights once again the differences between the structural information captured by the Laplacian and the normalized Laplacian entropy, as well as their quadratic approximations. The tendency of the process which maximizes the exact Laplacian entropy to connect low degree nodes is particularly evident in the plots of the index of dispersion. As explained above, maximizing the (approximated) Laplacian entropy tends to create connections between low degree nodes. This in turn tends to create a regular structure where each node has the same degree, thus keeping the index of dispersion of the degree distribution low throughout the graph evolution. Note that this does not happen when maximizing or minimizing the (approximated) normalized Laplacian entropy. The difference between the exact Laplacian entropy and its approximated version is instead clear by looking at the average clustering coefficient. By connecting nodes that have both low degree sum and high distance, maximizing the exact Laplacian entropy keep the clustering coefficient as it effectively attempts to avoid creating triangles, at least in the first stages of the evolution. This however does not happen for the approximated Laplacian entropy, where the connection of two nodes with a common neighbour introduces a new triangle and thus increases the value of the average clustering coefficient.

\begin{figure}[t!]
\centering
\includegraphics[width=0.15\textwidth]{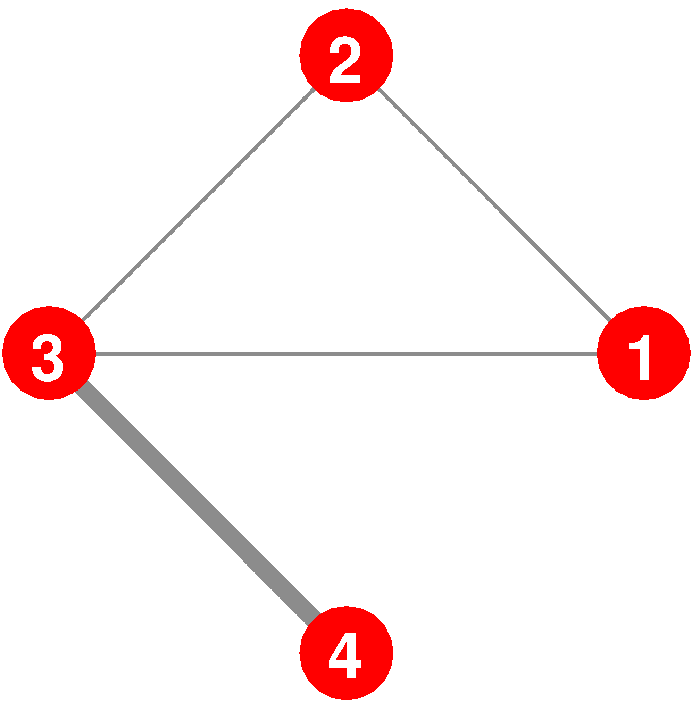}
\hspace{0.15in}
\includegraphics[width=0.15\textwidth]{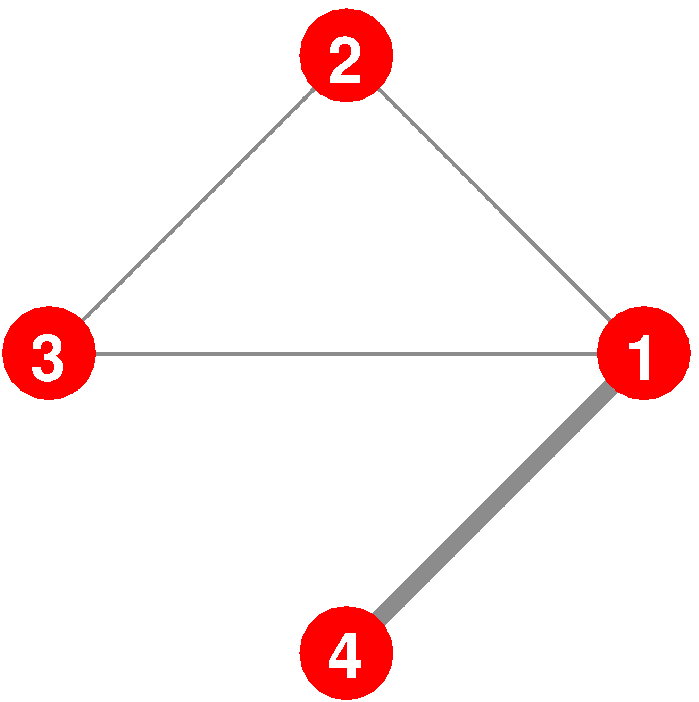} 
\hspace{0.2in}
\includegraphics[width=0.15\textwidth]{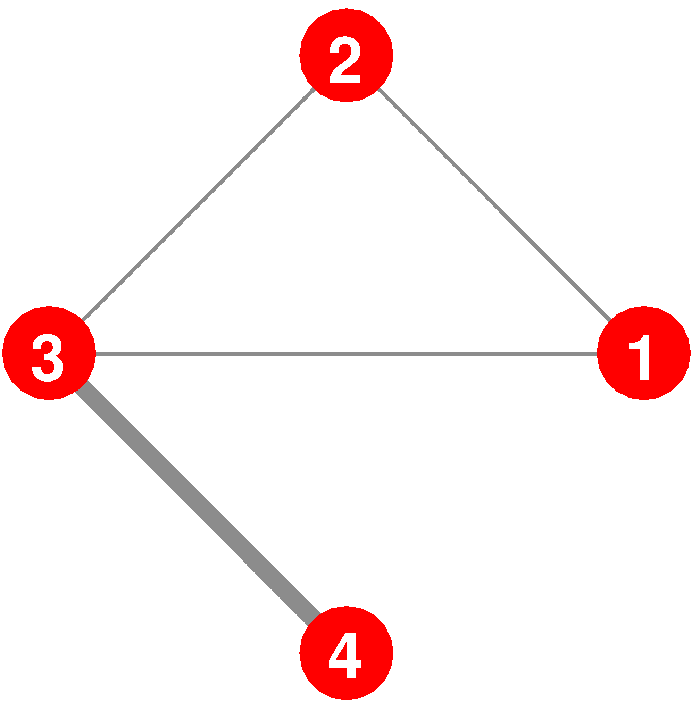}
\hspace{0.15in}
\includegraphics[width=0.15\textwidth]{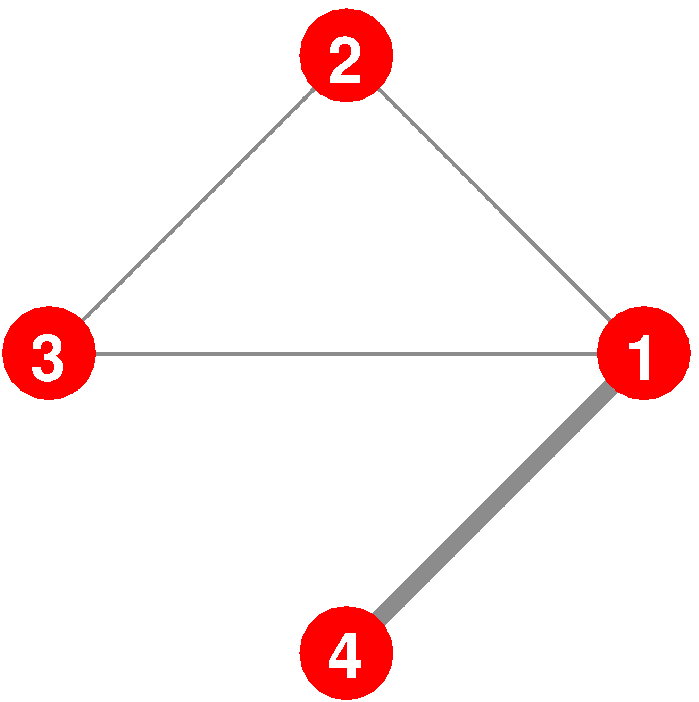}\\

\vspace{0.2in}
\includegraphics[width=0.15\textwidth]{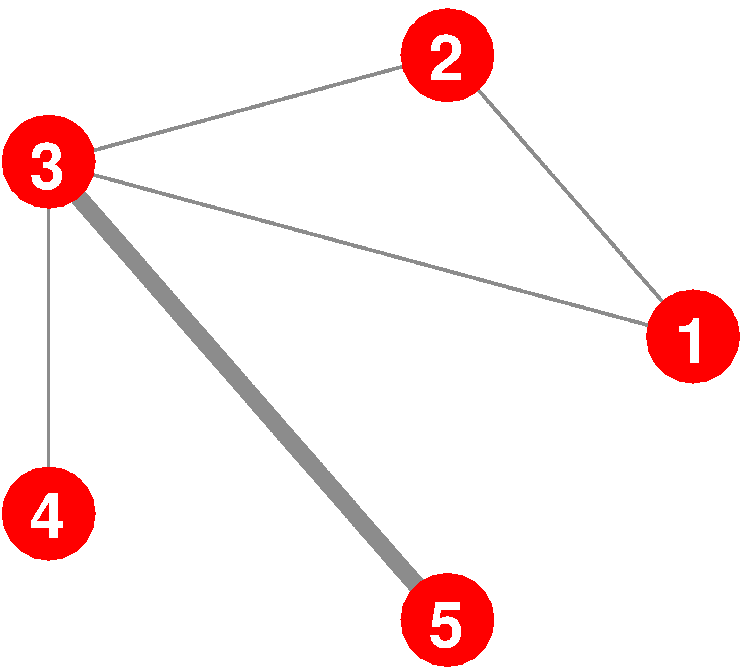}
\hspace{0.15in}
\includegraphics[width=0.15\textwidth]{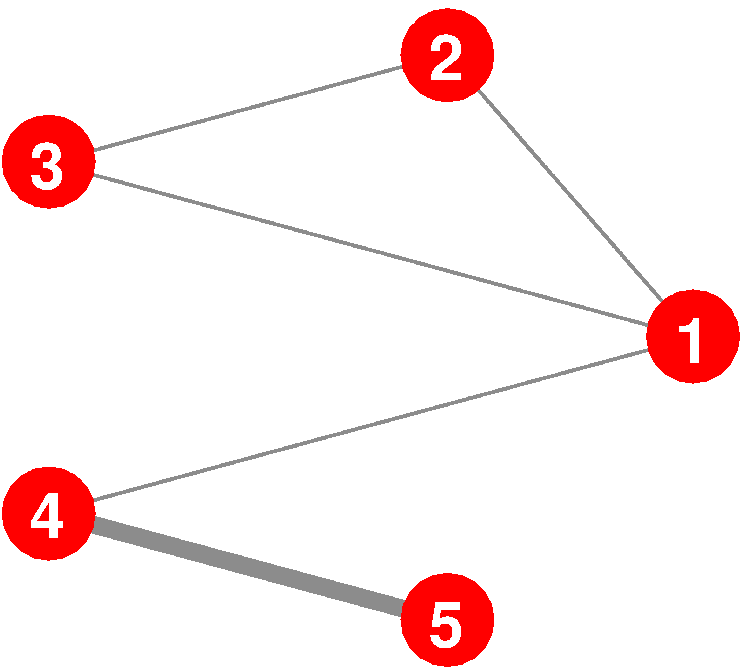}
\hspace{0.15in}
\includegraphics[width=0.15\textwidth]{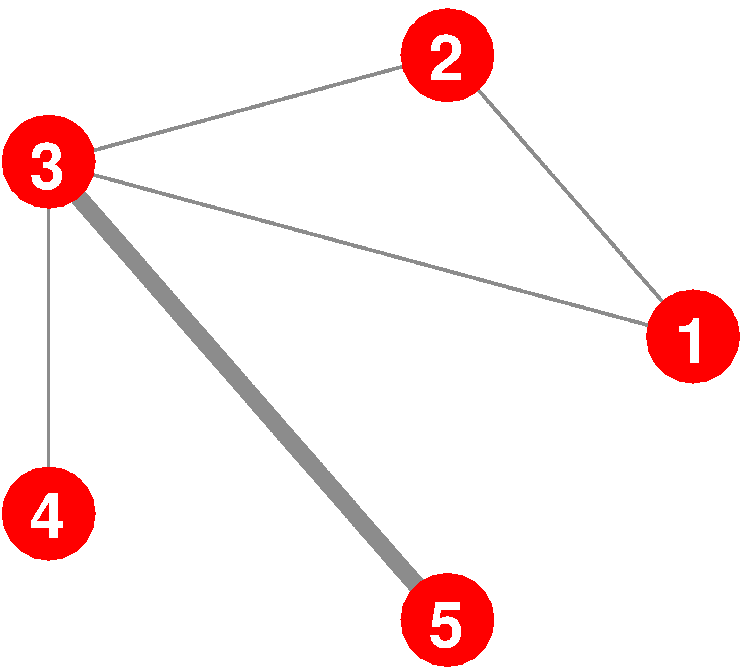}
\hspace{0.15in}
\includegraphics[width=0.15\textwidth]{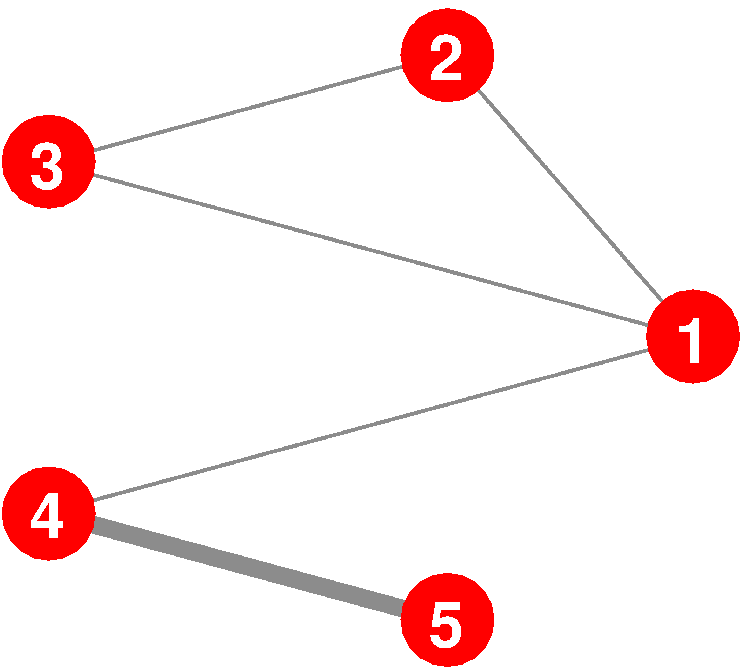}\\

\vspace{0.2in}
\includegraphics[width=0.15\textwidth]{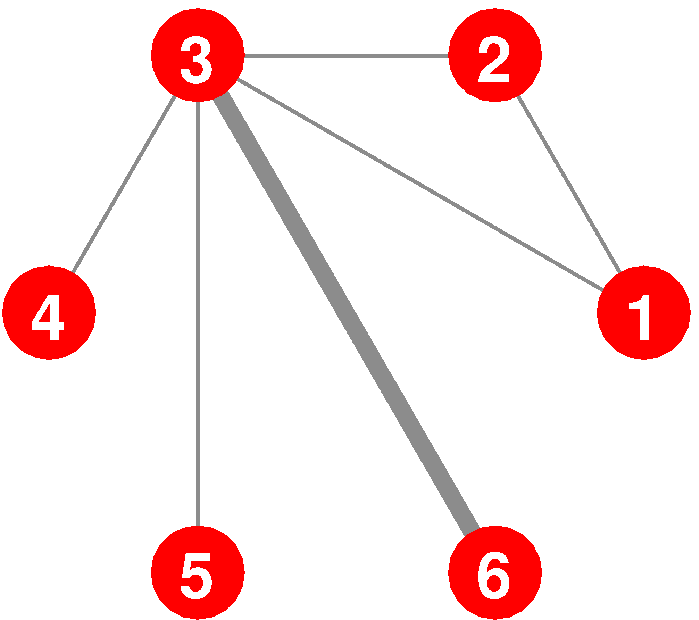}
\hspace{0.15in}
\includegraphics[width=0.15\textwidth]{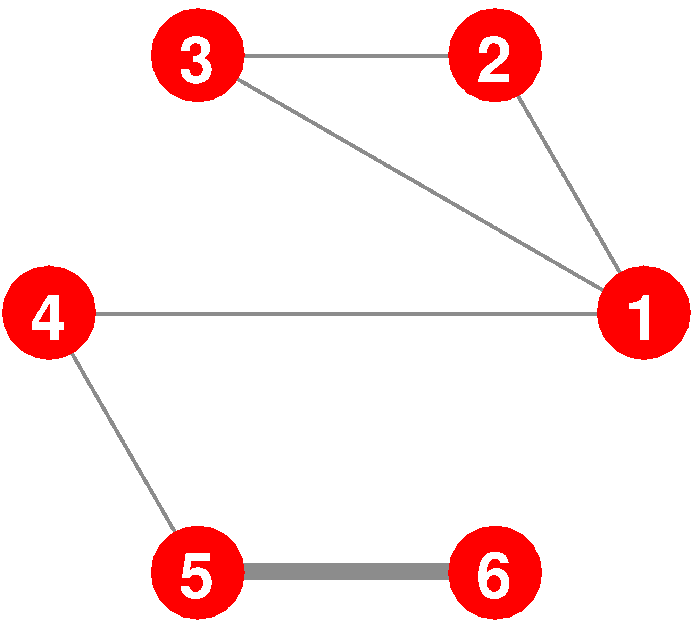}
\hspace{0.15in}
\includegraphics[width=0.15\textwidth]{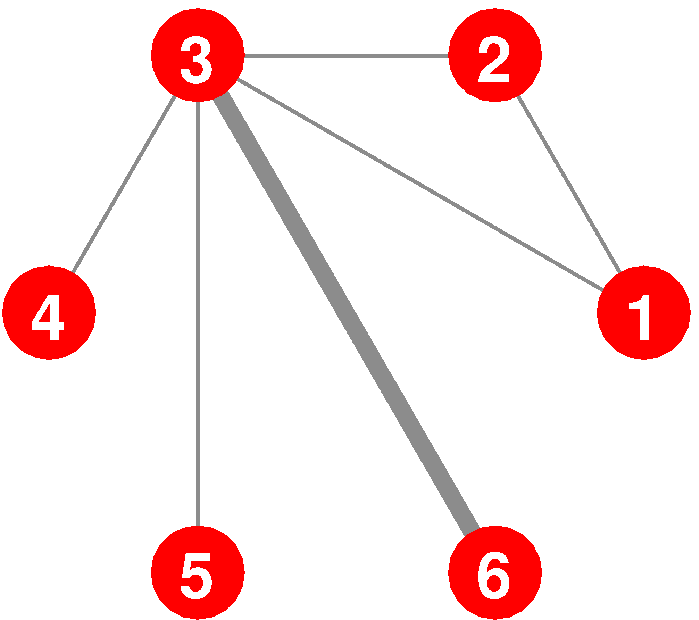}
\hspace{0.15in}
\includegraphics[width=0.15\textwidth]{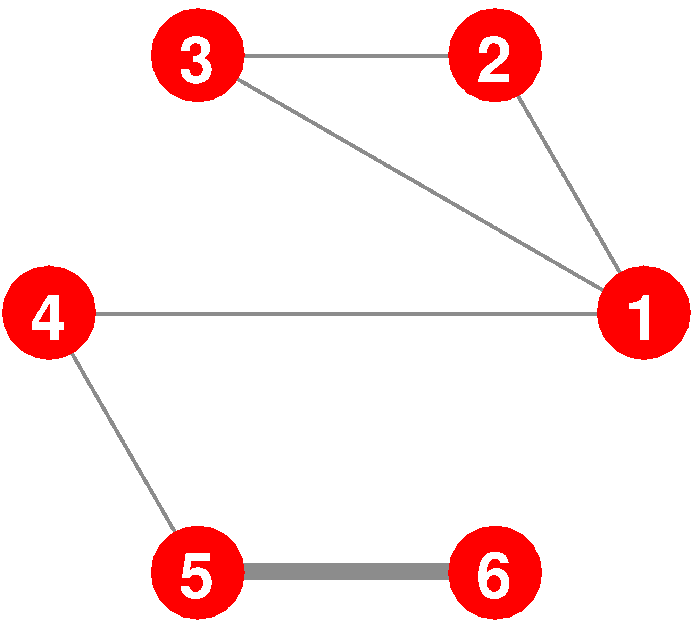}\\

\vspace{0.2in}
\subfloat[Min LE]{\includegraphics[width=0.15\textwidth]{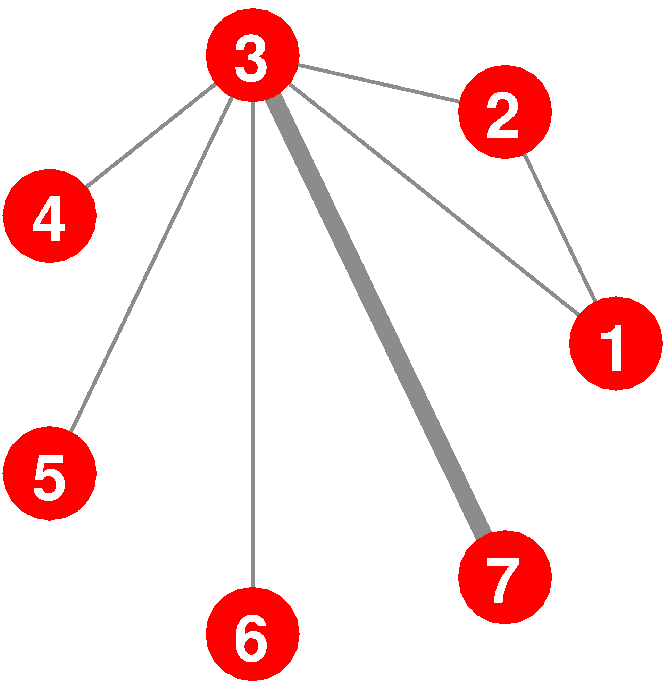}}
\hspace{0.15in}
\subfloat[Max LE]{\includegraphics[width=0.15\textwidth]{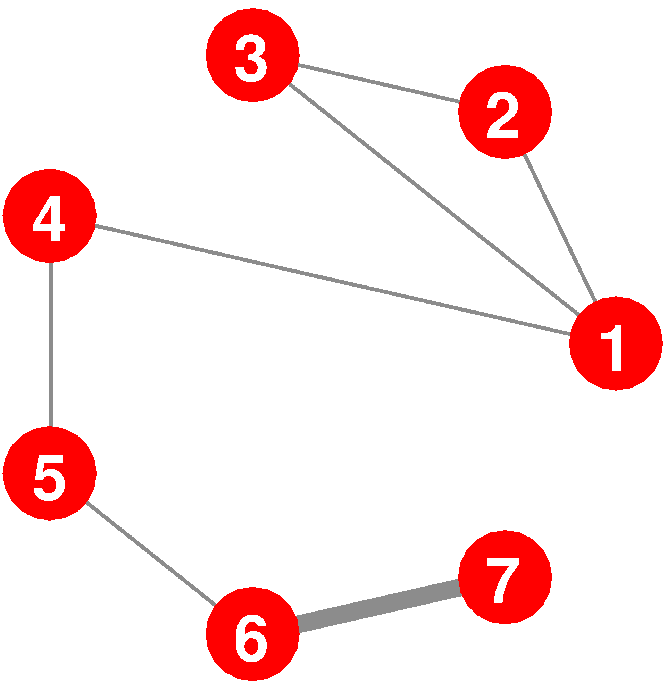}}
\hspace{0.15in}
\subfloat[Min ALE]{\includegraphics[width=0.15\textwidth]{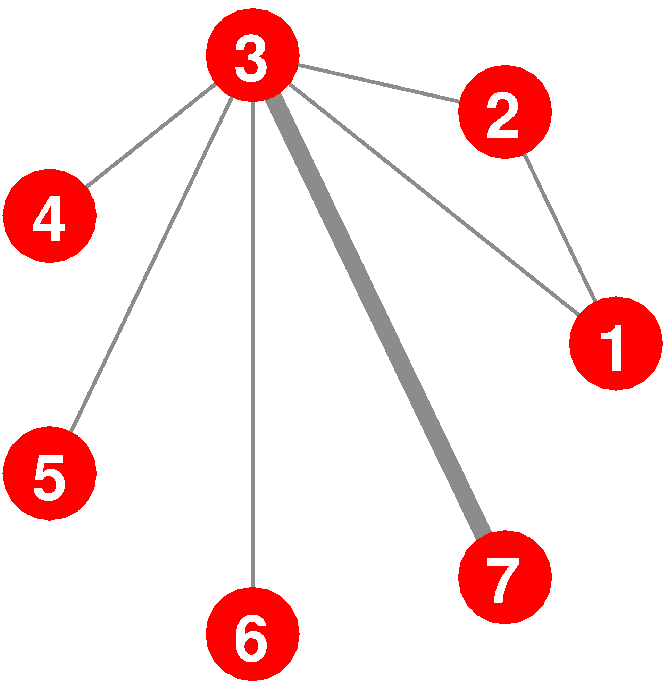}}
\hspace{0.15in}
\subfloat[Max ALE]{\includegraphics[width=0.15\textwidth]{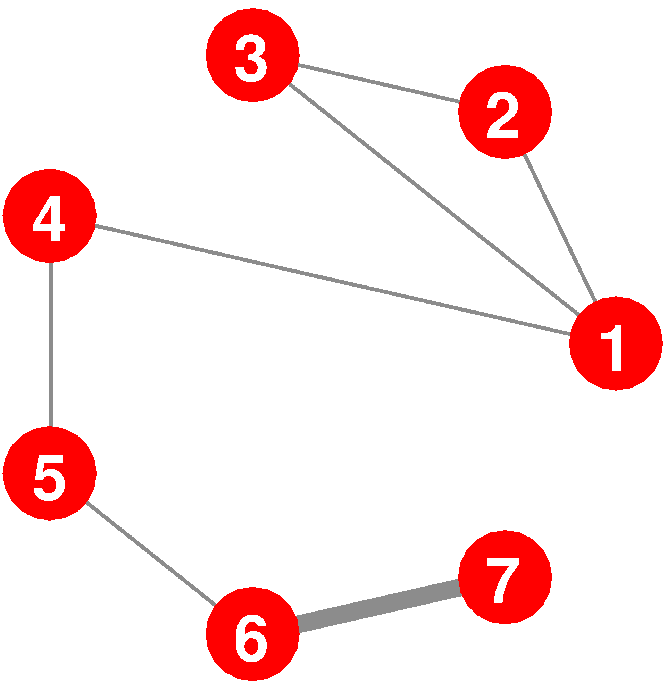}}
\caption{Evolution of the graph structure when we iteratively add a new node and we connect it to the graph so as to (a) minimize and (b) maximize the Laplacian entropy (LE). (c) and (d) show similar results for the approximate Laplacian entropy (ALE). Here the seed graph is a clique over 3 nodes.}
\label{fig:lucatoynode1}
\end{figure}

\subsubsection{Node Growth Model}
\begin{figure}[t!]
\centering
\includegraphics[width=0.15\textwidth]{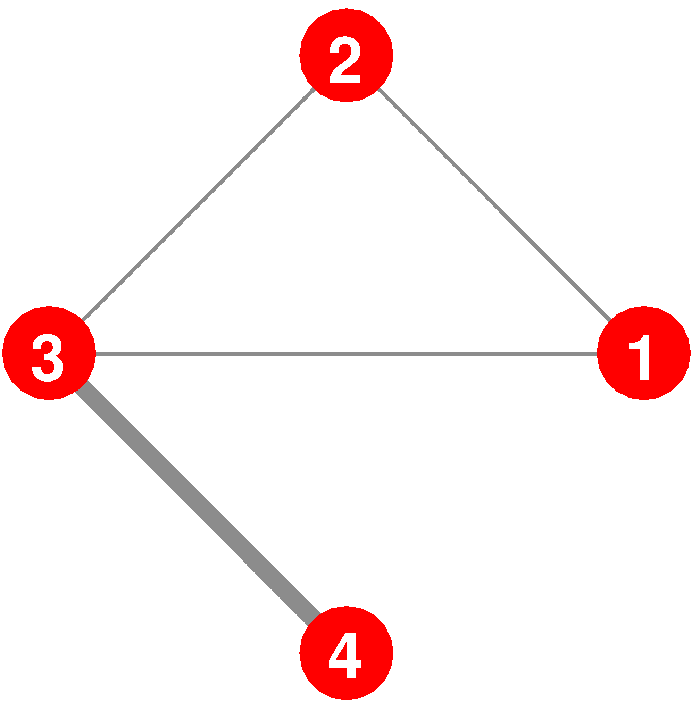}
\hspace{0.15in}
\includegraphics[width=0.15\textwidth]{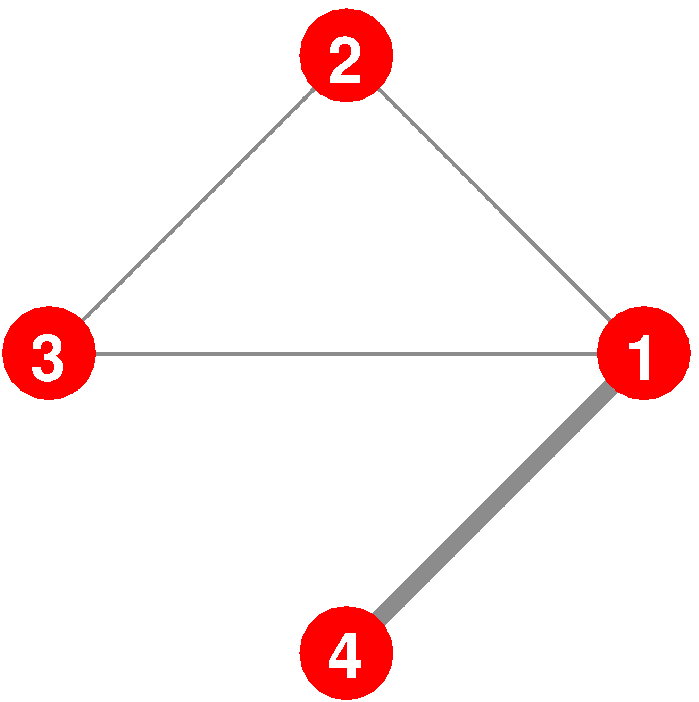}
\hspace{0.15in}
\includegraphics[width=0.15\textwidth]{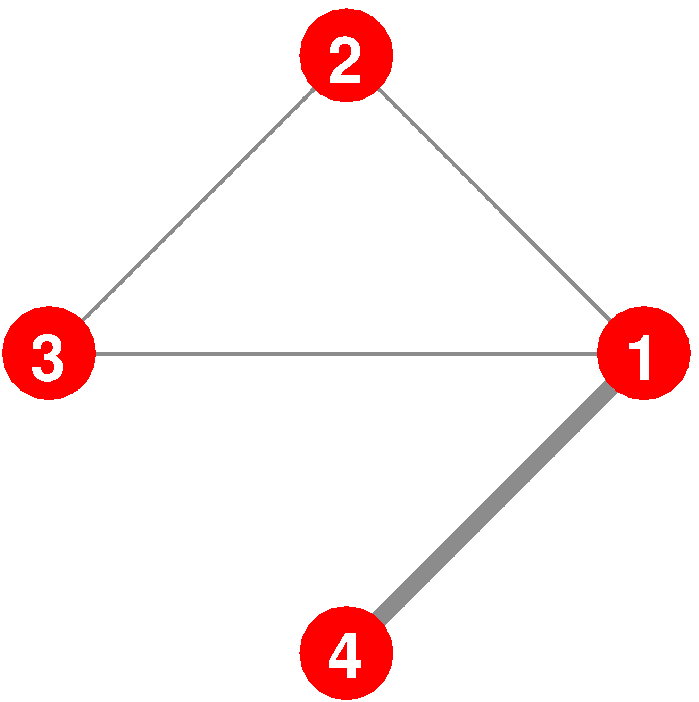}
\hspace{0.15in}
\includegraphics[width=0.15\textwidth]{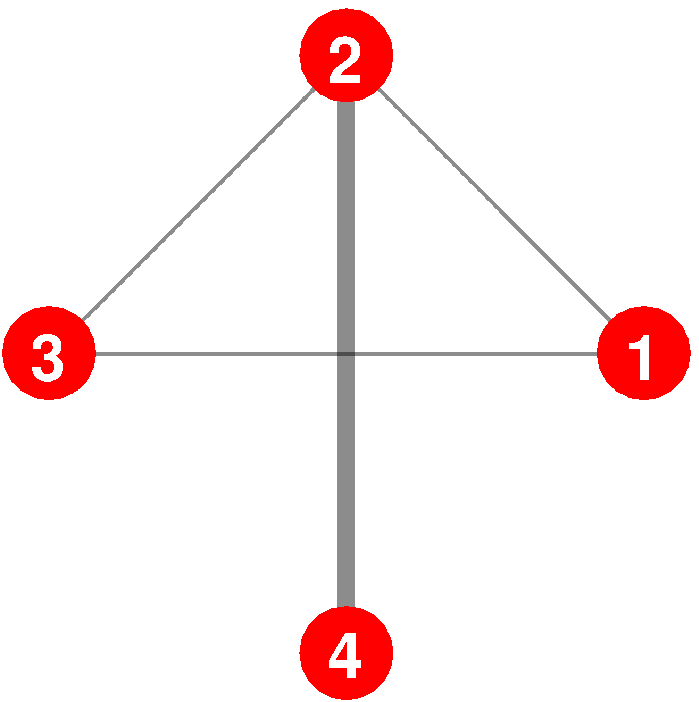}\\

\vspace{0.2in}
\includegraphics[width=0.15\textwidth]{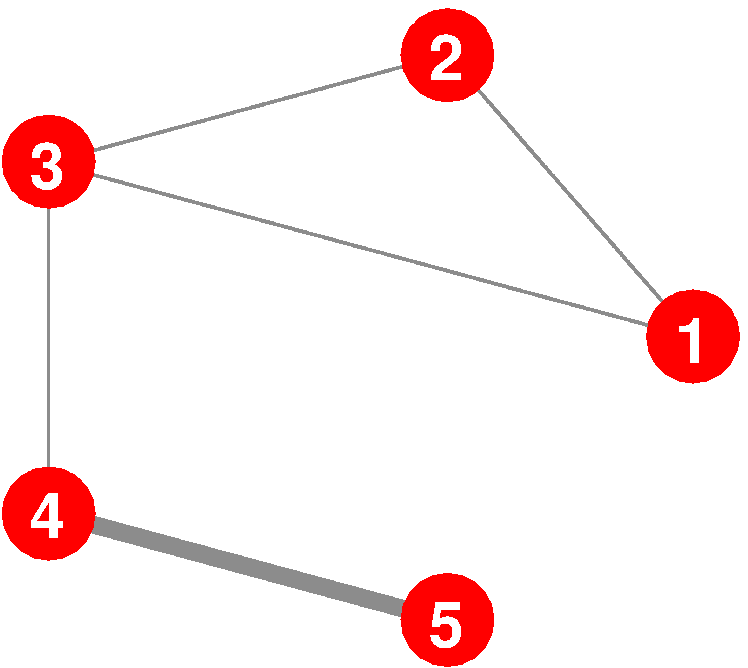}
\hspace{0.15in}
\includegraphics[width=0.15\textwidth]{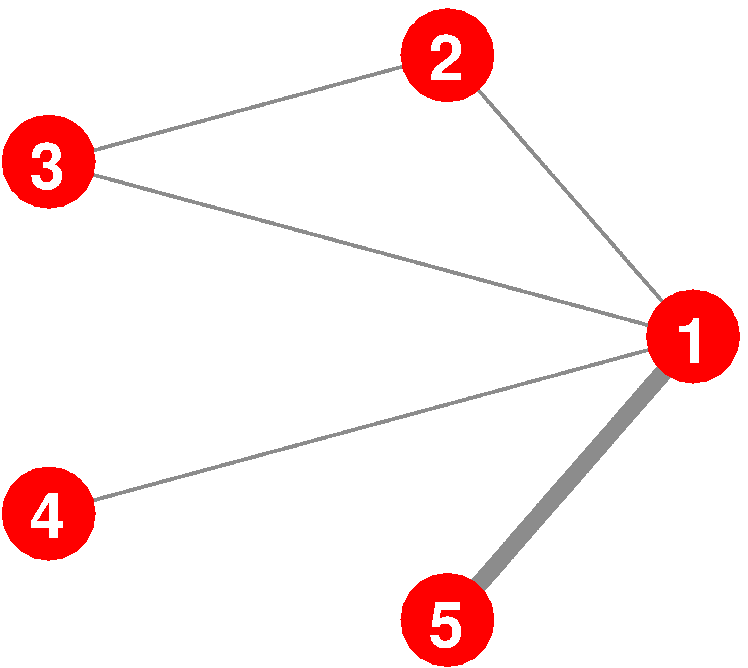}
\hspace{0.15in}
\includegraphics[width=0.15\textwidth]{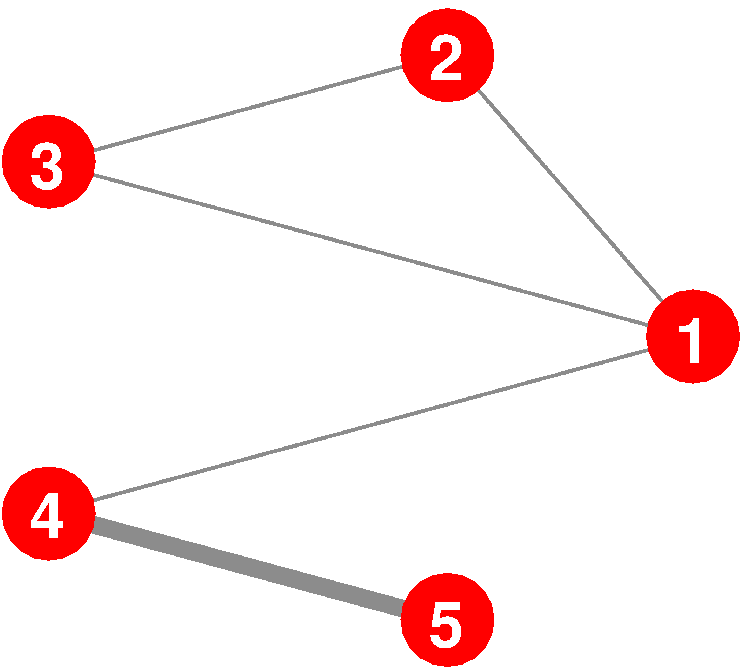}
\hspace{0.15in}
\includegraphics[width=0.15\textwidth]{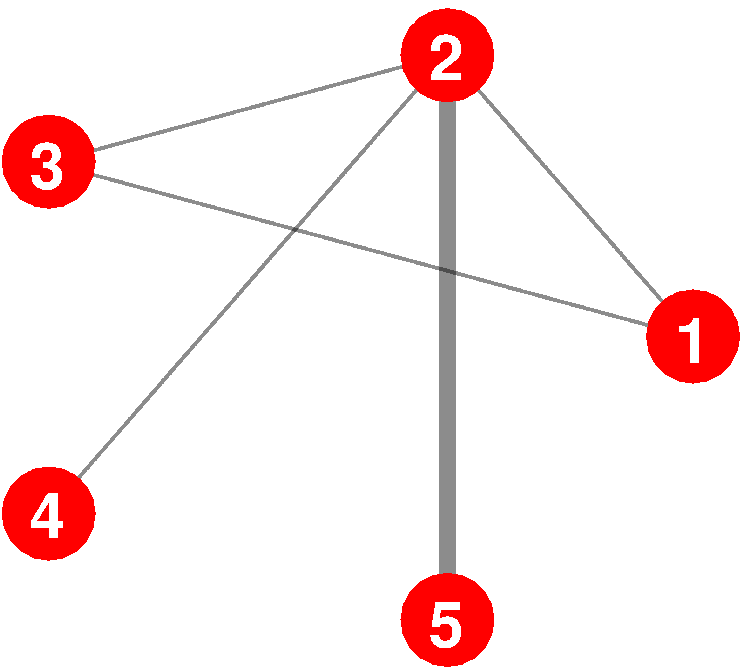}\\

\vspace{0.2in}
\includegraphics[width=0.15\textwidth]{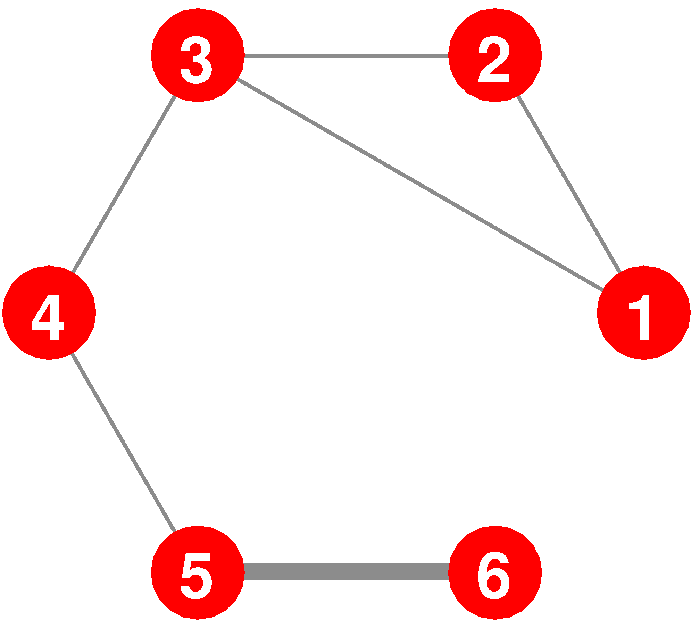}
\hspace{0.15in}
\includegraphics[width=0.15\textwidth]{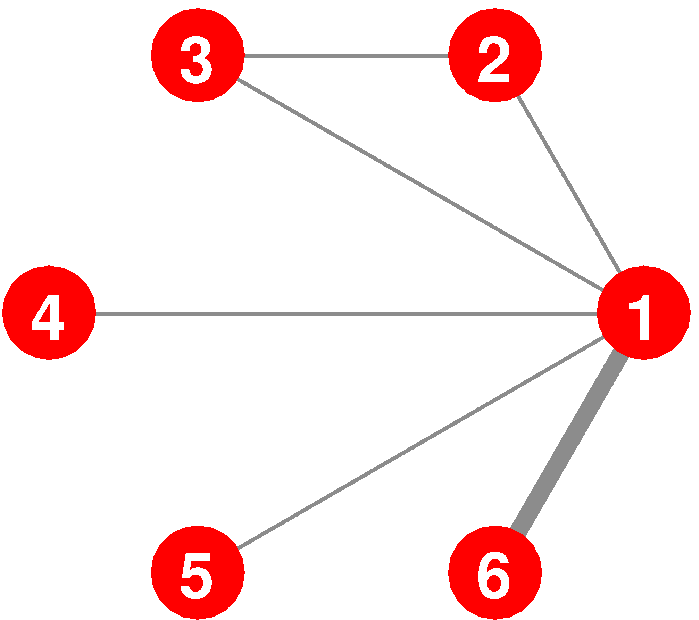}
\hspace{0.15in}
\includegraphics[width=0.15\textwidth]{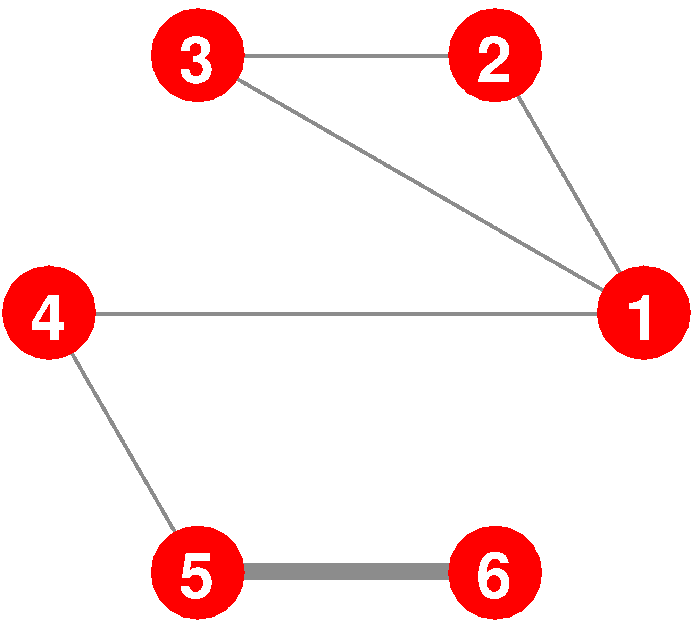}
\hspace{0.15in}
\includegraphics[width=0.15\textwidth]{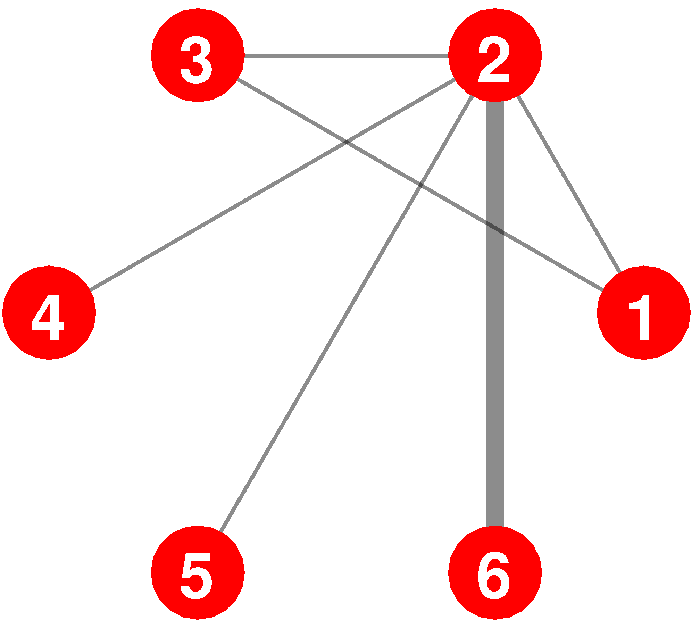}\\

\vspace{0.2in}
\subfloat[Min NLE]{\includegraphics[width=0.15\textwidth]{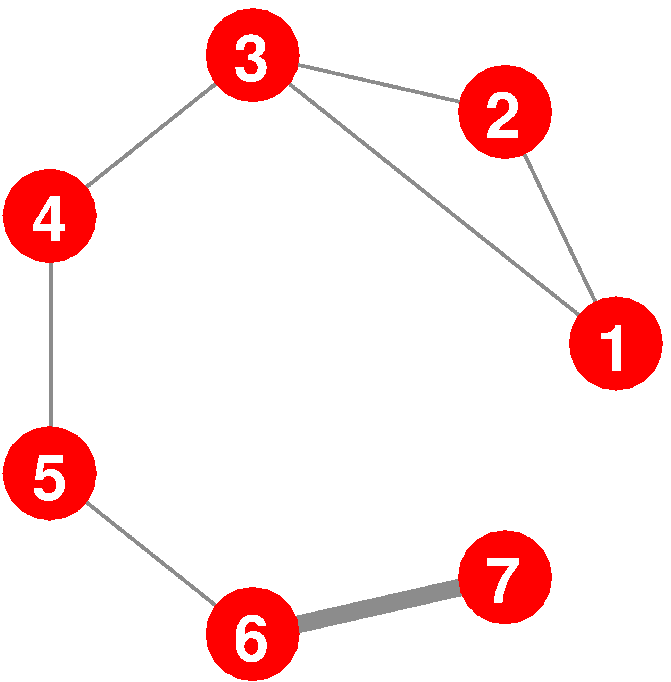}}
\hspace{0.15in}
\subfloat[Max NLE]{\includegraphics[width=0.15\textwidth]{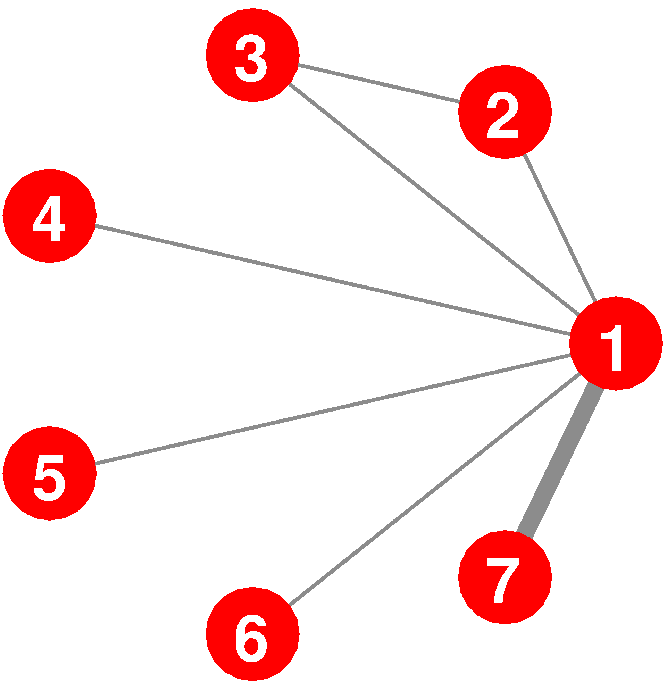}}
\hspace{0.15in}
\subfloat[Min ANLE]{\includegraphics[width=0.15\textwidth]{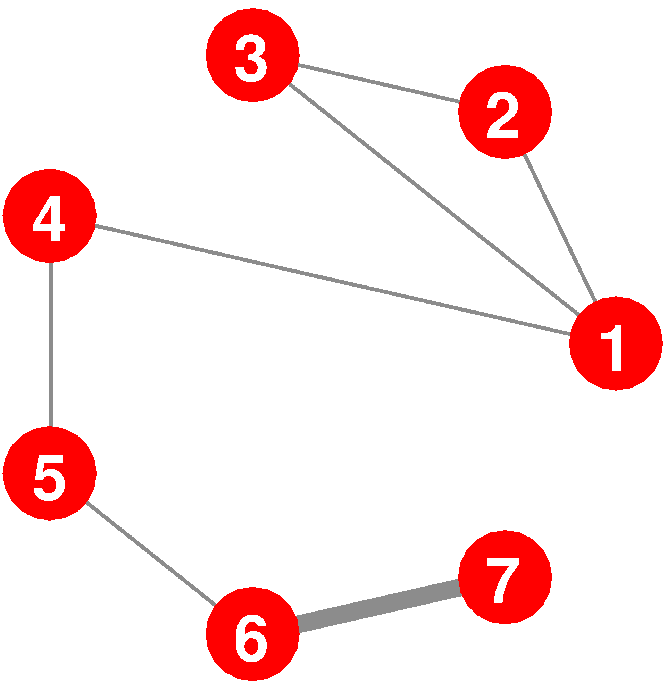}}
\hspace{0.15in}
\subfloat[Max ANLE]{\includegraphics[width=0.15\textwidth]{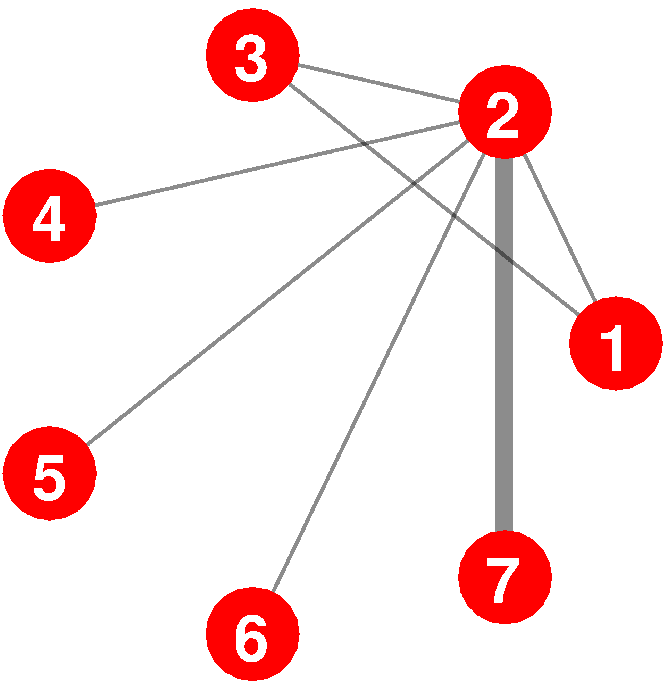}}
\caption{Evolution of the graph structure when we iteratively add a new node and we connect it to the graph so as to (a) minimize and (b) maximize the normalized Laplacian entropy (NLE). (c) and (d) show similar results for the approximate normalized Laplacian entropy (ANLE). Here the seed graph is a clique over 3 nodes.}\label{fig:giorgiatoynode1}
\end{figure}

We also consider the case of graph where the both the number of nodes is not fixed over time. Instead, at each time step we add a new node and we connect it to the $m$ nodes that lead to a maximal increment of the entropy. Each column in Fig.~\ref{fig:lucatoynode1} corresponds to a different choice of the process (maximization or minimization) and entropy (exact or approximated). Similarly, we show the results of the same experiment for the normalized Laplacian entropy in Fig.~\ref{fig:giorgiatoynode1}. In both cases we start from a clique over three nodes. We only show the results for $m=1$, as we observe the same behaviour for larger values of $m$. In contrast to the edge growth model, here maximizing (minimizing) the exact and approximated entropies yields the same structural evolution. Interestingly, while minimizing the (approximated) Laplacian entropy yields the formation of hubs, minimizing the (approximated) normalized Laplacian entropy leads to the formation of a long tail of low degree nodes. We observe opposite behaviour when maximizing the (approximated) Laplacian and normalized Laplacian entropy, respectively.

\subsection{Experiments on Random Graphs}
While the previous experiments gave us some first interesting insights in the nature of the structural pattern captured by the (approximated) Laplacian and normalized Laplacian entropies, in this section we aim to perform a more thorough analysis of the two entropies and their approximated versions on a large set of synthetically generated graphs.

\begin{figure}[t!]
\centering
\includegraphics[width=0.8\textwidth]{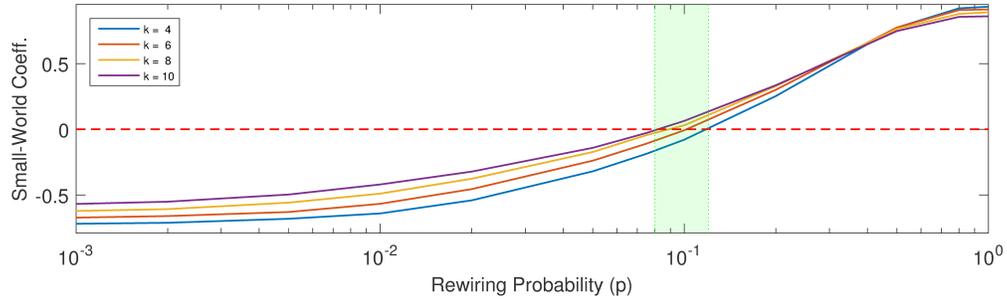}
\caption{We choose the optimal value of $p$ in order to generate graphs displaying the small-world property~\cite{telesford2011ubiquity}.}
\label{fig:SMALLness}
\end{figure}

\subsubsection{Datasets}\label{sec:graph_model}
We perform our experiments on synthetic networks generated by three well-known random graph models: 1) the Erd\"os-R\'enyi model, 2) the Watts-Strogatz model and 3) the Preferential Attachment model~\cite{barabasi1999emergence}. For each model we vary its parameters as explained below, except from the number of nodes, which is fixed to $n=100$ in all three cases. This is to control for the well-known dependency between the value of the von Neumann entropy of a graph and its vertex set size, which would otherwise skew the results of our correlation study. This is particularly evident when our results are compared to those of Han et al.~\cite{han2012graph}, where no such control was introduced.
\\
\\
\noindent\textbf{Erd\"os-R\'enyi  model}: the graphs in this dataset are generated by varying the parameter $p$, namely the probability of connecting two nodes, between $0.1$ and $0.9$. Unless otherwise stated, for each choice of $p$ we generate 100 instances, for a total of 900 graphs.\\
\\
\\
\noindent\textbf{Preferential Attachment model}: the parameter of this model is $m$, i.e. the number of edges to add from a new node to the existing nodes, at each temporal iteration. We let $m$ vary from 2 to 10, and, unless otherwise stated, we generate 100 instances for each choice of $m$, for a total of 900 graphs.
\\
\\
\noindent\textbf{Watts-Strogatz model}: here the model parameters are $k$ and $p$. Starting from a ring graph where each node is connected to its $k$ nearest neighbours, we rewire each edge with probability $p$. When $p=0$, the graph is regular. As $p$ increases the graph structure becomes more random. We follow the quantitative metric presented in~\cite{telesford2011ubiquity} to measure the small-worldness of a graph and we select the value of $p=0.1$, as shown in Fig.~\ref{fig:SMALLness}. More precisely, in~\cite{telesford2011ubiquity} the authors propose a way to measure the small-worldness of a network based on the original model described by Watts and Strogatz, comparing the network clustering coefficient to an equivalent lattice network and the path length to a random network. This in turn ensure that the generated graphs display the small-worldness property~\cite{telesford2011ubiquity}, i.e., they simultaneously have high clustering coefficient and low path length. As for the parameter $k$, we let it vary from 2 to 10. Unless otherwise stated, we generate 100 instances for each choice of $k$, for a total 900 graphs.

\begin{figure}[t!]
{\includegraphics[width=0.22\textwidth]{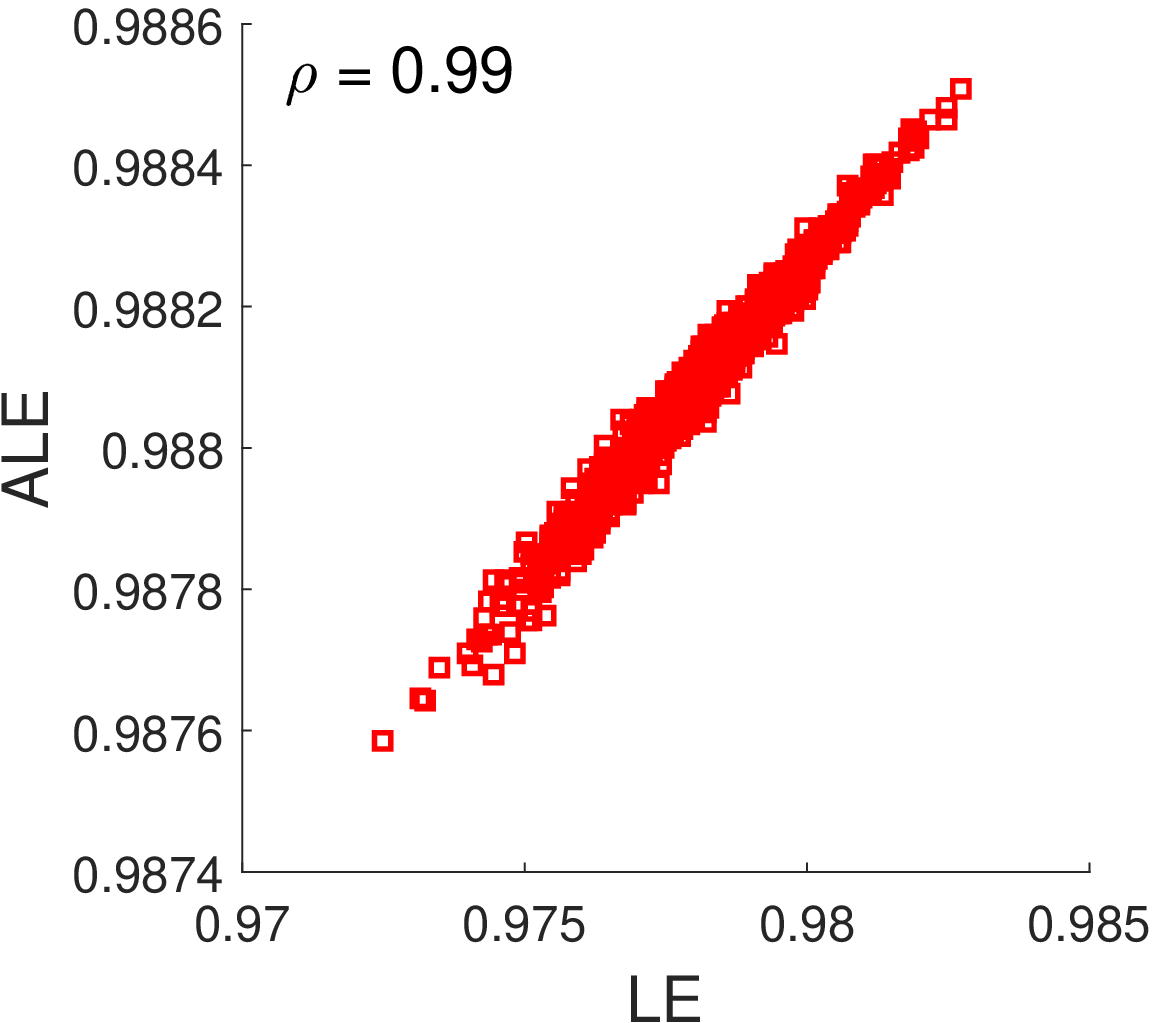}}
{\includegraphics[width=0.22\textwidth]{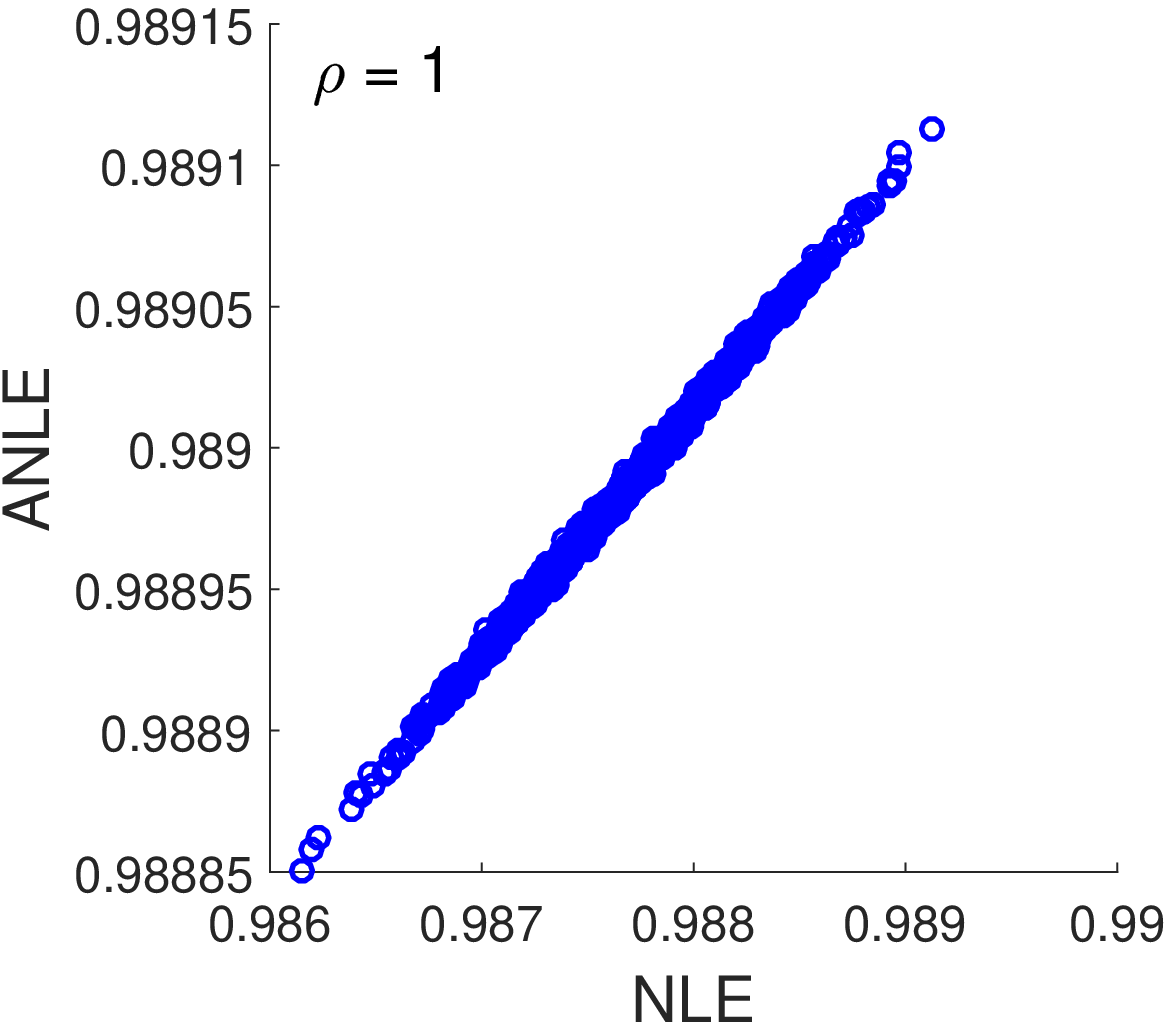}}
{\includegraphics[width=0.22\textwidth]{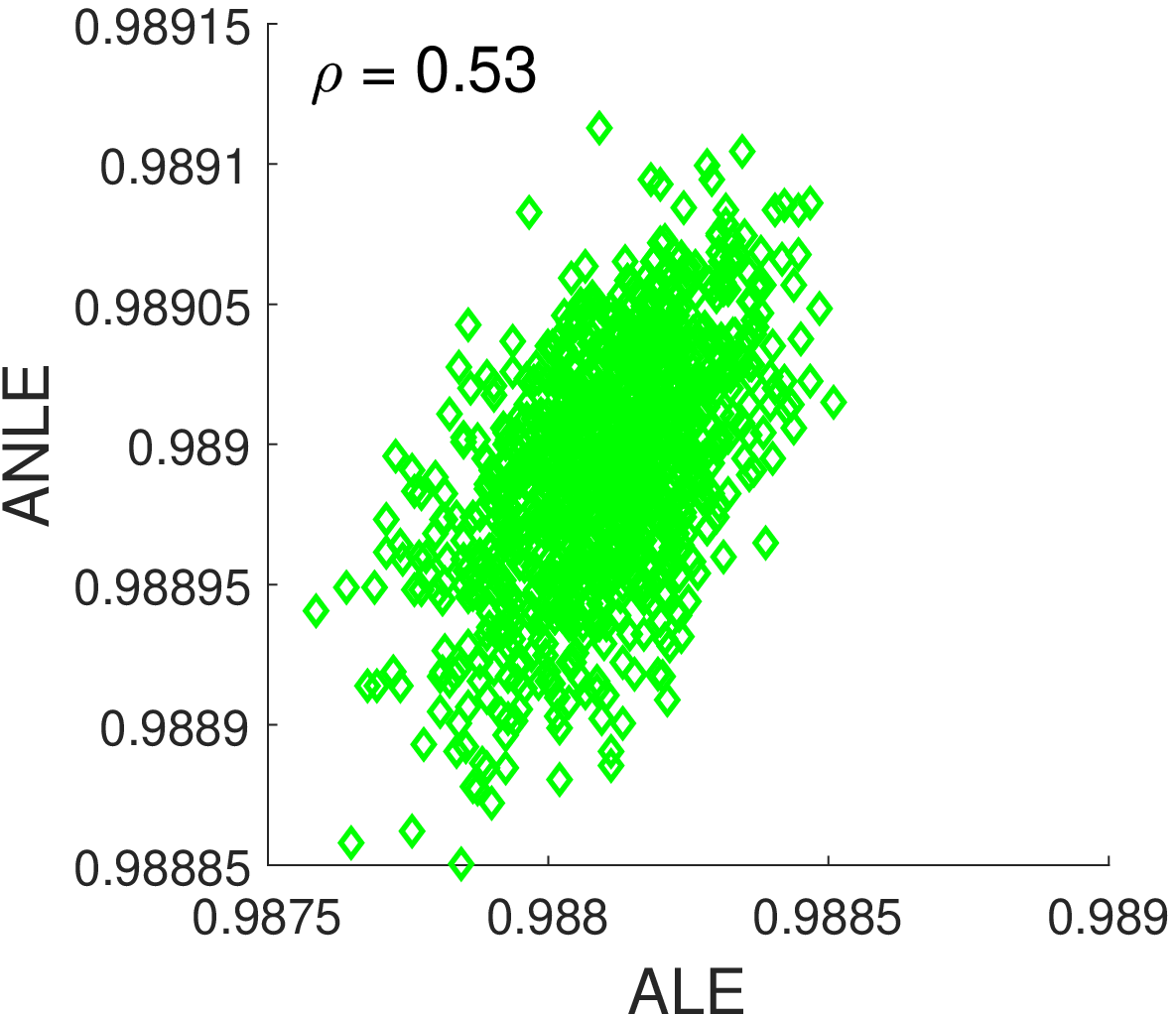}}
{\includegraphics[width=0.22\textwidth]{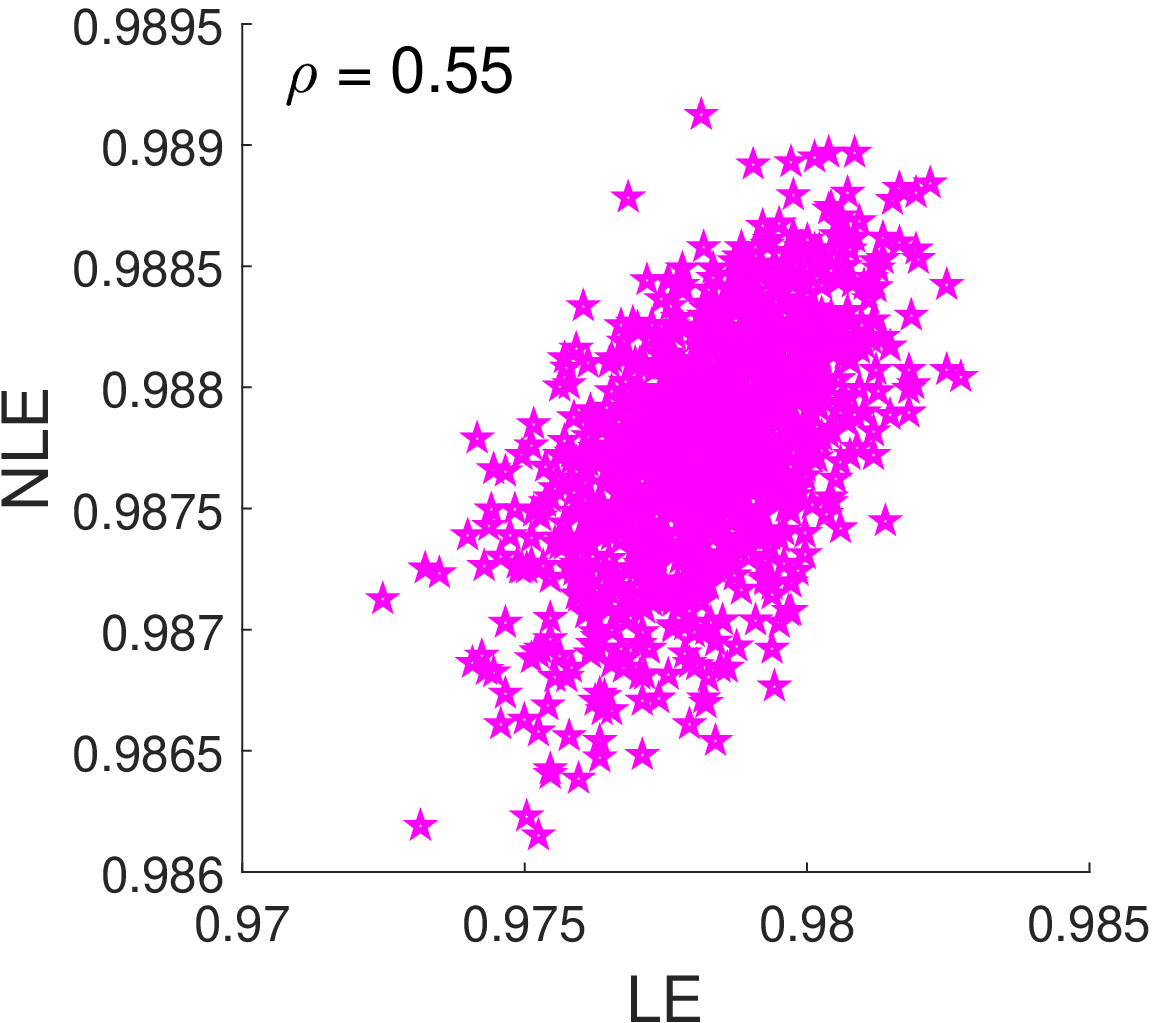}}\\

{\includegraphics[width=0.22\textwidth]{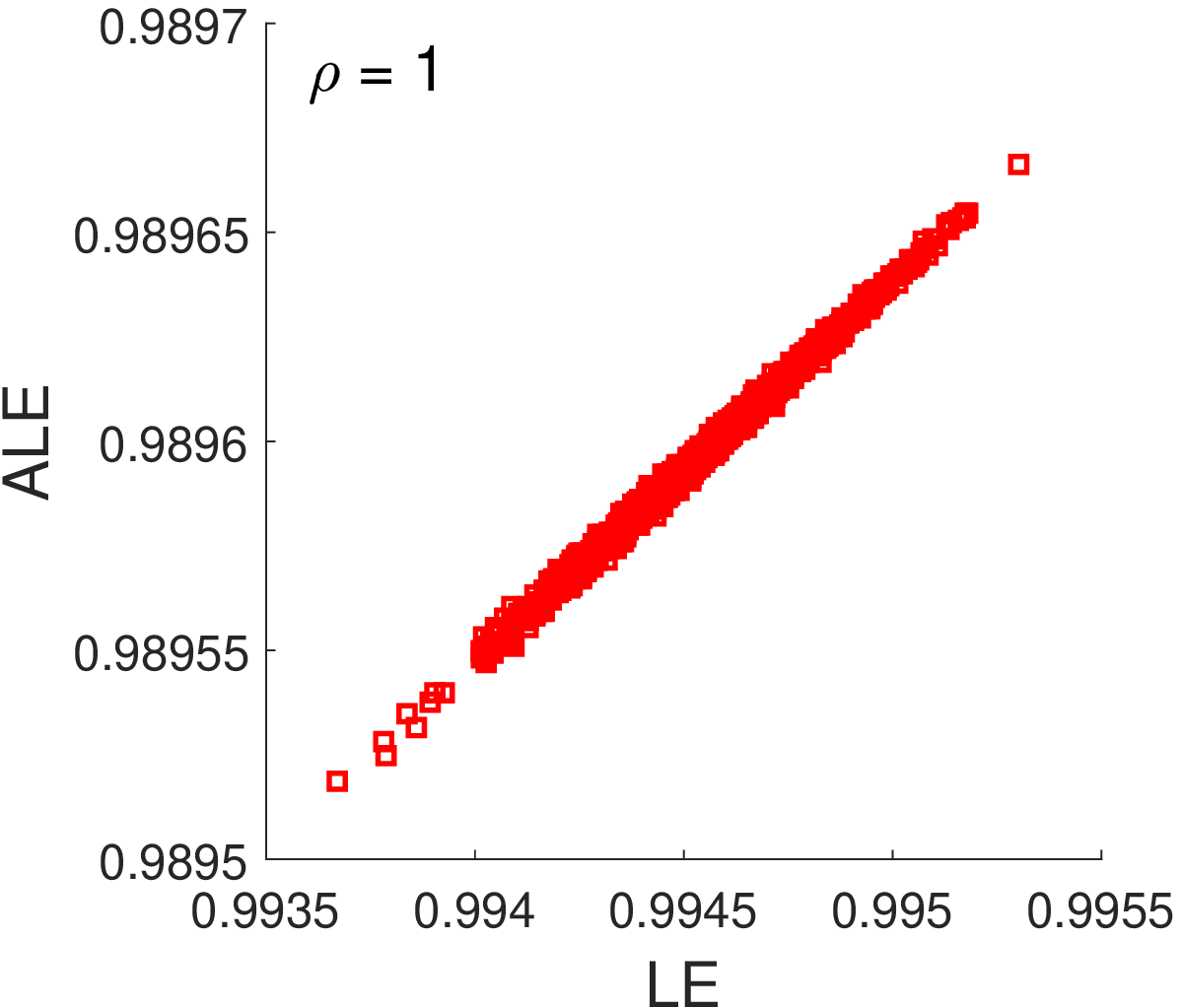}}
{\includegraphics[width=0.22\textwidth]{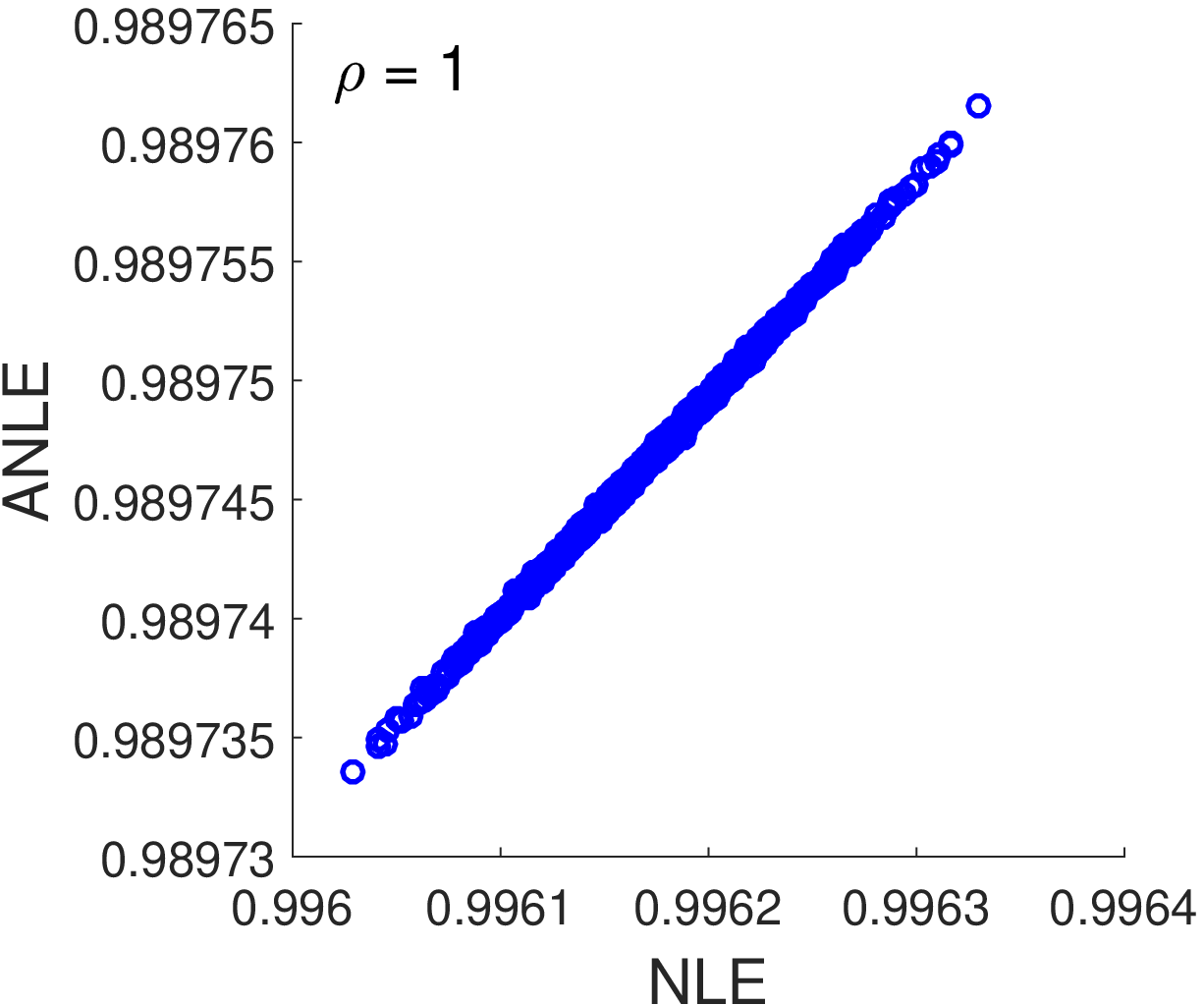}}
{\includegraphics[width=0.22\textwidth]{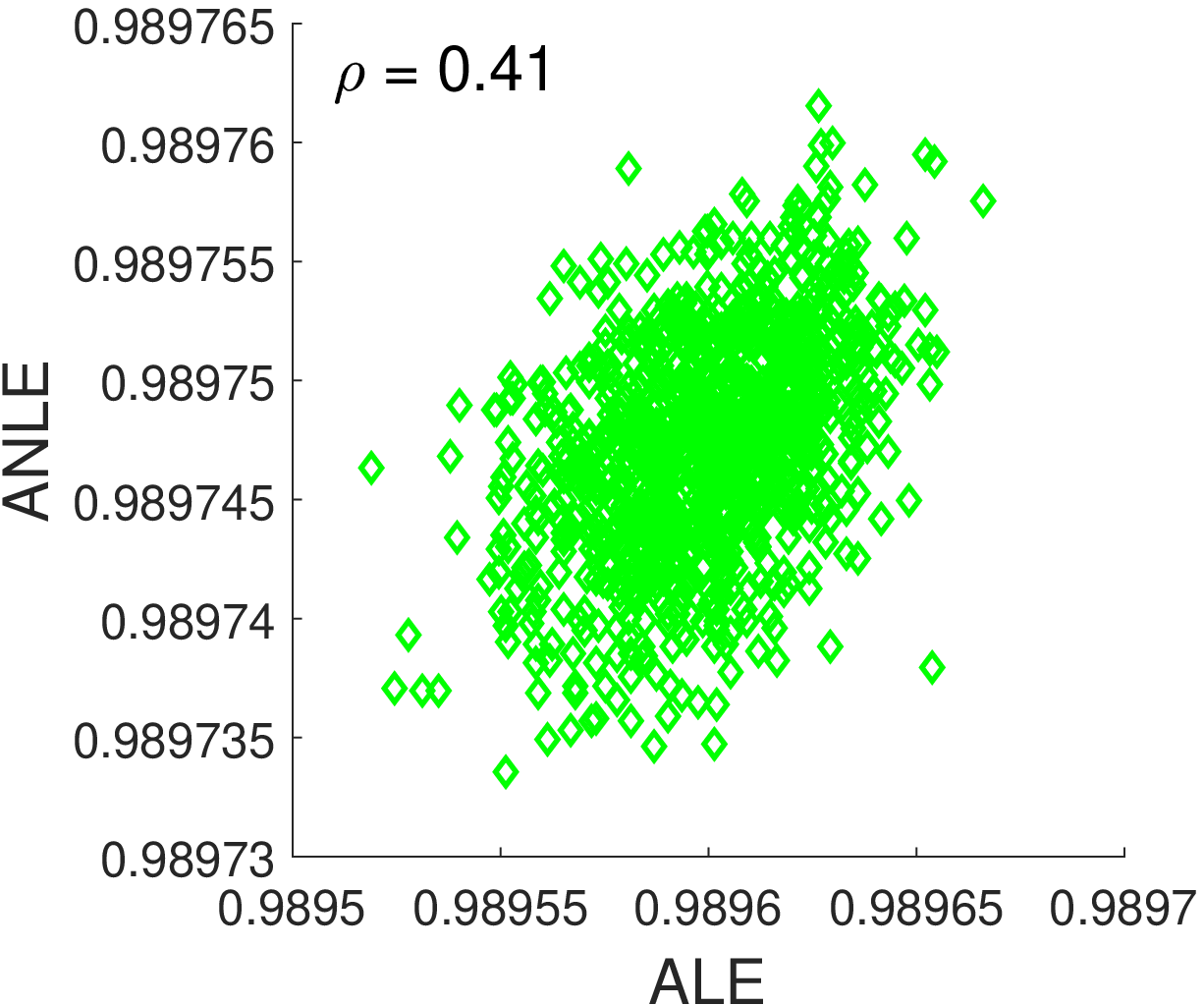}}
{\includegraphics[width=0.22\textwidth]{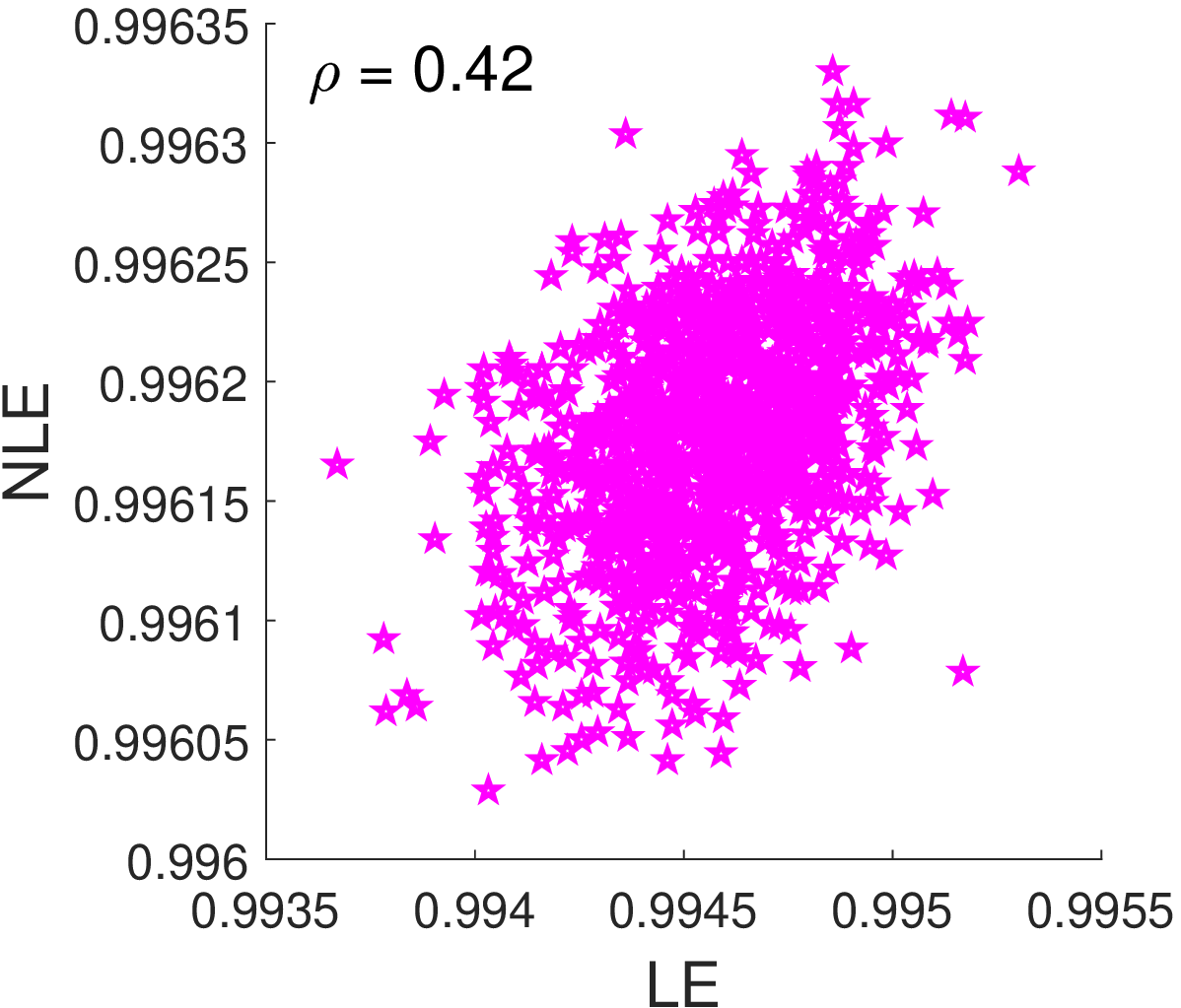}}\\

{\includegraphics[width=0.22\textwidth]{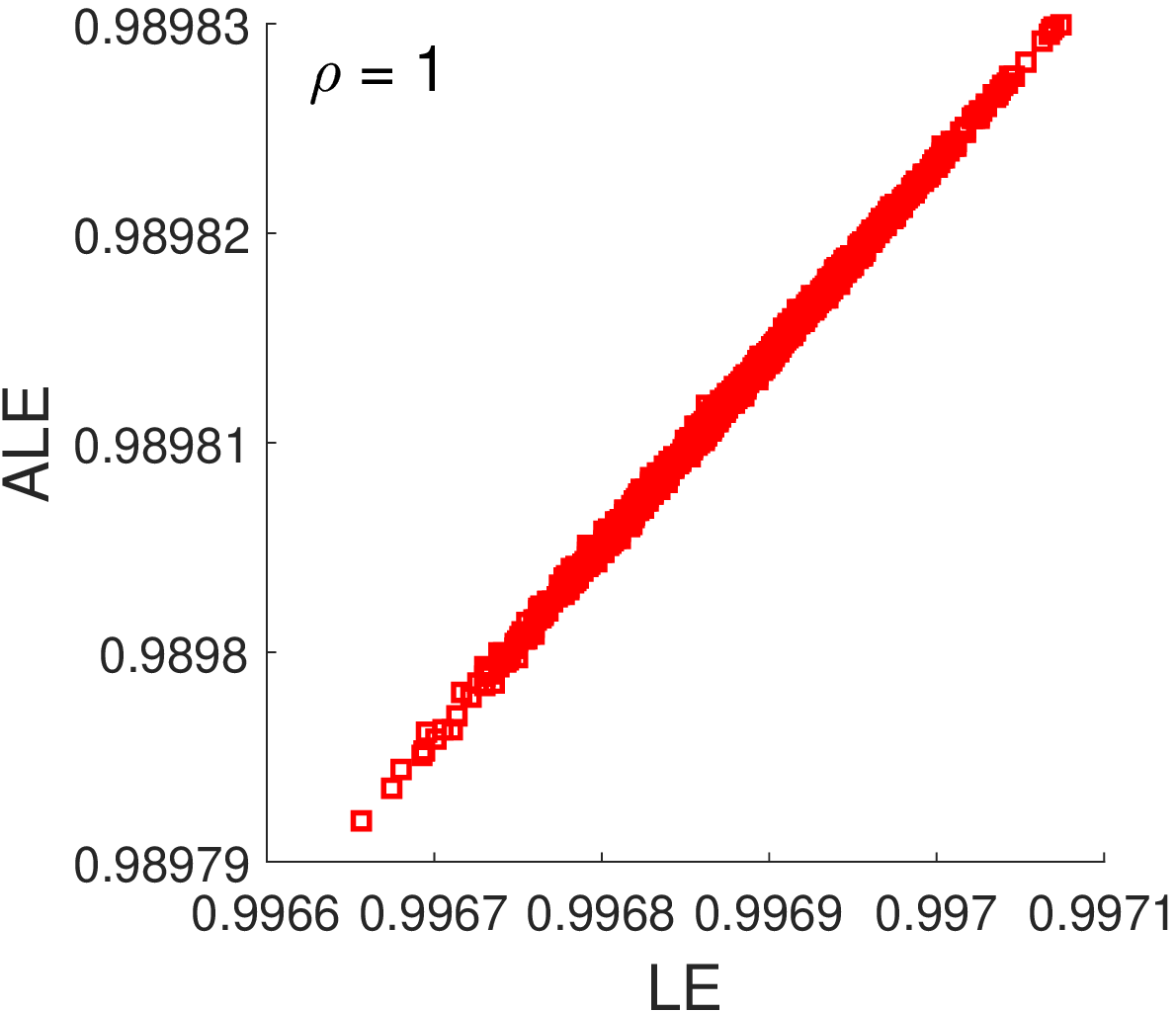}}
{\includegraphics[width=0.22\textwidth]{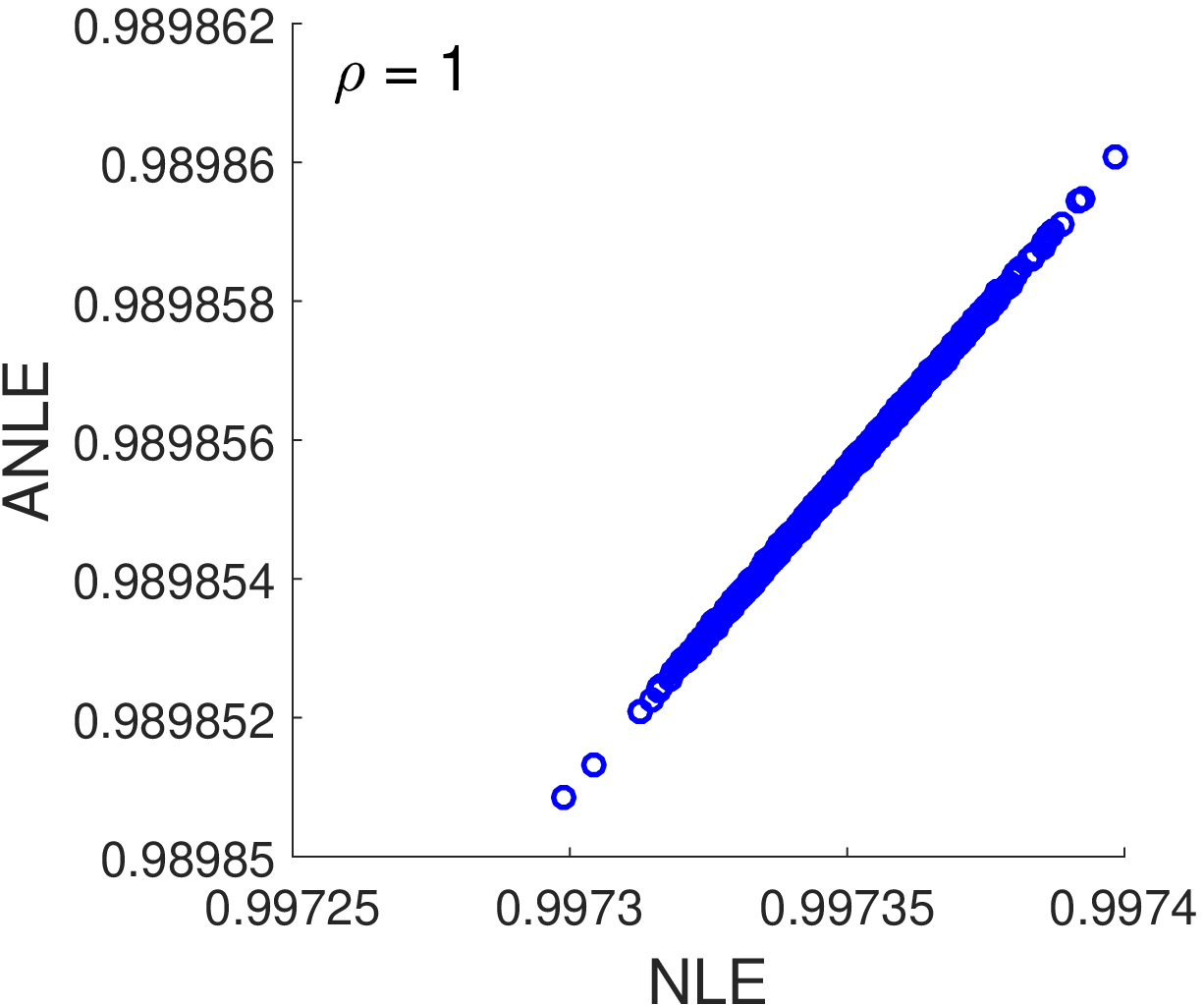}}
{\includegraphics[width=0.22\textwidth]{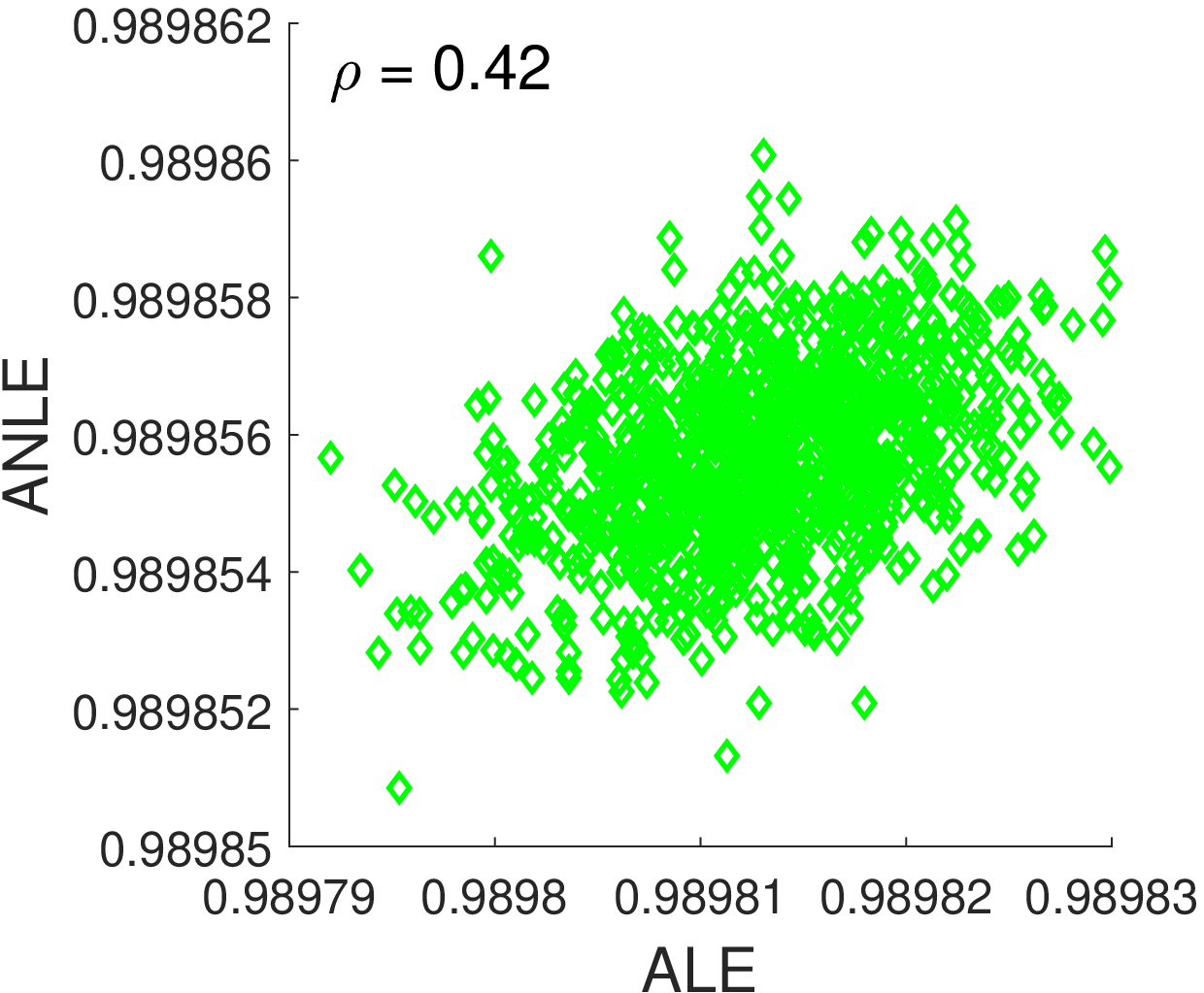}}
{\includegraphics[width=0.22\textwidth]{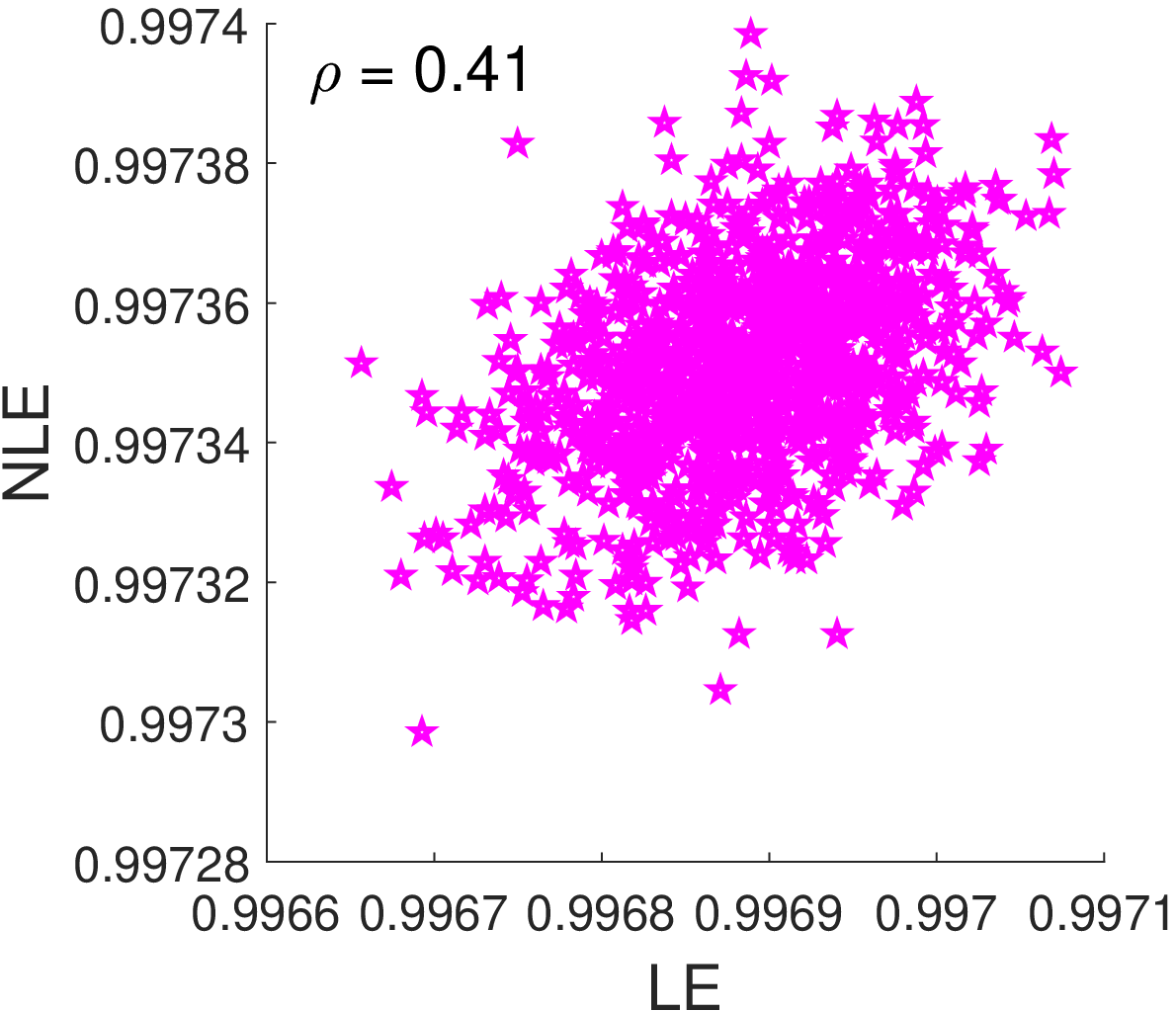}}\\
\caption{Entropies correlations on the Erd\"os-R\'enyi graphs for $p=0.1$ (top), $p=0.4$ (middle), and $p=0.7$ (bottom).}
\label{fig:corr_erdos}
\end{figure}

\begin{figure}[!t]
{\includegraphics[width=0.22\textwidth]{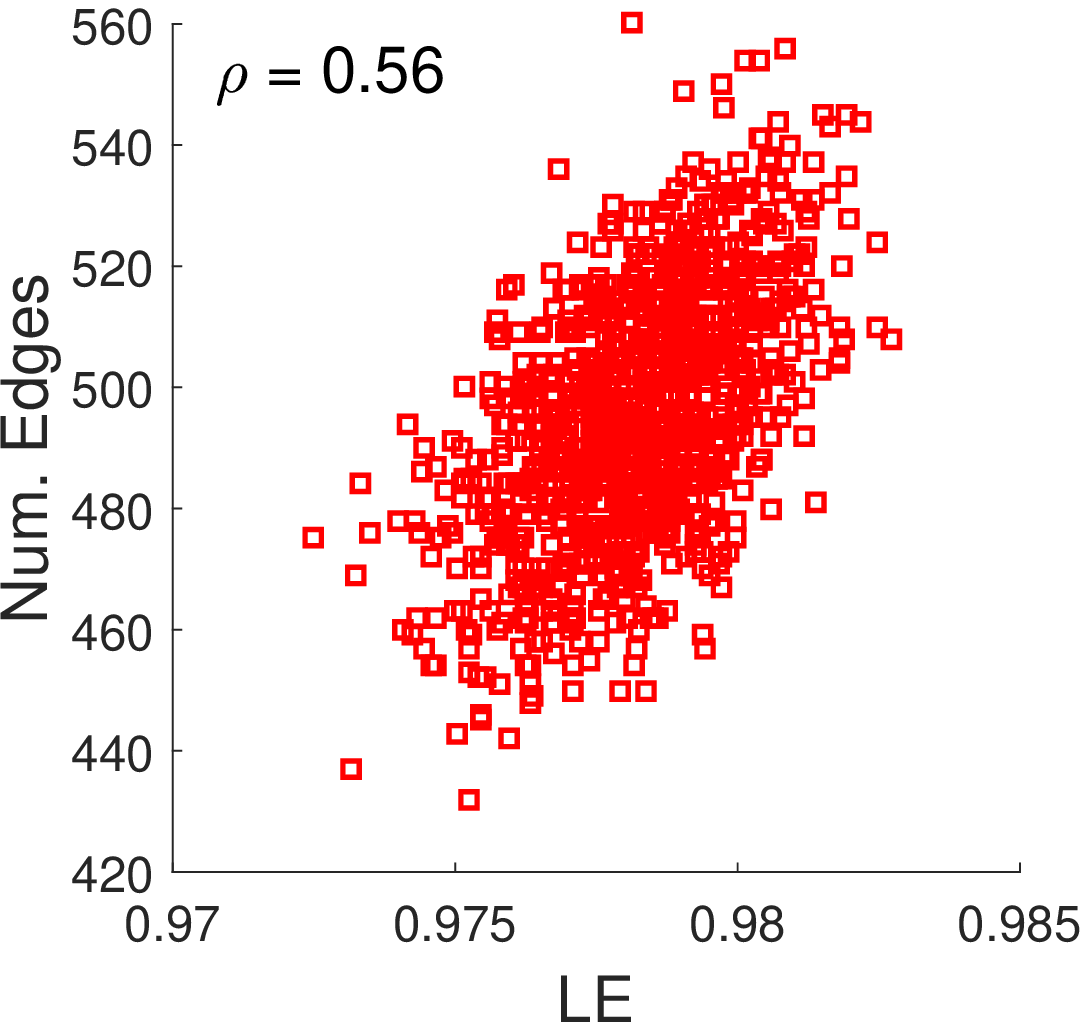}}
{\includegraphics[width=0.22\textwidth]{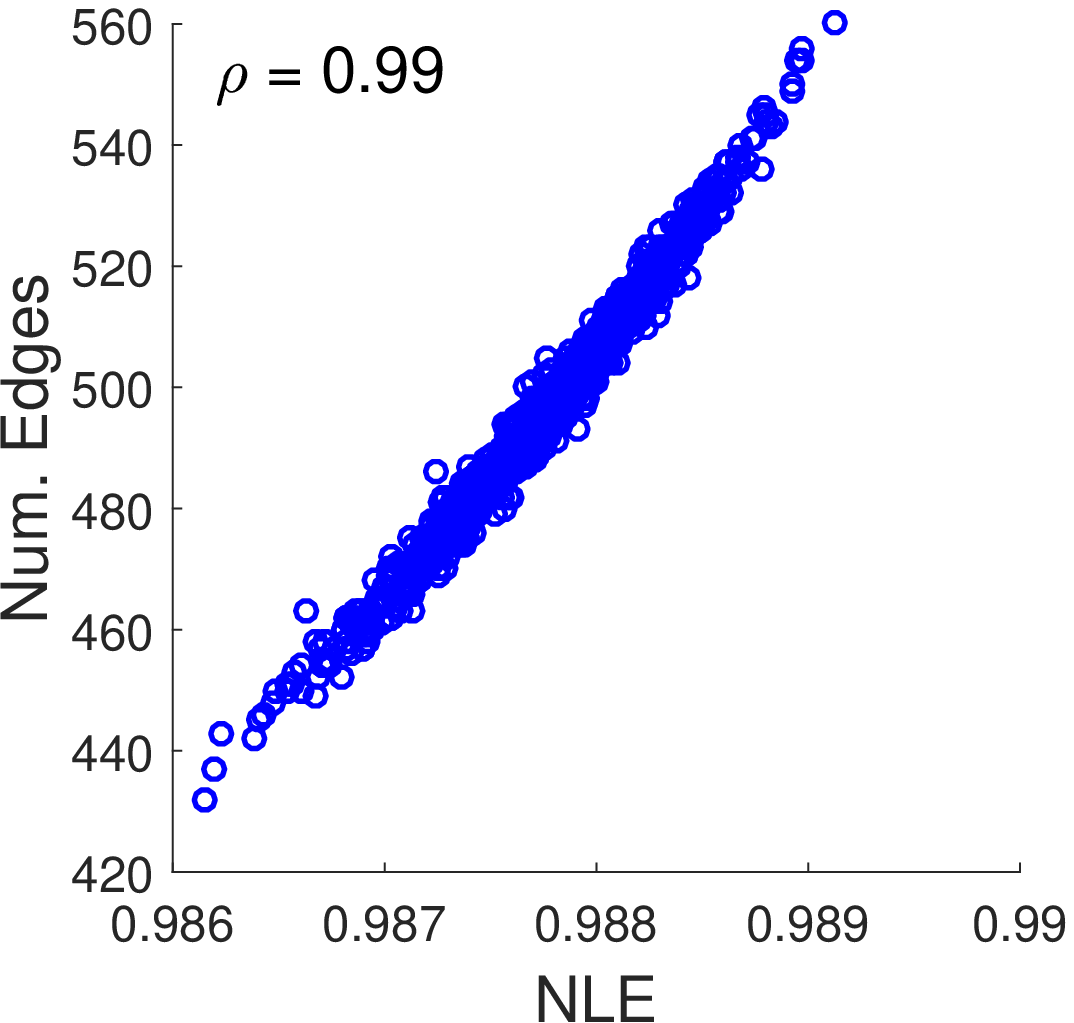}}
{\includegraphics[width=0.22\textwidth]{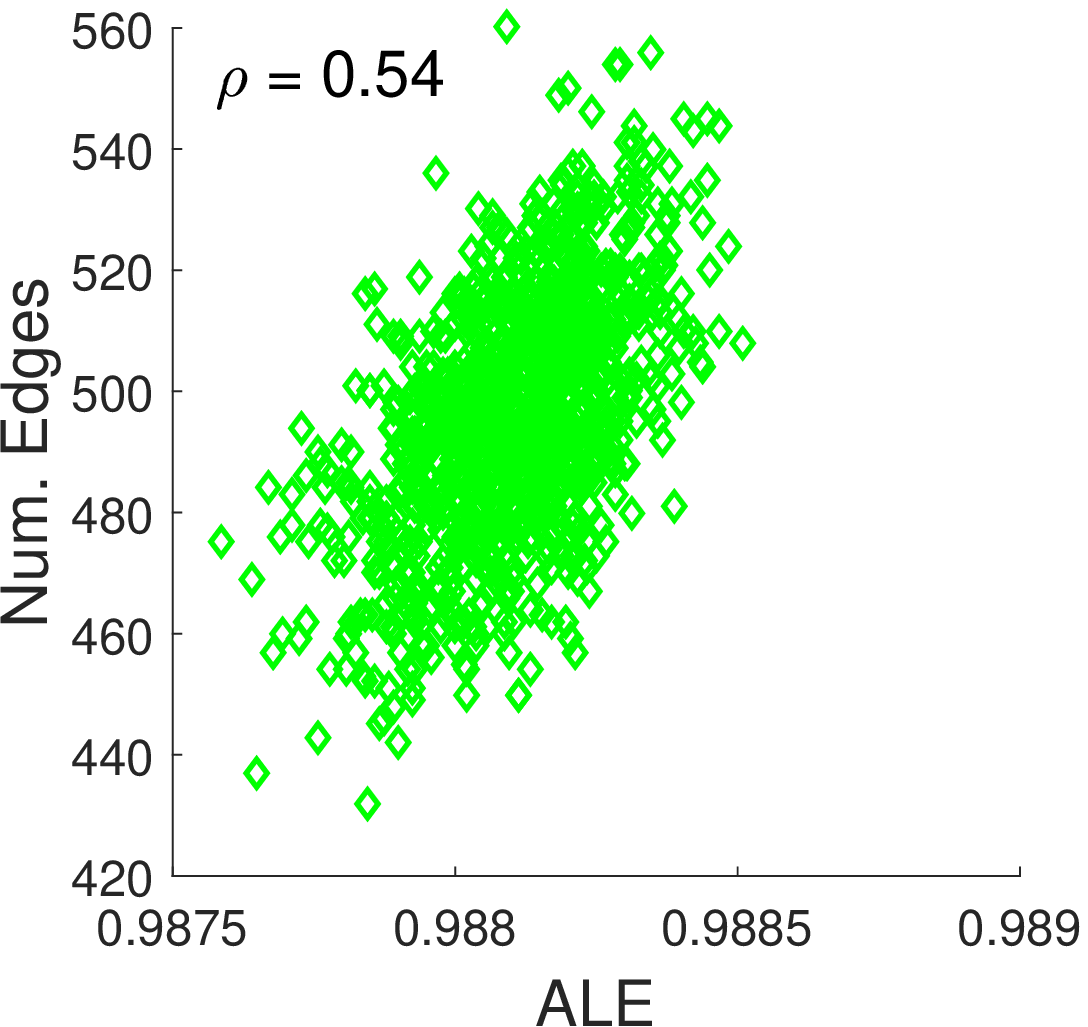}}
{\includegraphics[width=0.22\textwidth]{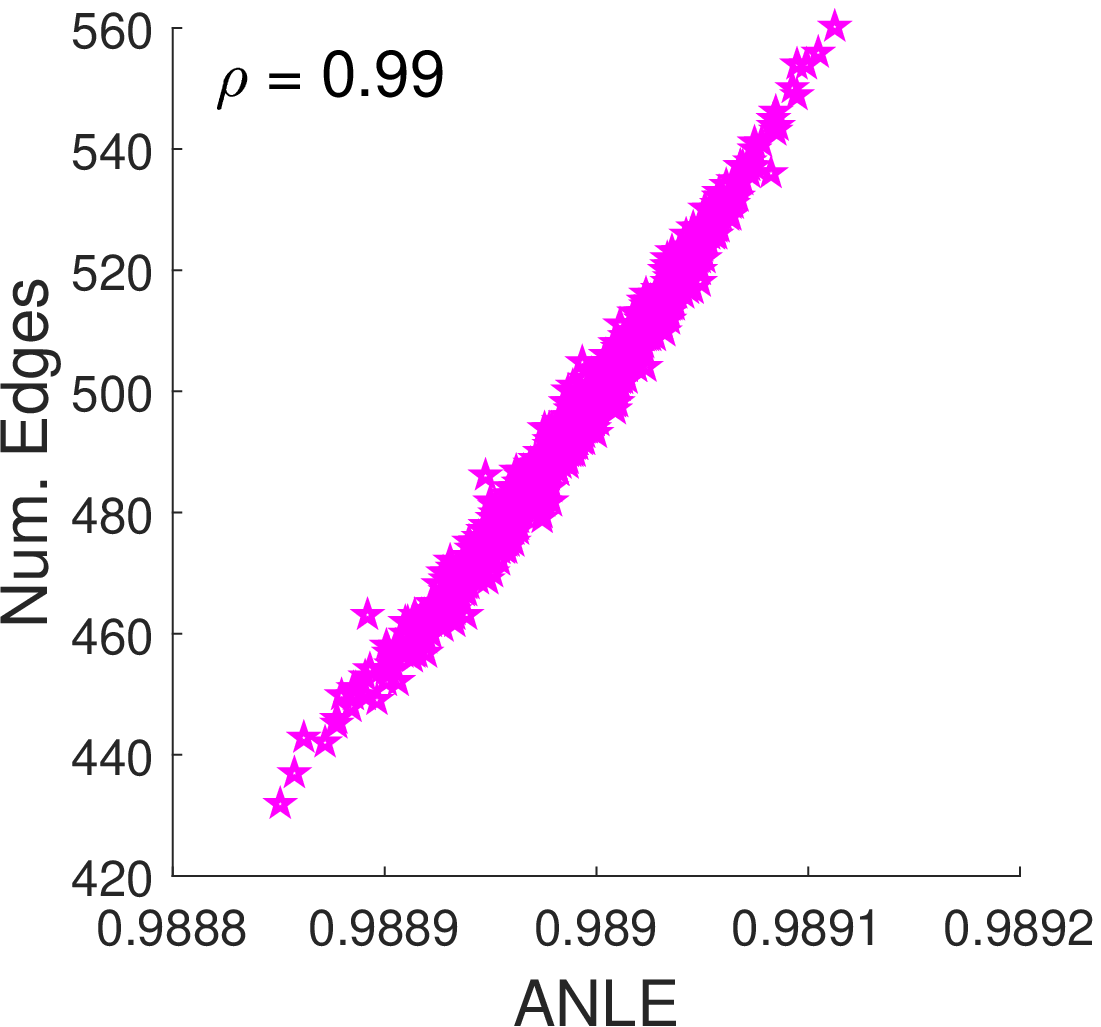}}\\
\caption{Correlation between entropy and number of edges for Erd\"os-R\'enyi graphs with $p=0.1$.}
\label{fig:corr_edges_erdos}
\end{figure}

\begin{figure}[t!]
{\includegraphics[width=0.22\textwidth]{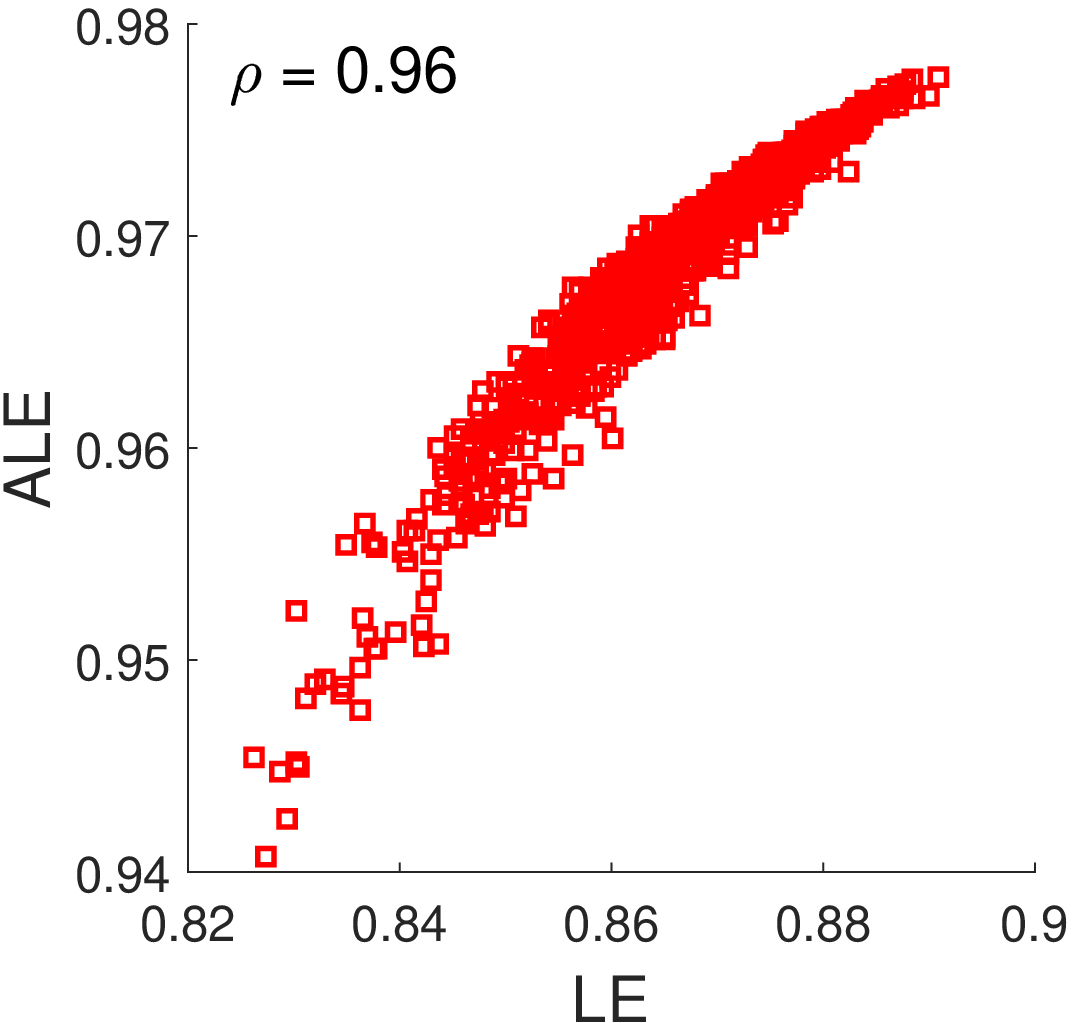}}
{\includegraphics[width=0.22\textwidth]{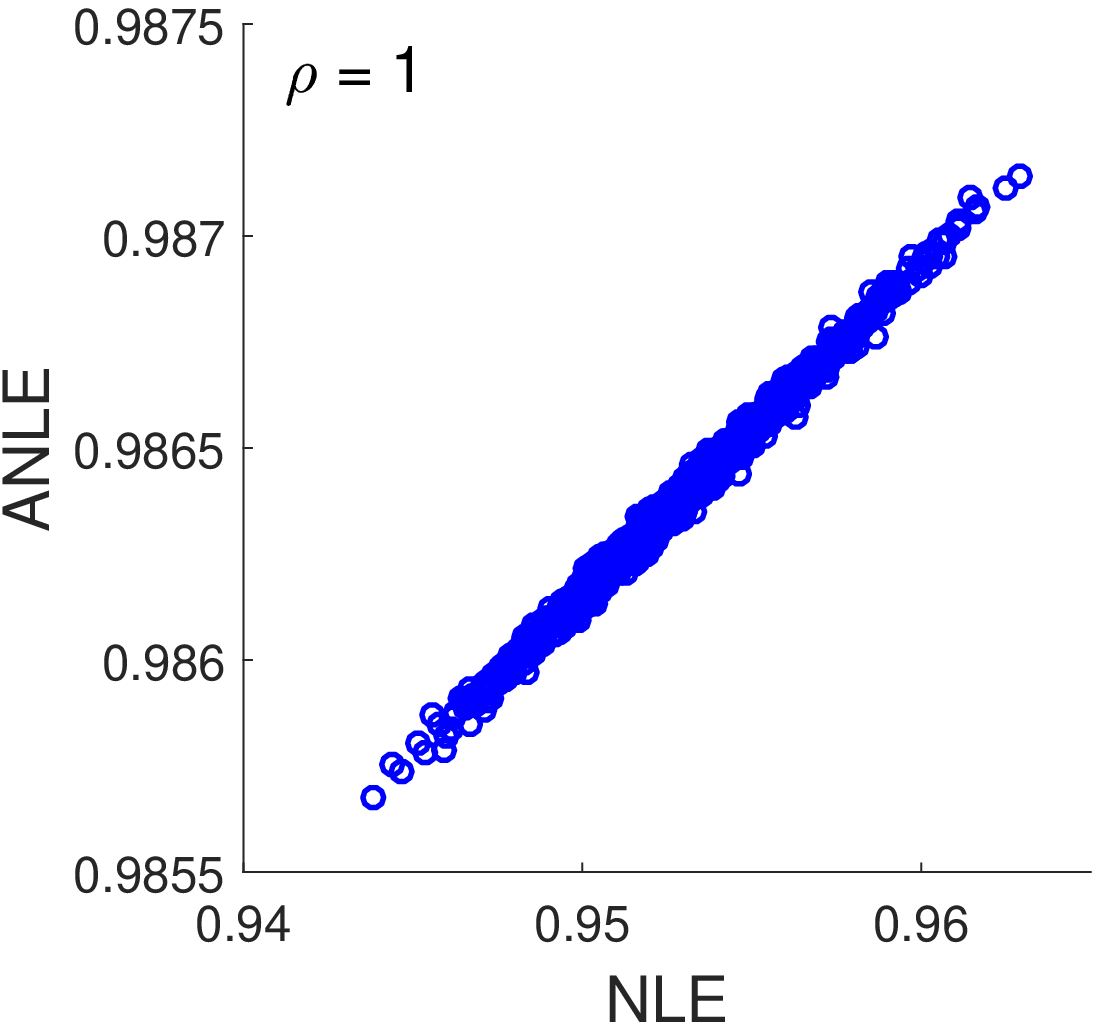}}
{\includegraphics[width=0.22\textwidth]{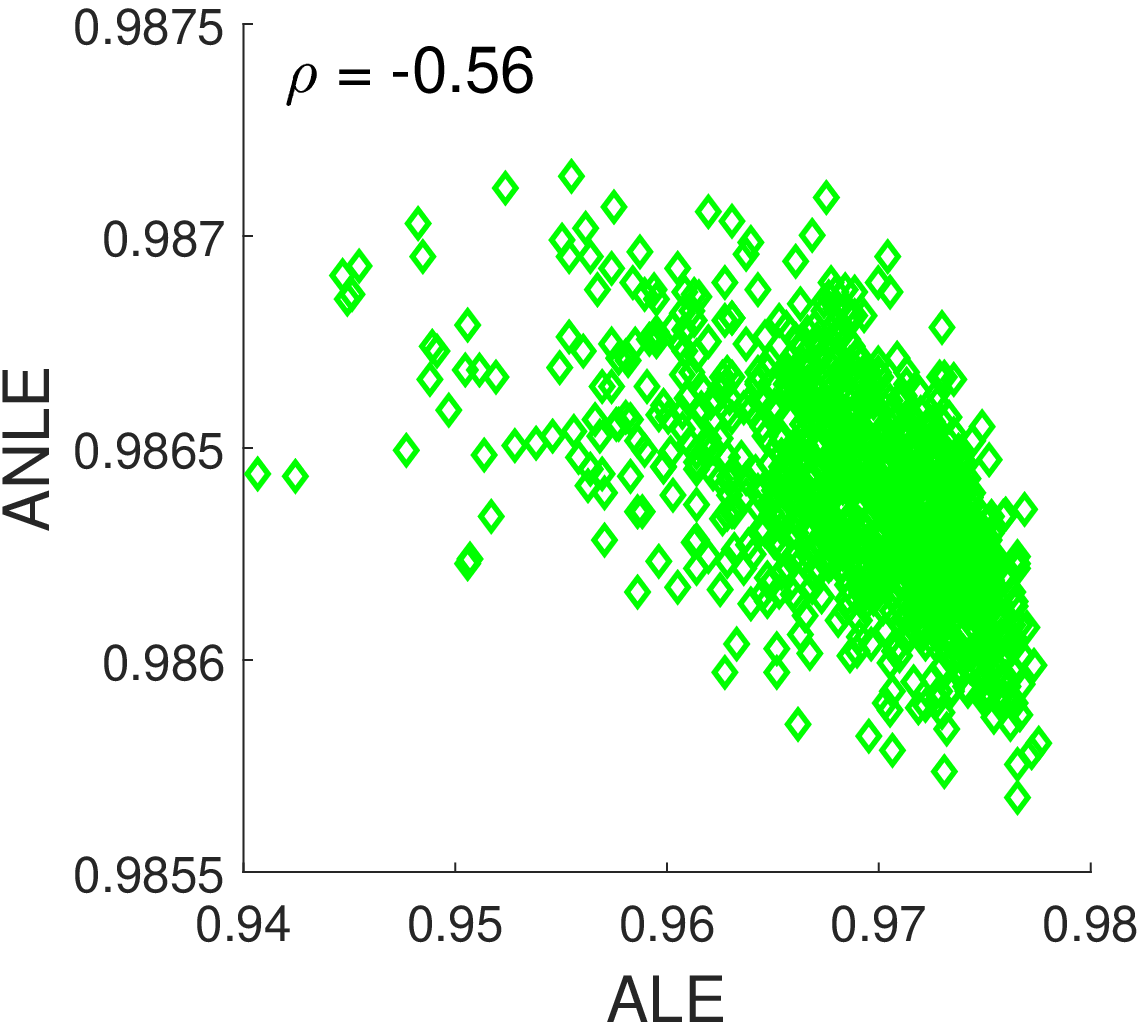}}
{\includegraphics[width=0.22\textwidth]{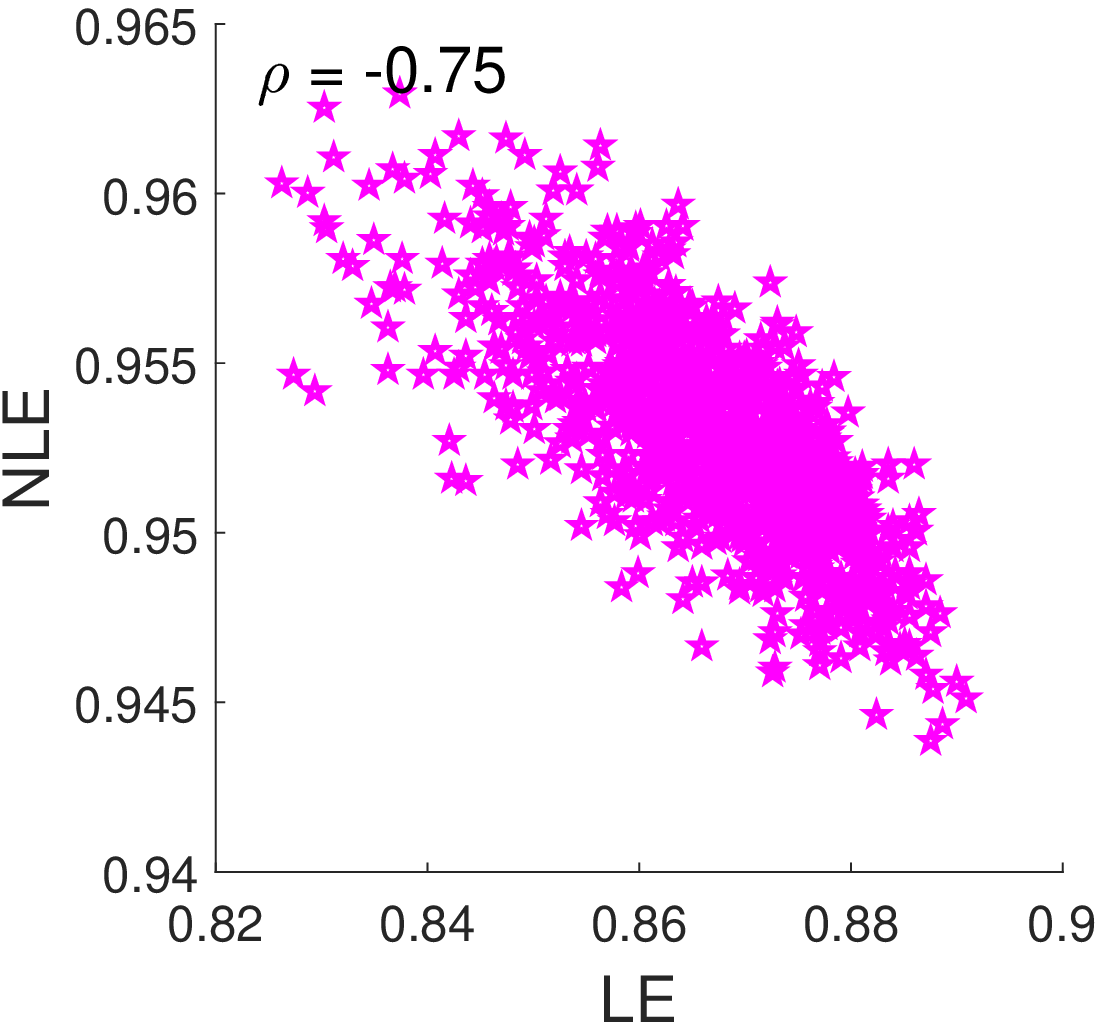}}\\

{\includegraphics[width=0.22\textwidth]{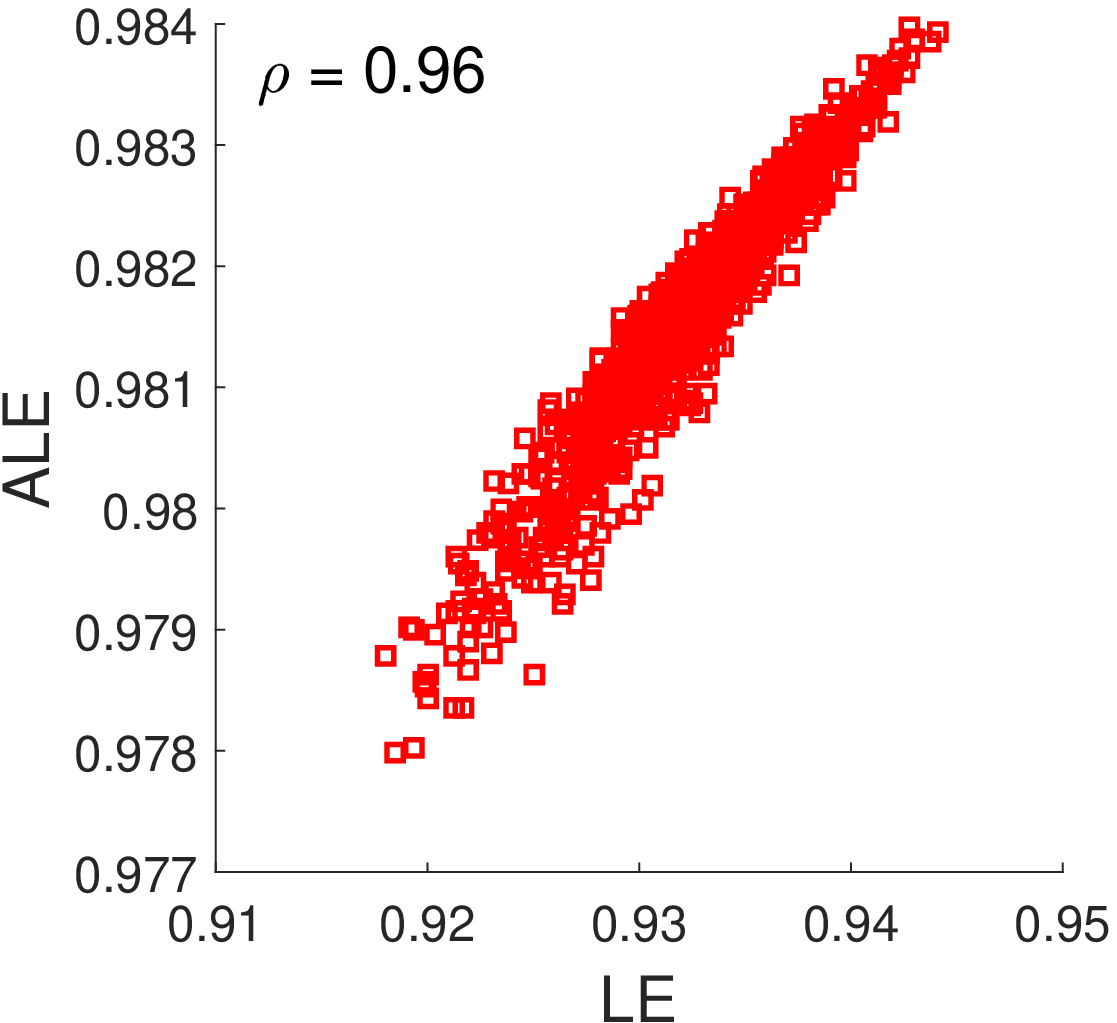}}
{\includegraphics[width=0.22\textwidth]{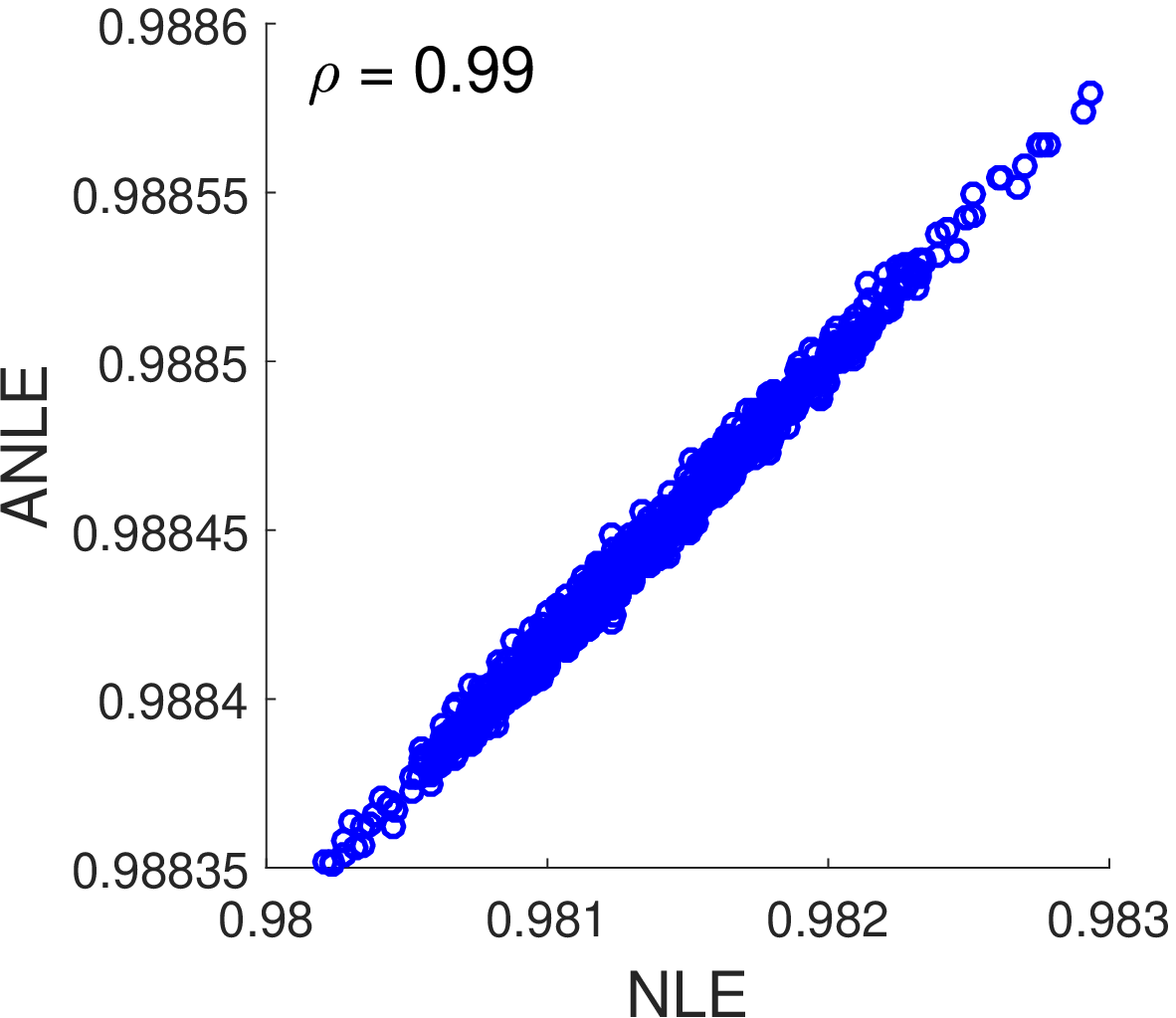}}
{\includegraphics[width=0.22\textwidth]{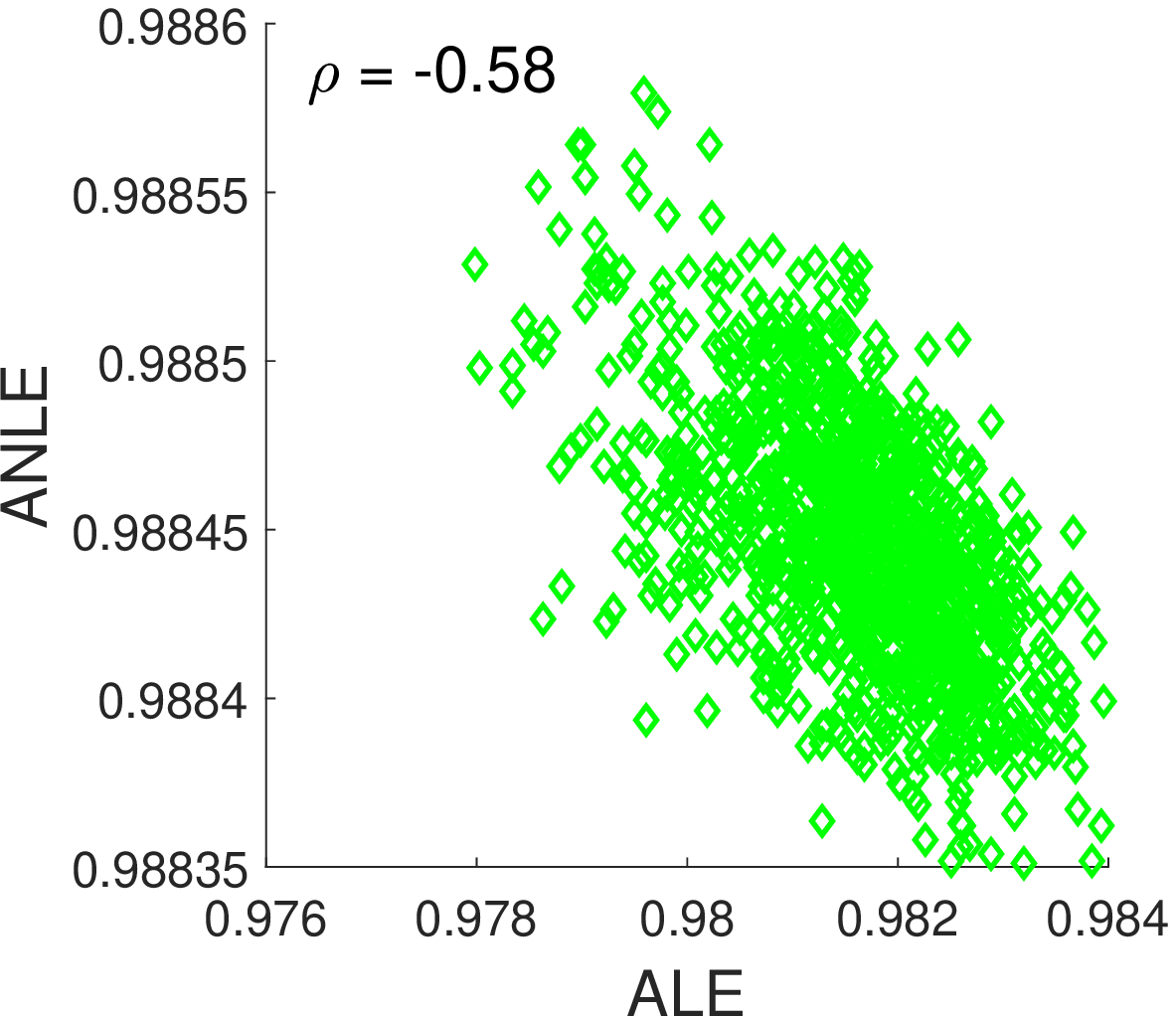}}
{\includegraphics[width=0.22\textwidth]{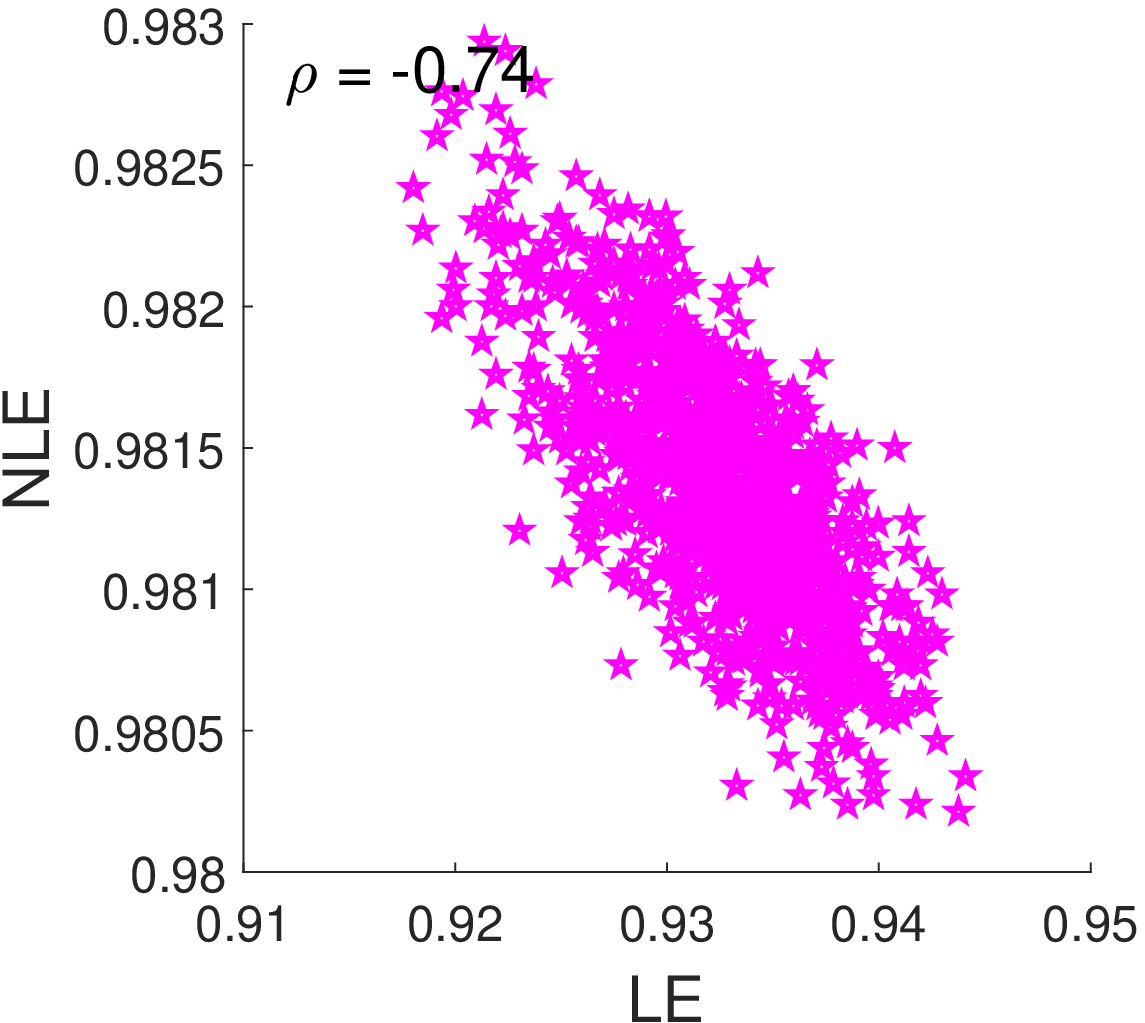}}\\

{\includegraphics[width=0.22\textwidth]{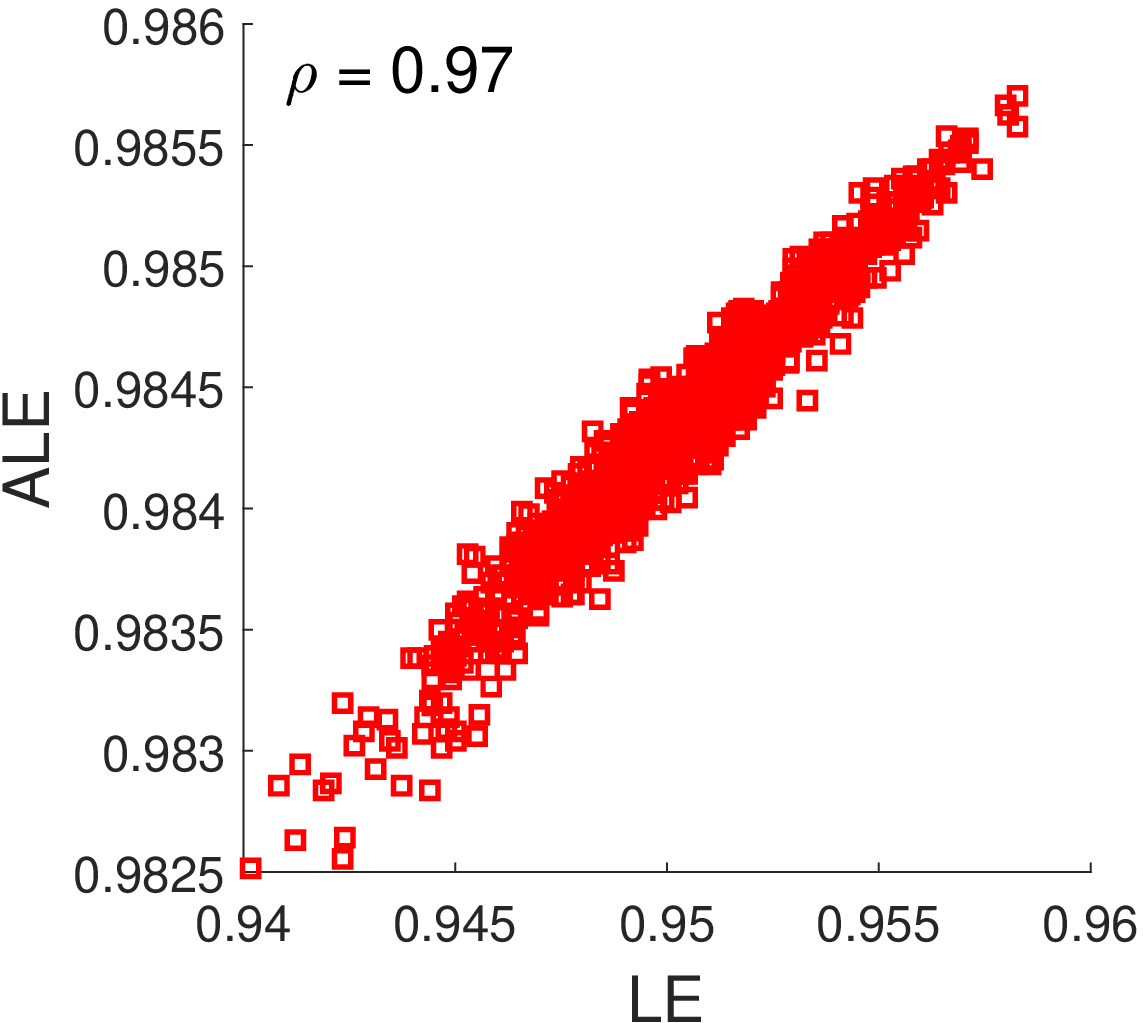}}
{\includegraphics[width=0.22\textwidth]{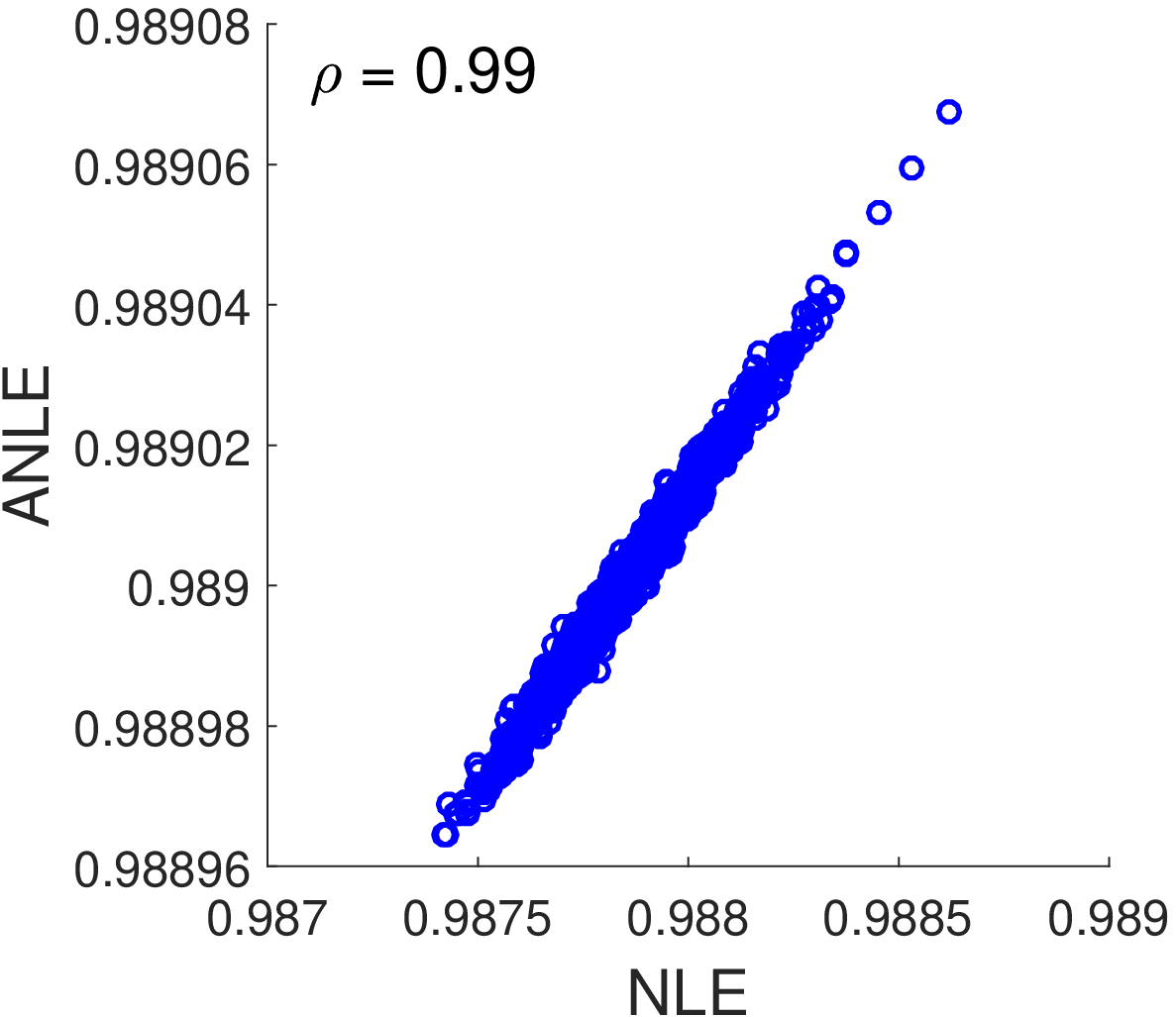}}
{\includegraphics[width=0.22\textwidth]{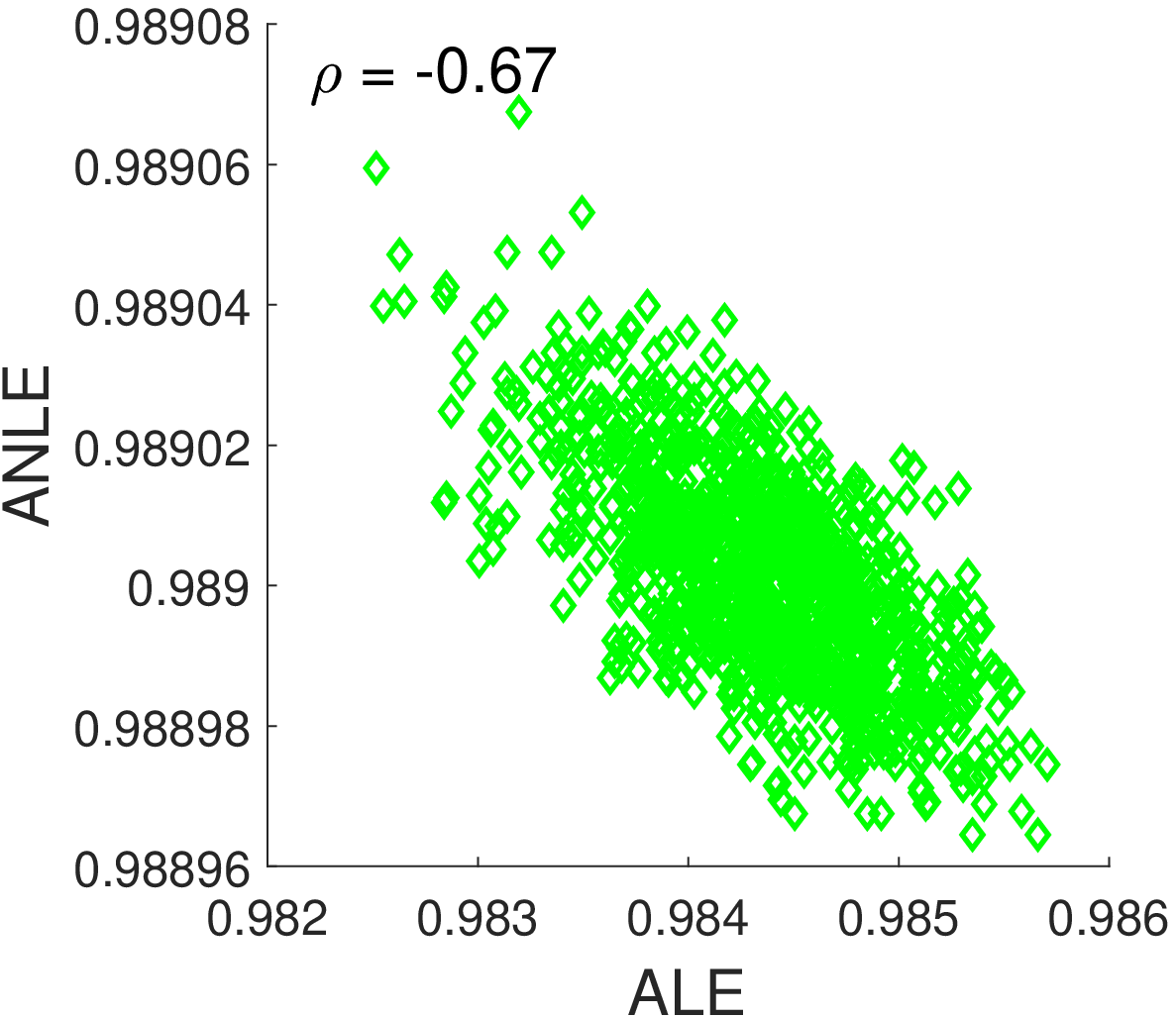}}
{\includegraphics[width=0.22\textwidth]{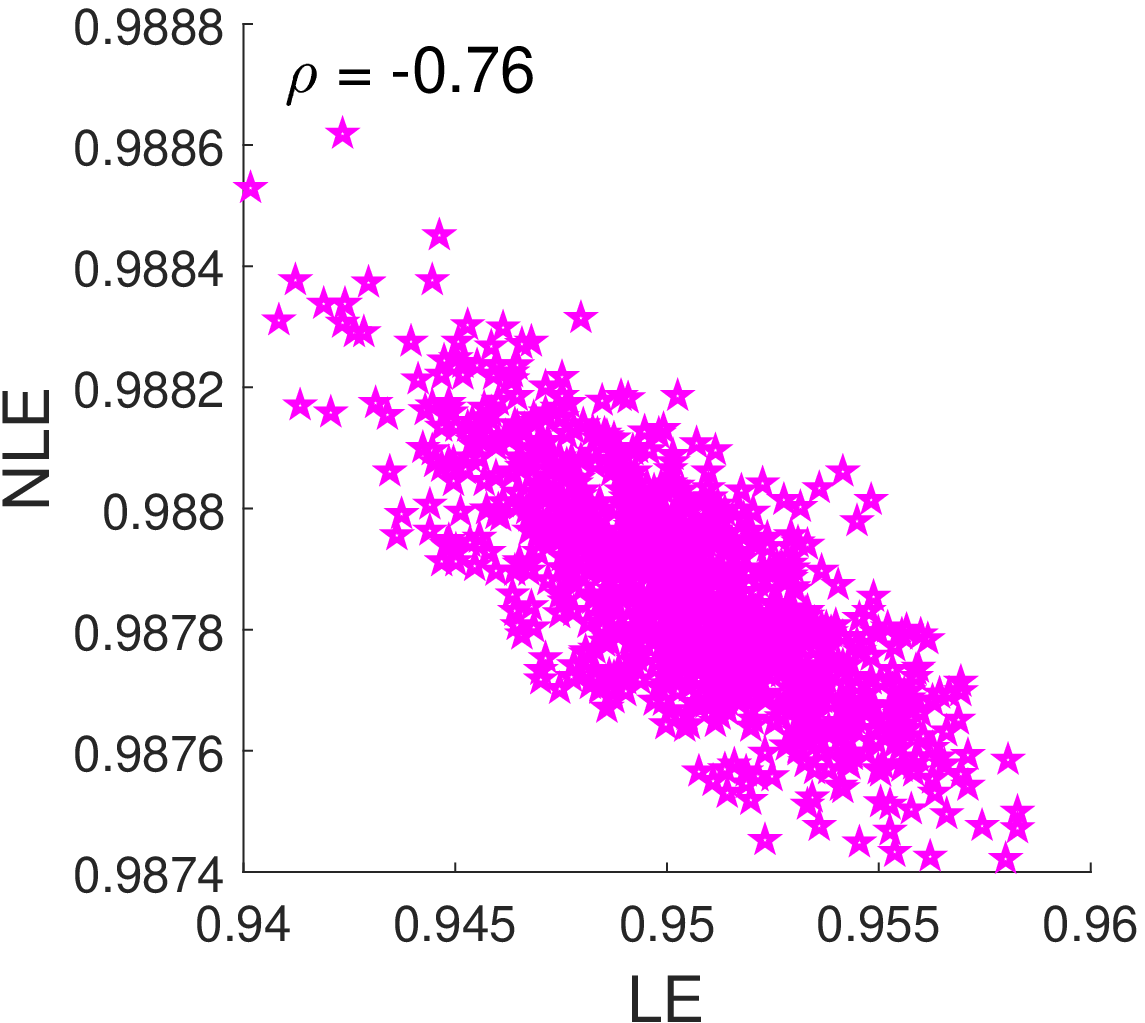}}\\
\caption{Entropies correlation on the scale-free graphs for $m=1$ (top), $m=3$ (middle), and $m=5$ (bottom).}
\label{fig:corr_scale}
\end{figure}

\begin{figure}[!t]
{\includegraphics[width=0.22\textwidth]{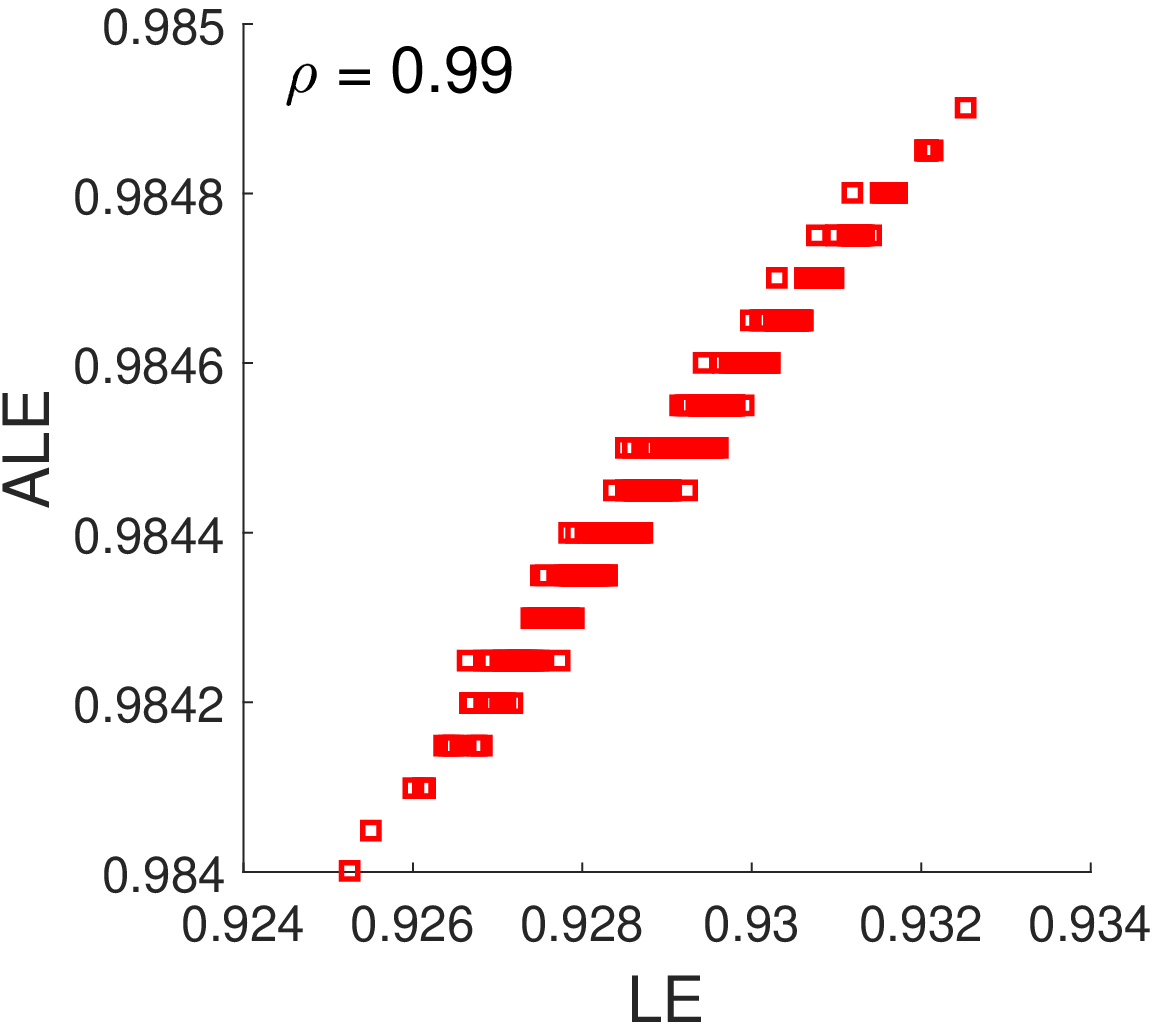}}
{\includegraphics[width=0.22\textwidth]{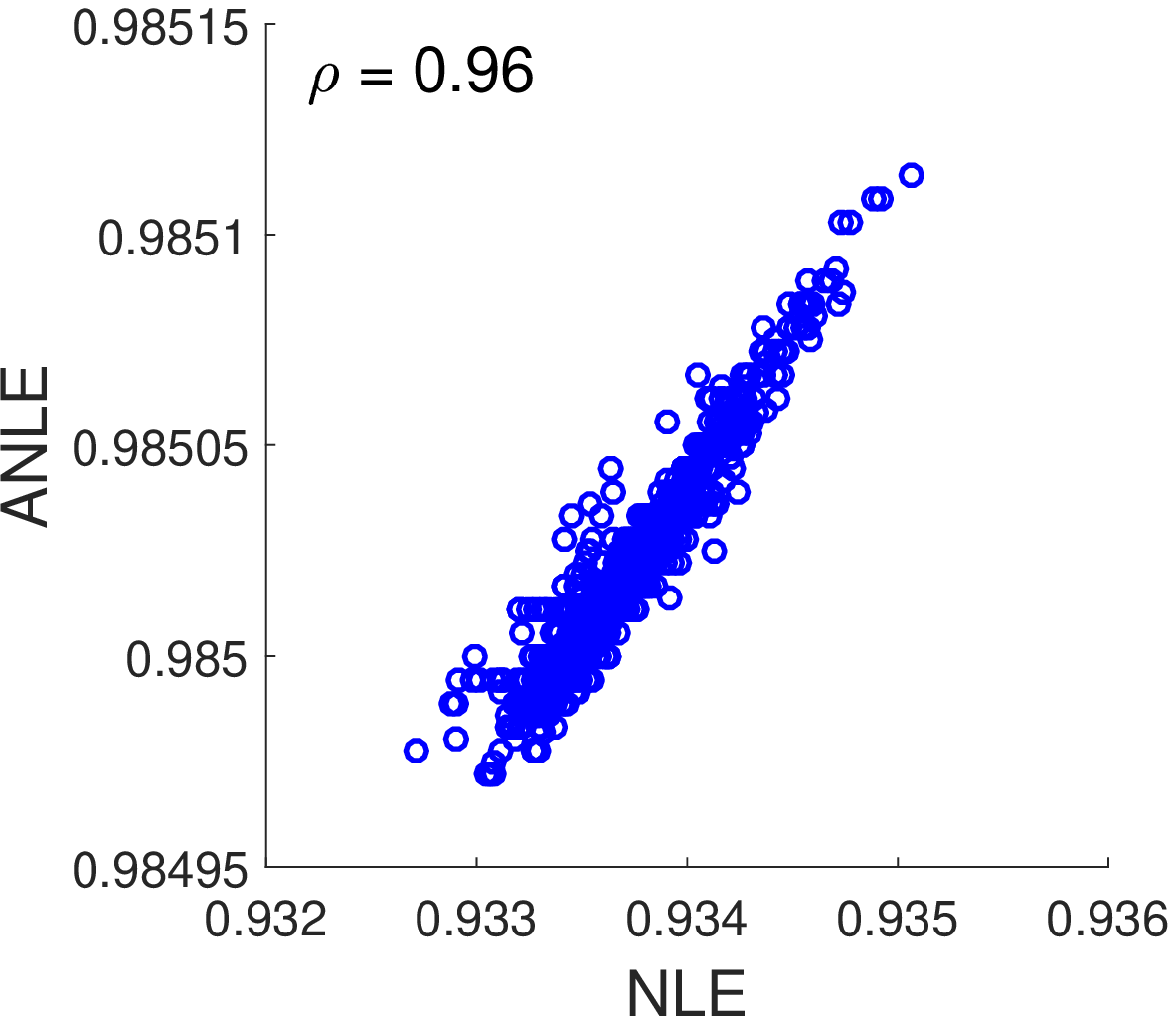}}
{\includegraphics[width=0.22\textwidth]{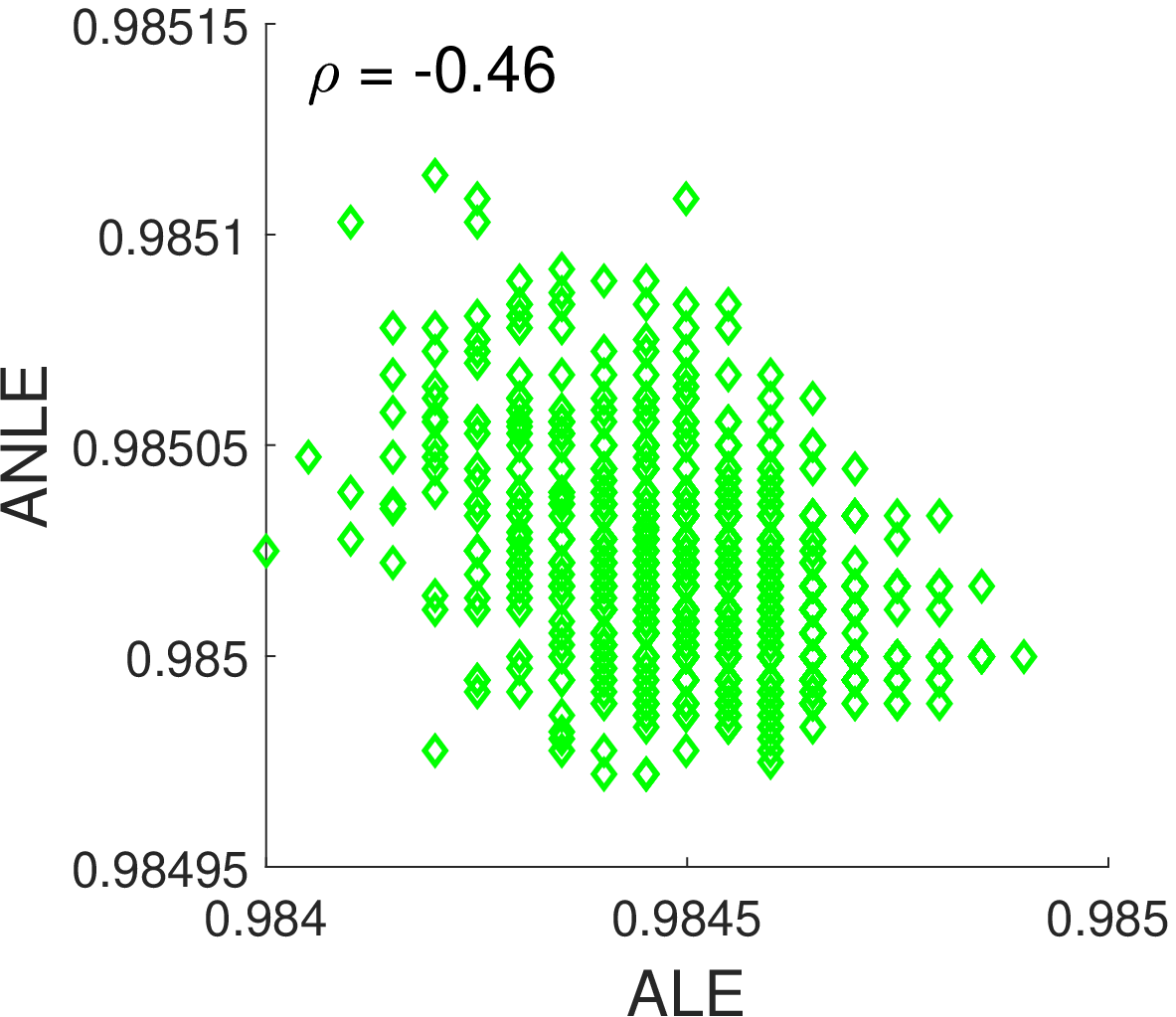}}
{\includegraphics[width=0.22\textwidth]{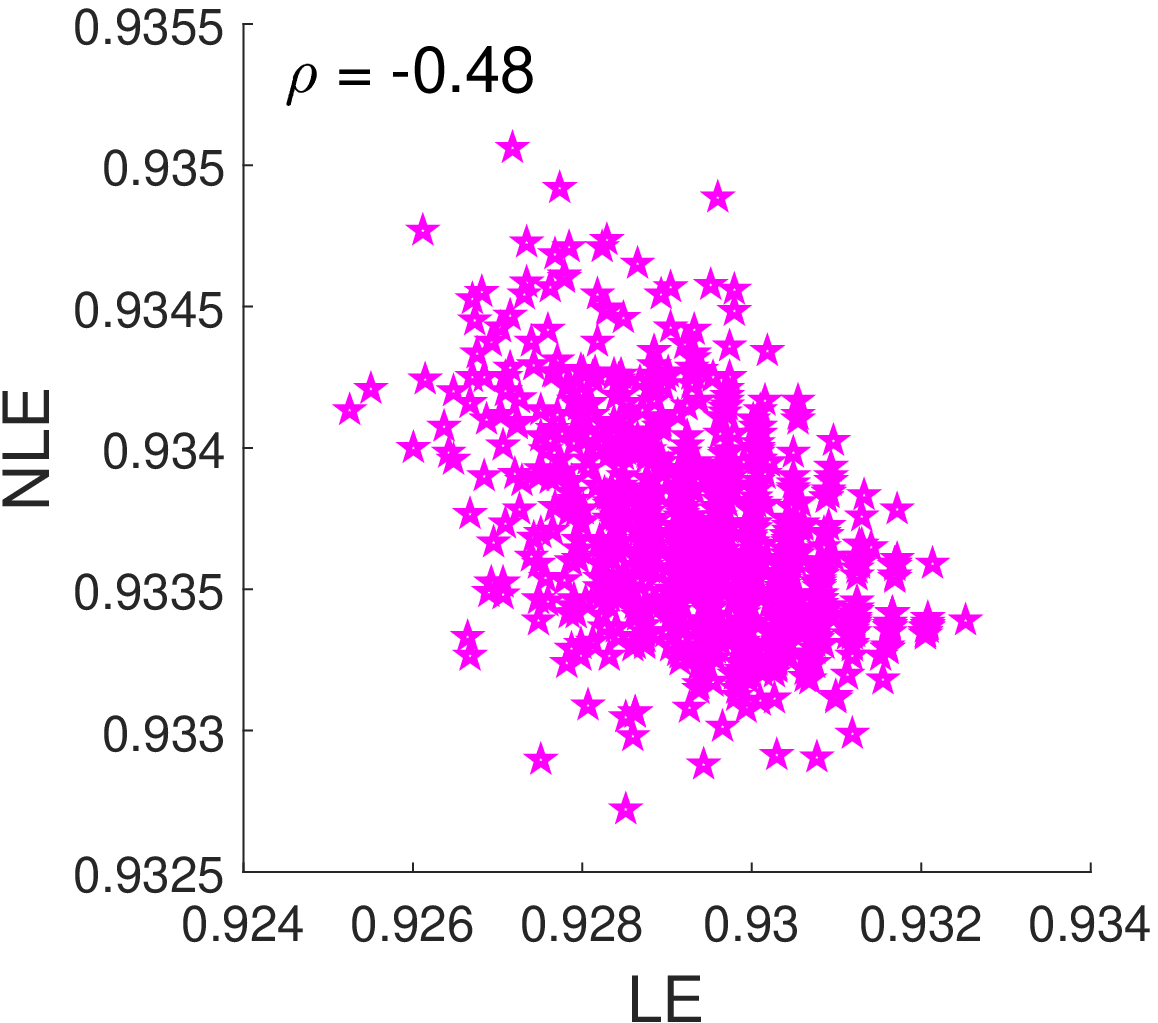}}\\

{\includegraphics[width=0.22\textwidth]{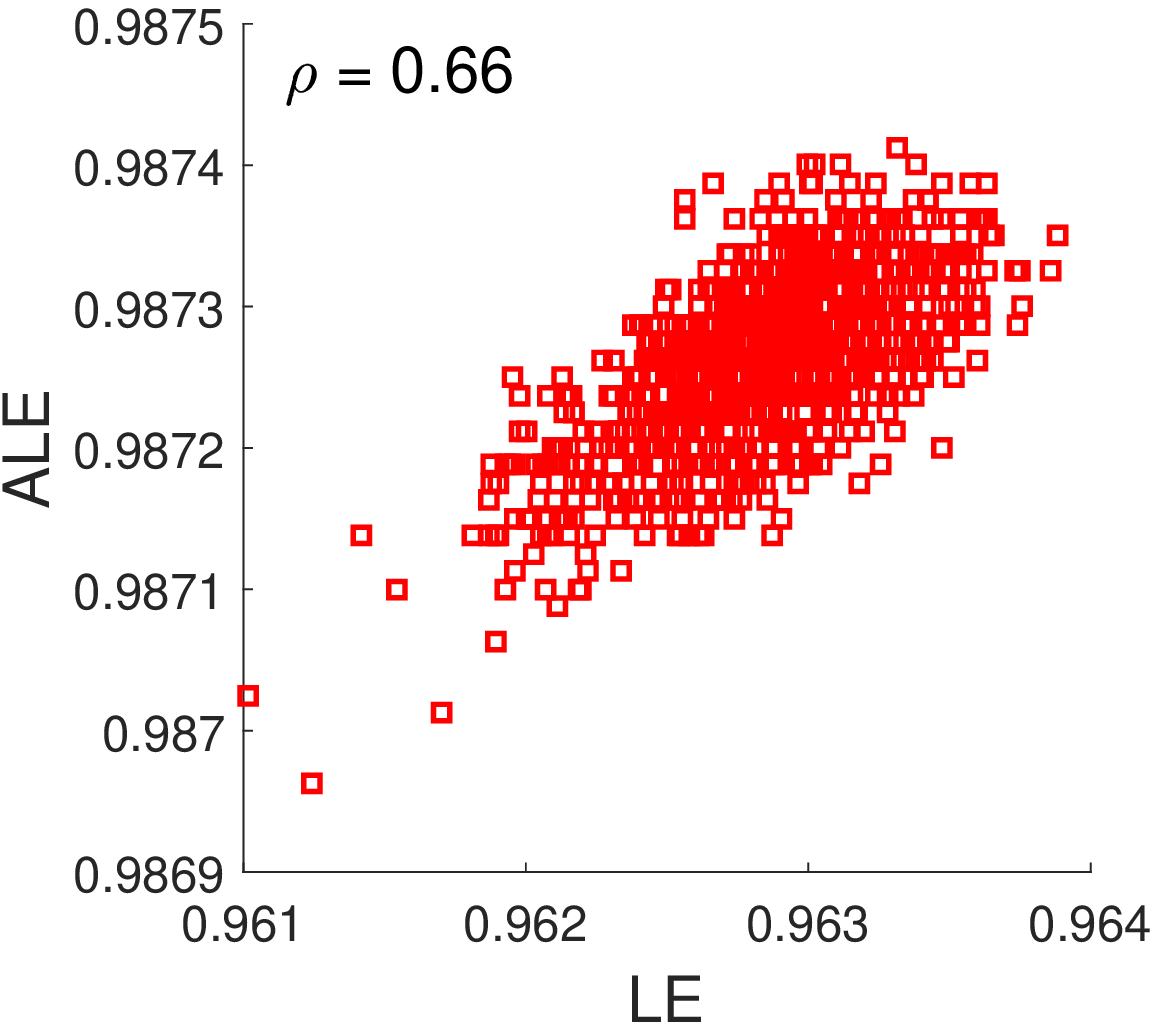}}
{\includegraphics[width=0.22\textwidth]{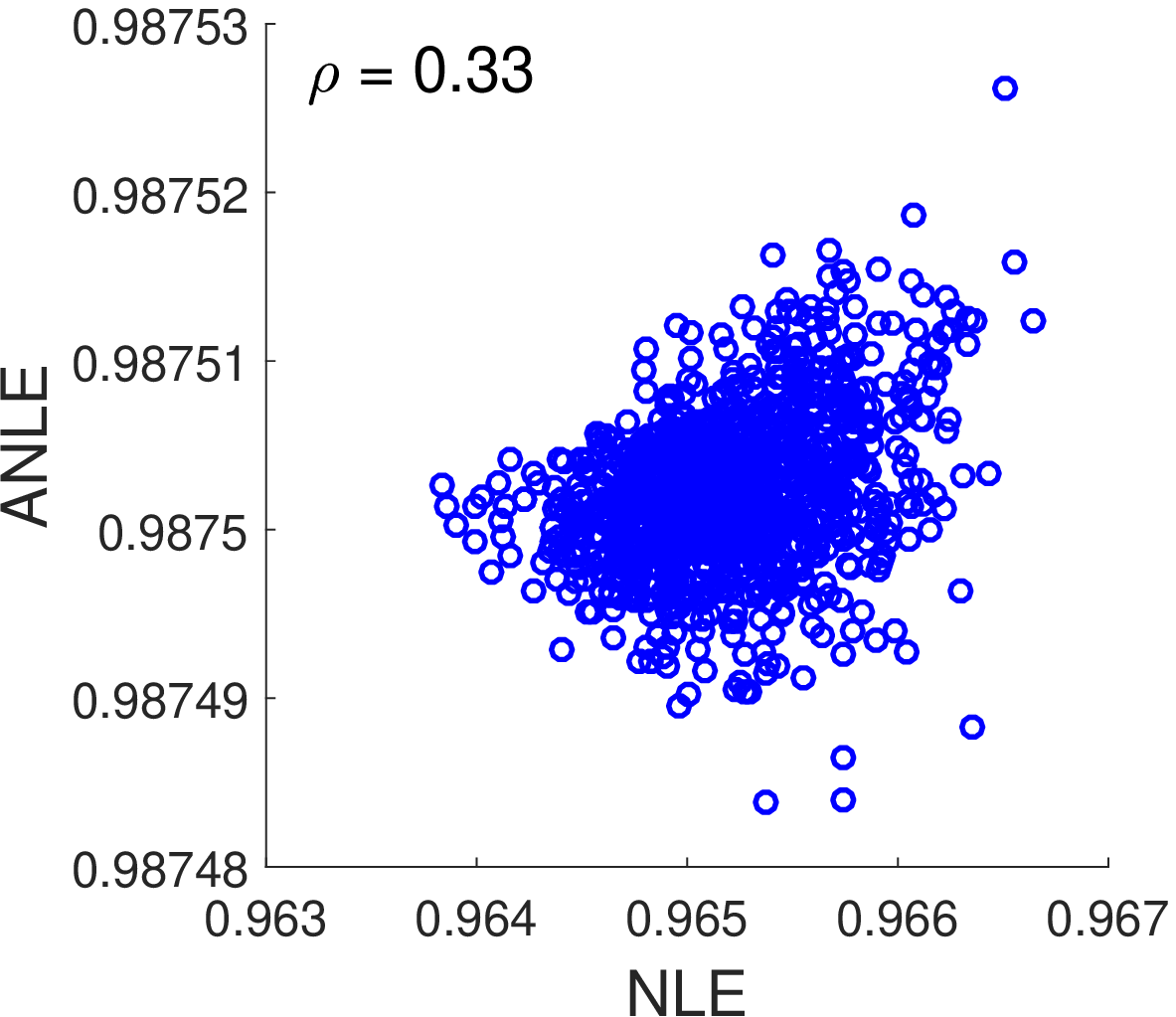}}
{\includegraphics[width=0.22\textwidth]{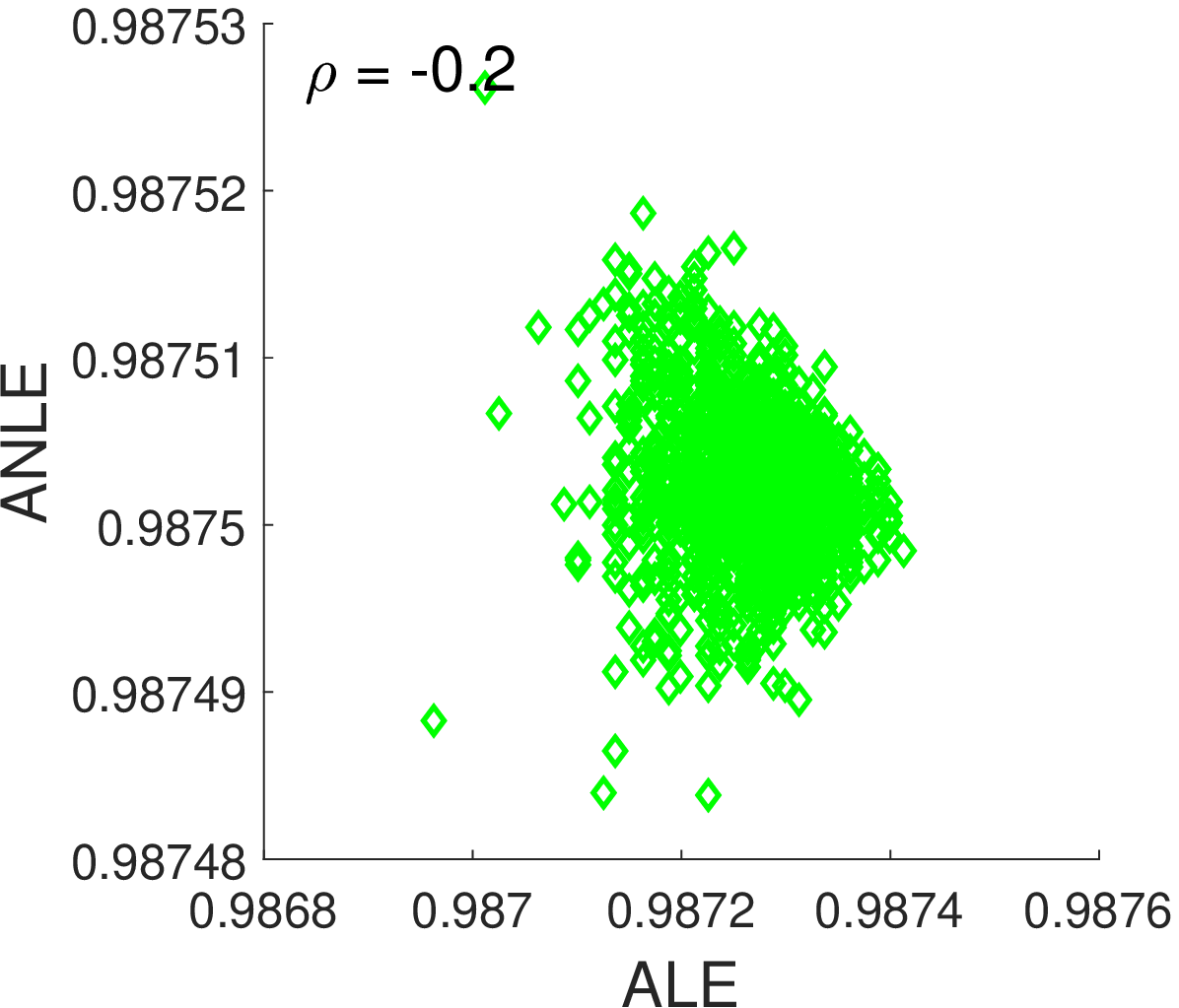}}
{\includegraphics[width=0.22\textwidth]{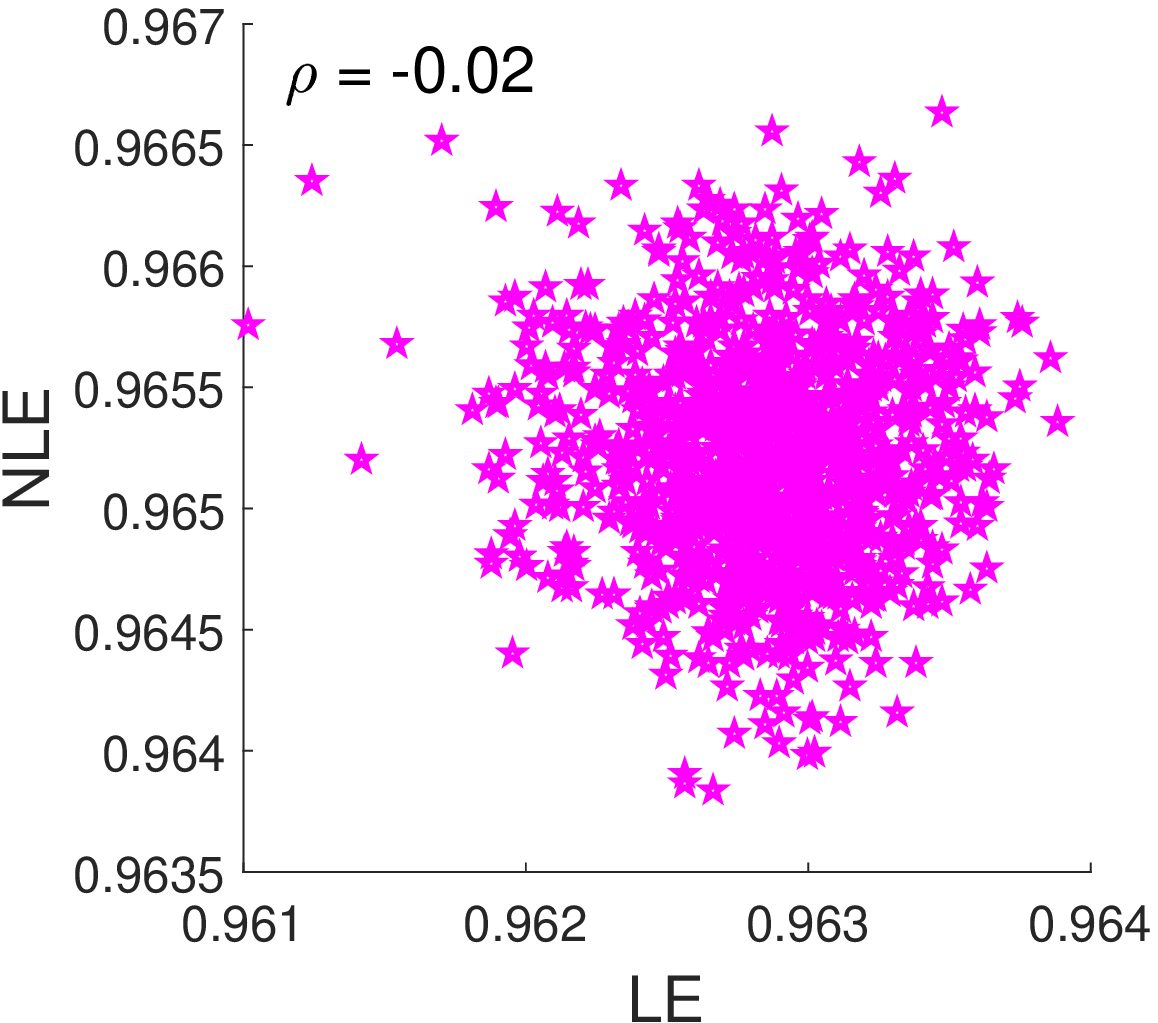}}\\

{\includegraphics[width=0.22\textwidth]{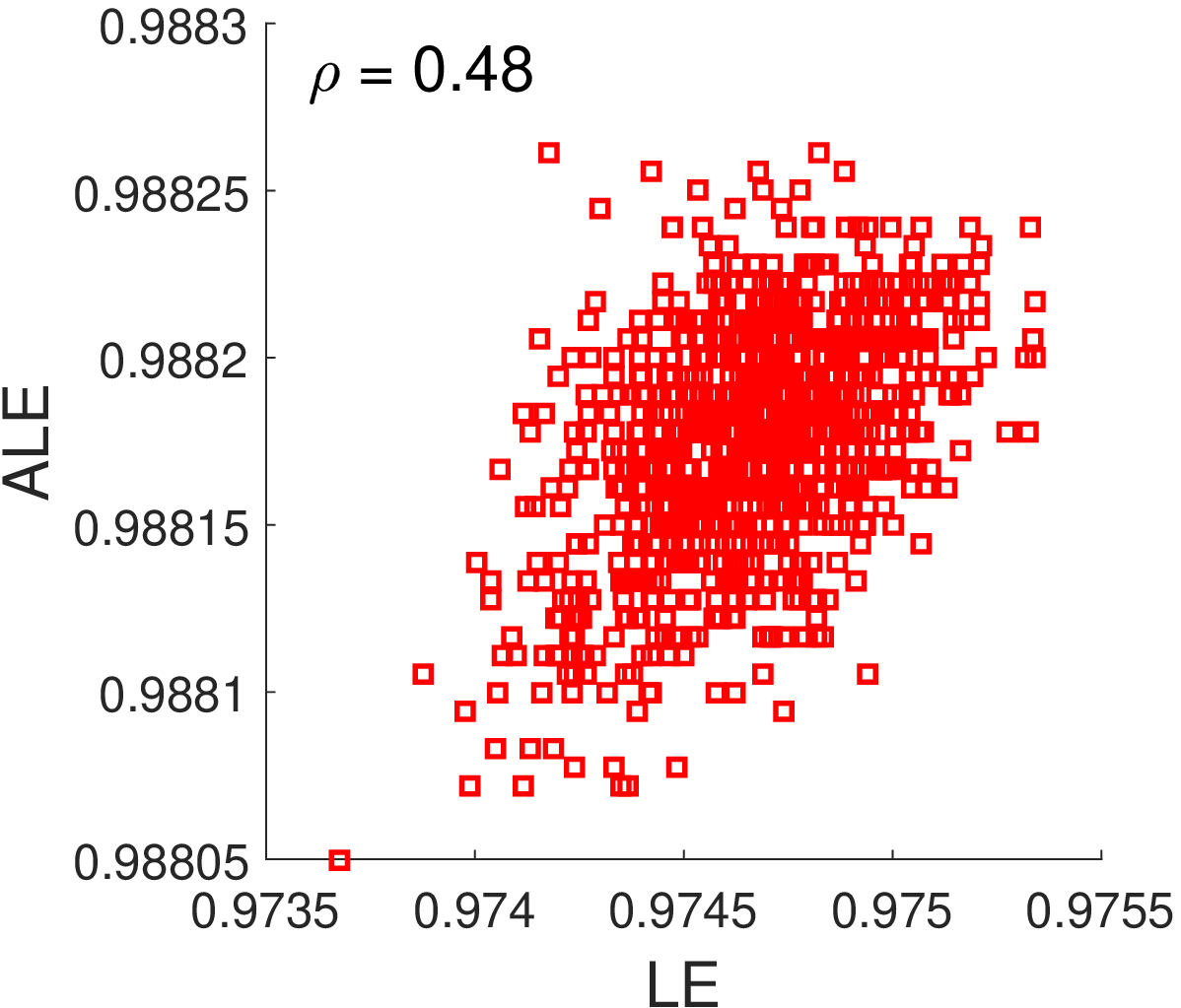}}
{\includegraphics[width=0.22\textwidth]{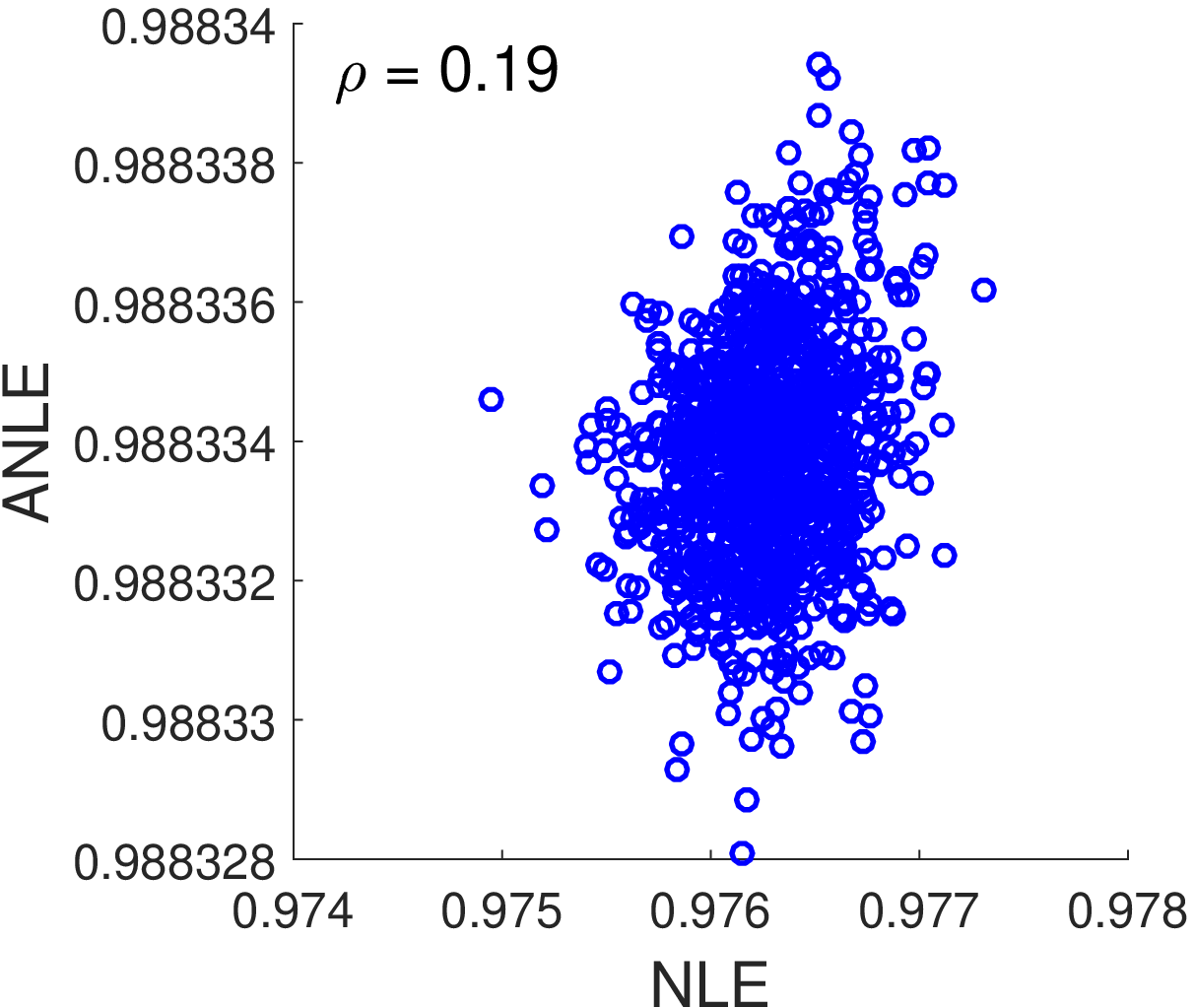}}
{\includegraphics[width=0.22\textwidth]{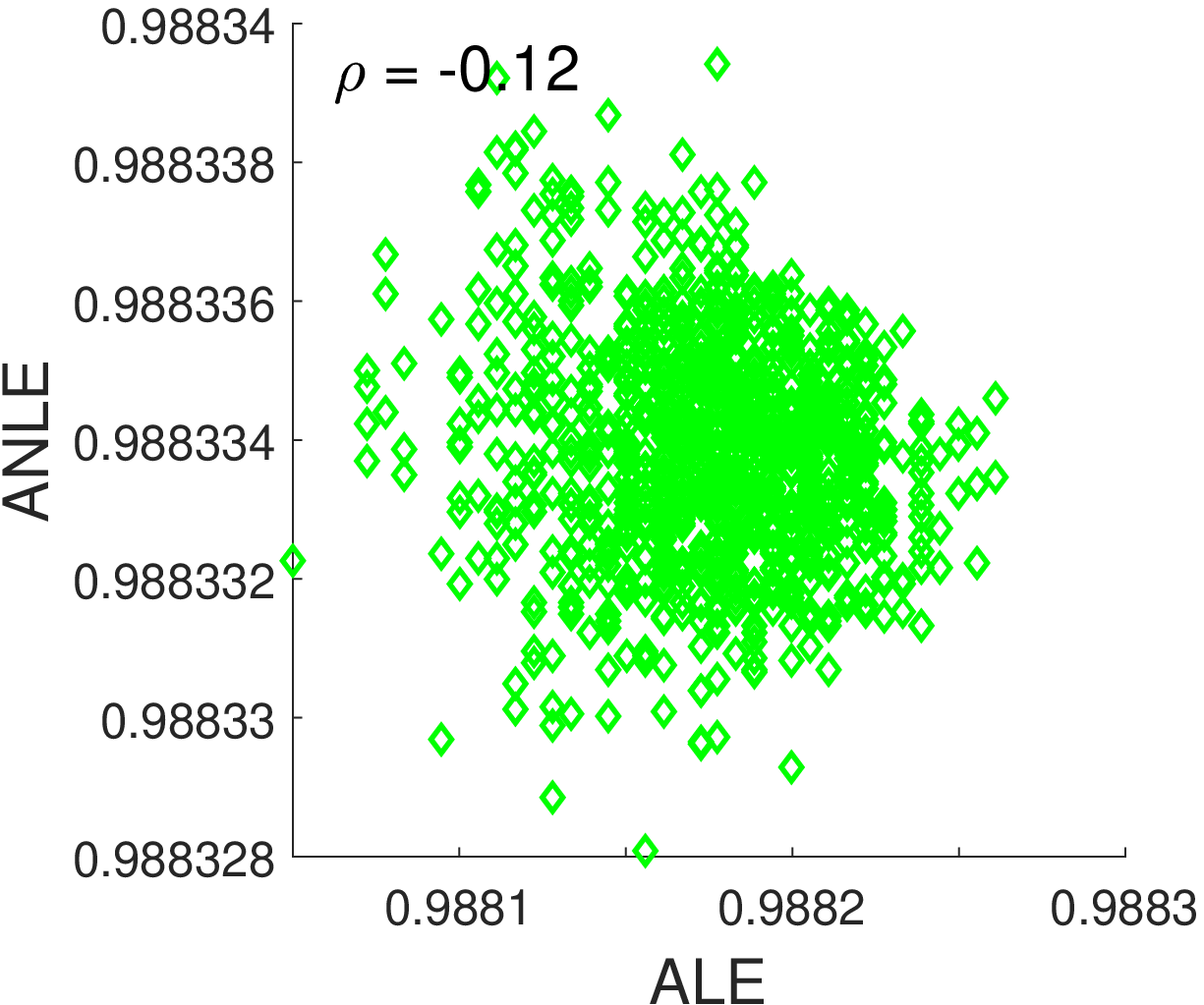}}
{\includegraphics[width=0.22\textwidth]{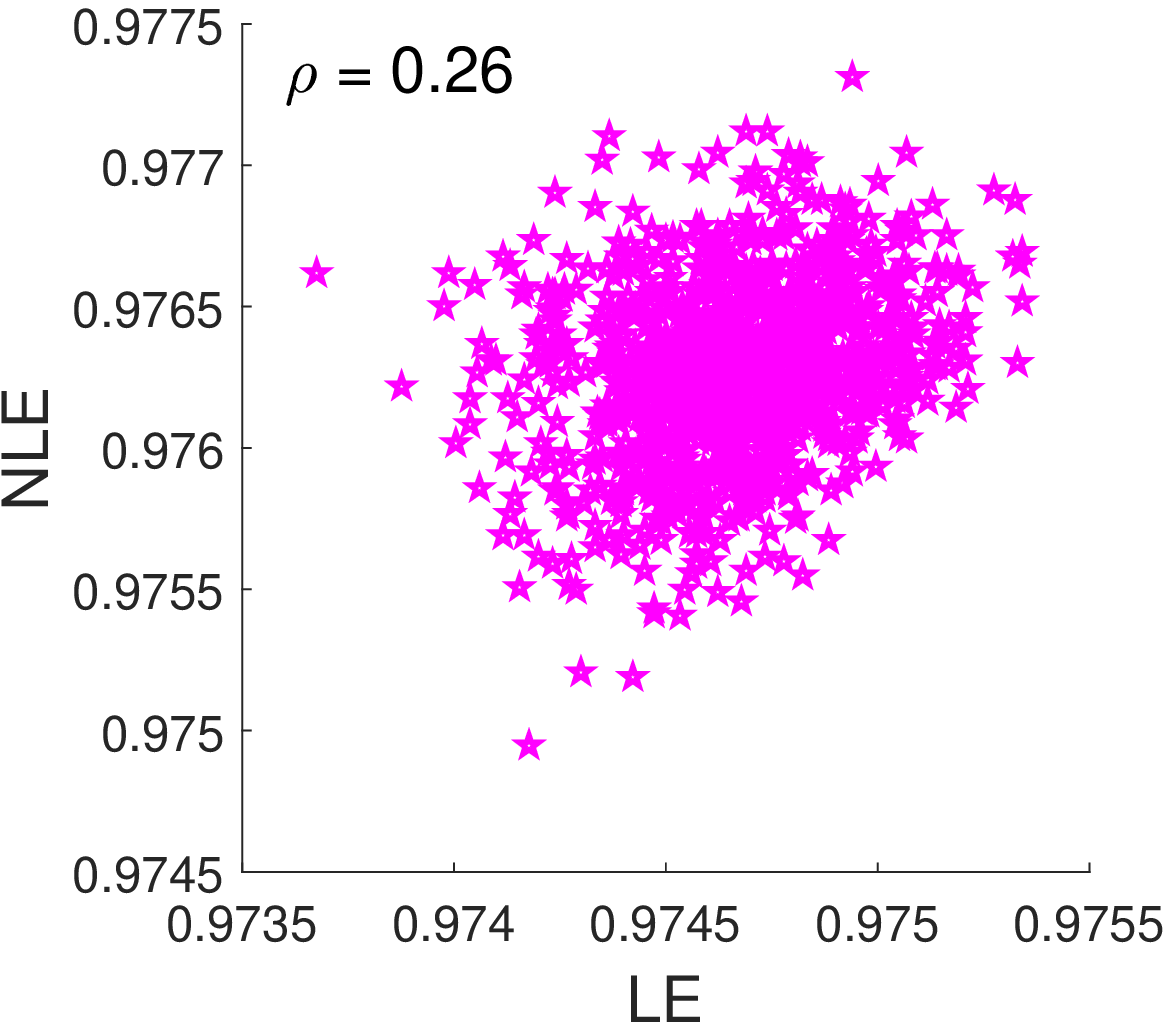}}\\
\caption{Entropies correlation on the small-world graphs for $k=2$ (top), $k=4$ (middle), and $k=6$ (bottom).}
\label{fig:corr_small}
\end{figure}

\subsubsection{Correlation Analysis}
With the synthetic graph datasets to hand, we perform a correlation study between the various version of the von Neumann entropy considered so far. More specifically, we measure the Pearson correlation coefficient (denoted as $\rho$ in Figs.~\ref{fig:corr_erdos}-\ref{fig:corr_small}) between 1) the approximate and exact Laplacian entropy, 2) the approximate and exact normalized Laplacian entropy, 3) the exact Laplacian entropy and the exact normalized Laplacian entropy, and 4) the approximate Laplacian entropy and the approximate normalized Laplacian entropy. Note that, for each model and each choice of the model parameters (see Section~\ref{sec:graph_model}), we generate 1000 graphs.

Fig.~\ref{fig:corr_erdos} shows the results of the correlation analysis on the Erd\"os-R\'enyi graphs. The first column refers to the Laplacian entropy, the second one to the normalized Laplacian entropy, whereas the third and fourth columns concern the approximate and exact formulation. Here, we consider three choices of $p$. We first observe that there exists a strong correlation between the exact and approximate versions of each entropy. On the other hand, when we compare the normalized Laplacian entropy and unnormalized Laplacian entropy, both in their exact and approximate forms, the correlation becomes weaker. Indeed, as observed in the previous section, we expect the quadratic approximations of the two entropies to show a weak positive, or potentially negative, correlation. We posit that the weak positive correlation observed for Erd\"os-R\'enyi graphs is a consequence of the degree distribution of the nodes neighbourhoods being close to uniform. Interestingly, here we observe that this result holds also for the exact versions of the entropies. This in turn suggests that the structural patterns captured by the two entropies are not necessarily the same. This is an important observation, as it implies, for example, that when using the von Neumann entropy in pattern recognition applications swapping one entropy for the other is likely not to give the same result.

As for the strong correlation observed between the exact and approximate version of the normalized Laplacian entropy, this is likely to be due to the tight relationship with the number of edges of a graph and its normalized Laplacian entropy, as shown in Fig.~\ref{fig:corr_edges_erdos}. More precisely, in Fig.~\ref{fig:corr_edges_erdos} we show the correlation between the number of edges of a graph and its entropy for Erd\"os-R\'enyi graphs with $p =0.1$. Indeed, Eq.~\ref{eq:nlaplacian_approximation} suggests a strong dependency between the number of edges of a graph and the quadratic approximation of the normalized Laplacian entropy. Note however that we do not observe a strong correlation between the (approximate) Laplacian entropy and the number of edges.

\begin{figure}[!t]
\centering
\subfloat[\hspace{0.1cm} Maximization]{\includegraphics[width=.49\textwidth]{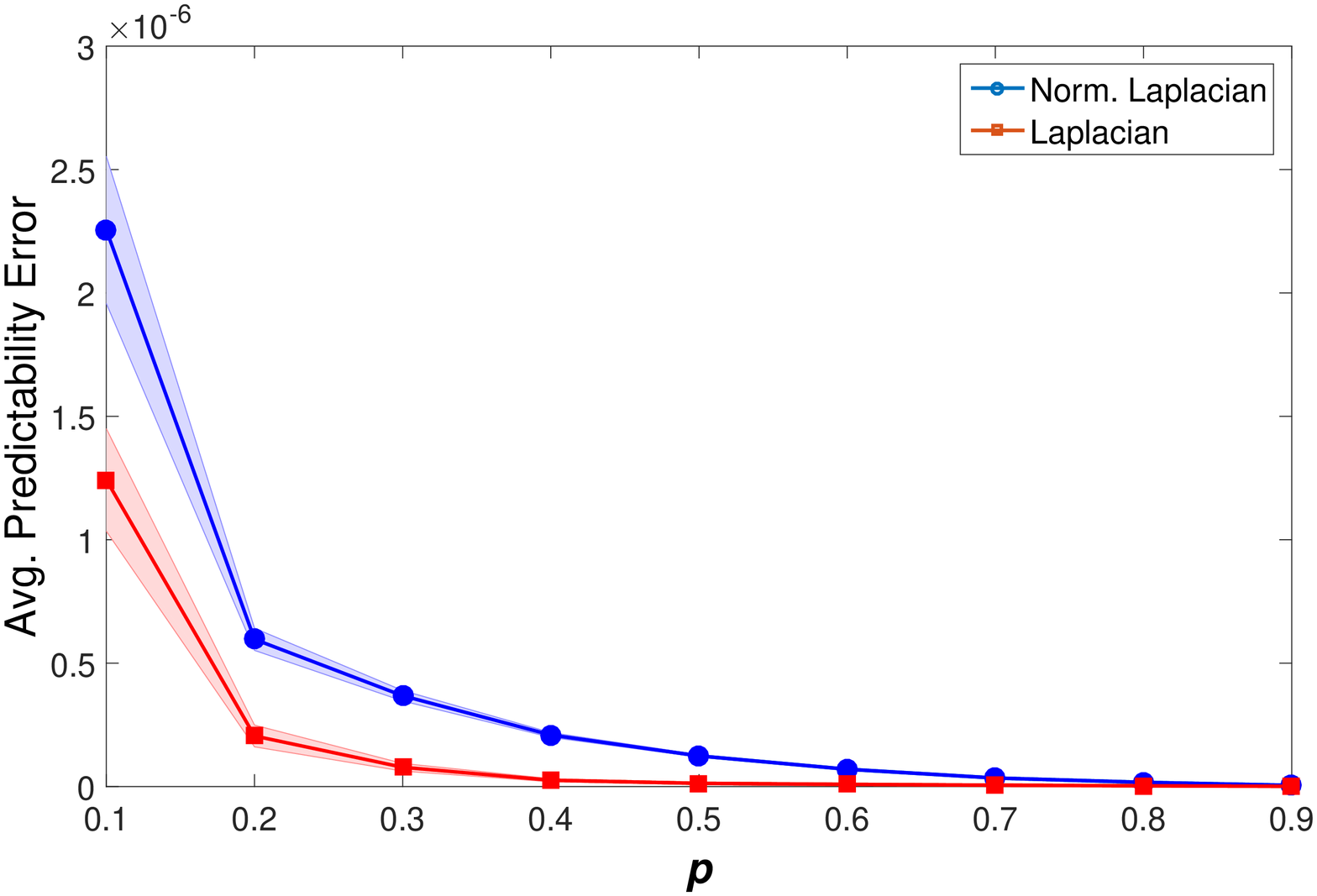}}
\hspace{0.01in}
\subfloat[\hspace{0.1cm}  Minimization]{\includegraphics[width=.49\textwidth]{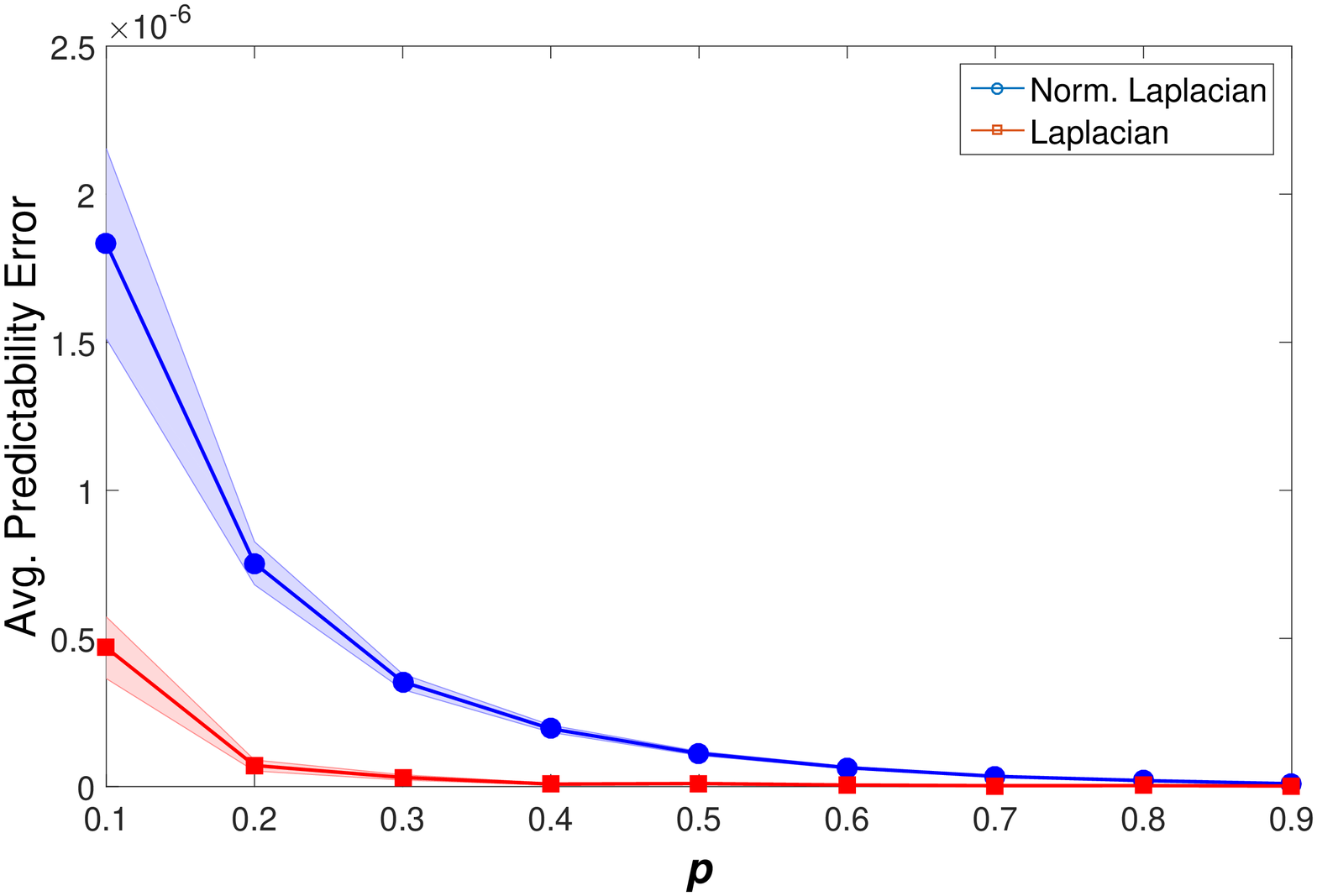}}
\hspace{0.01in}
\caption{Average predictability error ($\pm$ standard error) on Erd\"os-R\'enyi graphs.}
\label{fig:ERDOS}
\end{figure}

We then continue the correlation study on the set of scale-free graphs generated by the Preferential Attachment model. The results are shown in Fig.~\ref{fig:corr_scale}. On the one hand, when we consider the relationship between the approximate and the exact versions of the entropies, we observe a similar behaviour to that seen for the Erd\"os-R\'enyi graphs, with a strong correlation for both the Laplacian and the normalized Laplacian entropy with their quadratic approximations. Note that in this case, given a pair of values for $n$ and $m$, the number of edges of the generated graphs does not vary, so the observed effect cannot be explained by a varying edge set size.

On the other hand, in this case we observe a negative correlation between the two entropies, both in their exact and approximated versions. The correlation is stronger between the approximated entropies. Indeed, in Eq.~\ref{eq:laplacian_approximation} the term $-\sum_{u\in V} d_u^2$ prevails, whereas in Eq.~\ref{eq:nlaplacian_approximation} the leading term is $-\sum_{(u,v)\in E} \frac{1}{d_u\,d_v}$ . In other words, when a graph contains very high degree nodes, the Laplacian entropy becomes very small while the normalized counterpart tends to increase.

We conclude this correlation study on the small-world graphs generated by the Watts-Strogatz model. As for the scale-free graphs, note for a choice of the parameters $n$ and $k$ the number of edges in the generate graphs does not vary. Fig.~\ref{fig:corr_small} shows a stark contrast between the results obtained for $k>2$ and those obtained for $k=2$. To understand why this happens, recall that $k$ controls the number of neighbours for each node in the initial ring graph. The higher the value of $k$, the more robust the graph structure is to the edge flips that turn the ring into a small-world graph by reducing the average path length. The result is a quasi-regular ring lattice structure with relatively uniform degree where the approximate entropy. Since the approximated entropies only capture simple degree statistics, they are unable to capture the structural differences observed by the exact entropies, which go beyond structural information at degree level. As a consequence, the correlation between the exact entropies and the approximated ones decreases as $k$ increases, until the two are practically uncorrelated. However this does not happen when $k=2$. In fact, in this case the regularity of the initial ring graph is easily disrupted by the noise addition process, with the removal and addition of a few edges causing significant deviations from the initial lattice structure.

\subsubsection{Edge Predictability}
From the previous analyses it is clear that the quality of the quadratic approximations of exact entropies depend on the topology of the underlying graphs. We have also seen that the Laplacian and the normalized Laplacian entropies generally capture different types of structural information. We now take our investigation one step further and we look at entropic contribution of a single edge, when either the approximate entropy or the exact entropy ares used. Previous works~\cite{lockhart2016edge} have looked at the entropic content of an edge as a way to measure its centrality. More in general, our interest is again to understand how well the quadratic approximations of the Laplacian and normalized Laplacian entropies are able to capture the contributions of single edges to the overall graph entropy.

We generate three sets of synthetic graphs as described in \ref{sec:graph_model}. Let $G = (V,E)$ be one such graph with edge set $E \subseteq V \times V $, where $V$ denotes the node set. For each edge not in $E$, we calculate the increase in entropy obtained by adding that connection to the graph, both for the exact and the approximate form of the entropy. Let $list^{\mathcal{E}}$ and  $list^{\mathcal{A}}$ be two edge-indexed lists containing the values of the exact and the approximate entropic increases, respectively. Given $list^{\mathcal{A}}$, we choose the index $iM$ of the edge that leads to the maximum entropic increment, and the index $im$ of the edge that leads to the minimum entropic increment. With these indices to hand, we select the corresponding value of the entropic increment for the same edges in the exact case, i.e., $list^{\mathcal{E}}_{iM}$ and $list^{\mathcal{E}}_{im}$. Then, we define the predictability error for the edge that leads to the maximum entropic increment as
\begin{equation}
Prd_{M} = 1-\frac{list^{\mathcal{E}}_{iM}}{\max{(list^{\mathcal{E}})}}.
\end{equation}
Similarly, we define the predictability error for the edge that leads to the minimum entropic increment as
\begin{equation}
Prd_{m} = 1-\frac{\min{(list^{\mathcal{E}})}}{list{\mathcal{E}_{im}}}
\end{equation}
\noindent In both cases, the smaller the predictability error the better we are able to approximate the exact maximum using the quadratic entropy.

\begin{figure}[!t]
\centering
\subfloat[\hspace{0.1cm} Maximization]{\includegraphics[width=.49\textwidth]{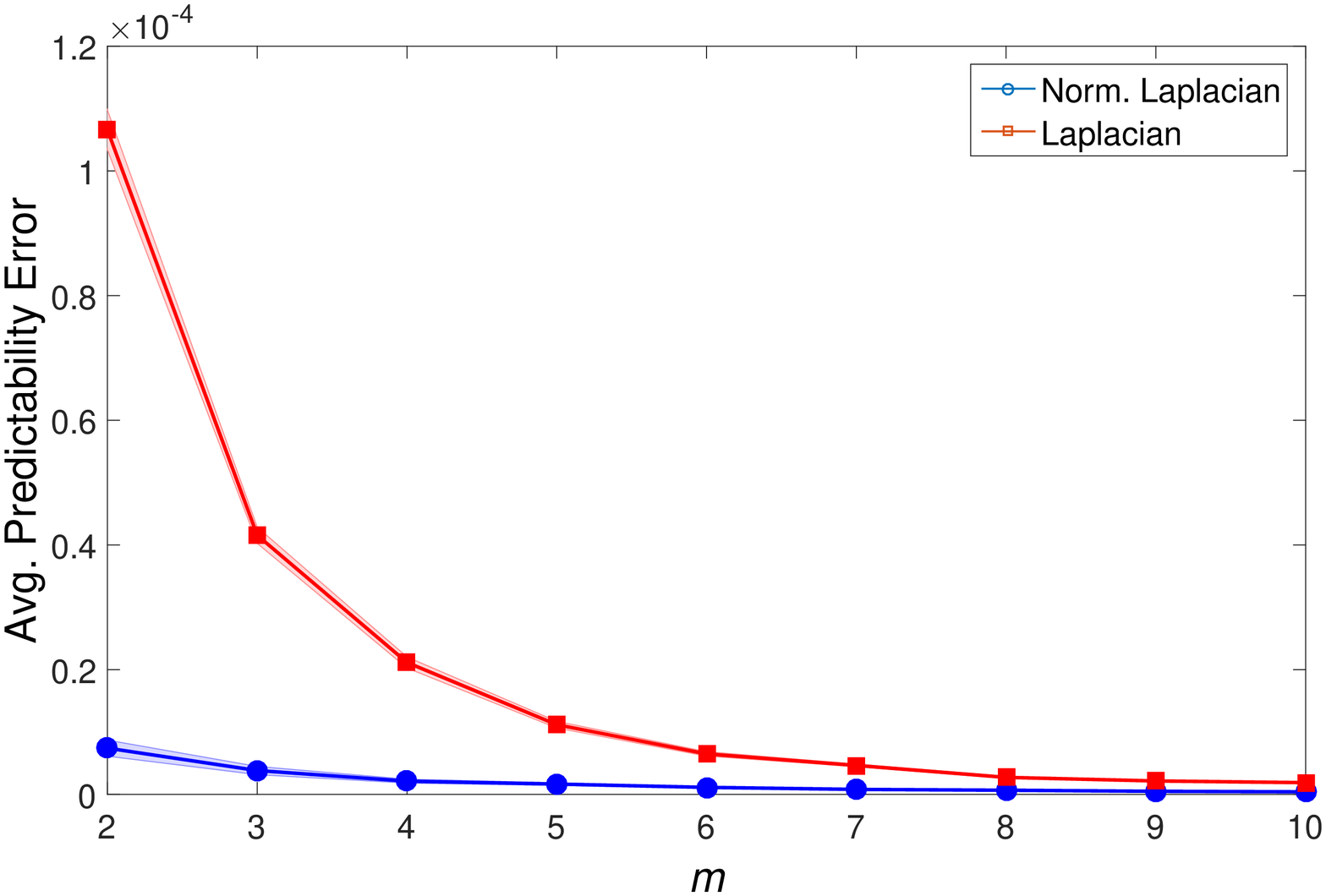}\label{fig:ratio_max_scale}}
\hspace{0.01in}
\subfloat[\hspace{0.1cm}  Minimization]{\includegraphics[width=.49\textwidth]{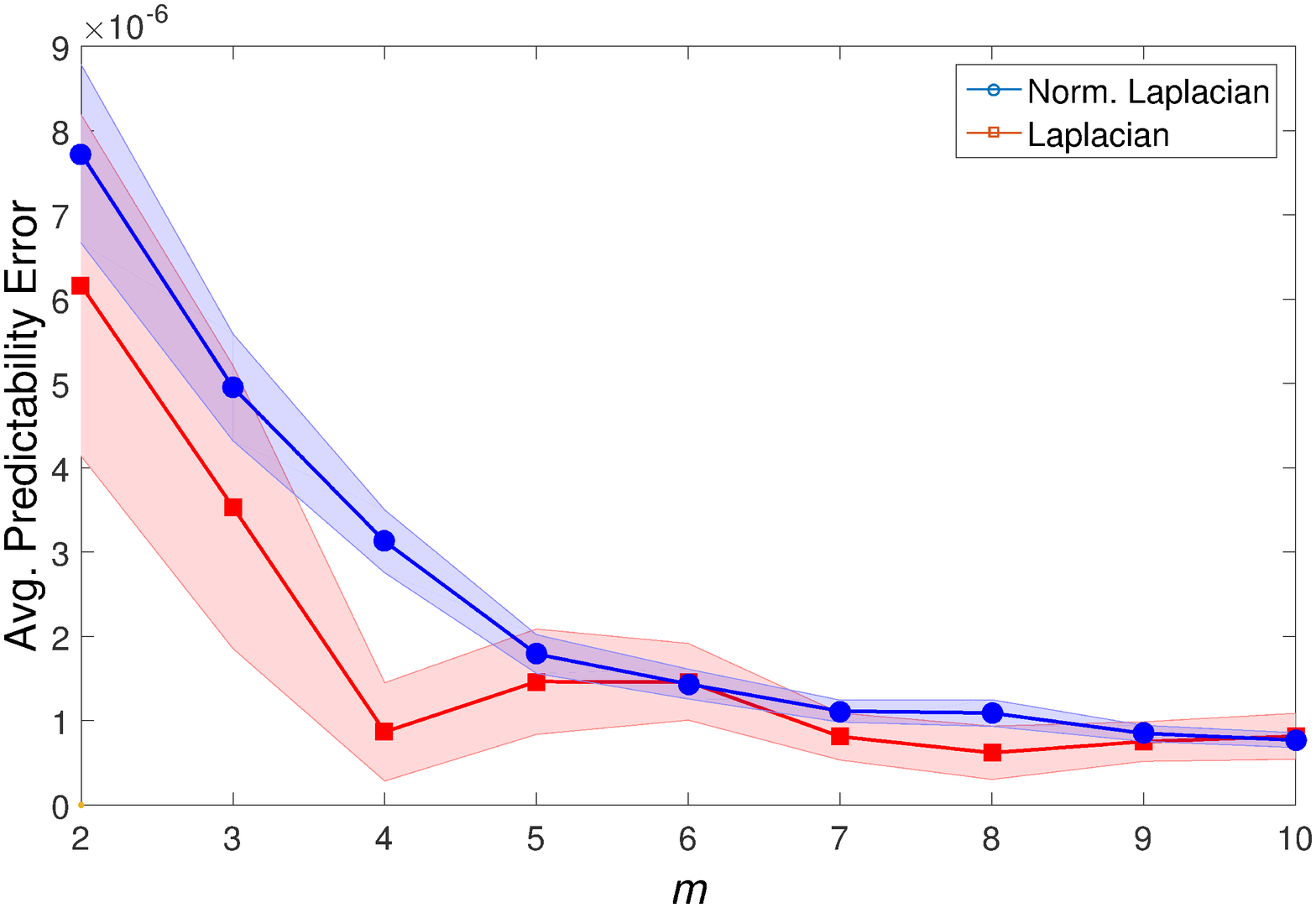}\label{fig:ratio_min_scale}}
\hspace{0.01in}
\caption{Average predictability error ($\pm$ standard error) on scale-free graphs.}
\label{fig:SCALEFREE}
\end{figure}

\begin{figure}[!t]
\centering
\subfloat[\hspace{0.1cm} Maximization]{\includegraphics[width=.49\textwidth]{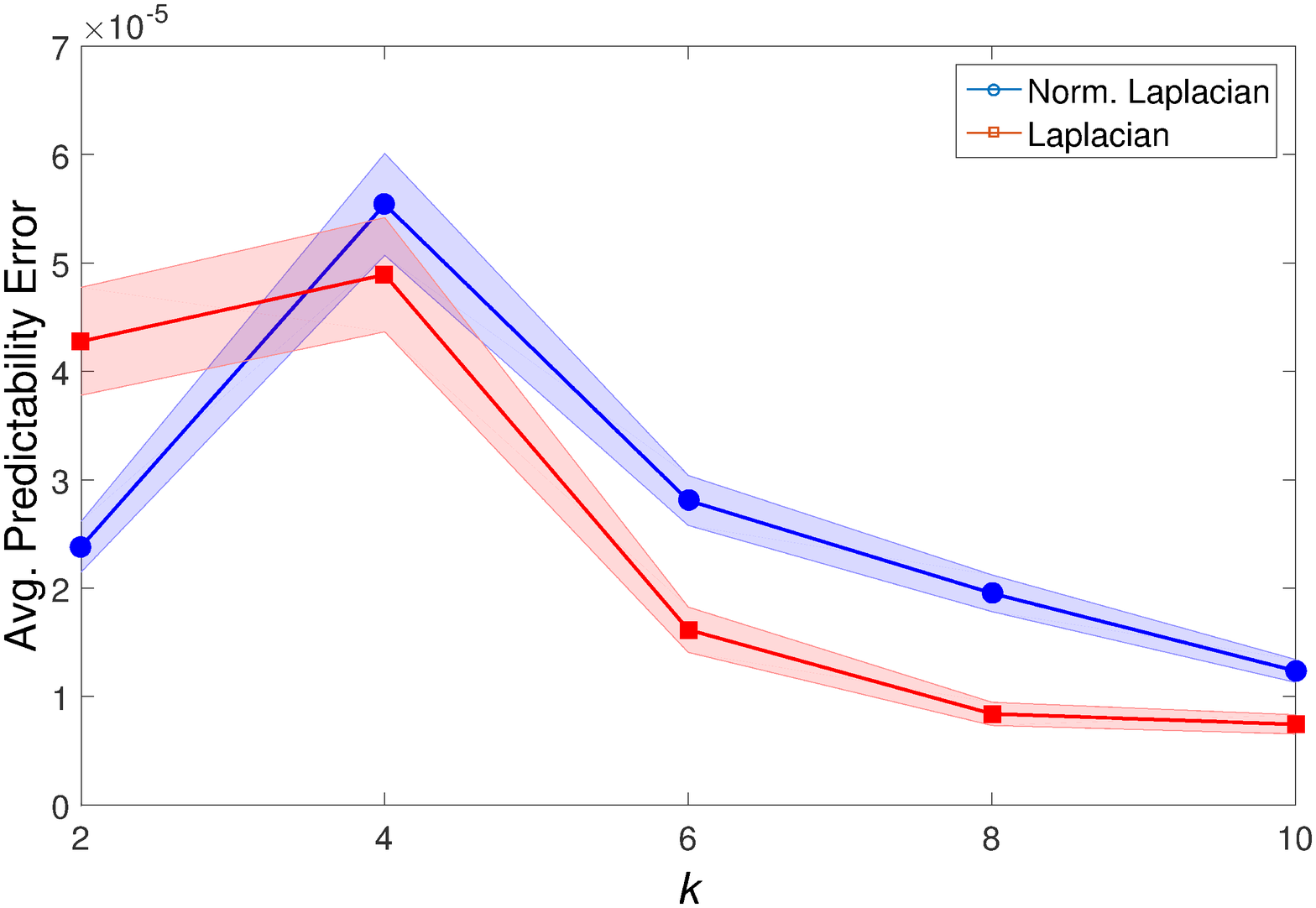}\label{fig:ratio_max_smallw}}
\hspace{0.01in}
\subfloat[\hspace{0.1cm}  Minimization]{\includegraphics[width=.49\textwidth]{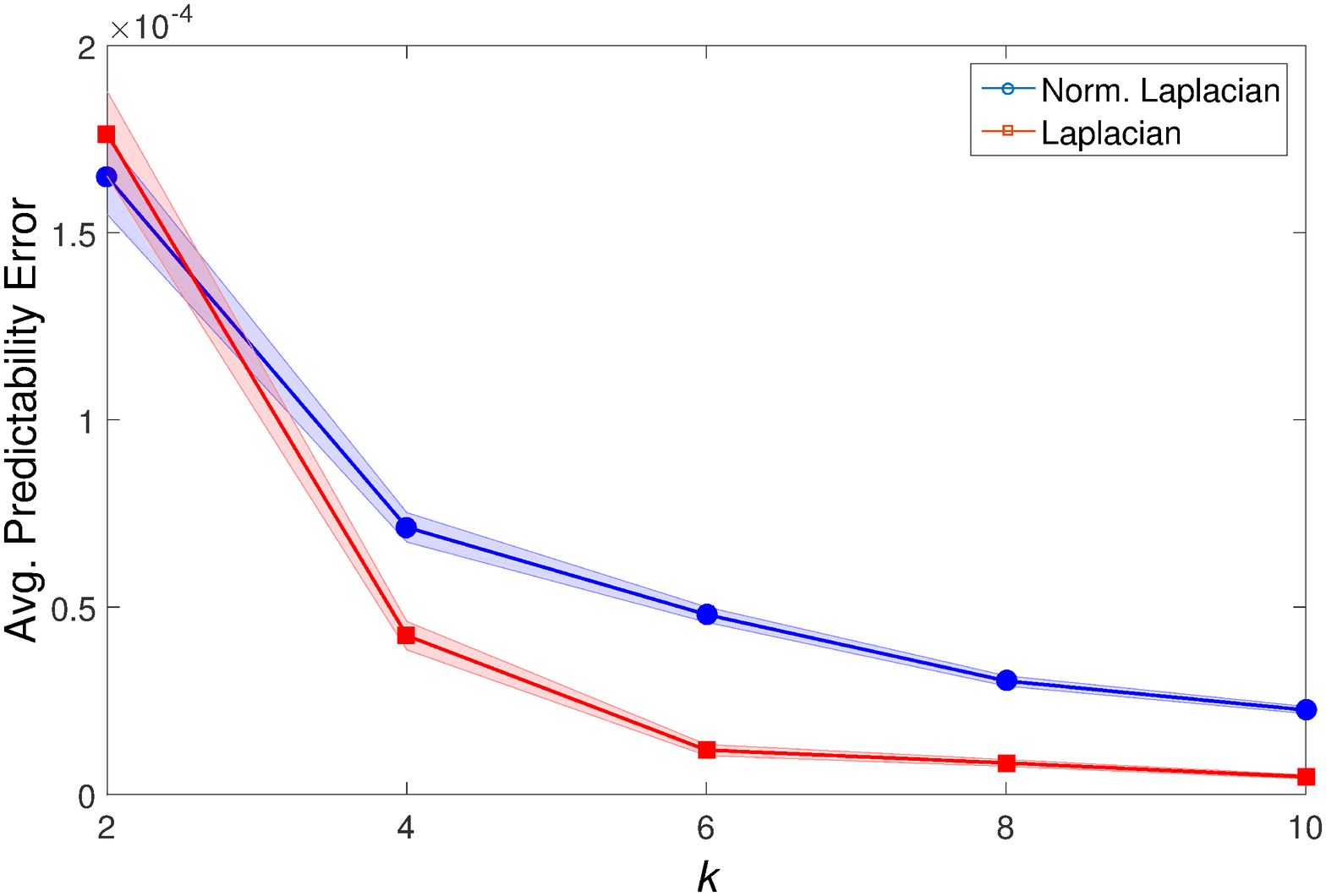}\label{fig:ratio_min_smallw}}
\hspace{0.01in}
\caption{Average predictability error ($\pm$ standard error) on small-world graphs.}
\label{fig:SMALL}
\end{figure}

Figs.~\ref{fig:ERDOS},~\ref{fig:SCALEFREE}, and~\ref{fig:SMALL}  show how the error changes as we vary the model parameters of the Erd\"os-R\'enyi, Preferential Attachment  and Watts-–Strogatz model, respectively. We observe that in general, regardless of the model, the error tends to decrease. In other words, as graphs become denser the number of non-existing edges decreases and thus it becomes easier to correctly identify the edges associated to the maximum (minimum) entropic increment. The only exception is that of the Watts-–Strogatz model, where the error first increases and then decreases, as shown in Fig.~\ref{fig:ratio_max_smallw}. Note that this fits with our previous observation of a higher correlation between the approximate and exact entropies for this type of graphs when $k=2$. However while in the correlation study we observe that for $k>2$ the correlation decreases, in this case the graph densification appears to dominate and to lead to a decrease of the observed predictability error.

Interestingly, we see that the predictability error is significantly lower when we are trying to approximate the Laplacian entropy as opposed to the normalized Laplacian entropy, with the only exception being that of scale free graphs. This is probably due to an ambiguity created by the approximate version. Indeed, according to Eq. \ref{eq:laplacian_increment}, in order to maximize the approximate Laplacian entropy, two nodes with low degree should be linked. However, when we take into account networks whose degree distribution follows a power law, the choice gets nearly random. More specifically, the formula does not consider the node neighbourhood and thus two nodes belonging to the same hub or two nodes belonging to different hubs may be indistinctly connected. However, the exact version could be making a distinction and favouring the connection between nodes belonging to different hubs. This in turn could be explained as an effort to connect distant nodes, already observed in Fig.~\ref{fig:heuristic}. Such an ambivalence between the approximate and exact Laplacian entropy eventually leads to a poor predictability. On the other hand, Fig.~\ref{fig:ratio_min_scale} shows no substantial difference between the two entropies because there is less uncertainty in choosing which pair of nodes (with high degree) to connected.

\subsection{Experiments on Real-world Networks}
We conclude our analysis by considering networks extracted from real-world complex systems.

\subsubsection{Datasets}

{\bfseries \textit{Dataset} 1:} the \textsf{USSM} dataset is extracted from a database consisting of the daily prices of 431 companies in 8 different sectors from the New York Stock Exchange (NYSE) and the Nasdaq Stock Market (NASDAQ).  To construct the dynamic network, 431 stocks with historical data from January 1995 to December 2016 are selected. The dataset is arranged to be around 5500 trading days.  In order to build an evolving network, a time window of 28 days is used and it is moved along time to obtain a sequence (from day 29 to day 5500); in this way, each temporal window contains a time-series of the daily return stock values over a 28 day period. Afterward, trades among the different stocks are set as a network. For each time window, we compute the cross correlation coefficients between the time-series for each pair of stocks and create connections between them if the absolute value of the correlation coefficient exceeds a threshold. The result is a stock market network which changes over the time, with a fixed number of 431 nodes and varying edge structure for each of trading days.
\\
{\bfseries \textit{Dataset} 2:} this dataset collects protein-protein interaction \texttt{PPI} networks related to histidine kinase \footnote{Lars J Jensen, Michael Kuhn, Manuel Stark, Samuel Chaffron, Chris Creevey, Jean Muller, Tobias Doerks, Philippe Julien, Alexander Roth, Milan Simonovic, et al. String global view on proteins and their functional interactions in 630 organisms. Nucleic acids research 2009}. Histidine kinase is a key protein in the development of signal transduction. The graphs describe the interaction relationships between histidine kinase in different species of bacteria. If two proteins (graph nodes) have direct (physical) or indirect (functional) association, they are connected by an edge. PPIs are collected from 5 different kinds of bacteria with the following evolution order (from older to more recent): 4 PPIs from \textit{Aquifex aelicus} and 4 PPIs from \textit{Thermotoga maritima}, 52 PPIs from Gram-Positive \textit{Staphylococcus aureus}, 73 PPIs from Cyanobacteria \textit{Anabaena variabilis} and 40 PPIs from Proteobacteria \textit{Acidovorax avenae}. 

\begin{figure}[t!]
\subfloat{\includegraphics[width=0.22\textwidth]{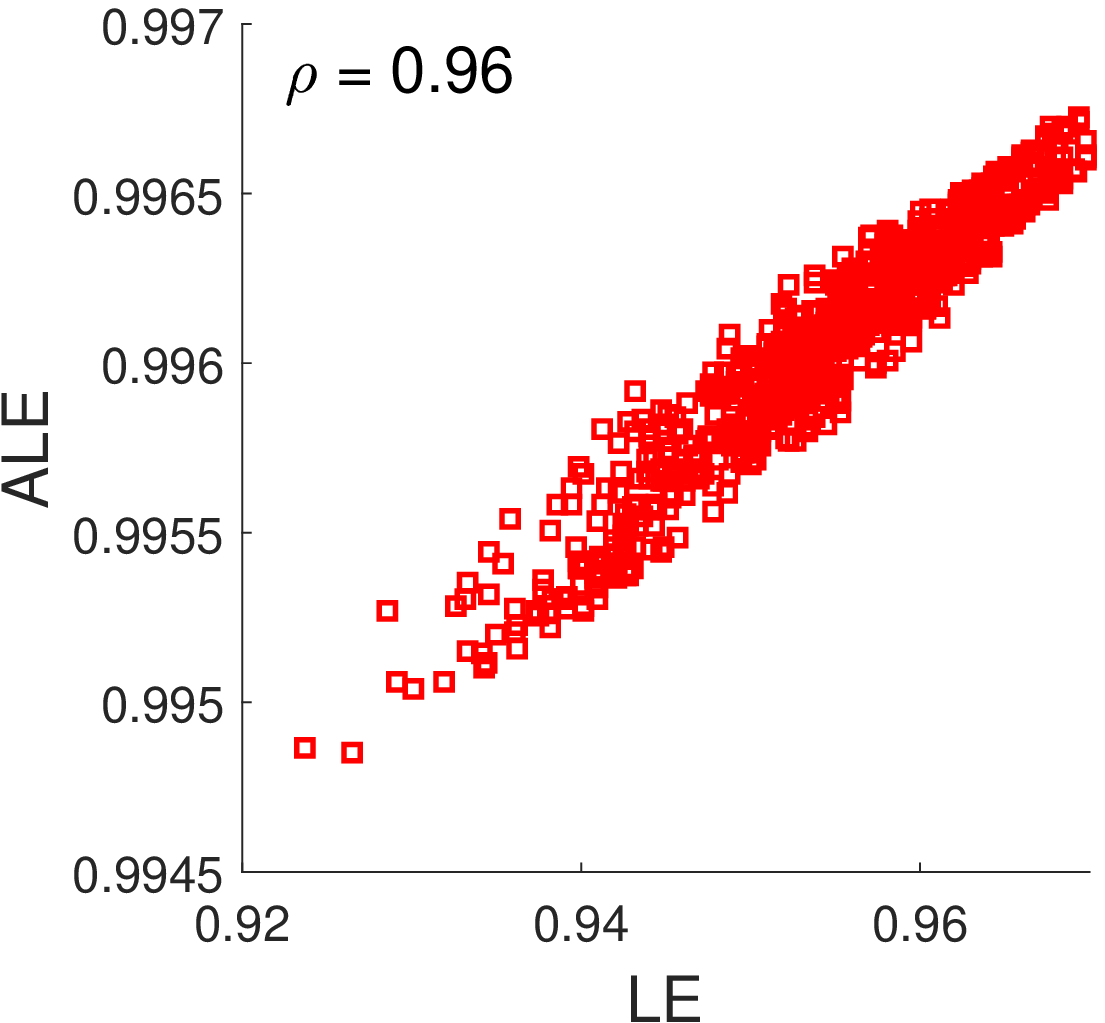}}
\subfloat{\includegraphics[width=0.22\textwidth]{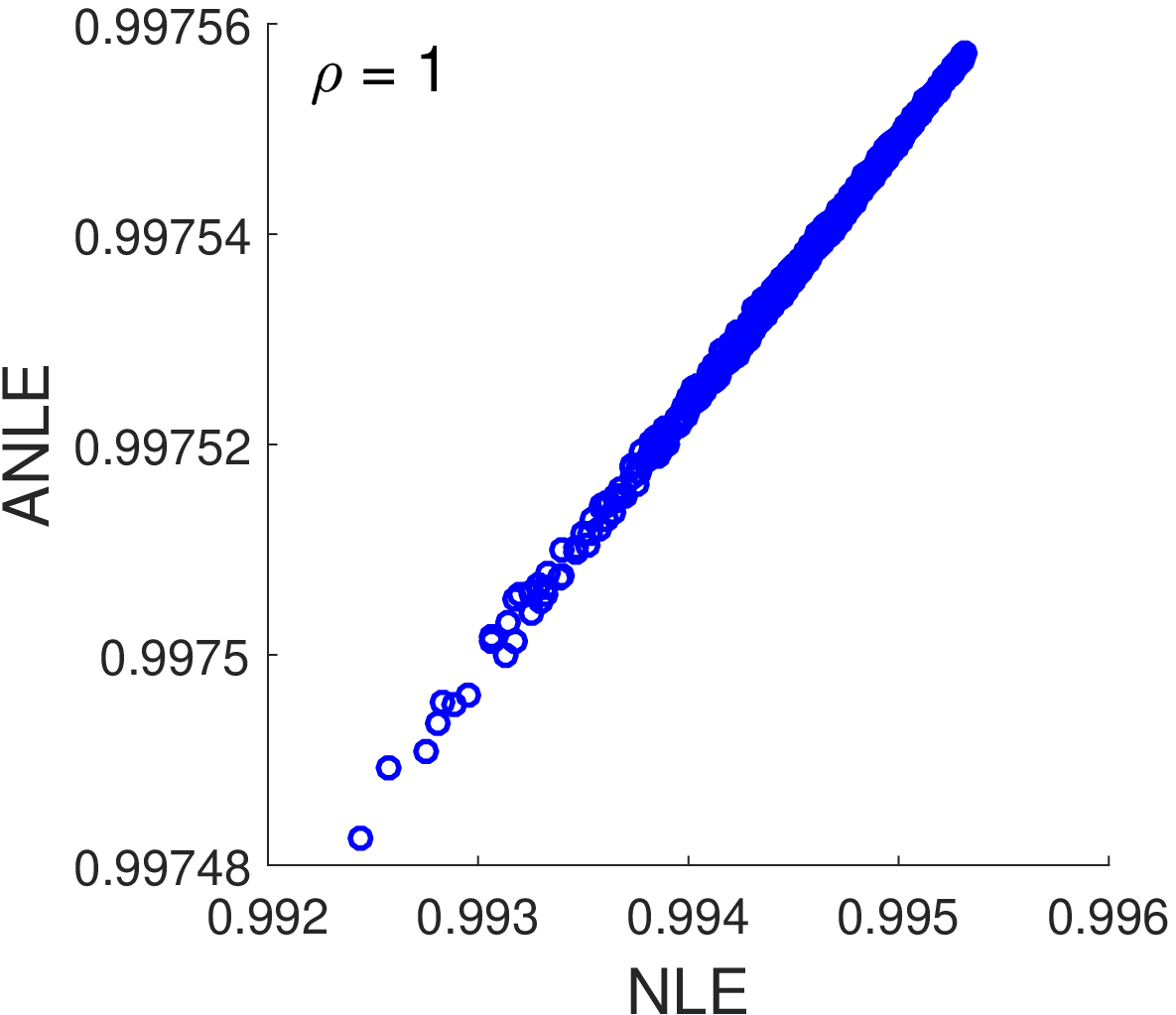}}
\subfloat{\includegraphics[width=0.22\textwidth]{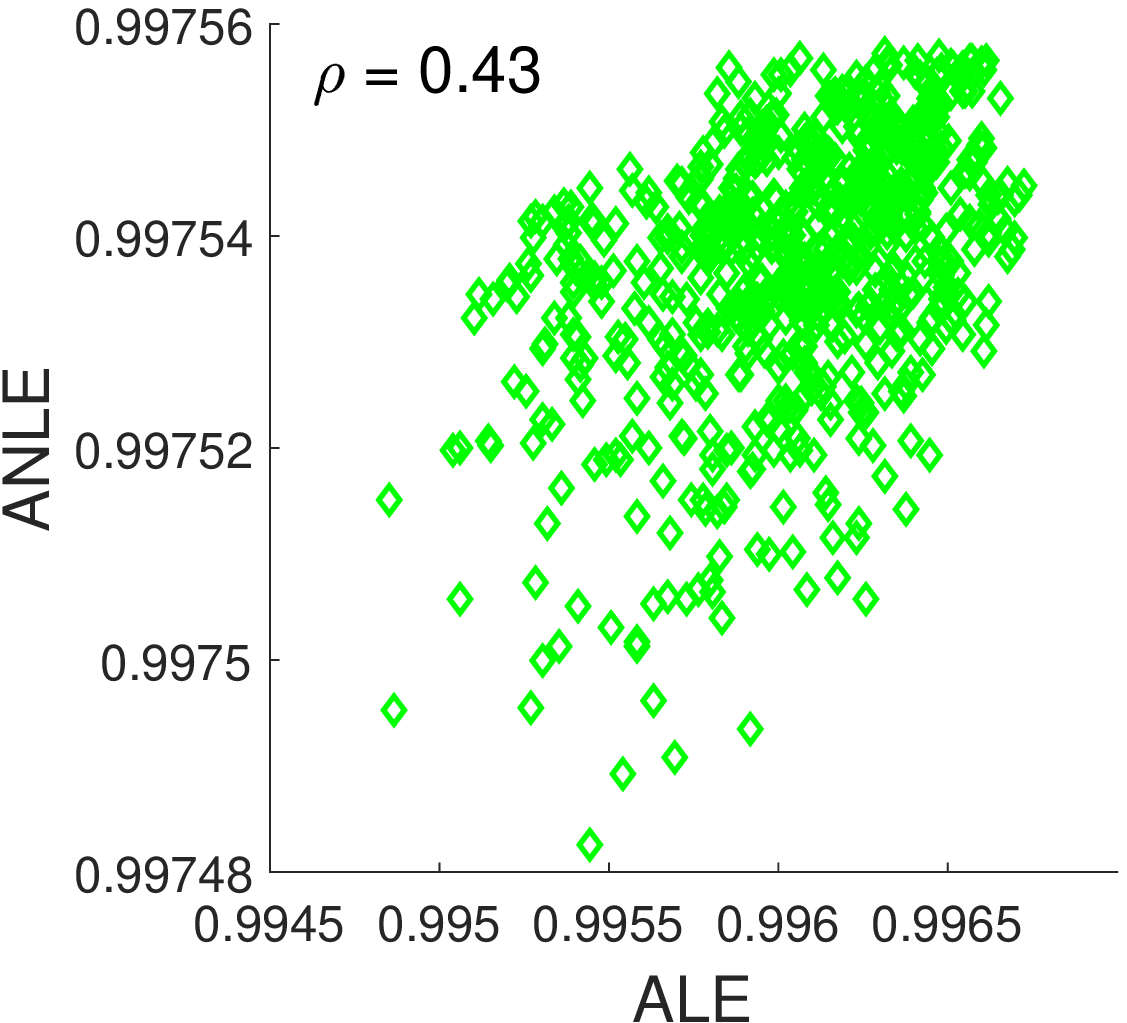}}
\subfloat{\includegraphics[width=0.22\textwidth]{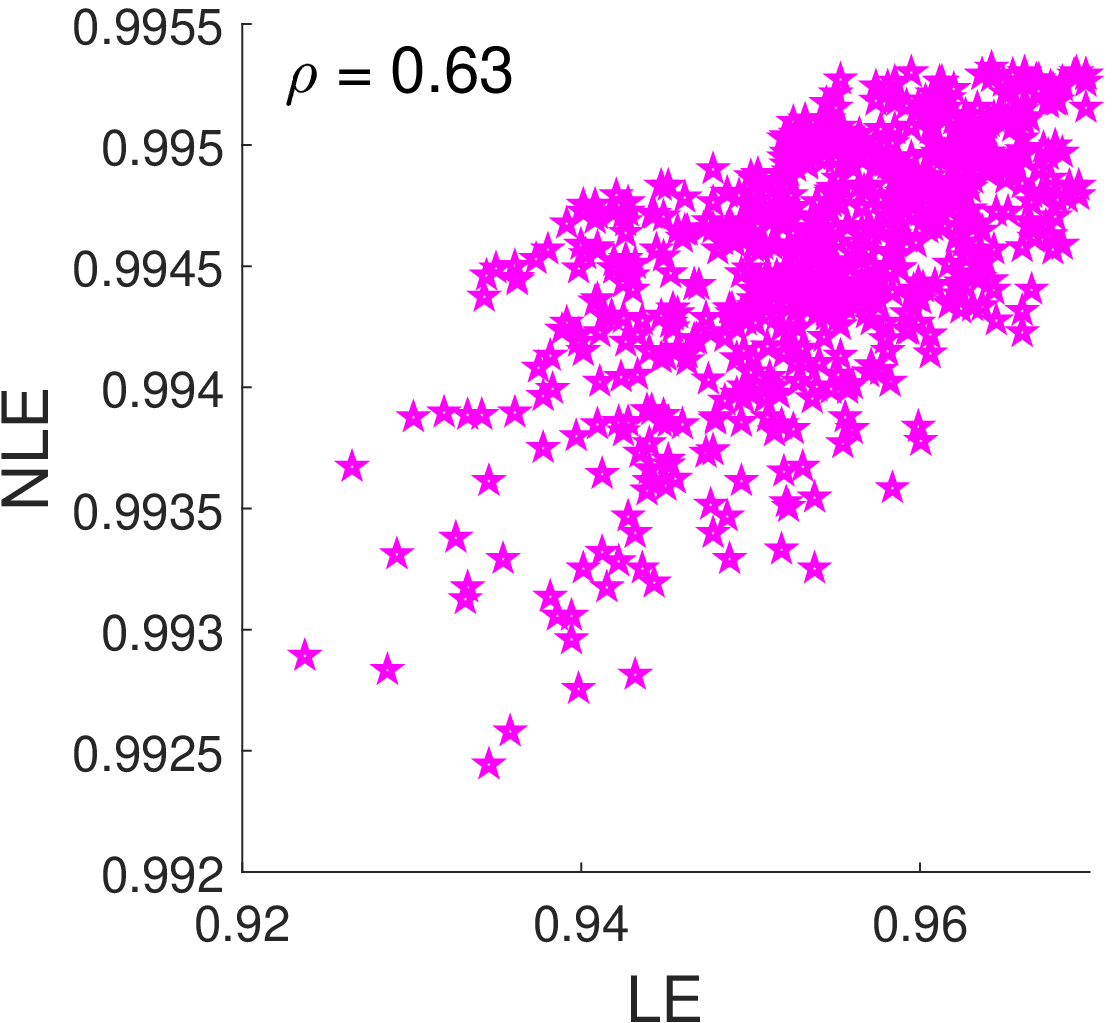}}

\subfloat{\includegraphics[width=0.22\textwidth]{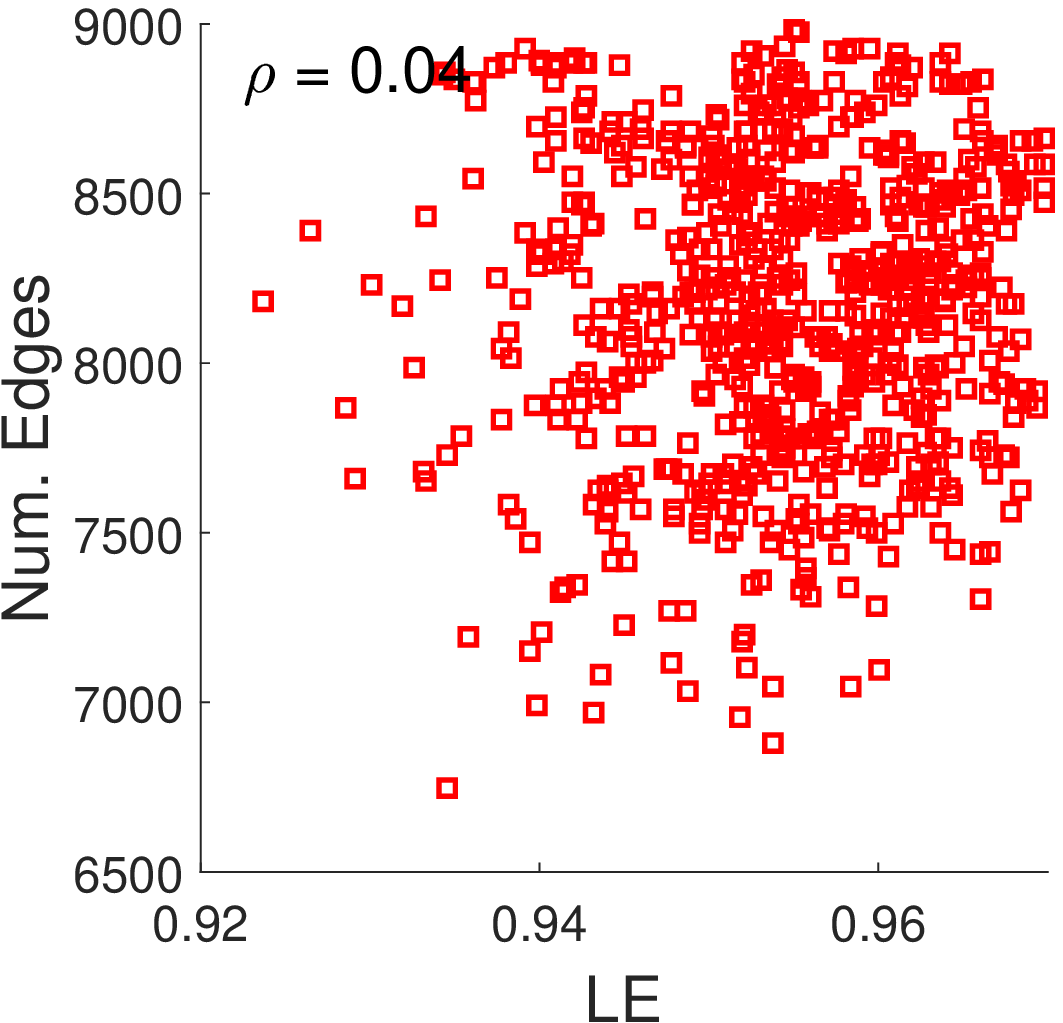}}
\subfloat{\includegraphics[width=0.22\textwidth]{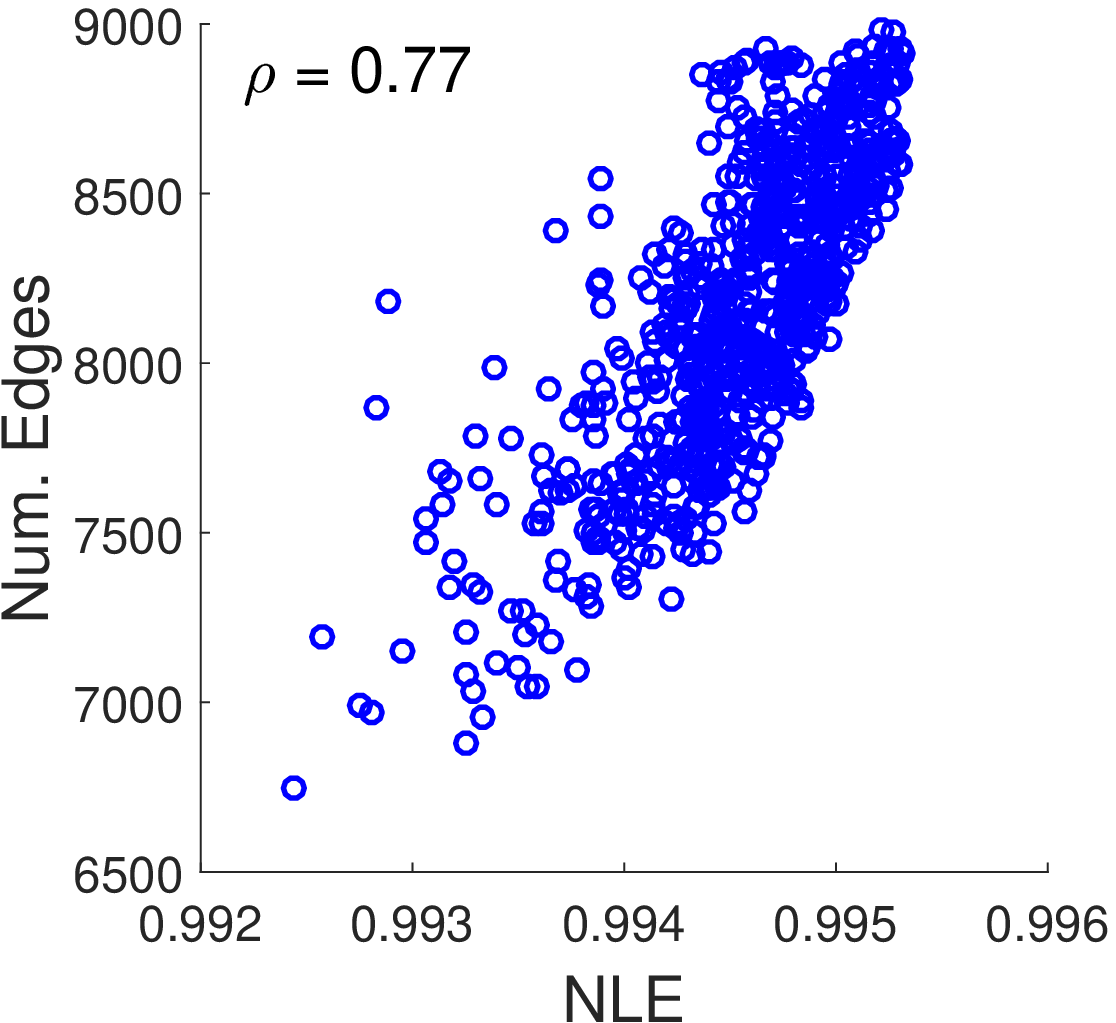}}
\subfloat{\includegraphics[width=0.22\textwidth]{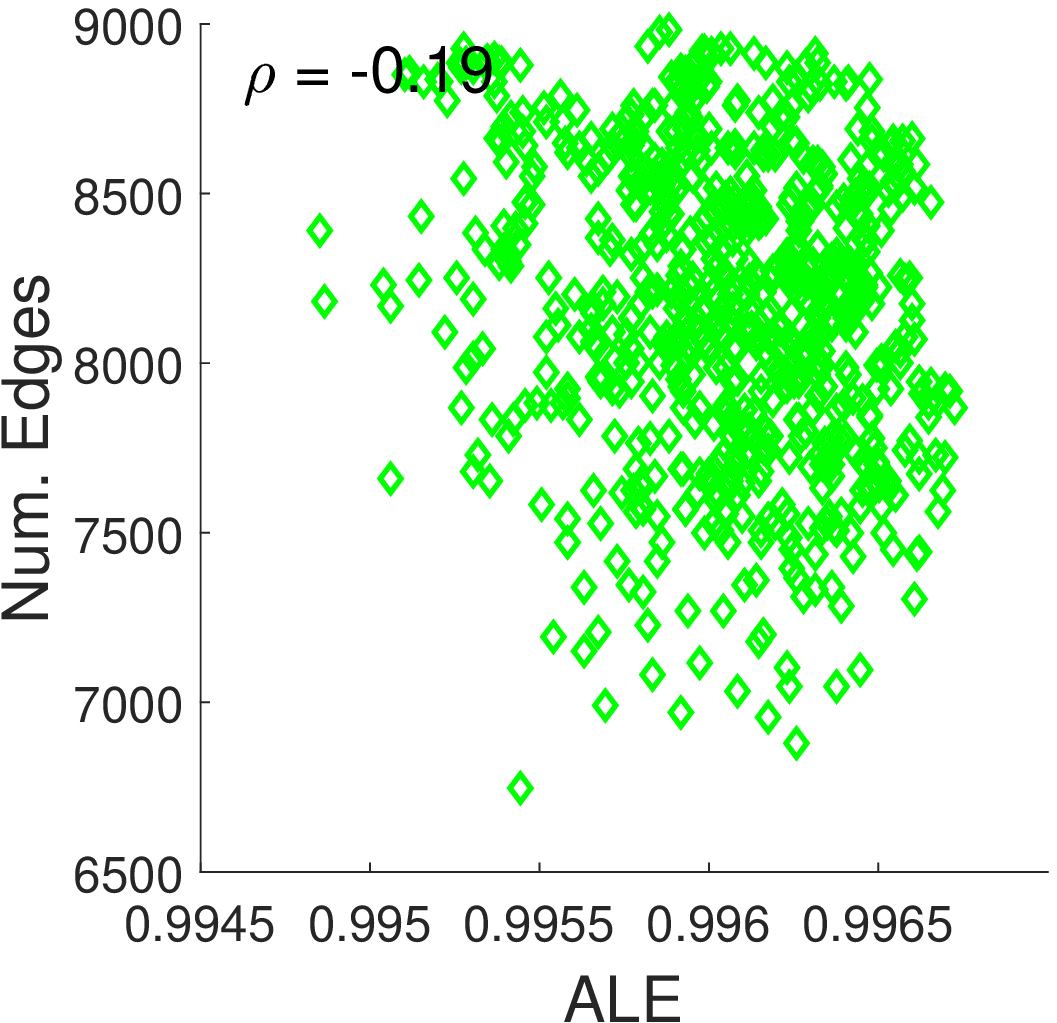}}
\subfloat{\includegraphics[width=0.22\textwidth]{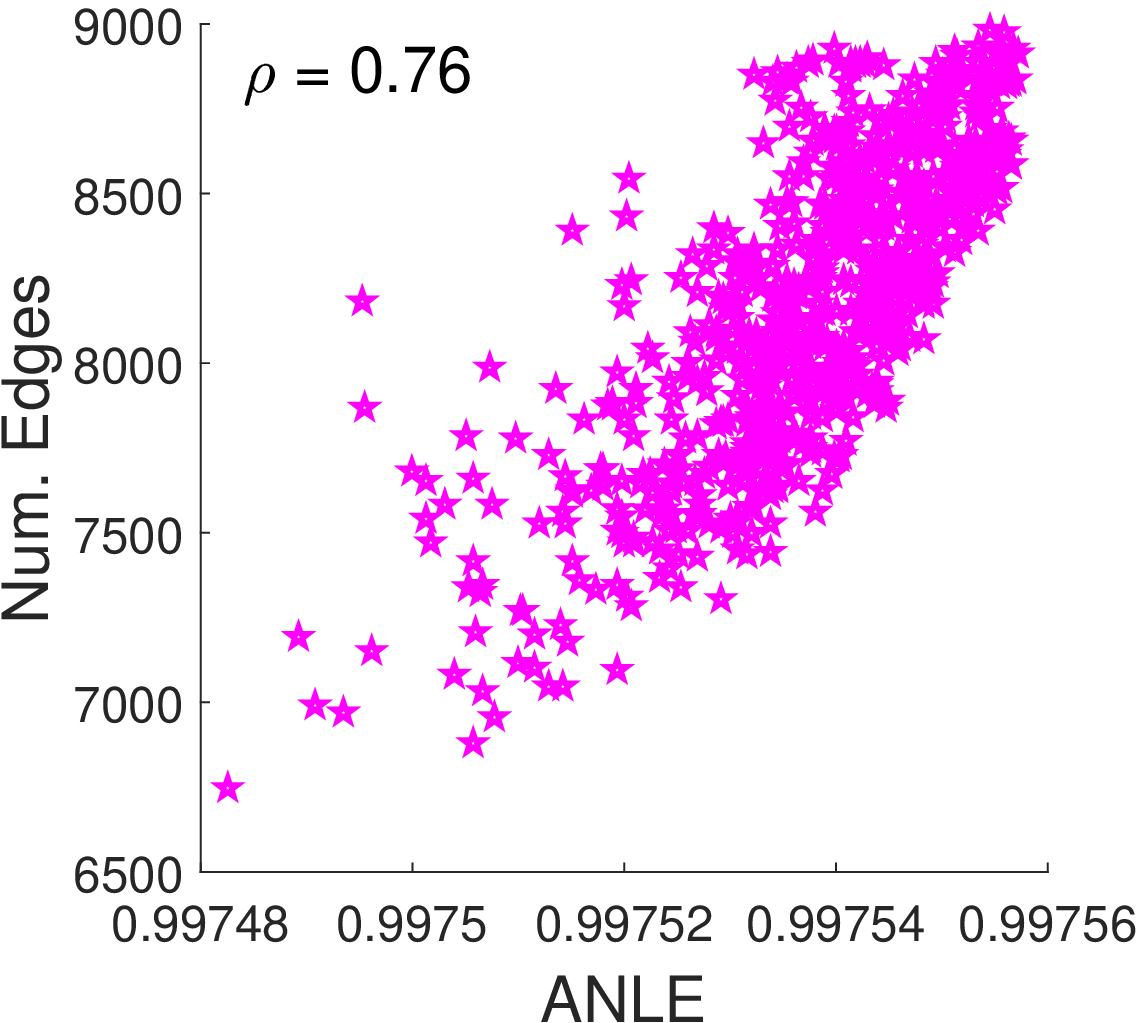}}
\caption{Correlation between the entropies (top) and entropy and number of edges (bottom) on the \textsf{USSM} dataset.}
\label{fig:ussm_corr}
\end{figure}

\subsubsection{Correlation Analysis}
The \textsf{USSM} dataset contains a time-evolving complex network  consisting of graphs having components of different sizes. Thus, we selected only some defined instances among the available ones. Specifically, we chose 707 samples. Each sample has to satisfy two requirements: a) being a connected graph, b) having maximum size (431 nodes). The correlation plots between the entropies are shown in Fig.~\ref{fig:ussm_corr} (top), where $\rho$ denotes the Pearson correlation coefficient. We first observe that the correlation is always strong. This is presumably due to the fact all instances belong to the same time-varying process, making them intrinsically correlated to each other. However, it is worth recalling that another factor may be causing this dependence. We have already stressed that the entropy can be influenced by the volume of a graph (i.e., the number of node it contains) as well as by its density (i.e., the number of edges it contains). While in this case the volume is fixed, the density changes over time. Indeed, in Fig.~\ref{fig:ussm_corr} (bottom) we see that the normalized Laplacian entropy of is highly correlated with the graphs density, in accordance to what already observed in Fig.~\ref{fig:corr_erdos}.

\begin{figure}[t!]  
\subfloat{\includegraphics[width=0.22\textwidth]{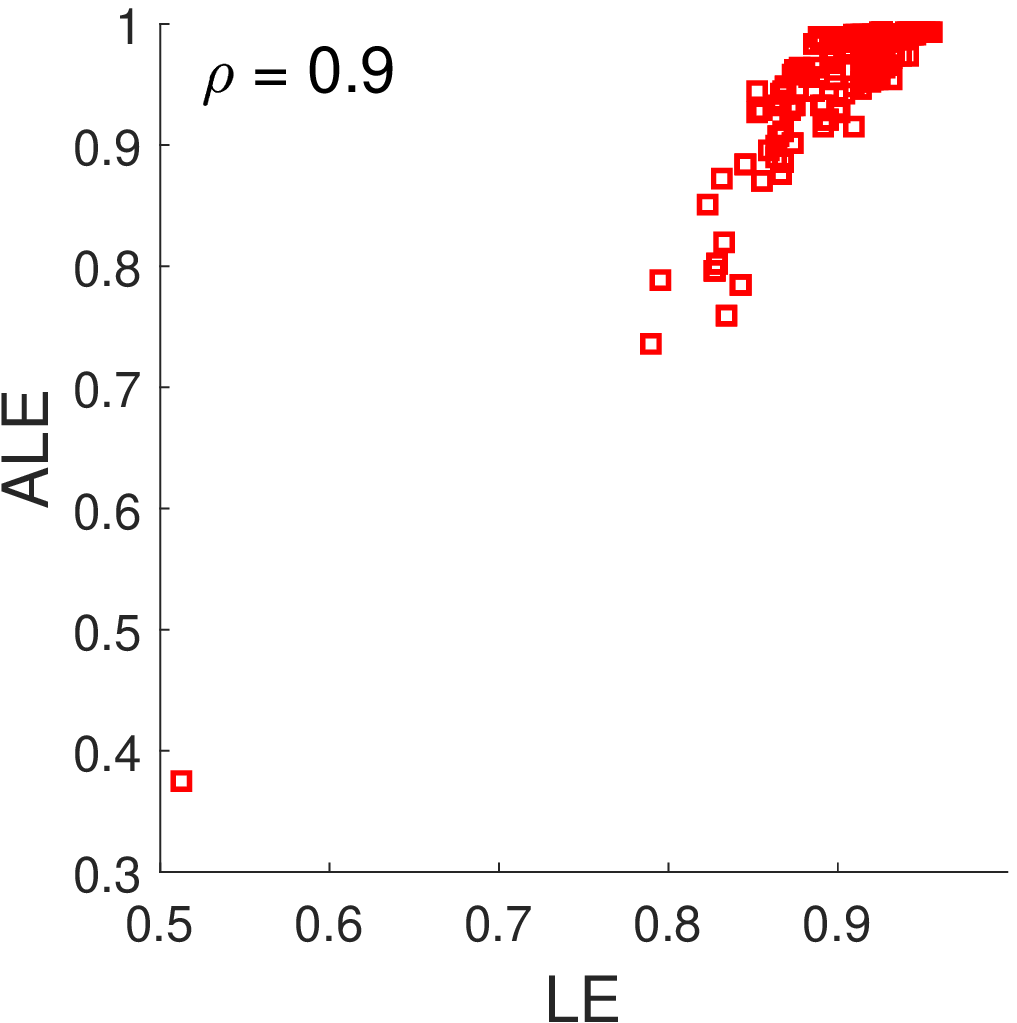}}
\subfloat{\includegraphics[width=0.22\textwidth]{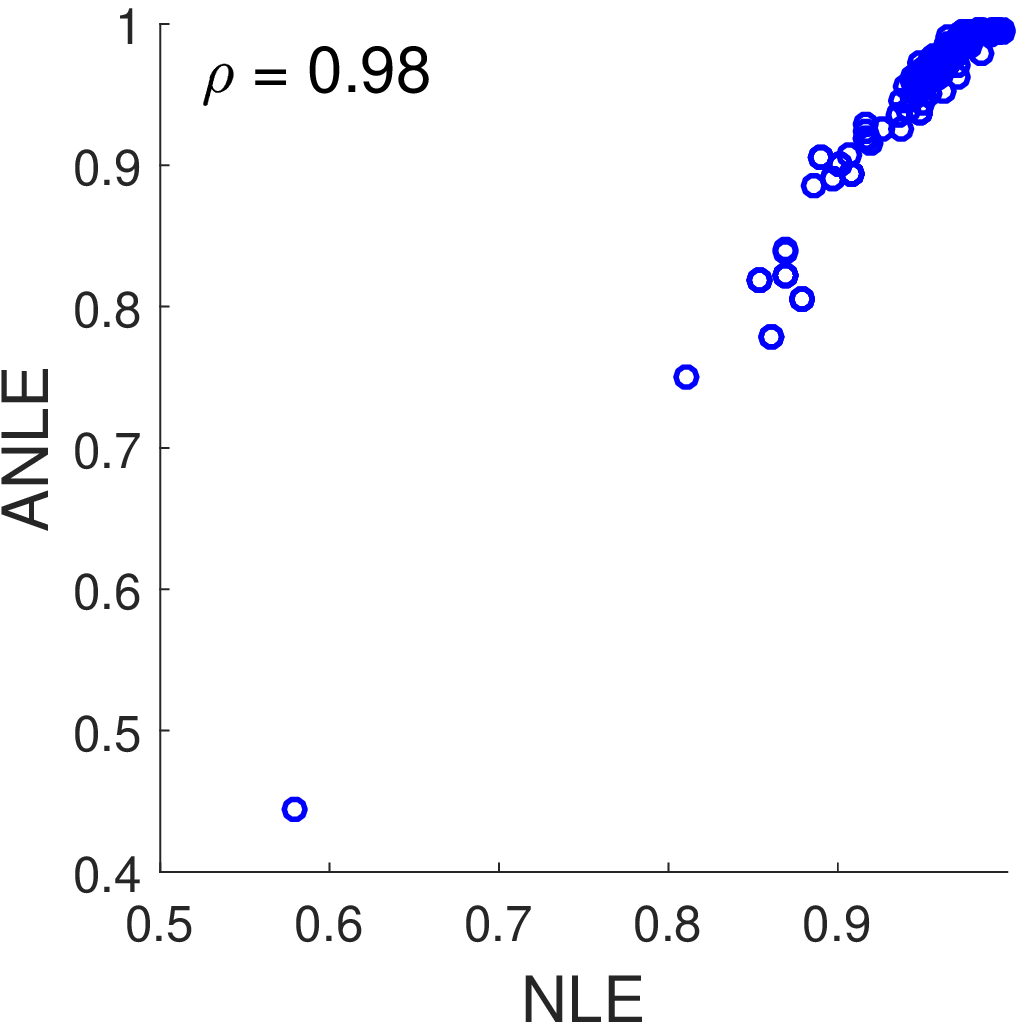}}
\subfloat{\includegraphics[width=0.22\textwidth]{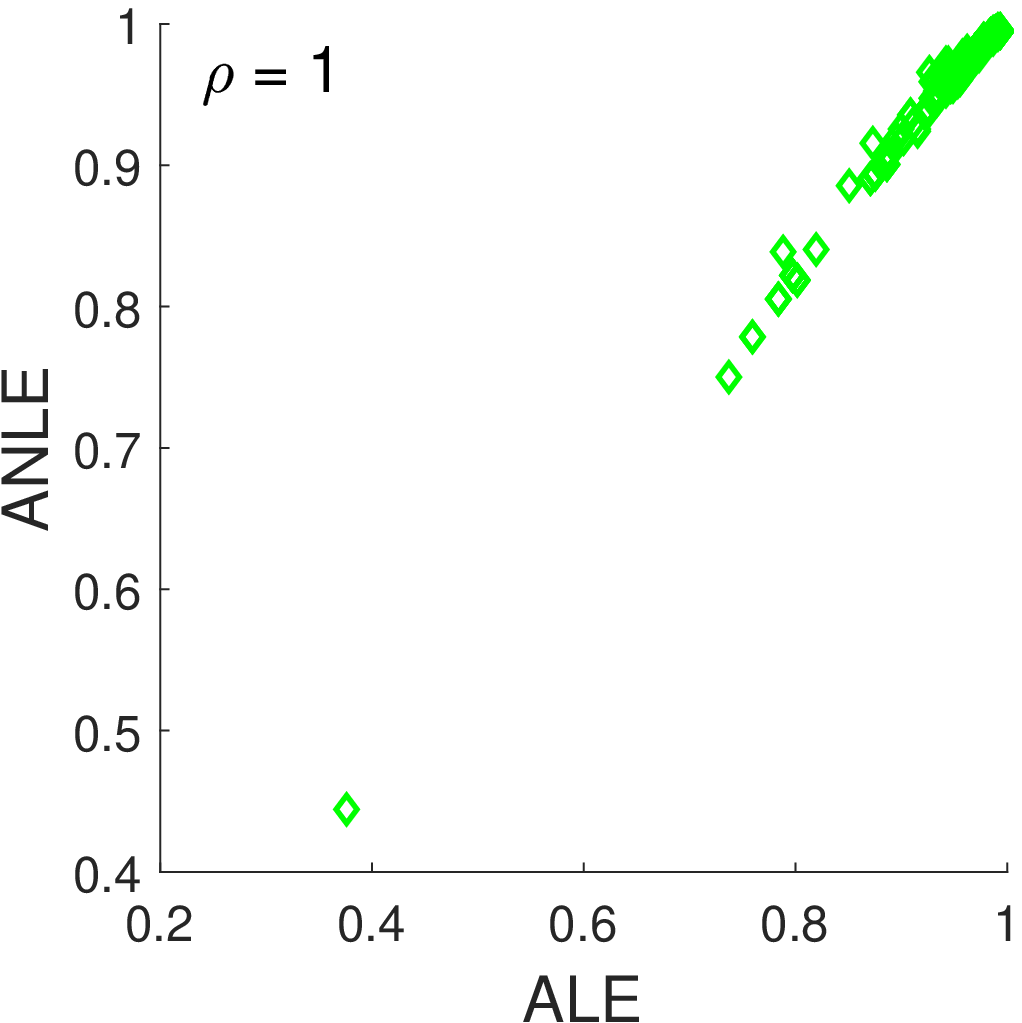}}
\subfloat{\includegraphics[width=0.22\textwidth]{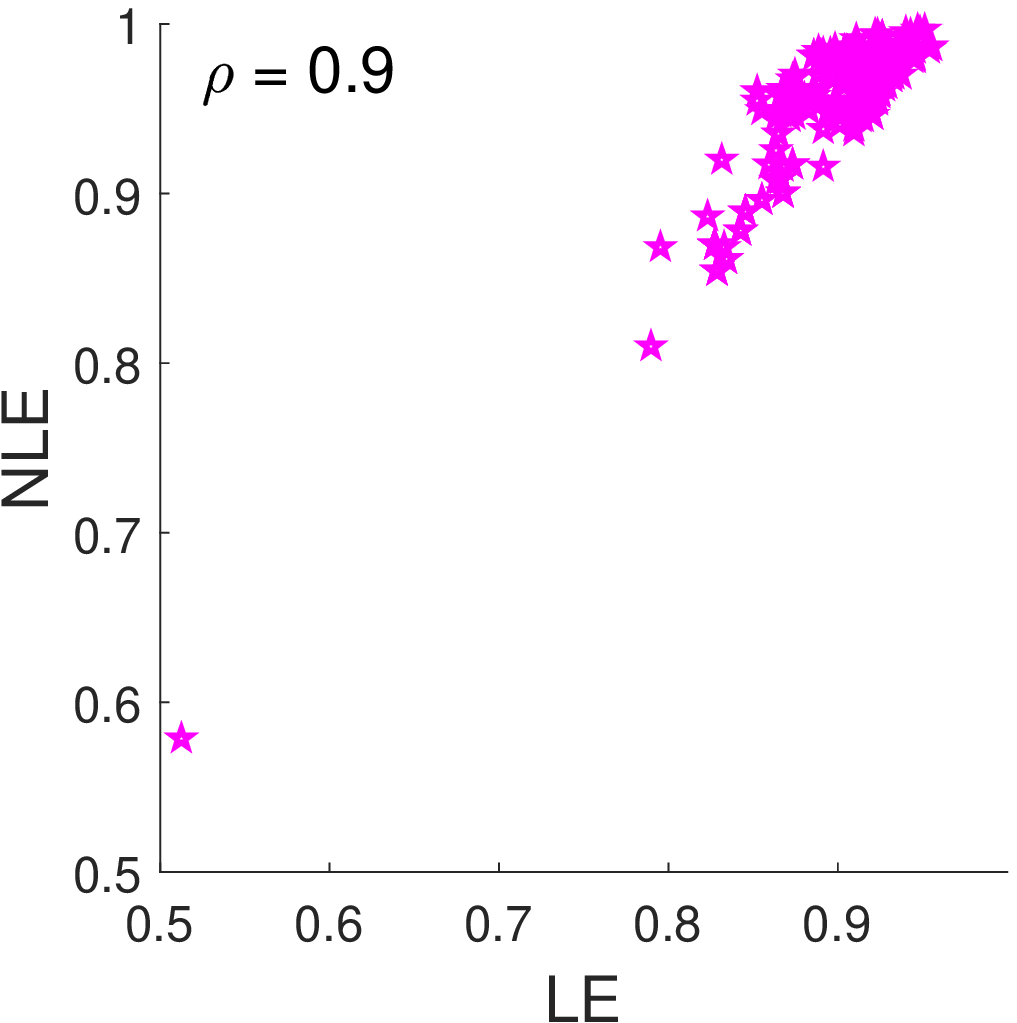}}

\subfloat{\includegraphics[width=0.22\textwidth]{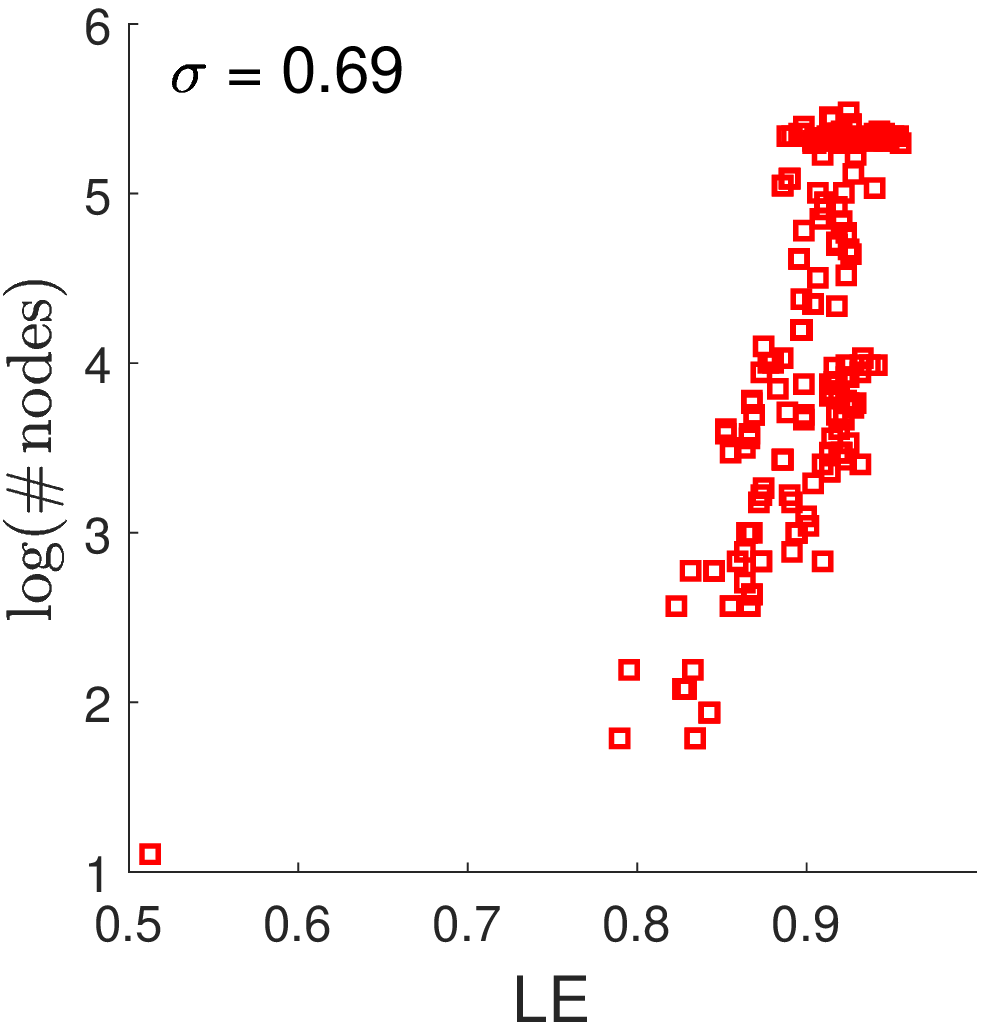}}
\subfloat{\includegraphics[width=0.22\textwidth]{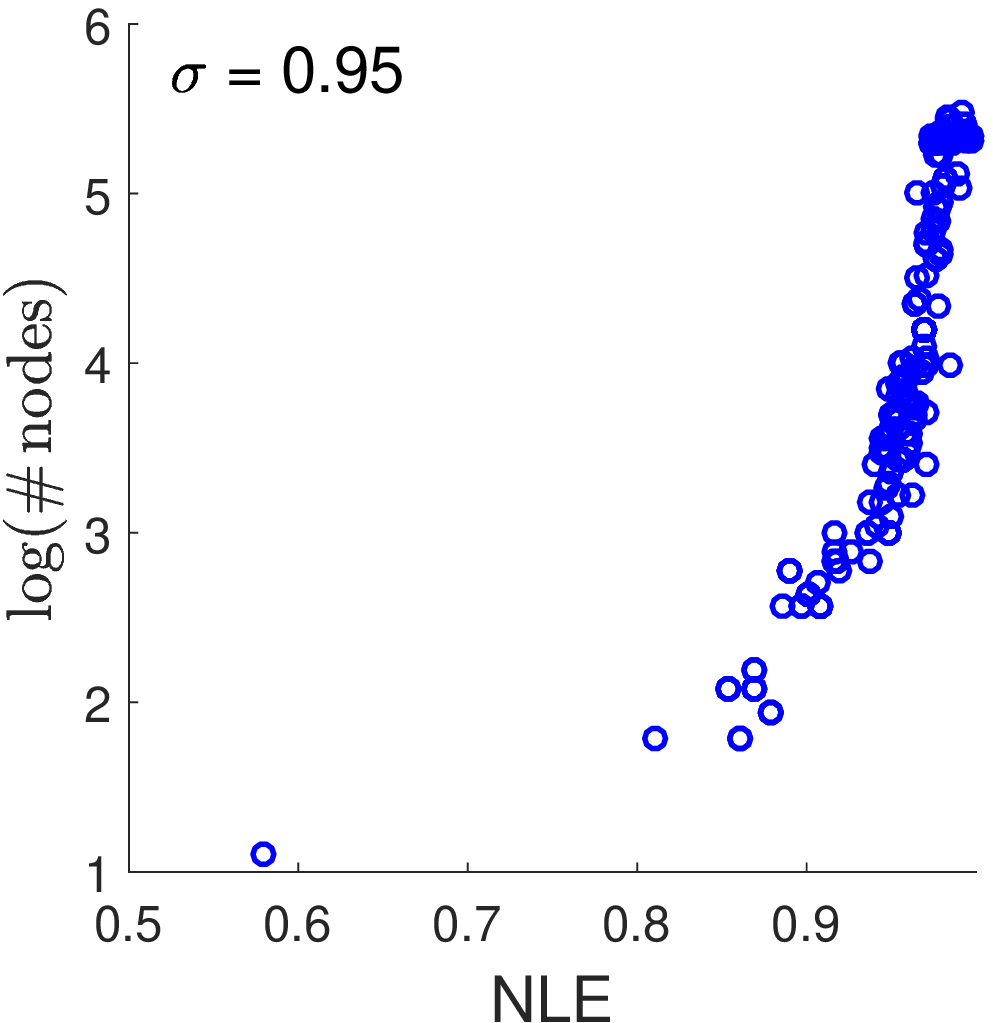}}
\subfloat{\includegraphics[width=0.22\textwidth]{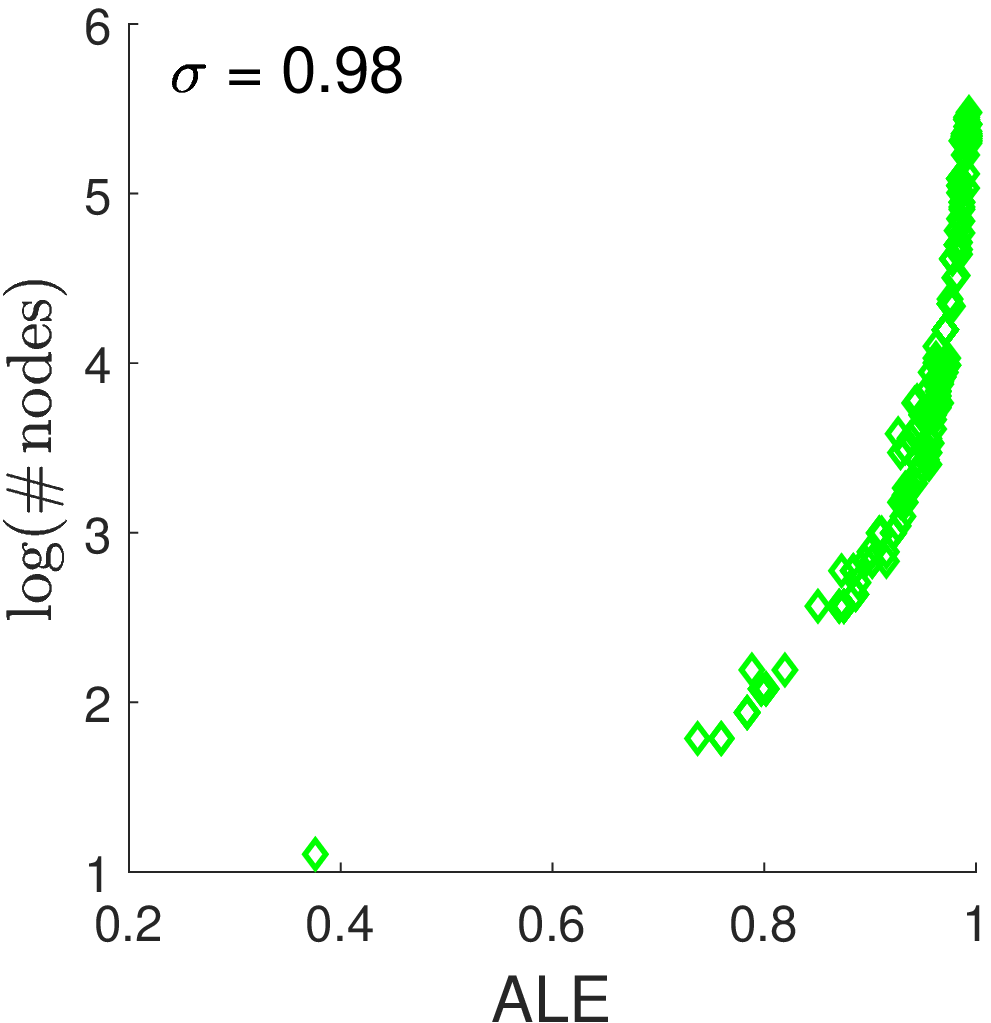}}
\subfloat{\includegraphics[width=0.22\textwidth]{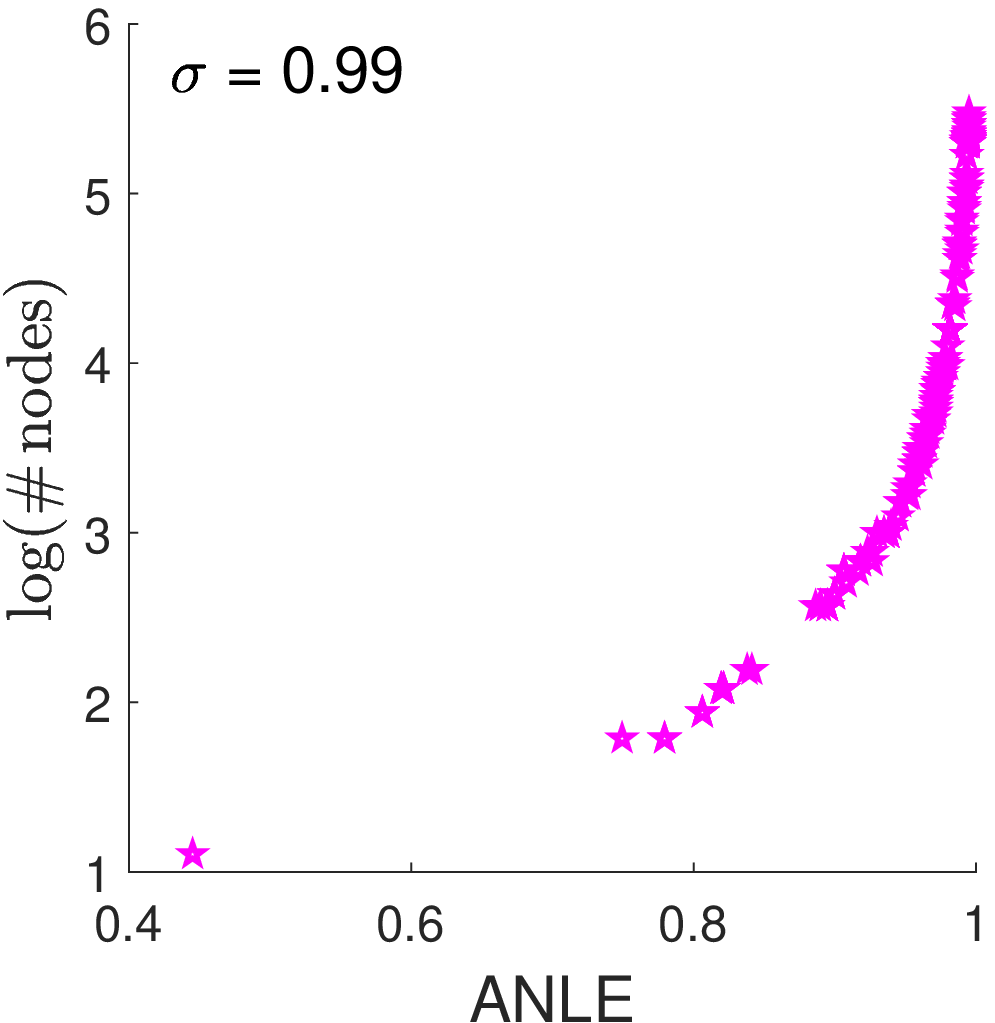}}

\subfloat{\includegraphics[width=0.22\textwidth]{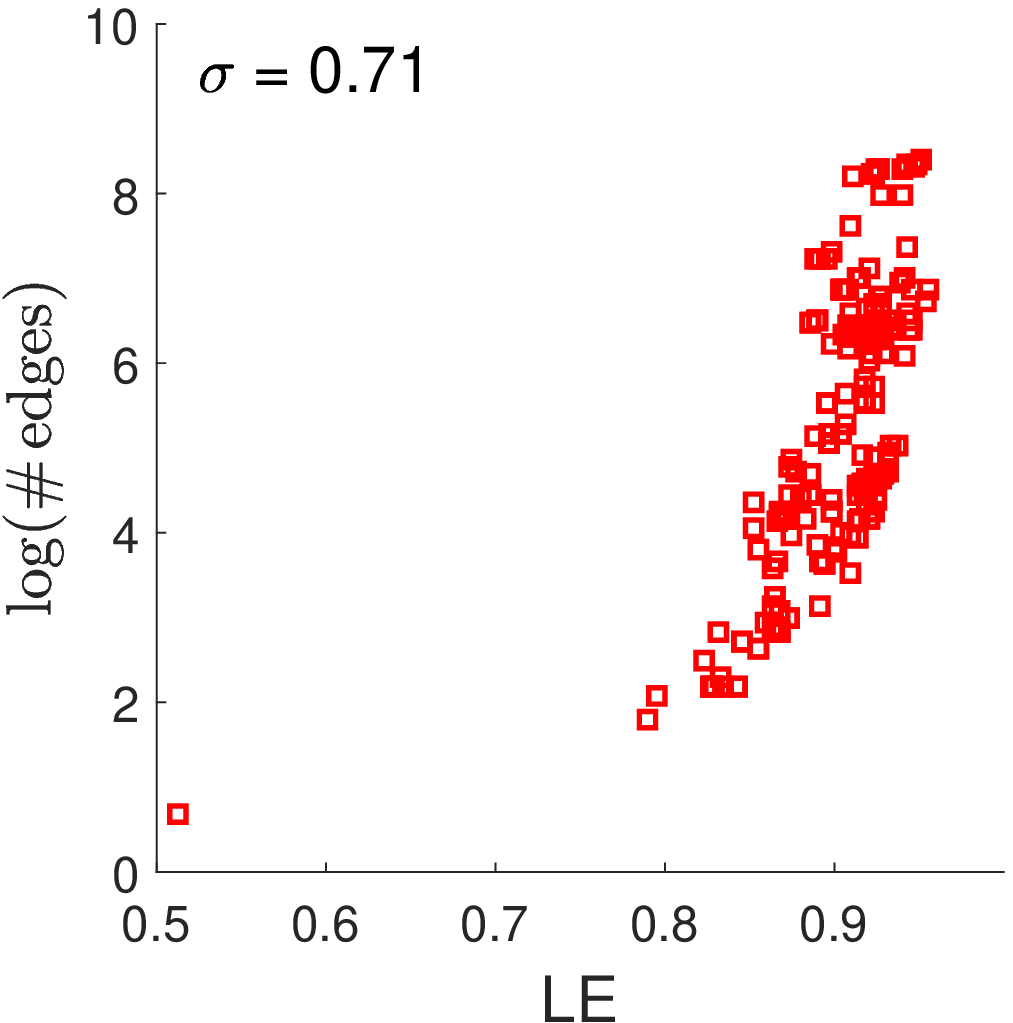}}
\subfloat{\includegraphics[width=0.22\textwidth]{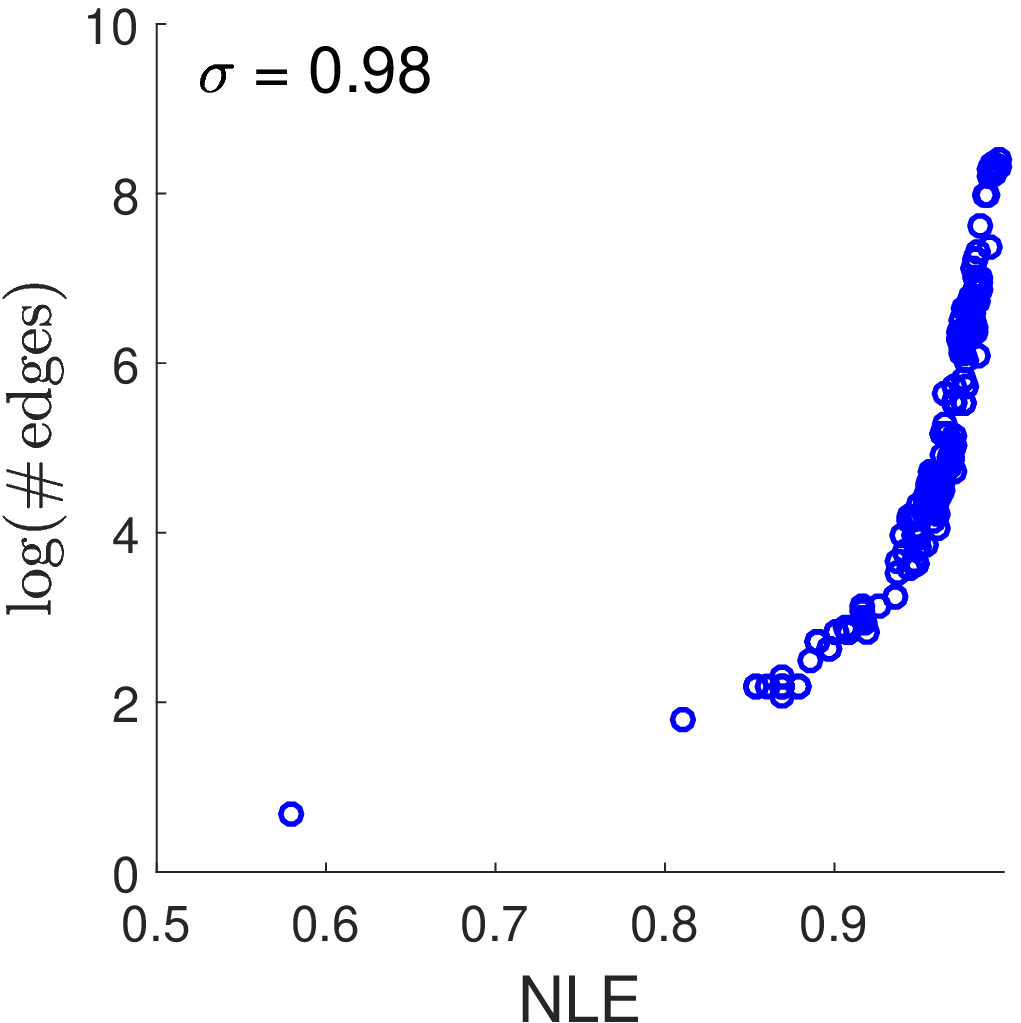}}
\subfloat{\includegraphics[width=0.22\textwidth]{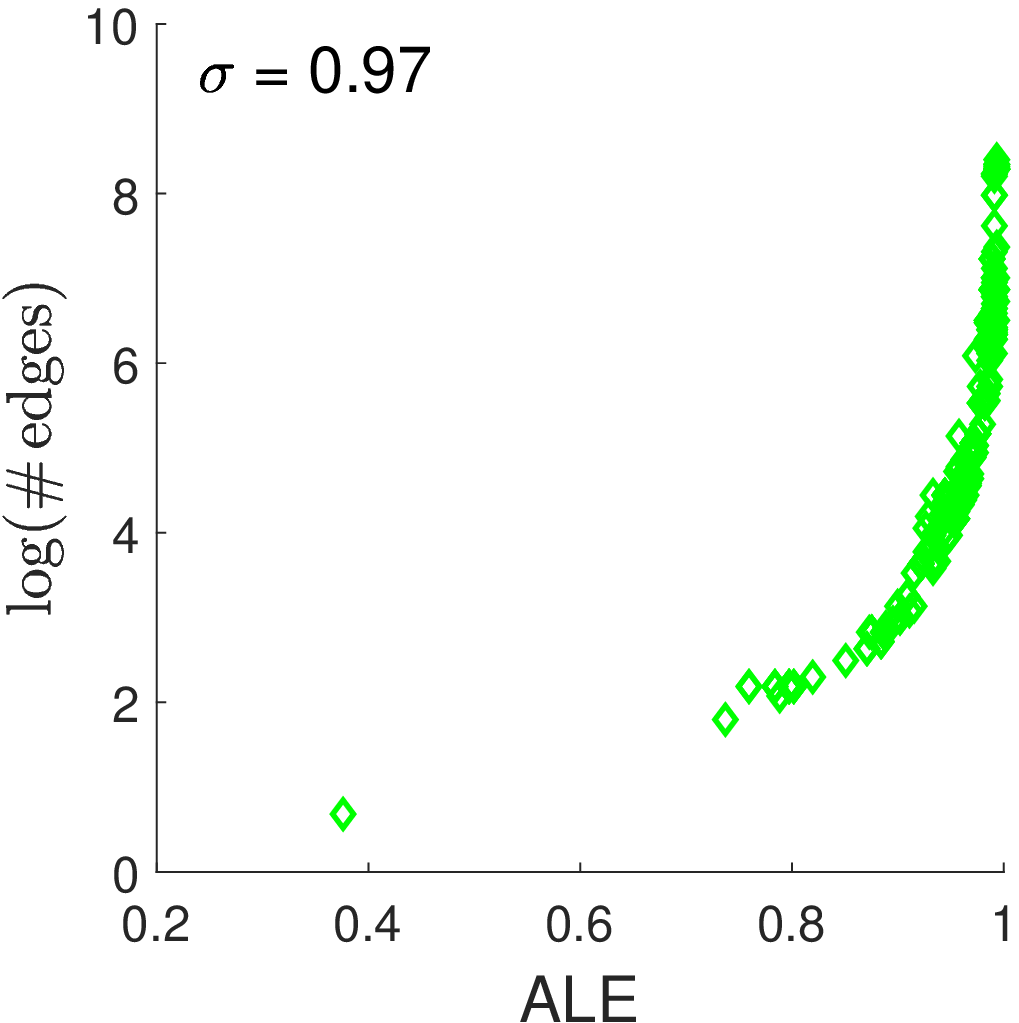}}
\subfloat{\includegraphics[width=0.22\textwidth]{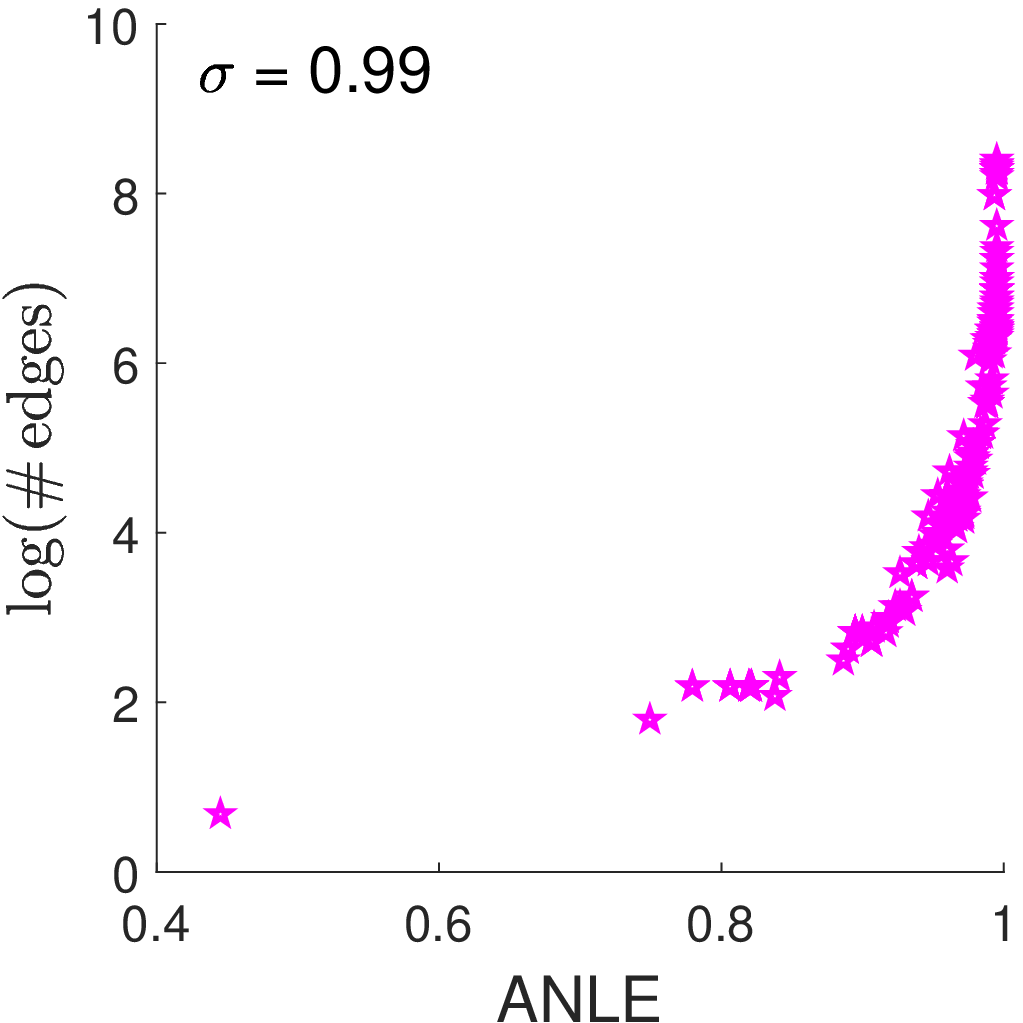}}
\caption{Correlation between entropies (top), entropy and number of nodes (middle) and edges (bottom) on the \textsf{PPI} dataset.}
\label{fig:ppi_corr}
\end{figure}

The \textsf{PPI} dataset consists of connected graphs with varying number of nodes. Due to the limited number of graphs in the \textsf{PPI} dataset, we prefer not to restrict our analysis to same size graphs. Fig.~\ref{fig:ppi_corr} (top) shows the correlation plots for the \textsf{PPI} dataset. Once again, we observe a high (Pearson) correlation between all pairs of entropies. With the exception of the Laplacian entropy, this appears to be largely due to the correlation between the entropy of a graph and its number of edges and nodes (Fig.~\ref{fig:ppi_corr}, middle and bottom). Finally, note that in Fig.~\ref{fig:ppi_corr} $\sigma$ denotes the Spearman rank correlation coefficient. Indeed, we observed the existence of a non-linear relation between the entropy and the number of edges (nodes).

\section{Conclusion}\label{sec:conclusion}
In this paper we have investigated two variants of the von Neumann entropy of a graph, based on the normalized and unnormalized Laplacian, respectively. With their quadratic approximations to hand, we have studied the entropic change as the new edges are added to the graph, giving new insight in the type of structural patterns that influence the value of the (approximated) entropy.

We performed an extensive set of experiments which showed that 1) the Laplacian and the normalized Laplacian entropies capture the presence of related yet different structural patterns, 2) the quadratic approximation fail to explain the emergence of non-trivial structures, in particular for the case of the Laplacian entropy, and that in general 3) the quality of the quadratic approximation, as well as which variant of the von Neumann entropy is better approximated, depends on the topology of the underlying graph. Our results suggest that the quadratic approximation of the von Neumann entropy can be an efficient way to measure the complexity of large networks, however the quality of this approximation depends on the topology of the network being studied. In particular, with the exception of small world networks, we find that the Laplacian entropy is easier to approximate. The normalized Laplacian entropy, on the other hand, can be approximated better for Erd\"os-R\'enyi and scale-free networks with low edge density.
 

\end{document}